\documentclass[prl,aps,twocolumn,superscriptaddress,showpacs,showkeys]{revtex4-1}
\usepackage{amsmath,amsthm,amssymb,graphicx,subfigure,xcolor}
\usepackage{esint}
\usepackage[unicode=true]{hyperref}
\hypersetup{
     colorlinks=true,       		
     linkcolor=blue,          	
     citecolor=red,            
     urlcolor=magenta,           	
 }

\newtheorem*{theorem*}{Theorem}

\newtheorem*{corollary*}{Corollary}

\newtheorem*{lemma*}{Lemma}

\newtheorem*{proposition*}{Proposition}
\theoremstyle{definition}

\newtheorem*{definition*}{Definition}
\theoremstyle{remark}

\newtheorem*{remark*}{Remark}

\begin{document}

\title{Spin Vector Potential and Spin Aharonov-Bohm Effect}

	\author{Jing-Ling Chen}
	\email{chenjl@nankai.edu.cn}
	\affiliation{Theoretical Physics Division, Chern Institute of Mathematics, Nankai University, Tianjin 300071, People's Republic of
	 	China}

	\author{Xing-Yan Fan}
   	\affiliation{Theoretical Physics Division, Chern Institute of Mathematics, Nankai University, Tianjin 300071, People's Republic of
	 	China}

	\author{Xiang-Ru Xie}
	\affiliation{School of Physics, Nankai University, Tianjin 300071, People's Republic of China}

	\date{\today}
	\begin{abstract}
		The Aharonov-Bohm (AB) effect is an important discovery of quantum theory. It serves as a surprising quantum phenomenon in which an electrically charged particle can be affected by an electromagnetic potential, despite being confined to a region in which both the magnetic field and electric field are zero. This fact gives the electromagnetic potentials greater significance in quantum physics than in classical physics.
The original AB effect belongs to an ``electromagnetic type". A certain vector potential is crucial for building a certain type of AB effect.
In this work, we focus on the ``spin", which is an intrinsic property of microscopic particles that has been widely accepted nowadays. First, we propose the hypothesis of \emph{spin vector potential} by considering a particle with a spin operator. Second, to verify the existence of such a spin vector potential, we present a gedanken double-slit interference experiment (i.e., the spin AB effect), which is possible to be observed in the lab. Third, we apply the spin vector potential to naturally explain why there were the Dzyaloshinsky-Moriya-type interaction and the dipole-dipole interaction between spins, and also predict a new type of spin-orbital interaction.
	\end{abstract}
\maketitle

	\emph{Introduction.}---The concept of \emph{spin} is an ancient subject. The famous Stern-Gerlach experiment, together with other experiments referring to the magnetic moment, have exhibited the quantization phenomena without satisfactory interpretation \cite{2001GreinerSpin}, until Uhlenbeck and Goudsmit
		proposed the hypothesis of \emph{electronic spin} \cite{1925Uhlenbeck,1926UHLENBECK}. Thereafter, this gradually orthodox concept has
		attracted much attention in quantum community. For instances, Pauli introduced it into a nonrelativistic field equation
		\cite{1927Pauli}, and Dirac proved the electronic spin as a corollary of the relativistic quantum theory for electron
		\cite{1928Dirac}. Recently, the spin has been not only an indispensable chapter in all textbooks of quantum mechanics, but
		also a newfound degree of freedom, which motivates scientists to extend some electromagnetic conceptions to the category of it, e.g.,
		the spintronics \cite{2001WolfS.A.,2004Zutic}, the spin Hall effect \cite{2008HostenOnur,2015Sinova}, and the spin Seebeck
		effect \cite{2008Uchida}.

		Meanwhile, the original (i.e., the electromagnetic-type) AB effect showcases the influences of an
        electromagnetic \emph{potential} on charged particles unaffected by the corresponding electromagnetic \emph{field} \cite{1959Aharonov}, which produces the so-called AB phase holding geometric significance \cite{2019Cohen}. In analogy with the complex numbers in quantum realm
		\cite{2021Renou,2022Chena,2022Li}, the AB effect emphasizes the indispensability of potentials therein, instead of working as a
		convenient mathematical tool. Now that an electromagnetic potential acts as a pivot in the electromagnetic AB effect, if one
		replaces it with other potentials, the primary AB effect may be realized in other areas. Very recently, a type of gravitational AB
		effect has been observed experimentally \cite{2022OverstreetChris}, which motivates us to explore the spin-type AB effect in the spin area.

From the perspectives of quantum theory, the electron is a very interesting matter. In the history of quantum mechanics, there have been several important hypotheses with respect to the electron, and surprisingly all of them have been either confirmed by the ingenious experiments, or reformulated by quantum theories with a firm foundation. For instances, (i) de Broglie boldly postulated that an electron (and other matters) with momentum $p$ should exhibit the wave properties and its wavelength equals to $\lambda=h/p$, with $h$ being Planck's constant \cite{deBroglie1,deBroglie2}. Later on, Davisson and Germer experimentally confirmed de Broglie' hypothesis on the wave-particle duality of an electron \cite{Davisson1927}; (ii) as mentioned above, Uhlenbeck and Goudsmit boldly postulated that an electron should have a \emph{spin} in order to account for the splitting of some optical lines seen in atomic spectra. Later on Dirac confirmed this hypothesis by formulating a relativistic Hamiltonian for a free electron, since the total angular momentum of an electron should be a conservation quantity, thus there must be an intrinsic spin-1/2 operator attached to the orbital angular momentum operator.

The purpose of this work is to advance the study of the \emph{potential theory} as well as the AB effect. The organization of the work is as follows: First, we present the hypothesis of \emph{spin vector potential} by considering a particle with a spin operator. A derivation for such a spin vector potential is also provided from the physical insight. We then exactly solve the eigen-problem for the corresponding spin AB Hamiltonian system, i.e., an electron moving in the $xyz$-space in the presence of the spin vector potential. Second, in order to verify the existence of such a spin vector potential, we present a gedanken double-slit interference experiment (i.e., the spin AB effect). Third, we discuss some applications of the spin vector potential. We use it to naturally explain why there were the Dzyaloshinsky-Moriya-type interaction and the dipole-dipole interaction between spins in the literatures, and also predict some new types of spin interactions.

\onecolumngrid
\begin{widetext}
\begin{figure}[t]
\includegraphics[width=160mm]{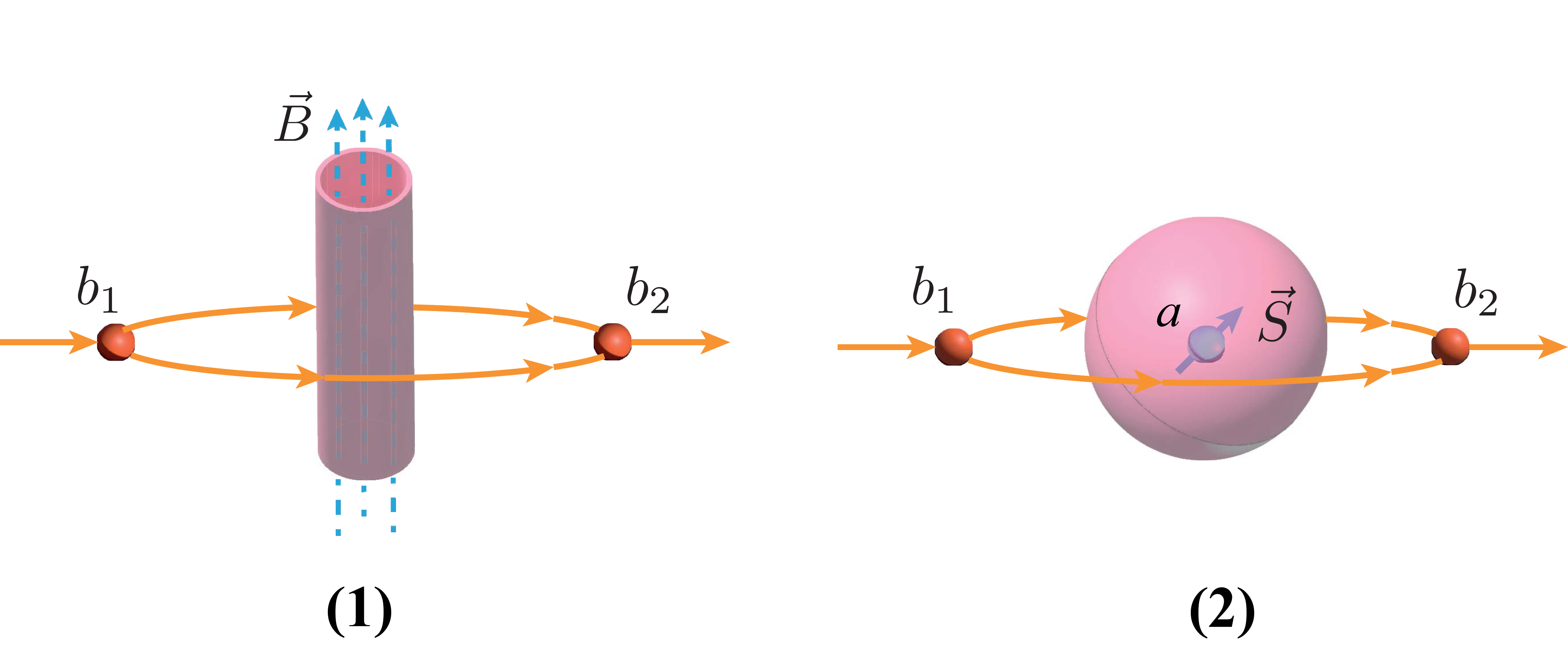}\\
	\caption{Two types of AB effects: (1) The magnetic AB effect. An electron is split into two parts at point
				$b_1$, then moves to the opposite point $b_2$ around a solenoid, where the external magnetic field $\vec{B}$ is
				confined; however the two split electrons are still affected by the magnetic vector potential $\vec{A}_{\rm M}$ induced via $\vec{B}$
				outside the solenoid. (2) The spin AB effect. Analogously, an electron is split at point $b_1$, then moves to the opposite point $b_2$ around another electron placed at point $a$ with a spin $\vec{S}$. Note that
				the electric field excited by the ``source'' electron is restricted inside the pink spherical district, consequently the electrons moving outside the pink area are influenced by the
				spin vector potential induced via the electronic spin $\vec{S}$ merely.}\label{fig:ABEffects}
\end{figure}
\end{widetext}
\twocolumngrid

\emph{Spin Vector Potential.---}Let us consider a particle with a spin $\vec{S}$ and a charge $q$, then we present a hypothesis that the spin could induce a spin vector potential as
	\begin{eqnarray}\label{eq:v-1}
				\vec{A} &=& \frac{c}{q}\frac{\vec{r}\times\vec{S}}{r^2},
			\end{eqnarray}
where $c$ is the speed of light in vacuum. In the following, based on the angular momentum operator, we would provide an alternative derivation from physical insight.

Let us denote $\vec{\ell}=\vec{r}\times \vec{p}$ as the orbital angular momentum operator, then the total angular momentum operator of the particle is given by $\vec{J}=\vec{\ell}+\vec{S}$, which can be recast to the form $ \vec{J}  = \vec{r} \times [\vec{p}-(q/c) \vec{A}]+ S_r \hat{e}_r$.
Thus, when one views $[\vec{p}-(q/c) \vec{A}]$ as the canonical momentum, then $\vec{A}$ is accordingly a kind of spin vector potentials. Here $S_r$ is one of the components of $\vec{S}=S_r \hat{e}_r+ S_\theta \hat{e}_\theta+S_\phi\hat{e}_\phi$ expanded in the spherical coordinates, $\hat{e}_r=\vec{r}/r$, and $r=|\vec{r}|$. In this work, for simplicity we just take this particle as a usual electron, then $q=-e$, with $-e$ being the electric charge of an electron. In this case, the spin operator reads $\vec{S}=(\hbar/2)\vec{\sigma}$, with $\vec{\sigma}=(\sigma_x, \sigma_y, \sigma_z)$ being the vector of Pauli matrices and $\hbar=h/{2\pi}$. In other words, the spin vector potential $\vec{A}$ is induced by the spin of an electron. For the detail proof, one may refer to Supplemental Material (SM) \cite{SM}. In the SM, by using the similar approach we have also naturally derived the well-known vector potential of Wu-Yang monopole \cite{1976Wub}.

\emph{The Hamiltonian System.---}We now consider a particle with mass $M$ and charge $Q=g(-e)$, $g\in\mathbb{Z}$ (when $g=1$, the particle is the usual electron). When the particle moves in the $xyz$-space in the presence of the ``spin" vector potential, the corresponding Hamiltonian of the particle is given by
\begin{eqnarray}\label{eq:v-3}
			H_{\rm S}&=&\dfrac{1}{2M} \left(\vec{p}-\frac{Q}{c}\vec{A}\right)^2= \dfrac{1}{2M} \left(\vec{p}-\vec{\mathcal{A}}\right)^2,
\end{eqnarray}
with
\begin{eqnarray}\label{eq:v-4}
			\vec{\mathcal{A}}&=&g \frac{\vec{r}\times\vec{S}}{r^2},
\end{eqnarray}
which does not depend on charge $e$ and satisfies $\vec{\nabla}\cdot\vec{\mathcal{A}}=0$.

The eigen-problem is given by
\begin{eqnarray}\label{eq:v-5}
			H_{\rm S} \Psi_{\rm S}(\vec{r}) &=& E  \Psi_{\rm S}(\vec{r}).
		\end{eqnarray}
Actually, the spin AB Hamiltonian system can be exactly solved in the common set $\{H, \vec{\ell}^2, \vec{\ell}\cdot\vec{S}\}$ \cite{SM}. One finds that the energy spectrum is continuous ($E\geq 0$), and the wavefunction reads
\begin{eqnarray}\label{eq:v-7}
\Psi_{\rm S}(\vec{r})=R(r) \; \Phi_{lm}(\theta, \phi),
\end{eqnarray}
where the radial wavefunction is given by the Bessel function as $R(r)= (1/{\sqrt{r}})\; J_{\nu}(\sqrt{\epsilon} r)$,
and the angular wavefunctions are
\begin{eqnarray}\label{eq:v-8}
\Phi^A_{lm}(\theta, \phi)&=& \frac{1}{\sqrt{2l+1}}\begin{bmatrix}
                                                     \sqrt{l+m+1} \;Y_{lm}(\theta, \phi) \\
                                                       \sqrt{l-m} \;Y_{l,m+1}(\theta, \phi)\\
                                                    \end{bmatrix},\nonumber\\
\Phi^B_{lm}(\theta, \phi)&=& \frac{1}{\sqrt{2l+1}}\begin{bmatrix}
                                                       -\sqrt{l-m} \;Y_{lm}(\theta, \phi) \\
                                                       \sqrt{l+m+1} \;Y_{l,m+1}(\theta, \phi)\\
                                                     \end{bmatrix},\nonumber
		\end{eqnarray}
for $K=l/2$ or $-(l+1)/2$ ($l\neq 0$), respectively. Here $K\hbar^2$ is the eigenvalue of $\vec{\ell}\cdot\vec{S}$, $\epsilon = 2ME/\hbar^2$,
$\nu=\sqrt{1+4\kappa}/2$, $\kappa = l(l+1) +2gK +g^2/2 \geq 0$, and
$Y_{lm}(\theta, \phi)$ is the spherical harmonics, with quantum numbers $l=0, 1, 2, \cdots$, and $m=0, \pm 1, \pm 2, \cdots, \pm l$.


\emph{Spin AB Effect.---}Due to the spin vector potential, we now propose a new physical effect: the \emph{spin AB effect}. In Fig. 1, we have made a sketch on the original \emph{magnetic} AB effect and the new proposed \emph{spin} AB effect. For the former, the magnetic field $\vec{B}$ is confined inside a solenoid with radius $r_0$, whose corresponding magnetic vector potential is given by \cite{2005QParadox}
\begin{equation}\label{eq:v-10}
   \vec{A}_{\rm M}= \begin{cases}
        & \frac{B \sqrt{x^2+y^2}}{2} \hat{e}_\phi,  \; \;\; (\rho< r_0)\\
        &  \frac{\Phi_{\rm M}}{2\pi\sqrt{x^2+y^2}} \hat{e}_\phi,\;\; (\rho> r_0)
    \end{cases}
  \end{equation}
where $\vec{B}=B \hat{z}$ is the magnetic field, $\Phi_{\rm M}=B\pi r_0^2$ is the magnetic flux, $\rho=\sqrt{x^2+y^2}$, and $\hat{e}_\phi = (-\sin\phi, \cos\phi, 0)$.  Accordingly, the magnetic Aharonov-Bohm Hamiltonian is given by
\begin{eqnarray}\label{eq:v-11}
			H_{\rm M}&=&\dfrac{1}{2M} \left(\vec{p}+\frac{e}{c}\vec{A}_{\rm M}\right)^2.
\end{eqnarray}
For the latter, we have used the spin $\vec{S}$ to replace the solenoid. Accordingly, the spin vector potential and the spin Aharonov-Bohm Hamiltonian are given in Eq. (\ref{eq:v-1}) and Eq. (\ref{eq:v-3}), respectively.

To demonstrate the AB effect, one usually adopts a gedanken double-slit experiment. In Fig. 2, a solenoid (or a spin) is placed behind the double-slit plate, which contributes a magnetic (or a spin) vector potential. Electrons are emitted from the electron source $O$, which travel to the point $D$ on the screen along two different paths 1 and 2. Under the influence of the magnetic (or the spin) vector potential, one will observe the interference patterns produced on the screen. The interference patterns with vector potentials are generally different from that of without a vector potential. In such a way, one demonstrates the AB effect.

Now we come to calculate the interference patterns. We shall make a unified treatment for both the magnetic and the spin AB effects. For the magnetic AB Hamiltonian, the eigen-equation reads $H_{\rm M} \Psi_{\rm M}(\vec{r}) = E_{\rm M}  \Psi_{\rm M}(\vec{r})$,
where $E_{\rm M}$ is the energy, and $\Psi_{\rm M}(\vec{r})$ is the eigenfunction.
The exact solution reads $E_{\rm M}=\hbar^2 k^2/{2M}\geq 0$, and
\begin{eqnarray}\label{eq:v-13}
&&\Psi_{\rm M}(\vec{r})=\mathcal{N}\; \xi_0(\vec{r}) \xi(\vec{r}),\nonumber\\
&&\xi_0(\vec{r}) = c_1 e^{{\rm i} \vec{k}\cdot \vec{r}} + c_2 e^{-{\rm i} \vec{k}\cdot \vec{r}}, \nonumber\\
&& \xi(\vec{r}) = e^{-\frac{ \mathrm{i} e\,\Phi_{\rm M}}{2\hbar c} \left[\arctan(\frac{y}{x})\right]},
\end{eqnarray}
where $\mathcal{N}$ is the normalized constant, $c_1, c_2$ are some complex numbers, $k=|\vec{k}|$, and $\xi_0(\vec{r})$ is just the eigenfunction of a free electron with the Hamiltonian equals to $\vec{p}^2/2M$.

To connect the result (\ref{eq:v-13}) with the observable double-slit interference experiment, we recast the wavefunction to the following integral form
\begin{eqnarray}\label{eq:v-14}
\Psi_{\rm M}(\vec{r}) & = & \mathcal{N}\; e^{\int^{\mathcal{L}(\vec{r})} \vec{F}(\vec{r}')\cdot d\vec{r}'},
\end{eqnarray}
where ${\mathcal{L}(\vec{r})}$ represents the path that the electron moves, and the vector $\vec{F}=(F_x, F_y, F_z)$ is determined by
\begin{eqnarray}\label{eq:v-15}
 &&F_x = \frac{ \frac{d \Psi_{\rm M}(\vec{r})}{d x}}{\Psi_{\rm M}(\vec{r})},\;\; F_y = \frac{ \frac{d \Psi_{\rm M}(\vec{r})}{d y}}{\Psi_{\rm M}(\vec{r})},\;\;
   F_z = \frac{ \frac{d \Psi_{\rm M}(\vec{r})}{d z}}{\Psi_{\rm M}(\vec{r})}.
\end{eqnarray}
For simplicity, let $c_1=1, c_2=0$, then we have $ \vec{F} = {\rm i} \vec{k}- ({\rm i} e/\hbar c) \vec{A}_{\rm M}$.

\begin{figure}[t]
\includegraphics[width=80mm]{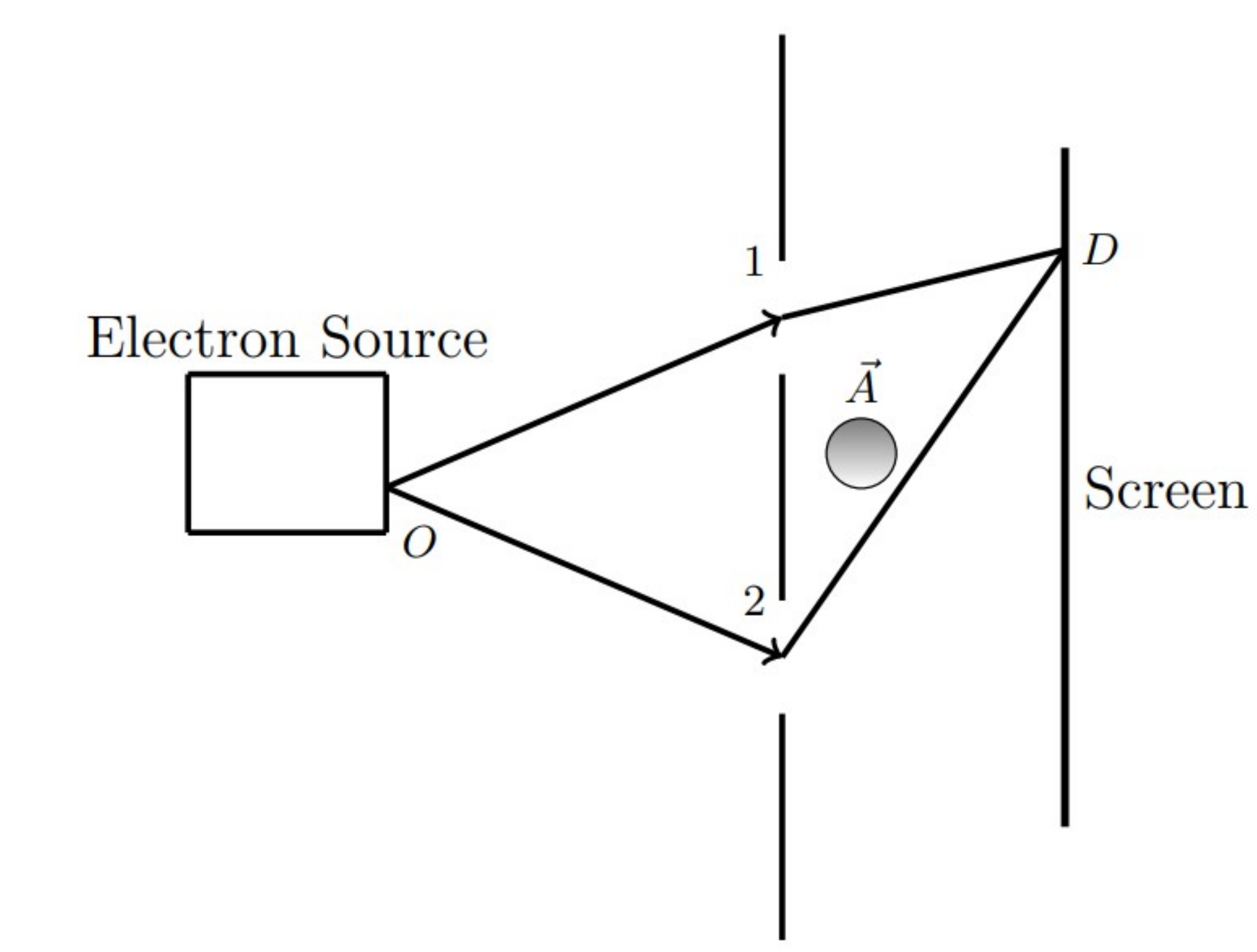}\\
	\caption{Illustration of the gedanken double-slit experiment.
}\label{fig:SpinAB}
\end{figure}

Let us focus on Fig. 2. The electron is initially at point $O$ and finally at point $D$, there are two different paths $\mathcal{L}_1$, $\mathcal{L}_2$ between them. Therefore, when the electron arrives at the screen, it is in the following superposition state
\begin{eqnarray}\label{eq:v-17}
\Psi(\vec{r})&=& \Psi^1_{\rm M}(\vec{r})+\Psi^2_{\rm M}(\vec{r})\nonumber\\
&=& \mathcal{N}\left(e^{\int^{\mathcal{L}_1(\vec{r})} \vec{F}(\vec{r}')\cdot d\vec{r}'}+e^{\int^{\mathcal{L}_2(\vec{r})} \vec{F}(\vec{r}')\cdot d\vec{r}'}\right).
\end{eqnarray}
Then the probability of finding the electron at the point $D$ is given by \cite{SM}
\begin{eqnarray}\label{eq:v-18}
			P_{\rm M}&=&|\Psi(\vec{r})|^2= \mathcal{N}^2 \left| 1+e^{\frac{\mathrm{i}}{\hbar} \oint (\hbar \vec{k}-\frac{e}{c} \vec{A}_{\rm M}(\vec{r}'))\cdot d \vec{r}'}\right|^2\nonumber\\
&=& 2 \mathcal{N}^2 [1+\cos(\delta_1+\delta_2)],
		\end{eqnarray}
with
\begin{eqnarray*}\label{eq:v-19}
&& \delta_1=\frac{1}{\hbar} \oint (\hbar \vec{k})\cdot d \vec{r}'=\oint \vec{k}\cdot d \vec{l},\nonumber\\
&& \delta_2=-\frac{e}{\hbar c} \oint \vec{A}_{\rm M}\cdot d \vec{l}=-\frac{e \Phi_{\rm M}}{\hbar c}.
		\end{eqnarray*}
In the ``ordinary" double-slit experiment, i.e., there is no any vector potential, it is just the phase $\delta_1$ that causes the interference fringes, i.e., $P_{\rm O}=2 \mathcal{N}^2 [1+\cos(\delta_1)]$. In the magnetic AB effect, the interference fringes are shifted by an additional phase $\delta_2$, which has been observed and confirmed in experiments \cite{1960Chambers,1986Tonomura}.

Similarly, for the spin AB wavefunction (\ref{eq:v-7}), one has
\begin{eqnarray}\label{eq:v-20}
		 \Psi_{\rm S}(\vec{r}) &=& \begin{bmatrix} c_1 \chi_1 (\vec{r}) \\ c_2 \chi_2 (\vec{r}) \end{bmatrix}
  =\mathcal{N}  \begin{bmatrix} c_1 e^{\int^{\mathcal{L}(\vec{r})} \vec{F}_1(\vec{r}')\cdot d\vec{r}'} \\ c_2 e^{\int^{\mathcal{L}(\vec{r})} \vec{F}_2(\vec{r}')\cdot d\vec{r}'} \end{bmatrix},
		\end{eqnarray}
where $\chi_1 (\vec{r})$ and $\chi_2 (\vec{r})$ are the normalized wavefunctions, $|c_1|^2+|c_2|^2=1$, and the vectors $\vec{F}_{i}  = \left( \vec{\nabla} \chi_i (\vec{r}) \right)/\chi_i (\vec{r})$, $(i=1, 2)$.
By moving along two different paths, the electron is in a superposition state $\Psi(\vec{r})
= \Psi^1_{\rm S}(\vec{r})+\Psi^2_{\rm S}(\vec{r})$,
and the probability of finding the electron at the point $D$ on the screen as
\begin{eqnarray}\label{eq:v-23}
	P_{\rm S} &&= \mathcal{N}^2\; \biggr\{|c_1|^2 \;e^{2 \int^{\mathcal{L}_1(\vec{r})}\,{\rm Re}\vec{F}_1 \cdot d\vec{r}'}
        \times \nonumber\\
        &&\left[1+e^{2 \oint{\rm Re}\vec{F}_1\cdot d\vec{r}'}
      +2 \; e^{\oint {\rm Re}\vec{F}_1\cdot d\vec{r}'}\; \cos\biggr(\oint {\rm Im}\vec{F}_1 \cdot d\vec{r}'\biggr) \right]\nonumber\\
    &&\qquad \;\;+\; |c_2|^2 \; e^{2 \int^{\mathcal{L}_1(\vec{r})}\,{\rm Re}\vec{F}_2 \cdot d\vec{r}'}
        \times \nonumber\\
    &&\left[1+e^{2 \oint {\rm Re}\vec{F}_2\cdot d\vec{r}'}
      +2 \; e^{\oint {\rm Re}\vec{F}_2\cdot d\vec{r}'}\; \cos\biggr(\oint {\rm Im}\vec{F}_2 \cdot d\vec{r}'\biggr) \right]\biggr\},\nonumber\\
\end{eqnarray}
where ${\rm Re}\vec{F}$ and ${\rm Im}\vec{F}$ are the real part and the imaginary part of $\vec{F}={\rm Re}\vec{F}+{\rm i}\; {\rm Im}\vec{F}$, respectively.

In SM \cite{SM}, we have performed the numerical simulations to show that the interference patterns of the spin AB effect are indeed different from that of the ordinary double-slit experiment. Since $P_{\rm S}\neq P_{\rm O}$, it is very possible to observe such a spin AB effect by performing the double-slit experiment, thus confirming the existence of the spin vector potential.

\emph{Application.---}Here we discuss a physical application of the spin vector potential. We find that it can be used to naturally explain the origination of the Dzyaloshinsky-Moriya (DM) interaction \cite{1958Dzyaloshinsky,1960Moriya}, the dipole-dipole interaction, and the tensor force operator \cite{1999Flugge}. The Hamiltonian $H_{\rm S}$ in Eq. (\ref{eq:v-3}) is written in a non-relativistic version. If we consider the Hamiltonian in its relativistic version, we find that some important types of spin interactions emerge naturally. Explicitly, let us consider a Dirac particle (marked `2') with charge $-ge$ and mass $M$, which moves under the spin vector potential induced by the spin $\vec{S}_1:=\hbar\vec{\sigma}_1 /2$ of an electron (marked `1') with charge $-e$. Namely, $\vec{\mathcal{A}}=g(\vec{r}\times\vec{S}_1)/{r^2}$, where $\vec{r}$ depicts the distance vector between the Dirac particle and the electron. Then the Dirac Hamiltonian reads $(c=1)$
\begin{eqnarray}\label{eq:v-24}
 \mathcal{H}_{\rm Dirac} &=&\ \vec{\alpha}_2 \cdot\left(\vec{p}-\vec{\mathcal{A}}\right)+\beta M,
\end{eqnarray}
 where $\vec{\alpha}_2=\sigma_x\otimes \vec{\sigma}_2$, $\beta=\sigma_z\otimes \openone$ are the Dirac matrices, and $\openone$ is the $2\times 2$ identity matrix.

\begin{table}[!h]
	\centering
\caption{The types of spin interactions involving in the Dirac Hamiltonian $\mathcal{H}_{\rm Dirac} =
\sigma_x \otimes\bigl[\vec{\sigma}_2 \cdot\vec{p} -(g\hbar/2)\,\vec{r}\cdot\bigl(\vec{\sigma}_1 \times\vec{\sigma}_2\bigr)/r^2\bigr]+\beta M$ and its square operator $\mathcal{H}^2_{\rm Dirac}$.}
\begin{tabular}{lr}
\hline\hline
 & {\rm The types of interactions} \\
  \hline
	$\vec{\sigma}_1 \cdot\vec{\sigma}_2$ & {\rm The spin-spin exchange interaction} \\
  \hline
	$\vec{\sigma}_1 \cdot{\vec{\ell}}$ & {\rm The spin-orbital interaction} \\
  \hline
  $\vec{r}\cdot\bigl(\vec{\sigma}_1 \times\vec{\sigma}_2\bigr)$ &{\rm The DM-type interaction} \\
   \hline
  \vspace{0.5mm}
  $\dfrac{(\vec{r}\cdot\vec{\sigma}_1)(\vec{r}\cdot\vec{\sigma}_2)}{r^2}$ & {\rm The dipole-dipole interaction} \\
  \hline
$\vec{\ell}\cdot(\vec{\sigma}_1 \times\vec{\sigma}_2)$ & {\rm New spin-orbital interaction} \\
  \hline
   \vspace{0.5mm}
$\dfrac{3(\vec{r}\cdot\vec{\sigma}_1)(\vec{r}\cdot\vec{\sigma}_2)}{r^2} -\vec{\sigma}_1 \cdot\vec{\sigma}_2$ & {\rm The tensor force operator} \\
 \hline\hline
\end{tabular}\label{tab:interactions}
\end{table}

In SM \cite{SM}, we have expanded the Dirac Hamiltonian $\mathcal{H}_{\rm Dirac}$ and its square operator $\mathcal{H}^2_{\rm Dirac}$, and we find that, besides the well-known spin-spin exchange interaction (i.e., the Heisenberg exchange interaction) $\vec{\sigma}_1 \cdot\vec{\sigma}_2$ and the spin-orbital interaction $\vec{\sigma}_1 \cdot{\vec{\ell}}$, there are additionally the DM-type interaction, the dipole-dipole interaction, and further a new-type of generalized spin-orbital interaction $\vec{\ell}\cdot(\vec{\sigma}_1 \times\vec{\sigma}_2)$. These spin interactions are listed in Table \ref{tab:interactions}.

It is worth mentioning the DM interaction, which is a significant interaction in condensed matter physics. The DM interaction was first proposed by Dzyaloshinsky \cite{1958Dzyaloshinsky} to explain phenomenologically the ``weak" ferromagnetism of antiferromagnetics in some crystals, such as $\alpha$-$\text{Fe}_2 \text{O}_3$, $\text{MnCO}_3$ and $\text{CoCO}_3$. Later on, Moriya \cite{1960Moriya} gave an insight into it from the perspective of spin-orbit coupling, together with Anderson's superexchange formula \cite{1959Anderson}. However, the origination of the DM interaction is still lacked. In Table \ref{tab:interactions}, we show that the DM interaction is a natural corollary from the Dirac Hamiltonian by considering the spin vector potential. Similarly, the dipole-dipole interaction is also a natural corollary of the spin vector potential.

Moreover, in the textbook of quantum mechanics, one may meet the following the tensor force operator between two spin-1/2 particles \cite{1999Flugge}:
\begin{eqnarray}\label{eq:v-25}
 \mathcal{T}_{12} &=& \frac{3}{r^2} (\vec{r}\cdot\vec{\sigma}_1)(\vec{r}\cdot\vec{\sigma}_2)-\vec{\sigma}_1 \cdot\vec{\sigma}_2.
\end{eqnarray}
The tensor operator $\mathcal{T}_{12}$ is a linear combination of the dipole-dipole interaction and the spin-spin exchange interaction, thus one may say that it is also a natural outcome of the spin vector potential.

\emph{Conclusion.---}Instead of the electric filed $\vec{E}$ and the magnetic field $\vec{B}$, the scalar and vector potentials $(\varphi , \vec{A})$ appear in Schr{\" o}dinger's equation. Thus, quantum mechanics needs $(\varphi , \vec{A})$ in a way that classical physics does not. From the viewpoint of classical physics, the physical quantities $\vec{E}$ and  $\vec{B}$ are more fundamental than $(\varphi , \vec{A})$, for the latter is merely a convenient mathematical tool. However, this situation is dramatically changed when the AB effect appears. After obtaining the experimental verifications, the assertion of Aharonov and Bohm that the potentials are essential for expressing the laws of physics is therefore founded. To advance the importance of the potential theory as well as the AB effect,  we turn to the domain of ``spin". To reach this purpose, we have proposed a hypothesis of \emph{spin vector potential} by considering a particle with a spin operator. To directly verify the existence of such a spin vector potential, we have presented the spin AB effect by suggesting a gedanken double-slit interference experiment, which could be observed in the lab. Moreover, we find that the Heisenberg exchange interaction, the spin-orbital interaction, the Dzyaloshinsky-Moriya-type interaction, the dipole-dipole interaction, and the tensor force operator can be naturally derived from the Dirac Hamiltonian with the spin vector potential. These facts strongly hint that the spin vector potential does exist and is quite profound. Eventually, due to the spin vector potential we have predicted a new type of spin-orbital interaction $\vec{\ell}\cdot(\vec{\sigma}_1 \times\vec{\sigma}_2)$, and we expect that it can also play a significant role in condensed matter physics as well as atomic physics.

\begin{acknowledgments}
This work was supported by the National Natural Science Foundations of China (Grant Nos. 12275136, 11875167).
\end{acknowledgments}


\end{document}


\title{Supplemental Material for\\
``Spin Vector Potential and Spin Aharonov-Bohm Effect''}

\author{Jing-Ling Chen}
      \email{chenjl@nankai.edu.cn}
      \affiliation{Theoretical Physics Division, Chern Institute of Mathematics, Nankai University, Tianjin
      300071, People's Republic of China}

\author{Xing-Yan Fan}
\affiliation{Theoretical Physics Division, Chern Institute of Mathematics, Nankai University, Tianjin
      300071, People's Republic of China}

\author{Xiang-Ru Xie}
\affiliation{School of Physics, Nankai University, Tianjin 300071, People's Republic of China}

\date{\today}
\maketitle

\tableofcontents

\newpage

\part{The Derivation of the Vector Potentials $\vec{A}$}

In this section, we present a significant approach to derive the vector potentials, such as the Wu-Yang monopole vector potential \cite{1976Wub,1931DiracMonopole} and the novel \emph{spin vector potential}, which we have used to show the spin AB effect in the main text. This approach is based on \emph{the angular momentum operator} in quantum mechanics.

Angular momentum is very important in physics. In quantum mechanics, the angular momentum operators $\{J_x, J_y, J_z\}$ are defined through the following commutation relations:
\begin{eqnarray}\label{angularm}
&& [J_\alpha, J_\beta]=J_\alpha J_\beta-J_\beta J_\alpha= \mathrm{i} \hbar\; \epsilon_{\alpha\beta\gamma} J_{\gamma},
\end{eqnarray}
with $\mathrm{i}=\sqrt{-1}$, $\hbar=h/{2\pi}$, $h$ is Planck's constant, $\alpha, \beta, \gamma\in \{x, y, z \}$, $\epsilon_{\alpha\beta\gamma}$ is the Levi-Civita symbol, for instances, $\epsilon_{xyz}=\epsilon_{yzx}=\epsilon_{zxy}=1$, $ \epsilon_{xzy}=\epsilon_{yxz}=\epsilon_{zyx}=-1$. Explicitly, from Eq. (\ref{angularm}) one has
\begin{eqnarray}\label{angularm1}
&& [J_x, J_y]= \mathrm{i} \hbar\; J_z, \nonumber\\
&& [J_y, J_z]= \mathrm{i} \hbar\; J_x,\nonumber\\
&& [J_z, J_x]= \mathrm{i} \hbar\; J_y.
\end{eqnarray}
Alternatively, one may introduce the vector form of the angular momentum operator as
\begin{eqnarray}\label{angularm2}
&& \vec{J}=(J_x, J_y, J_z),
\end{eqnarray}
then Eq. (\ref{angularm1}) can be equivalently expressed as
\begin{eqnarray}\label{angularm3}
&& \vec{J} \times \vec{J} = \mathrm{i} \hbar\; \vec{J}.
\end{eqnarray}
If a vector operator $\vec{J}$ satisfies Eq. (\ref{angularm3}), then we say that it is an angular momentum operator. In the following, let us provide three concrete examples.

\emph{Example 1.}---The typical example is the orbital angular momentum operator, which is given by
\begin{eqnarray}\label{angularm3-a}
&& \vec{\ell} = \vec{r} \times \vec{p},
\end{eqnarray}
where
\begin{eqnarray}\label{coord}
&& \vec{r} = (x, y, z)
\end{eqnarray}
is the coordinate operator of the particle, and
\begin{eqnarray}\label{linearp}
&& \vec{p} = (p_x, p_y, p_z)
\end{eqnarray}
is the linear momentum operator of the particle.
In quantum mechanics, the linear momentum operators are defined by differential operators with respect to coordinates, i.e.,
\begin{eqnarray}\label{linearp1}
&& p_x=-\mathrm{i}\hbar \frac{\partial}{\partial x}, \;\;\;\;\; p_y=-\mathrm{i}\hbar \frac{\partial}{\partial y}, \;\;\;\;\; p_z=-\mathrm{i}\hbar \frac{\partial}{\partial z},
\end{eqnarray}
or
\begin{eqnarray}\label{linearp2}
&& \vec{p} = -\mathrm{i}\hbar \vec{\nabla},\;\;\;\;\;  \vec{\nabla}= (\frac{\partial}{\partial x}, \frac{\partial}{\partial y}, \frac{\partial}{\partial z}).
\end{eqnarray}
Then, based on the basic commutation relation
\begin{eqnarray}\label{linearp3}
&& [r_\alpha, p_\beta]= \mathrm{i}\hbar \delta_{\alpha\beta},
\end{eqnarray}
with $\delta_{\alpha\beta}$ being the Kronecker delta function, or explicitly
\begin{eqnarray}\label{linearp3-a}
&& [x, p_x]= \mathrm{i}\hbar, \;\; [y, p_y]= \mathrm{i}\hbar, \;\; [z, p_z]= \mathrm{i}\hbar, \nonumber\\
&& [x, p_y]= [x, p_z]= 0, \;\;\;  [y, p_x]= [y, p_z]= 0, \;\;\;  [z, p_x]= [z, p_y]= 0,
\end{eqnarray}
one can directly verify that three components $\{\ell_x, \ell_y, \ell_z\}$ satisfy the definition of angular momentum operator, i.e.,
\begin{eqnarray}\label{orbit1}
&& [\ell_x, \ell_y]= \mathrm{i} \hbar\; \ell_z, \;\;\;\;\; [\ell_y, \ell_z]= \mathrm{i} \hbar\; \ell_x,\;\;\;\;\;
[\ell_z, \ell_x]= \mathrm{i} \hbar\; \ell_y,
\end{eqnarray}
or
\begin{eqnarray}\label{angularm3-d}
&& \vec{\ell} \times \vec{\ell}  = \mathrm{i} \hbar\; \vec{\ell} .
\end{eqnarray}

\emph{Example 2.}---The orbital angular momentum operator associated with $U(1)$ monopole was
introduced by Wu and Yang ~\cite{1976Wub}, which takes the following form (the speed of light $c=1$)
\begin{equation}
 \label{mo-eq3a}
 \vec{L}=\vec{r}\times \vec{\pi}-q\frac{\vec{r}}{r},
\end{equation}
with
\begin{equation}
 \label{mo-eq3b}
 {\vec \pi}=\vec{p}-Ze\vec{A}
\end{equation}
being the canonical momentum. The vector $\vec{A}$ attached to the linear momentum $\vec{p}$ is the so-called Wu-Yang vector-potential, which satisfies the following relation
\begin{equation}
 \label{mo-eq3c}
\vec{\nabla} \times \vec{A}=g\frac{\vec{r}}{r^3},
\end{equation}
with $g$ being the strength of the monopole, and
\begin{equation}
 \label{mo-eq4}
 q=Zeg=\frac{1}{2}\times{\rm integer}.
\end{equation}
In Ref. \cite{1976Wub}, the Wu-Yang vector potential $\vec{A}$ are defined in two regions $a$ and $b$, and the expressions of $\vec{A}_a$ and $\vec{A}_b$ are given by
\begin{eqnarray}
\label{commu-6}
 \vec{A}_a &=& \frac{g}{r} \frac{1-\cos\theta}{\sin\theta}\; \hat{e}_\phi,\nonumber\\
  \vec{A}_b &=& \frac{-g}{r} \frac{1+\cos\theta}{\sin\theta}\; \hat{e}_\phi,
\end{eqnarray}
here $\hat{e}_\phi$ is one of the unit vectors in the spherical coordinates $\{\hat{e}_r, \hat{e}_\theta, \hat{e}_\phi\}$.

When $g=0$ and $q=0$, it is easy to observe from (\ref{mo-eq3a}) that $\vec{L}$ reduces to the usual orbital angular momentum $\vec{\ell}$.
By using the basic relation (\ref{linearp3}), one can directly verify that the vector $\vec{L}$ satisfies the definition of the angular momentum operator
\begin{eqnarray}\label{angularm3b}
&& \vec{L} \times \vec{L}  = \mathrm{i} \hbar\; \vec{L},
\end{eqnarray}
thus showing that $\vec{L}$ is a kind of angular momentum operators.

\emph{Example 3.}---The spin-1/2 operator is given by
 \begin{eqnarray}
 \label{S-1}
  && \vec{S}=\frac{\hbar}{2} \vec{\sigma},
  \end{eqnarray}
where $\vec{\sigma}$ is the vector of Pauli matrices, whose three components read
 \begin{eqnarray}
 \label{S-3}
  && \sigma_x=
  \left(
    \begin{array}{cc}
      0 & 1 \\
      1 & 0 \\
    \end{array}
  \right),\;\;\;\sigma_y=
  \left(
    \begin{array}{cc}
      0 & -{\rm i} \\
      {\rm i} & 0 \\
    \end{array}
  \right),\;\;\;  \sigma_z=
  \left(
    \begin{array}{cc}
      1 & 0 \\
      0 & -1 \\
    \end{array}
  \right).
   \end{eqnarray}
one can directly check that
   \begin{eqnarray}
 \label{S-2}
  && \vec{S} \times \vec{S}=\mathrm{i} \hbar \vec{S},
  \end{eqnarray}
thus $\vec{S}$ is an angular momentum operator.

\section{The Derivation of the Wu-Yang Monopole Vector Potential}

People may ask a curious question: Why does the Wu-Yang monopole vector potential take a form as shown in Eq. (\ref{commu-6})? Here we would like to provide a natural derivation.

We begin the derivation with the usual orbital angular momentum operator $\vec{\ell}=\vec{r}\times \vec{p}$. Let us define a new vector
\begin{eqnarray}
 \label{M-3}
 \vec{L}&=&\vec{\ell}+q \vec{G},
\end{eqnarray}
i.e., by shifting the vector $\vec{\ell}$ by a vector $q \vec{G}$ we obtain the new vector $\vec{L}$, and $q$ is a certain parameter. We have known that $\vec{\ell}$ is the angular momentum, however, by performing such a shift, we require that the resultant vector $\vec{L}$ is still an angular momentum operator. According to the definition as shown in Eq. (\ref{angularm3}), one must have
\begin{eqnarray}
 \label{M-4}
 \vec{L}\times \vec{L}=\mathrm{i} \hbar \vec{L},
 \end{eqnarray}
which leads to
\begin{eqnarray}
 \label{M-6}
&& \vec{L}\times \vec{L}=(\vec{\ell}+q \vec{G})\times (\vec{\ell}+q \vec{G})=\vec{\ell}\times \vec{\ell}+ q(\vec{\ell} \times \vec{G}+\vec{G}\times \vec{\ell}) +q^2\vec{G}\times\vec{G}.
 \end{eqnarray}

Now, for simplicity, let us consider the following case
\begin{eqnarray}
 \label{M-5}
 \vec{G}\times \vec{G}=0,
 \end{eqnarray}
i.e., the vector $\vec{G}$ is an Abelian operator, for its three components satisfy
\begin{eqnarray}
&& [G_i, G_j]=0, \;\;\;\;\;(i, j=1, 2, 3).
 \end{eqnarray}
After substituting Eq. (\ref{M-5}) into Eq. (\ref{M-6}), one obtains
\begin{eqnarray}
 \label{M-6b}
&& \vec{L}\times \vec{L}=\mathrm{i} \hbar \;\vec{\ell}+ q(\vec{\ell} \times \vec{G}+\vec{G}\times \vec{\ell}).
 \end{eqnarray}
By comparing Eq. (\ref{M-4}) and Eq. (\ref{M-6b}), we must have
\begin{eqnarray}
 \label{M-7}
&& \vec{\ell} \times \vec{G}+\vec{G}\times \vec{\ell} =\mathrm{i} \hbar \;\vec{G}.
 \end{eqnarray}

We now derive the conditions for $\vec{G}$, for which $\vec{L}$ is an angular momentum operator. For convenient, we first consider the $z$-component, i.e.,
\begin{eqnarray}
 \label{M-8a}
 (\vec{\ell} \times \vec{G})_z&=&\left[(\vec{r}\times \vec{p})\times \vec{G}\right]_z =\vec{r} \cdot (p_z \vec{G})-z (\vec{p}\cdot\vec{G})\nonumber\\
&=& x (p_z G_x)+y (p_z G_y)+z (p_z G_z)-z (p_x G_x)-z (p_y G_y)-z (p_z G_z)\nonumber\\
&=& p_z (x G_x+y G_y)-z (p_x G_x+p_y G_y),
 \end{eqnarray}
and
\begin{eqnarray}
 \label{M-8b}
 (\vec{G}\times \vec{\ell} )_z &=& \left[\vec{G}\times(\vec{r}\times \vec{p})\right]_z =\vec{G} \cdot(z \;\vec{p})-(\vec{G}\cdot\vec{r})p_z\nonumber\\
&=& G_x (z \;p_x)+G_y (z \;p_y)+G_z (z \;p_z) -(G_x x)p_z-(G_y y)p_z-(G_z z)p_z\nonumber\\
&=&z (G_x p_x+G_y p_y)-(G_x x+ G_y y)p_z\nonumber\\
&=& z (G_x p_x+G_y p_y)-(x G_x + y G_y )p_z.
\end{eqnarray}
Then we have
\begin{eqnarray}
 \label{M-9}
 (\vec{\ell} \times \vec{G})_z+(\vec{G}\times \vec{\ell} )_z &=&p_z (x G_x+y G_y)-z (p_x G_x+p_y G_y)+z (G_x p_x+G_y p_y)-(x G_x + y G_y )p_z\nonumber\\
&=&p_z (x G_x+y G_y+z G_z)-z (p_x G_x+p_y G_y+p_z G_z)+z (G_x p_x+G_y p_y+G_z p_z)\nonumber\\
&&  -(x G_x + y G_y +z G_z)p_z-p_z z G_z+ zp_z G_z-z G_zp_z+z G_zp_z\nonumber\\
&=& [p_z, \vec{r}\cdot \vec{G}]-z \;\left([p_x, G_x]+[p_y, G_y]+[p_z, G_z]\right)-[p_z, z] G_z\nonumber\\
&=&- \mathrm{i} \hbar \frac{\partial  (\vec{r}\cdot \vec{G})}{\partial z}+ z \;\mathrm{i} \hbar \left(\frac{\partial G_x}{\partial x}+\frac{\partial G_y}{\partial y}+\frac{\partial G_z}{\partial z}\right)+\mathrm{i} \hbar G_z\nonumber\\
&=& - \mathrm{i} \hbar \frac{\partial  (\vec{r}\cdot \vec{G})}{\partial z}+ z \;\mathrm{i} \hbar\left( \vec{\nabla}\cdot\vec{G}\right)+\mathrm{i} \hbar G_z.
 \end{eqnarray}
Thus, we have
\begin{eqnarray}
 \label{M-10}
&& (\vec{\ell} \times \vec{G})_x+(\vec{G}\times \vec{\ell} )_x=- \mathrm{i} \hbar \frac{\partial  (\vec{r}\cdot \vec{G})}{\partial x}+ x\; \mathrm{i} \hbar \left(\vec{\nabla}\cdot\vec{G}\right)+\mathrm{i} \hbar G_x,\nonumber\\
&& (\vec{\ell} \times \vec{G})_y+(\vec{G}\times \vec{\ell} )_y=- \mathrm{i} \hbar \frac{\partial  (\vec{r}\cdot \vec{G})}{\partial y}+ y\; \mathrm{i} \hbar \left(\vec{\nabla}\cdot\vec{G}\right)+\mathrm{i} \hbar G_y,\nonumber\\
&& (\vec{\ell} \times \vec{G})_z+(\vec{G}\times \vec{\ell} )_z=- \mathrm{i} \hbar \frac{\partial  (\vec{r}\cdot \vec{G})}{\partial z}+ z \;\mathrm{i} \hbar \left(\vec{\nabla}\cdot\vec{G}\right)+\mathrm{i} \hbar G_z,\nonumber
 \end{eqnarray}
or in a vector form as
\begin{eqnarray}
 \label{M-11}
 \vec{\ell} \times \vec{G}+\vec{G}\times \vec{\ell} =- \mathrm{i} \hbar \left[\vec{\nabla} (\vec{r}\cdot \vec{G})\right]+ \mathrm{i} \hbar \vec{r}  \; \left(\vec{\nabla}\cdot\vec{G}\right)+\mathrm{i} \hbar \vec{G},
 \end{eqnarray}
with the gradient operator
\begin{eqnarray}
 \label{M-12}
&&  \vec{\nabla} f={\rm grad} f= \hat{e}_x \frac{\partial f}{\partial x} + \hat{e}_y\frac{\partial f}{\partial y} +\hat{e}_z\frac{\partial f}{\partial z} .
 \end{eqnarray}
By comparing Eq. (\ref{M-7}) and Eq. (\ref{M-11}), we have the condition for $\vec{G}$ as
\begin{eqnarray}
 \label{M-13}
&&  \vec{\nabla} (\vec{r}\cdot \vec{G})= \vec{r} \left(\vec{\nabla}\cdot\vec{G}\right).
 \end{eqnarray}

We may express the above result as the following theorem:

\begin{theorem}
If a vector $\vec{G}$ satisfies
\begin{eqnarray}
 \label{G-1}
&&\vec{G}\times \vec{G}=0,\nonumber\\
&&\vec{\nabla} (\vec{r}\cdot \vec{G})= \vec{r} \left(\vec{\nabla}\cdot\vec{G}\right),
\end{eqnarray}
then the vector
\begin{eqnarray}\label{G-2}
\vec{L}&=&\vec{r}\times\vec{p}+q \vec{G}
\end{eqnarray}
is an angular momentum operator.
\end{theorem}

\begin{remark}
If $\vec{G}$ is zero vector $\vec{0}$, i.e.,
\begin{eqnarray}\label{G-4}
\vec{G}&=& \vec{0},
\end{eqnarray}
then Eq. (\ref{G-1}) is automatically satisfied. However, this case is trivial, because the vector $\vec{L}$ reduces to the usual angular momentum $\vec{\ell}$, and the parameter $q$ becomes useless.
\end{remark}

We shall study the nontrivial vectors $\vec{G}$ that can satisfy Theorem 1. Firstly let us list some useful formulas, which are used for the calculation. In the rectangular coordinate system $\{\hat{e}_x, \hat{e}_y, \hat{e}_z\}$, the vector $\vec{G}$, the nabla operator $\vec{\nabla}$, and the divergence $\vec{\nabla}\cdot\vec{G}$ are given by
\begin{eqnarray}\label{G-4-a}
\vec{G}=G_x \hat{e}_x+ G_y \hat{e}_y+G_z \hat{e}_z,
\end{eqnarray}
\begin{eqnarray}\label{N-1a}
\vec{\nabla}=\hat{e}_x \frac{\partial }{\partial x}+\hat{e}_y \frac{\partial }{\partial y}+\hat{e}_z \frac{\partial }{\partial z},
\end{eqnarray}
\begin{eqnarray}
 \label{G-12a}
&&  \vec{\nabla}\cdot\vec{G}= \frac{\partial G_x}{\partial x}+\frac{\partial G_y}{\partial y}+\frac{\partial G_z}{\partial z}.
 \end{eqnarray}
And in the spherical coordinate system $\{\hat{e}_r, \hat{e}_\theta, \hat{e}_\phi\}$, they can be expressed as
\begin{eqnarray}\label{G-3}
\vec{G}=G_r \hat{e}_r+ G_\theta \hat{e}_\theta+G_\phi \hat{e}_\phi,
\end{eqnarray}
\begin{eqnarray}\label{N-1b}
\vec{\nabla}=\hat{e}_r \frac{\partial }{\partial r}+\hat{e}_\theta \frac{1}{r}\frac{\partial }{\partial \theta}+\hat{e}_\phi \frac{1}{r\sin\theta}\frac{\partial }{\partial \phi},
\end{eqnarray}
\begin{eqnarray}
 \label{G-12}
&&  \vec{\nabla}\cdot\vec{G}=\frac{1}{r^2}\biggr[ \frac{\partial (r^2 G_r)}{\partial r}\biggr]+\frac{1}{r\sin\theta}\biggr[ \frac{\partial (\sin\theta G_\theta)}{\partial \theta}\biggr]+\frac{1}{r\sin\theta}\biggr[ \frac{\partial G_\phi}{\partial \phi}\biggr].
 \end{eqnarray}
The coordinate transformation between the basis $\{\hat{e}_r, \hat{e}_\theta, \hat{e}_\phi\}$ and the basis $\{\hat{e}_x, \hat{e}_y, \hat{e}_z\}$ is \begin{eqnarray}
\label{G-5}
&&  \hat{e}_r =\sin\theta\cos\phi \; \hat{e}_x+ \sin\theta\sin\phi \; \hat{e}_y +\cos\theta \; \hat{e}_z, \nonumber\\
&&  \hat{e}_\theta= \cos\theta\cos\phi \; \hat{e}_x+ \cos\theta\sin\phi \; \hat{e}_y -\sin\theta \; \hat{e}_z, \nonumber\\
&&  \hat{e}_\phi=-\sin\phi \; \hat{e}_x+ \cos\phi \; \hat{e}_y,
\end{eqnarray}
or in a matrix form as
\begin{eqnarray}
\label{G-5aa}
\left(
  \begin{array}{c}
    \hat{e}_r \\
    \hat{e}_\theta \\
    \hat{e}_\phi \\
  \end{array}
\right)
=\left(
  \begin{array}{ccc}
    \sin\theta\cos\phi & \sin\theta\sin\phi & \cos\theta \\
    \cos\theta\cos\phi & \cos\theta\sin\phi & -\sin\theta \\
    -\sin\phi & \cos\phi & 0 \\
  \end{array}
\right)
\left(
  \begin{array}{c}
    \hat{e}_x \\
    \hat{e}_y \\
    \hat{e}_z \\
  \end{array}
\right).
\end{eqnarray}

\begin{remark}
Let us consider the nonzero vector $\vec{G}$. In the following, the investigation is divided into two different cases: (i) $G_r=0$; (ii) $G_r\neq 0$.
\end{remark}

\subsection{Case (i). $G_r=0$}

In this case, we have the vector $\vec{G}$ as
\begin{eqnarray}\label{G-7}
\vec{G}&=&G_\theta \hat{e}_\theta+G_\phi \hat{e}_\phi.
\end{eqnarray}
Because
\begin{eqnarray}\label{G-7a}
\hat{e}_r=\frac{\vec{r}}{r},
\end{eqnarray}
we have
\begin{eqnarray}\label{G-9}
(\vec{r}\cdot \vec{G})=r \; (\hat{e}_r\cdot \vec{G})=r \;G_r=0,
\end{eqnarray}
thus Eq. (\ref{M-13}) becomes
\begin{eqnarray}
 \label{G-10}
&&  \vec{0}= \vec{r} \left(\vec{\nabla}\cdot\vec{G}\right),
 \end{eqnarray}
which leads to
\begin{eqnarray}
 \label{G-11}
&&  \vec{\nabla}\cdot\vec{G}=0.
 \end{eqnarray}
From Eq. (\ref{G-12}) we have the condition as
\begin{eqnarray}
 \label{G-13}
&&  \vec{\nabla}\cdot\vec{G}=\frac{1}{r\sin\theta}\biggr[ \frac{\partial (\sin\theta G_\theta)}{\partial \theta}\biggr]+\frac{1}{r\sin\theta}\biggr[ \frac{\partial G_\phi}{\partial \phi}\biggr]=0,
 \end{eqnarray}
which yields
\begin{eqnarray}
 \label{G-13b}
&&  \biggr[ \frac{\partial (\sin\theta G_\theta)}{\partial \theta}\biggr]=0, \;\;\;\biggr[ \frac{\partial G_\phi}{\partial \phi}\biggr]=0.
 \end{eqnarray}
Thus, it is easy to have the solution of $\vec{G}$ as
\begin{eqnarray}
\label{solution-G-1}
 &&\vec{G}= \frac{W_1(r, \phi)}{\sin\theta} \hat{e}_\theta+  W_2(r, \theta) \hat{e}_\phi,
  \end{eqnarray}
where $W_1(r,\phi)$ is function depending only on $r$ and $\phi$, and $W_2(r,\theta)$ is function depending only on $r$ and $\theta$.

By using the following relation
\begin{eqnarray}
 \label{G-13c}
&&  \hat{e}_\theta=-\hat{e}_r\times \hat{e}_\phi, \;\;\;\;\;  \hat{e}_\phi=\hat{e}_r\times \hat{e}_\theta,
 \end{eqnarray}
from Eq. (\ref{solution-G-1}) we have
\begin{eqnarray}
\label{solution-G-1-aa}
 &&\vec{G}= \hat{e}_r\times\left(-\frac{W_1(r, \phi)}{\sin\theta} \hat{e}_\phi+  W_2(r, \theta) \hat{e}_\theta\right)=\vec{r}\times\left(-\frac{W_1(r, \phi)}{r\sin\theta} \hat{e}_\phi+  \frac{W_2(r, \theta) }{r}\hat{e}_\theta\right).
  \end{eqnarray}
After substituting Eq. (\ref{solution-G-1-aa}) into Eq. (\ref{M-3}), one obtains
\begin{eqnarray}
 \label{M-3a}
 \vec{L}&=&\vec{\ell}+q \vec{G}=\vec{r}\times \vec{p}+q\vec{r}\times\left(-\frac{W_1(r, \phi)}{r\sin\theta} \hat{e}_\phi+  \frac{W_2(r, \theta) }{r}\hat{e}_\theta\right)\nonumber\\
 &=&\vec{r}\times \left[\vec{p}+q\left(-\frac{W_1(r, \phi)}{r\sin\theta} \hat{e}_\phi+  \frac{W_2(r, \theta) }{r}\hat{e}_\theta\right)\right]\nonumber\\
  &:=&\vec{r}\times \left[\vec{p}-\frac{q}{c} \vec{A}\right].
\end{eqnarray}
In Eq. (\ref{M-3a}), if one views the term $\left[\vec{p}-(q/c) \vec{A}\right]$ as the canonical momentum, then he obtains a kind of vector potential as follows
\begin{eqnarray}
 \label{M-3b}
\vec{A}&=&c\left(\frac{W_1(r, \phi)}{r\sin\theta} \hat{e}_\phi-  \frac{W_2(r, \theta) }{r}\hat{e}_\theta\right).
\end{eqnarray}
Let us compare Eq. (\ref{M-3b}) with the magnetic vector potential adopted in the magnetic AB effect \cite{2005QParadox}
\begin{equation}\label{eq:v-10}
   \vec{A}_{\rm M}= \begin{cases}
        & \dfrac{B \sqrt{x^2+y^2}}{2} \hat{e}_\phi,  \; \;\; (\rho< r_0)\\
        & \\
        &  \dfrac{\Phi_{\rm M}}{2\pi\sqrt{x^2+y^2}} \hat{e}_\phi,\;\; (\rho> r_0)
    \end{cases}
  \end{equation}
where $\hat{e}_\phi= (-\sin\phi, \cos\phi, 0)$, $\rho=r\sin\theta= \sqrt{x^2+y^2}$, $r_0$ is the radius of the solenoid, and
\begin{eqnarray}
 \label{M-3d-a}
\Phi_{\rm M}=B\pi r_0^2
\end{eqnarray}
is the magnetic flux. For Eq. (\ref{M-3b}), after selecting
\begin{eqnarray}
 \label{M-3c}
W_1(r, \phi)= \frac{\Phi_{\rm M}}{2\pi\,c},\;\;\;\;\;  W_2(r, \theta)=0,
\end{eqnarray}
one obtains
\begin{eqnarray}\label{eq:v-10b}
  \vec{A}= \vec{A}_{\rm M},\;\; (\rho > r_0).
  \end{eqnarray}
Namely, the magnetic vector potential $\vec{A}_{\rm M}$, $(\rho > r_0)$, can be derived from the approach based on the angular momentum operator. Certainly, we admit that such an approach is not omnipotent. However, it can indeed derive some significant vector potentials in quantum physics. As we shall show behind, the Wu-Yang monopole vector potential can be naturally derived by this approach.

\subsection{Case (ii). $G_r\neq 0$}

Let us now consider the case with $G_r \neq 0$.

\emph{Case (ii-1).} We consider the most simple case with $G_r\neq 0, G_\theta=G_\phi=0$, i.e.,
\begin{eqnarray}
\label{Gr-1}
 &&\vec{G}= G_r \hat{e}_r.
  \end{eqnarray}
By substituting Eq. (\ref{Gr-1}) into Eq. (\ref{M-13}), we have
 \begin{eqnarray}
 \label{Gr-2}
&&  \vec{\nabla} (\vec{r}\cdot \vec{G})=\vec{r} \left(\vec{\nabla}\cdot\vec{G}\right),\nonumber\\
&& \Rightarrow \;\;\;\;{\rm grad}(r G_r)= \vec{r} \frac{1}{r^2}\biggr[ \frac{\partial (r^2 G_r)}{\partial r}\biggr],\nonumber\\
&&\Rightarrow \;\;\;\;\hat{e}_r \frac{\partial (r G_r)}{\partial r}+\hat{e}_\theta \frac{1}{r}\frac{\partial (r G_r)}{\partial \theta}+\hat{e}_\phi \frac{1}{r\sin\theta}\frac{\partial (r G_r)}{\partial \phi} = \hat{e}_r \frac{1}{r}\biggr[ \frac{\partial (r^2 G_r)}{\partial r}\biggr],
 \end{eqnarray}
which leads to
 \begin{eqnarray}
 \label{Gr-3}
&& \frac{\partial (r G_r)}{\partial r}= \frac{1}{r}\biggr[ \frac{\partial (r^2 G_r)}{\partial r}\biggr], \;\;\;\;\; \frac{\partial (r G_r)}{\partial \theta}=0,\;\;\;\;\; \frac{\partial (r G_r)}{\partial \phi}=0.
 \end{eqnarray}
The last two equations of (\ref{Gr-3}) implies that $G_r$ is a function depending only on $r$. We have from the first equation that
  \begin{eqnarray}
 \label{Gr-4-0}
&& \frac{\partial (r G_r)}{\partial r}= \frac{1}{r}\biggr[ \frac{\partial (r^2 G_r)}{\partial r}\biggr],\nonumber\\
&&\Rightarrow\;\;\;G_r+r \frac{\partial ( G_r)}{\partial r}=2G_r+r\frac{\partial ( G_r)}{\partial r},\nonumber\\
 &&\Rightarrow\;\;\;G_r=0.
 \end{eqnarray}
This implies that $\vec{G}=\vec{0}$ is a zero vector, which is a trivial solution.

\emph{Case (ii-2).} We consider the case with $G_r\neq 0, G_\theta\neq 0, G_\phi=0$, i.e.,
\begin{eqnarray}
\label{Gr-1aa}
 &&\vec{G}= G_r \hat{e}_r +G_\theta \hat{e}_\theta.
  \end{eqnarray}
By substituting Eq. (\ref{Gr-1aa}) into Eq. (\ref{M-13}), we have
 \begin{eqnarray}
 \label{Gr-2a}
&&  \vec{\nabla} (\vec{r}\cdot \vec{G})=\vec{r} \left(\vec{\nabla}\cdot\vec{G}\right), \nonumber\\
&& \Rightarrow \;\;{\rm grad}(r G_r)= \vec{r} \biggr(\frac{1}{r^2}\biggr[ \frac{\partial (r^2 G_r)}{\partial r}\biggr]+\frac{1}{r\sin\theta}\biggr[ \frac{\partial (\sin\theta G_\theta)}{\partial \theta}\biggr]\biggr), \nonumber\\
&&\Rightarrow \;\;\hat{e}_r \frac{\partial (r G_r)}{\partial r}+\hat{e}_\theta \frac{1}{r}\frac{\partial (r G_r)}{\partial \theta}+\hat{e}_\phi \frac{1}{r\sin\theta}\frac{\partial (r G_r)}{\partial \phi} = \hat{e}_r \biggr(\frac{1}{r}\biggr[ \frac{\partial (r^2 G_r)}{\partial r}\biggr]+\frac{1}{\sin\theta}\biggr[ \frac{\partial (\sin\theta G_\theta)}{\partial \theta}\biggr]\biggr),
 \end{eqnarray}
which leads to
 \begin{eqnarray}
 \label{Gr-3a}
&& \frac{\partial (r G_r)}{\partial r}= \frac{1}{r}\biggr[ \frac{\partial (r^2 G_r)}{\partial r}\biggr]+\frac{1}{\sin\theta}\biggr[ \frac{\partial (\sin\theta G_\theta)}{\partial \theta}\biggr], \;\;\;\;\; \frac{\partial (r G_r)}{\partial \theta}=0,\;\;\;\;\; \frac{\partial (r G_r)}{\partial \phi}=0.
 \end{eqnarray}
The last two equations of (\ref{Gr-3a}) implies that $G_r$ is a function depending only on $r$. We have from the first equation that
  \begin{eqnarray}
 \label{Gr-4}
&& \frac{\partial (r G_r)}{\partial r}= \frac{1}{r}\biggr[ \frac{\partial (r^2 G_r)}{\partial r}\biggr]+\frac{1}{\sin\theta}\biggr[ \frac{\partial (\sin\theta G_\theta)}{\partial \theta}\biggr],\nonumber\\
&&\Rightarrow\;\;\;G_r+r \frac{\partial  G_r}{\partial r}=2G_r+r\frac{\partial  G_r}{\partial r}+\frac{1}{\sin\theta}\biggr[ \frac{\partial (\sin\theta G_\theta)}{\partial \theta}\biggr],\nonumber\\
 &&\Rightarrow\;\;\;G_r+\frac{1}{\sin\theta}\biggr[ \frac{\partial (\sin\theta G_\theta)}{\partial \theta}\biggr]=0,\nonumber\\
  &&\Rightarrow\;\;\; \frac{\partial (\sin\theta G_\theta)}{\partial \theta}=-G_r \sin\theta,\nonumber\\
  &&\Rightarrow\;\;\; \int \frac{\partial (\sin\theta G_\theta)}{\partial \theta} d\theta=-\int G_r \sin\theta d\theta,\nonumber\\
  &&\Rightarrow\;\;\; \sin\theta G_\theta=-G_r\int \sin\theta d\theta,\nonumber\\
  &&\Rightarrow\;\;\; \sin\theta G_\theta=G_r (\cos\theta+C), \nonumber\\
  &&\Rightarrow\;\;\;  G_\theta=G_r \frac{\cos\theta+C}{\sin\theta},
\end{eqnarray}
where $C$ is real constant number.

By substituting Eq. (\ref{Gr-4}) into Eq. (\ref{Gr-1aa}), we have
\begin{eqnarray}
\label{Gr-1ab}
 \vec{G}&=& G_r \hat{e}_r +G_\theta \hat{e}_\theta=G_r \hat{e}_r +G_\theta (-\hat{e}_r \times \hat{e}_\phi)\nonumber\\
 &=&G_r \hat{e}_r -G_r \frac{\cos\theta+C}{\sin\theta}(\hat{e}_r \times \hat{e}_\phi)
 =G_r \frac{\vec{r}}{r} -G_r \frac{\cos\theta+C}{r\sin\theta}(\vec{r} \times \hat{e}_\phi),
  \end{eqnarray}
therefore
\begin{eqnarray}
 \label{M-3d}
 \vec{L}&=&\vec{\ell}+q \vec{G}=\vec{r}\times \vec{p}+q\vec{r}\times\left(-G_r \frac{\cos\theta+C}{r\sin\theta} \hat{e}_\phi\right)+q G_r \frac{\vec{r}}{r}\nonumber\\
 &=&\vec{r}\times \left(\vec{p}-q G_r \frac{\cos\theta+C}{r\sin\theta} \hat{e}_\phi\right)+q G_r \frac{\vec{r}}{r}.
\end{eqnarray}
Let us compare Eq. (\ref{M-3d}) with the Wu-Yang angular momentum operator as shown in Eq. (\ref{mo-eq3a}), i.e.,
\begin{equation}
 \label{mo-eq3aa}
 \vec{L}=\vec{r}\times \left(\vec{p}-Ze \vec{A}\right)-q\frac{\vec{r}}{r},
\end{equation}
we easily find that for
\begin{eqnarray}
 \label{M-3e}
G_r=-1, \;\;\; q=Zeg, \;\;\; C=-1,
\end{eqnarray}
one immediately obtains the Wu-Yang monopole vector potential in the region $a$ as
\begin{eqnarray}
\label{commu-6-a}
 \vec{A}=\vec{A}_a &=& \frac{g}{r} \frac{1-\cos\theta}{\sin\theta}\; \hat{e}_\phi.
 \end{eqnarray}
Similarly, for
\begin{eqnarray}
 \label{M-3f}
G_r=-1, \;\;\; q=Zeg, \;\;\; C=1,
\end{eqnarray}
one immediately obtains the Wu-Yang monopole vector potential in the region $b$ as
\begin{eqnarray}
\label{commu-6-b}
 \vec{A}= \vec{A}_b &=& \frac{-g}{r} \frac{1+\cos\theta}{\sin\theta}\; \hat{e}_\phi.
\end{eqnarray}
Thus, we have naturally derived the Wu-Yang monopole vector potential based on the approach of the angular momentum operator.

\emph{Case (ii-3).} We consider the case with $G_r\neq 0, G_\phi\neq 0, G_\theta=0$, i.e.,
\begin{eqnarray}
\label{Gr-1b}
 &&\vec{G}= G_r \hat{e}_r +G_\phi \hat{e}_\phi.
  \end{eqnarray}
By substituting Eq. (\ref{Gr-1b}) into Eq. (\ref{M-13}), we have
 \begin{eqnarray}
 \label{Gr-2b}
&&  \vec{\nabla} (\vec{r}\cdot \vec{G})=\vec{r} \left(\vec{\nabla}\cdot\vec{G}\right),\nonumber\\
&& \Rightarrow \;\;{\rm grad}(r G_r)= \vec{r} \biggr(\frac{1}{r^2}\biggr[ \frac{\partial (r^2 G_r)}{\partial r}\biggr]+\frac{1}{r\sin\theta}\biggr[ \frac{\partial G_\phi}{\partial \phi}\biggr]\biggr),\nonumber\\
&&\Rightarrow \;\;\hat{e}_r \frac{\partial (r G_r)}{\partial r}+\hat{e}_\theta \frac{1}{r}\frac{\partial (r G_r)}{\partial \theta}+\hat{e}_\phi \frac{1}{r\sin\theta}\frac{\partial (r G_r)}{\partial \phi},\nonumber\\
&& \;\;\;\;\;\; = \hat{e}_r \biggr(\frac{1}{r}\biggr[ \frac{\partial (r^2 G_r)}{\partial r}\biggr]+\frac{1}{\sin\theta}\biggr[ \frac{\partial G_\phi}{\partial \phi}\biggr]\biggr),
 \end{eqnarray}
which leads to
 \begin{eqnarray}
 \label{Gr-3b}
&& \frac{\partial (r G_r)}{\partial r}= \frac{1}{r}\biggr[ \frac{\partial (r^2 G_r)}{\partial r}\biggr]+\frac{1}{\sin\theta}\biggr[ \frac{\partial G_\phi}{\partial \phi}\biggr], \;\;\;\;\; \frac{\partial (r G_r)}{\partial \theta}=0,\;\;\;\;\;
 \frac{\partial (r G_r)}{\partial \phi}=0.
 \end{eqnarray}
The last two equations of (\ref{Gr-3b}) implies that $G_r$ is a function depending only on $r$. We have from the first equation that
  \begin{eqnarray}
 \label{Gr-4a}
&& \frac{\partial (r G_r)}{\partial r}= \frac{1}{r}\biggr[ \frac{\partial (r^2 G_r)}{\partial r}\biggr]+\frac{1}{\sin\theta}\biggr[ \frac{\partial G_\phi}{\partial \phi}\biggr],\nonumber\\
&&\Rightarrow\;\;\;G_r+r \frac{\partial  G_r}{\partial r}=2G_r+r\frac{\partial  G_r}{\partial r}+\frac{1}{\sin\theta}\biggr[ \frac{\partial G_\phi}{\partial \phi}\biggr],\nonumber\\
 &&\Rightarrow\;\;\;G_r+\frac{1}{\sin\theta}\biggr[ \frac{\partial G_\phi}{\partial \phi}\biggr]=0,\nonumber\\
  &&\Rightarrow\;\;\; \frac{\partial G_\phi}{\partial \phi}=-G_r \sin\theta,\nonumber\\
  &&\Rightarrow\;\;\; \int \frac{\partial G_\phi}{\partial \phi} d\phi=-\int G_r \sin\theta d\phi, \nonumber\\
  &&\Rightarrow\;\;\; G_\phi=-\left(\phi+C\right) G_r \sin\theta,
\end{eqnarray}
where $C$ is real constant number.

\emph{Case (ii-D).} We consider the case with $G_r\neq 0, G_\theta\neq 0, G_\phi\neq 0$, i.e.,
\begin{eqnarray}
\label{Gr-1c}
 &&\vec{G}= G_r \hat{e}_r+G_r \hat{e}_r +G_\phi \hat{e}_\phi.
  \end{eqnarray}
By substituting Eq. (\ref{Gr-1c}) into Eq. (\ref{M-13}), we have
 \begin{eqnarray}
 \label{Gr-2c}
&&  \vec{\nabla} (\vec{r}\cdot \vec{G})=\vec{r} \left(\vec{\nabla}\cdot\vec{G}\right),\nonumber\\
&& \Rightarrow \;\;{\rm grad}(r G_r)= \vec{r} \biggr(\frac{1}{r^2}\biggr[ \frac{\partial (r^2 G_r)}{\partial r}\biggr]+\frac{1}{r\sin\theta}\biggr[ \frac{\partial (\sin\theta G_\theta)}{\partial \theta}\biggr]+\frac{1}{r\sin\theta}\biggr[ \frac{\partial G_\phi}{\partial \phi}\biggr]\biggr),\nonumber\\
&&\Rightarrow \;\;\hat{e}_r \frac{\partial (r G_r)}{\partial r}+\hat{e}_\theta \frac{1}{r}\frac{\partial (r G_r)}{\partial \theta}+\hat{e}_\phi \frac{1}{r\sin\theta}\frac{\partial (r G_r)}{\partial \phi} = \hat{e}_r \biggr(\frac{1}{r}\biggr[ \frac{\partial (r^2 G_r)}{\partial r}\biggr]+\frac{1}{\sin\theta}\biggr[ \frac{\partial (\sin\theta G_\theta)}{\partial \theta}\biggr]+\frac{1}{\sin\theta}\biggr[ \frac{\partial G_\phi}{\partial \phi}\biggr]\biggr),
 \end{eqnarray}
which leads to
 \begin{eqnarray}
 \label{Gr-3c}
&& \frac{\partial (r G_r)}{\partial r}= \frac{1}{r}\biggr[ \frac{\partial (r^2 G_r)}{\partial r}\biggr]+\frac{1}{\sin\theta}\biggr[ \frac{\partial (\sin\theta G_\theta)}{\partial \theta}\biggr]+\frac{1}{\sin\theta}\biggr[ \frac{\partial G_\phi}{\partial \phi}\biggr], \;\;\;\;\;
\frac{\partial (r G_r)}{\partial \theta}=0,\;\;\;\;\; \frac{\partial (r G_r)}{\partial \phi}=0.
 \end{eqnarray}
The last two equations of (\ref{Gr-3c}) implies that $G_r$ is a function depending only on $r$. We have from the first equation that
  \begin{eqnarray}
 \label{Gr-4b}
 && \frac{\partial (r G_r)}{\partial r}= \frac{1}{r}\biggr[ \frac{\partial (r^2 G_r)}{\partial r}\biggr]+\frac{1}{\sin\theta}\biggr[ \frac{\partial (\sin\theta G_\theta)}{\partial \theta}\biggr]+\frac{1}{\sin\theta}\biggr[ \frac{\partial G_\phi}{\partial \phi}\biggr],\nonumber\\
 &&\Rightarrow \;\;\;G_r+\frac{1}{\sin\theta}\biggr[ \frac{\partial (\sin\theta G_\theta)}{\partial \theta}\biggr]+\frac{1}{\sin\theta}\biggr[ \frac{\partial G_\phi}{\partial \phi}\biggr]=0.
\end{eqnarray}

\emph{Analysis (a):}  One solution is
 \begin{eqnarray}
 \label{Gr-4c}
  &&\frac{1}{\sin\theta}\biggr[ \frac{\partial (\sin\theta G_\theta)}{\partial \theta}\biggr]=-\mu G_r, \;\;\;\;\;\frac{1}{\sin\theta}\biggr[ \frac{\partial G_\phi}{\partial \phi}\biggr]=-(1-\mu)G_r,\nonumber\\
 &&\Rightarrow\;\;\; G_\theta=\mu G_r \frac{\cos\theta+C_1}{\sin\theta},\;\;\;\;\;\; G_\phi=(\mu-1)G_r \sin\theta(\phi+C_2),
 \end{eqnarray}
where $C_1, C_2$ are real constant numbers.

\emph{Analysis (b):}  If $G_\theta=T(r,\phi)$, which is a real constant number or a function depending only on $r$ and $\phi$, then we have
 \begin{eqnarray}
 \label{Gr-4d}
  &&G_r+\frac{1}{\sin\theta}\biggr[ \frac{\partial (\sin\theta G_\theta)}{\partial \theta}\biggr]+\frac{1}{\sin\theta}\biggr[ \frac{\partial G_\phi}{\partial \phi}\biggr]=0\nonumber\\
   &&\Rightarrow\;\;\; G_r\sin\theta+ \frac{\partial (\sin\theta G_\theta)}{\partial \theta}+ \frac{\partial G_\phi}{\partial \phi}=0\nonumber\\
   &&\Rightarrow\;\;\; G_r\sin\theta{+}\cos\theta T(r, \phi) + \frac{\partial G_\phi}{\partial \phi}=0\nonumber\\
  &&\Rightarrow\;\;\;\int \frac{\partial G_\phi}{\partial \phi}d \phi=-\int G_r\sin\theta d\phi{-}\int\cos\theta T(r, \phi)d\phi\nonumber\\
  &&\Rightarrow\;\;\;G_\phi=-G_r\sin\theta (\phi+C_1){-}\cos\theta\int T(r, \phi)d\phi.
 \end{eqnarray}
If $G_\theta=T(r)$, which does not depends on $\theta$ and $\phi$, then from above we may have a simple solution as follows
 \begin{eqnarray}
 \label{Gr-4e}
  &&G_\phi=-G_r\sin\theta (\phi+C_1){-}\cos\theta\int T(r)d\phi.\nonumber\\
  &&\Rightarrow\;\;\;G_\phi=-G_r\sin\theta (\phi+C_1){-}\cos\theta T(r)(\phi+C_2).\nonumber\\
  \end{eqnarray}

\emph{Analysis (c):}  If $G_\phi=W(r,\theta)$, which is a real constant number or a function depending only on $r$ and $\theta$, then we have
 \begin{eqnarray}
 \label{Gr-4f}
  &&G_r+\frac{1}{\sin\theta}\biggr[ \frac{\partial (\sin\theta G_\theta)}{\partial \theta}\biggr]+\frac{1}{\sin\theta}\biggr[ \frac{\partial G_\phi}{\partial \phi}\biggr]=0\nonumber\\
  &&\Rightarrow\;\;\; G_r+\frac{1}{\sin\theta}\biggr[ \frac{\partial (\sin\theta G_\theta)}{\partial \theta}\biggr]=0\nonumber\\
  &&\Rightarrow\;\;\; G_\theta=G_r \frac{\cos\theta+C}{\sin\theta}.
 \end{eqnarray}

\begin{remark}
Let us make a summary for this section, in which we have presented an approach to extract the vector potential from the angular momentum operator. Explicitly,  in Theorem 1, based on the orbital angular momentum $\vec{\ell}=\vec{r}\times \vec{p}$ and an appropriate vector $\vec{G}$, we can construct a new angular momentum operator as
\begin{eqnarray}\label{G-2m}
\vec{L}&=&\vec{r}\times\vec{p}+q \vec{G}.
\end{eqnarray}
We then recast Eq. (\ref{G-2m}) to the following form
\begin{eqnarray}\label{G-2n}
\vec{L}&=&\vec{r}\times\vec{p}+q \vec{G}=\vec{r}\times\vec{p}+q \left(G_r \hat{e}_r+G_\theta \hat{e}_\theta+G_\phi \hat{e}_\phi\right)\nonumber\\
&=&\vec{r}\times\vec{p}+q \left[G_r \hat{e}_r+G_\theta (-\hat{e}_r\times \hat{e}_\phi)+G_\phi (\hat{e}_r\times \hat{e}_\theta)\right]\nonumber\\
&=&\vec{r}\times\left[\vec{p}+\frac{q}{r}\left(-G_\theta \hat{e}_\phi+G_\phi  \hat{e}_\theta\right)\right]+q G_r \hat{e}_r,
\end{eqnarray}
and view the operator
\begin{eqnarray}\label{G-2p}
&& \vec{\Pi}:=\vec{p}+\frac{q}{r}\left(-G_\theta \hat{e}_\phi+G_\phi  \hat{e}_\theta\right)=\vec{p}-\frac{q}{c}\vec{A}
\end{eqnarray}
as the canonical momentum. Thus, from Eq. (\ref{G-2p}) one can extract a vector potential $\vec{A}$ as
\begin{eqnarray}\label{G-2r}
 \vec{A}&=&\frac{c}{r} \left(G_\theta \hat{e}_\phi - G_\phi  \hat{e}_\theta\right)\nonumber\\
&=& \dfrac{c}{r}\left|\begin{array}{ccc}
                  \hat{e}_r & \hat{e}_\theta & \hat{e}_\phi \\
                  1 & 0 & 0 \\
                  G_r & G_\theta & G_\phi
            \end{array}\right| = c \dfrac{\hat{e}_r\times\vec{G}}{r}\\
            &=& c \dfrac{\vec{r}\times\vec{G}}{r^2}.
\end{eqnarray}
Based on this approach, we have successfully derived the Wu-Yang monopole vector potential ($\vec{A}_a$ and $\vec{A}_b$) and the magnetic vector potential $\vec{A}_{\rm M}$ $(r_0>0)$ used in the magnetic AB effect. In the next section, we shall use this approach to establish the spin vector potential $\vec{A}_{\rm S}$.
\end{remark}

\section{The Derivation of the Spin Vector Potential}

In the above section, we have studied the Abelian case for the vector $\vec{G}$, i.e., it satisfies the relation $\vec{G}\times \vec{G}=0$. When the vector $\vec{G}$ is a non-Abelian one, the situation becomes complicate. However, in this section, we only focus on a simple case.

Let us consider the total angular momentum operator
\begin{equation}\label{eq:DefL}
      \vec{J}=\vec{\ell}+\vec{S},
      \end{equation}
where $\vec{S}$ is the spin operator, which satisfies the relation of the angular momentum operator, i.e.,
\begin{equation}\label{eq:Spin-1a}
      \vec{S}\times \vec{S}={\rm i} \hbar \vec{S}.
      \end{equation}
It is very easy to prove that
\begin{equation}\label{eq:Spin-1}
      \vec{J}\times \vec{J}={\rm i} \hbar \vec{J},
      \end{equation}
hence $\vec{J}$ is an angular momentum operator.

Similar to the previous section, the direct calculation shows
      \begin{eqnarray}\label{eq:SplitJ}
      \vec{J}  &=& \vec{\ell}+\vec{S} =\vec{r}\times\vec{p}+(S_r \hat{e}_r +S_\theta \hat{e}_\theta +S_\phi \hat{e}_\phi) \nonumber \\
      &=& \vec{r}\times\vec{p}+\Bigl[S_r \hat{e}_r +S_\theta (-\hat{e}_r \times\hat{e}_\phi)
            +S_\phi (\hat{e}_r \times\hat{e}_\theta)\Bigr] \nonumber \\
      &=& \vec{r}\times\vec{p}+\biggr[S_r \dfrac{\vec{r}}{r}
            -S_\theta \left(\dfrac{\vec{r}}{r} \times\hat{e}_\phi\right)
            +S_\phi \left(\dfrac{\vec{r}}{r} \times\hat{e}_\theta\right)\biggr] \nonumber \\
      &=& \vec{r}\times\Biggr[\vec{p}+\biggr(\dfrac{S_\phi}{r} \hat{e}_\theta
            -\dfrac{S_\theta}{r} \hat{e}_\phi\biggr)\Biggr]+S_r \dfrac{\vec{r}}{r} \nonumber \\
      &=& \vec{r}\times\left(\vec{p}-\dfrac{q}{c} \vec{A}\right)+S_r \dfrac{\vec{r}}{r},
      \end{eqnarray}
from which we extract the spin vector potential as
      \begin{eqnarray}
                 \vec{A} &=& \dfrac{c}{q\,r}\left(S_\theta\,\hat{e}_\phi-S_\phi\,\hat{e}_\theta\right)=\dfrac{c}{q} \dfrac{\vec{r}\times\vec{S}}{r^2}.
        \end{eqnarray}
\begin{remark}
      In Eq. (\ref{eq:DefL}), we have let the operator $q\,\vec{G}:=\vec{S}$, where $\vec{S}$ is an arbitrary spin-$s$ operator, with $s=1/2, 1, 3/2, \cdots$. Actually, the operator $q\,\vec{G}$ can be chosen as any angular momentum operator, such as (i) the isospin operator $\vec{\tau}=(\hbar/2)\sigma$ for the neutron and the proton; (ii) the orbital angular momentum operator $\vec{L}_2=\vec{r}_2\times \vec{p}_2$ of another particle. In this work, we merely focus on the simplest case with $\vec{S}=(\hbar/2)\sigma$, i.e., it is a spin-1/2 operator. In the future work, one may consider the spin-1 operator, which is given by
      \begin{eqnarray}
 \label{Spin1}
  && S_x=\frac{1}{\sqrt{2}}\left(
           \begin{array}{ccc}
             0 & 1 & 0 \\
             1 & 0 & 1 \\
             0 & 1 & 0 \\
           \end{array}
         \right),\;\;\;\;\;\;\;\;
  S_y=\frac{{\rm i}}{\sqrt{2}}\left(
           \begin{array}{ccc}
             0 & -1 & 0 \\
             1 & 0 & -1 \\
             0 & 1 & 0 \\
           \end{array}
         \right),\;\;\;\;\;\;\;\;
   S_z=\left(
           \begin{array}{ccc}
             1 & 0 & 0 \\
             0 & 0 & 0 \\
             0 & 0 & -1 \\
           \end{array}
         \right).
  \end{eqnarray}
\end{remark}

\newpage

      \part{The Eigen-Problem of the Hamiltonian $H_{\rm S}$}

      Consider a particle with mass $M$ and charge $Q=g(-e)$, where $-e$ is the electric charge of an electron, and  $g\in\mathbb{Z}$.  When the particle moves in the $xyz$-space in the presence of the ``spin'' vector potential, the corresponding Hamiltonian of the
      particle is given by
      \begin{eqnarray}\label{eq:v-3}
            H_{\rm S}=\dfrac{1}{2M} \left(\vec{p}-\dfrac{Q}{c}\vec{A}\right)^2
            =\dfrac{1}{2M} \left(\vec{p}-\vec{\mathcal{A}}\right)^2,
      \end{eqnarray}
      with
      \begin{eqnarray}\label{eq:v-4}
      \vec{\mathcal{A}}&=&g \dfrac{\vec{r}\times\vec{S}}{r^2},
      \end{eqnarray}
      and $\vec{S}=(\hbar/2)\vec{\sigma}$ is the spin-1/2 operator.  Because
      \begin{eqnarray}
                        \vec{\mathcal{A}}\cdot\vec{p} &=&\ \dfrac{g\,\hbar}{2}
                  \dfrac{(\vec{r}\times\vec{\sigma})\cdot\vec{p}}{r^2} = -\dfrac{g\,\hbar}{2} \dfrac{(\vec{\sigma}\times\vec{r})\cdot\vec{p}}{r^2} = -\dfrac{g\,\hbar}{2} \dfrac{\vec{\sigma} \cdot(\vec{r}\times\vec{p})}{r^2}
                  = -\dfrac{g\,\hbar}{2} \dfrac{\vec{\ell}\cdot\vec{\sigma}}{r^2},
            \end{eqnarray}
      and
      \begin{eqnarray}\label{eq:ASqure}
      \vec{\mathcal{A}}^2 &=& \left(\dfrac{g\,\hbar}{2} \dfrac{\vec{r}\times\vec{\sigma}}{r^2}\right)\cdot
            \left(\dfrac{g\,\hbar}{2} \dfrac{\vec{r}\times\vec{\sigma}}{r^2}\right) =\dfrac{g^2\,\hbar^2}{4\,r^4} (\vec{r}\times\vec{\sigma})\cdot(\vec{r}\times\vec{\sigma}) =\dfrac{g^2\,\hbar^2}{4\,r^4} \bigl[(\vec{r}\times\vec{\sigma})\times\vec{r}\bigr]\cdot\vec{\sigma} \nonumber\\
            &=& \dfrac{g^2\,\hbar^2}{4\,r^4} \bigl[(r^2 \vec{\sigma})-(\vec{\sigma} \cdot\vec{r})\vec{r}\bigr]\cdot\vec{\sigma}=\dfrac{g^2\,\hbar^2}{4\,r^4} \bigl(3r^2-r^2\bigr)= \dfrac{g^2\,\hbar^2}{2} \dfrac{1}{r^2},
            \end{eqnarray}
      we can expand the Hamiltonian in \Eq{eq:v-3} as follows
      \begin{equation}\label{eq:HSForm1}
            \begin{split}
            H_{\rm S} =&\ \dfrac{1}{2M} \left(
                  \vec{p}^2 -\vec{\mathcal{A}}\cdot\vec{p}-\vec{p}\cdot\vec{\mathcal{A}}
                  +\vec{\mathcal{A}}^2\right) \\
            =&\ \dfrac{1}{2M} \biggl[\vec{p}^2 -\vec{\mathcal{A}}\cdot\vec{p}
                  -\left(\vec{\mathcal{A}}\cdot\vec{p}-{\rm i}\hbar\vec{\nabla}\cdot\vec{\mathcal{A}}\right)
                  +\vec{\mathcal{A}}^2\biggr] \\
            =&\ \dfrac{1}{2M} \left(\vec{p}^2 -2\vec{\mathcal{A}}\cdot\vec{p}+\vec{\mathcal{A}}^2\right) \\
            =&\ \dfrac{1}{2M} \left(\vec{p}^2 +g\,\hbar\dfrac{\vec{\ell}\cdot\vec{\sigma}}{r^2}
                  +\dfrac{g^2\,\hbar^2}{2} \dfrac{1}{r^2}\right),
            \end{split}
      \end{equation}
where we have used $\vec{\nabla}\cdot\vec{\mathcal{A}}=0$.

      The eigen-problem is given by
      \begin{eqnarray}\label{eq:v-5}
      H_{\rm S}\,\Psi_{\rm S}(\vec{r})=E\,\Psi_{\rm S}(\vec{r}),
      \end{eqnarray}
      where $E$ is the energy, and $\Psi_{\rm S}(\vec{r})$ is the eigenfunction. In $\sigma_z$'s
      representation, because
      \begin{eqnarray}\label{eq:E-7}
            \vec{\ell}\cdot\vec{\sigma} &=& \begin{bmatrix}
                  \ell_z & \ell_x -\mathrm{i}\,\ell_y \\
                  \ell_x +\mathrm{i}\,\ell_y & -\ell_z
            \end{bmatrix},
      \end{eqnarray}
      then we have
      \begin{eqnarray}\label{eq:E-8}
            H_{\rm S} &=& \dfrac{1}{2M} \begin{bmatrix}
                  \vec{p}^2 +g\hbar\dfrac{\ell_z}{r^2} +\dfrac{g^2\hbar^2}{2} \dfrac{1}{r^2}
                        & g\hbar\dfrac{(\ell_x-\mathrm{i} \ell_y)}{r^2} \\
                  g\hbar\dfrac{(\ell_x+\mathrm{i} \ell_y)}{r^2}
                        & \vec{p}^2 -g\hbar\dfrac{\ell_z}{r^2} +\dfrac{g^2\hbar^2}{2} \dfrac{1}{r^2} \\
            \end{bmatrix}.
      \end{eqnarray}
      Because $ H_{\rm S}$ is a $2\times 2$ matrix, thus the structure of the wavefunction $\Psi(\vec{r})$ is of the following form
      \begin{eqnarray}\label{eq:E-9}
            \Psi(\vec{r}) &=& \begin{bmatrix}
                  \chi_1(\vec{r}) \\
                  \chi_2(\vec{r}) \\
            \end{bmatrix}.
      \end{eqnarray}

\begin{remark}
When there is no any vector potential (i.e., $g=0$), the corresponding Hamiltonian represents a free
      particle, which is given by
      \begin{equation}\label{eq:H0-b}
            H_0=\dfrac{1}{2M} \vec{p}^{\;2}.
      \end{equation}
      The eigen-equation hence reads
      \begin{equation}\label{eq:H0-f-b-01}
            H_0 \Psi_0(\vec{r}) =E_0 \Psi_0(\vec{r}),
      \end{equation}
      where $E_0$ is the eigen-energy of the particle, and $\Psi_0(\vec{r})$ is the eigenfunction.
      According to quantum mechanics, it is easy to have
      \begin{eqnarray}\label{eq:H0-P-b}
            && E_0=\dfrac{\hbar^2 \vec{k}^2 }{2M},\nonumber\\
            &&\Psi_0(\vec{r})= \mathcal{C}_1 {\rm e}^{\mathrm{i} \vec{k}\cdot\vec{r}}+\mathcal{C}_2 {\rm e}^{-\mathrm{i} \vec{k}\cdot\vec{r}},\;\;\;
            {\rm or}\;\;  \Psi_0(\vec{r})= \mathcal{D}_1 \cos\left(\vec{k}\cdot\vec{r}\right)
            +\mathcal{D}_2 \sin\left(\vec{k}\cdot\vec{r}\right),
      \end{eqnarray}
      here $\vec{k}$ represents the wave vector relating to the wave packet $\Psi_0(\vec{r})$, and the
      coefficients $\mathcal{C}_1$, $\mathcal{C}_2$, $\mathcal{D}_1$, $\mathcal{D}_2$ are some complex
      constants. On the other hand, we would like to see in the spherical coordinate
      system $(r, \theta, \phi)$, how the wavefunction $\Psi_0(\vec{r})$ looks like.

      Due to
      \begin{eqnarray}\label{eq:E-18}
            \vec{p}^{\;2} =-\hbar^2 \dfrac{\partial^2}{\partial\,r^2}
                  -2\hbar^2 \dfrac{1}{r} \dfrac{\partial}{\partial\,r} +\dfrac{\vec{\ell}^{\;2}}{r^2}.
      \end{eqnarray}
       we have
      \begin{eqnarray}\label{eq:J-4}
            H_0 &=& \dfrac{1}{2M} \left(-\hbar^2 \dfrac{\partial^2}{\partial\,r^2}
                  -2 \hbar^2 \dfrac{1}{r} \dfrac{\partial }{\partial\,r}
                  + \dfrac{\vec{\ell}^{\;2}}{ r^2}\right).
      \end{eqnarray}
      Since
      \begin{eqnarray}\label{eq:J-5}
            \left[H_0, \vec{\ell}^{\;2}\right]=0, \;\;\left[H_0, \ell_z\right]=0, \;\;
                  \left[\vec{\ell}^{\;2}, \ell_z\right]=0,
      \end{eqnarray}
      thus in the common set $\left\{H_0, \vec{\ell}^{\; 2}, \ell_z\right\}$, one may
      express the wavefunction $\Psi_0(\vec{r})$ as
      \begin{eqnarray}\label{eq:J-6}
            \Psi_0(\vec{r})=R_0(r)\,Y_{lm}(\theta, \phi),
      \end{eqnarray}
      which satisfies
      \begin{eqnarray}\label{eq:J-7}
            H_0\; \Psi_0(\vec{r})&=& E_0 \; \Psi_0(\vec{r}),\nonumber\\
            \vec{\ell}^{\;2}\; \Psi_0(\vec{r})&=& l(l+1) \hbar^2\; \Psi_0(\vec{r}),\nonumber\\
            \ell_z\; \Psi_0(\vec{r})&=& m\hbar \; \Psi_0(\vec{r}),
      \end{eqnarray}
      where $l=0, 1, 2, \cdots$, and $m=0, \pm 1, \pm 2, \cdots, \pm l$.

      By the way, if we only restrict the common eigenstates of
      $\left\{H_0, \vec{\ell}^{\;2}\right\}$, then the wavefunction is generally a
      superposition state as follows:
      \begin{eqnarray}\label{eq:J-8}
      \Psi_0(\vec{r})&=&R_0(r) \sum_{m=-l}^{+l} c_m Y_{lm}(\theta, \phi)= R_0(r) \sum_{m=-l}^{+l} c_m P_{l}^m(\cos\theta) {\rm e}^{{\rm i} m \phi}.
            \end{eqnarray}
      After replacing $\vec{\ell}^{\;2}$ by its eigenvalue $l(l+1) \hbar^2$, then we have
      the differential equation for $R_0(r)$ as
      \begin{eqnarray}\label{eq:J-9}
            \dfrac{1}{2M} \left[-\hbar^2 \dfrac{{\rm d}^2 R_0(r)}{d\,r^2}
                  -\dfrac{2 \hbar^2}{r} \dfrac{{\rm d}\,R_0(r) }{{\rm d}\,r}
                  + \dfrac{l(l+1) \hbar^2}{ r^2} R_0(r)\right]=E_0 R_0(r),
      \end{eqnarray}
      i.e.,
      \begin{eqnarray}\label{eq:J-11}
            \dfrac{{\rm d}^2 R_0(r)}{{\rm d}\,r^2}
                  +\dfrac{2 }{r} \dfrac{{\rm d}\,R_0(r) }{{\rm d}\,r}
                  +\biggr[\epsilon-\dfrac{l(l+1) }{ r^2}\biggr]R_0(r)=0,
      \end{eqnarray}
      with
      \begin{eqnarray}\label{eq:J-12}
            \epsilon= \dfrac{2M}{\hbar^2} E_0.
      \end{eqnarray}
      From Eq. (\ref{eq:J-11}) we have
      \begin{eqnarray}\label{eq:J-13}
            r^2 \dfrac{{\rm d}^2 R_0(r)}{{\rm d}\,r^2}
            +2r \dfrac{{\rm d}\,R_0(r) }{{\rm d}\,r}
            +\left[\epsilon r^2-l(l+1)\right]R_0(r)=0,
      \end{eqnarray}
      which can be transformed to the following standard Bessel equation
      \begin{eqnarray}\label{eq:J-14}
            t^2\dfrac{{\rm d}^2 w}{{\rm d}\,t^2}
            +t\dfrac{{\rm d}\,w}{{\rm d}\,t} +(t^2-\nu^2)w=0,
      \end{eqnarray}
      with
      \begin{eqnarray}\label{eq:J-15}
            w&=&\sqrt{r}\,R_0(r), \nonumber\\
            t&=&\sqrt{\epsilon}\,r,\nonumber\\
            \nu &=& l+\dfrac{1}{2}.
      \end{eqnarray}

      The solution of Eq. (\ref{eq:J-14}) is the Bessel function
      \begin{eqnarray}\label{eq:J-16}
      &&J_{\pm \nu}(t)=\sum_{k=0}^{\infty} \dfrac{(-1)^k}{k!}\dfrac{1}{\Gamma(\pm \nu+k+1)}\biggr(\dfrac{t}{2}\biggr)^{2k\pm \nu},
            \end{eqnarray}
      because $\nu = l+1/2$, which is not an integer, thus the two solutions
      $J_{\nu}(t)$ and $J_{-\nu}(t)$ are linear independent. Thus the general solution of  Eq. (\ref{eq:J-14}) is a superposition state
      \begin{eqnarray}\label{eq:J-17}
            w(t)=C_1\; J_{\nu}(t) +C_2 \;J_{-\nu}(t),
      \end{eqnarray}
      with $C_1$, and $C_2$ two constants independent of $t$, which yields the radial
      wavefunction
      \begin{eqnarray}\label{eq:J-18}
            R_0(r)= \dfrac{1}{\sqrt{r}}\;\biggr[C_1\; J_{\nu}(\sqrt{\epsilon} r) +C_2 \;J_{-\nu}(\sqrt{\epsilon} r) \biggr].
      \end{eqnarray}
      To guarantee the convergence of $R_0(r)$ when $r\rightarrow 0$ (or $\infty$),
      one has to set $C_2=0$. Thus we finally have
      \begin{eqnarray}\label{eq:J-19}
      &&R_0(r)= \dfrac{C_1}{\sqrt{r}}\; J_{\nu}(\sqrt{\epsilon} r)=\dfrac{C_1}{\sqrt{r}}\;\sum_{k=0}^{\infty} \dfrac{(-1)^k}{k!}\dfrac{1}{\Gamma(\nu+k+1)}\biggr(\dfrac{\sqrt{\epsilon} r}{2}\biggr)^{2k+ \nu}.
            \end{eqnarray}

%

      Hence, in the spherical coordinate system $(r, \theta, \phi)$, the wavefunction is given by
            \begin{eqnarray}\label{eq:J-21}
                  \Psi_0(\vec{r})&=&R_0(r) Y_{lm}(\theta, \phi)= \mathcal{N}\,\dfrac{1}{\sqrt{r}}\,J_{\nu}(\sqrt{\epsilon}\,r)\:
                        Y_{lm}(\theta, \phi).
            \end{eqnarray}		
      Notwithstanding the second equation in \Eq{eq:H0-P-b} is not equal directly to the one in Eq. (\ref{eq:J-21}). From the viewpoint of physics,
      for a fixed energy $E$, the former can be obtained by the superpositions of the later, i.e.,
            \begin{eqnarray}\label{eq:J-22}
                  \Psi_0(\vec{r})&=& \mathcal{C}_1 {\rm e}^{\mathrm{i} \vec{k}\cdot\vec{r}}+\mathcal{C}_2 {\rm e}^{-\mathrm{i} \vec{k}\cdot\vec{r}}= \sum_{\nu}\dfrac{1}{\sqrt{r}}\; J_{\nu}(\sqrt{\epsilon}\,r) \left(\sum_{l=0}^{\infty}\sum_{m=-l}^{+l} c_{lm} Y_{lm}(\theta, \phi)\right).
            \end{eqnarray}
\end{remark}

\begin{remark}
      For the general $g\neq 0$, first let us observe what physical quantities are commutative with the
      Hamiltonian $H_{\rm S}$. Because
      \begin{eqnarray}\label{eq:E-12}
            \left[\vec{\ell}^{\;2}, \vec{\ell}\right] =\left[\vec{\ell}^{\;2}, \vec{\sigma}\right]
            =\left[\vec{\ell}^{\;2}, \vec{p}^{\;2}\right]=\left[\vec{\ell}^{\;2}, r^2\right]=0,
      \end{eqnarray}
      thus $\left[\vec{\ell}^{\;2},\,\vec{\ell}\cdot\vec{\sigma}\right]=0$, and
      \begin{eqnarray}\label{eq:E-11}
            [H_{\rm S},\vec{\ell}^{\;2}] =0.
      \end{eqnarray}
      In addition,
      \begin{eqnarray}\label{eq:SquredJ}
            \vec{J}^2 &=& (\vec{\ell}+\vec{S})^2 \nonumber \\
            &=& \vec{\ell}^{\;2} +\dfrac{3\hbar^2}{4} +\hbar\left(\vec{\ell}\cdot\vec{\sigma}\right),
      \end{eqnarray}
      then
      \begin{eqnarray}\label{eq:E-13}
            [\vec{J}^2,\vec{\ell}^{\;2}]=[\vec{J}^2,\vec{\ell}\cdot\vec{\sigma}]=0.
      \end{eqnarray}
      besides, it is easy to check
      \begin{equation}
            \left[r^2,\ \vec{l}\cdot\sigma\right]=\left[\vec{p}^2,\ \vec{l}\cdot\sigma\right]=0,
      \end{equation}
      therefore,
      \begin{eqnarray}\label{eq:E-15}
            \left[H_{\rm S},\ \vec{\ell}\cdot\vec{\sigma}\right]
                  =\left[H_{\rm S},\ \vec{\ell}\cdot\vec{S}\right]=0.
      \end{eqnarray}
      By substituting Eq. (\ref{eq:E-18}) into Eq. (\ref{eq:HSForm1}), we obtain the Hamiltonian as
      \begin{eqnarray}\label{eq:E-19}
            H_{\rm S} &=& \dfrac{1}{2M} \left(\vec{p}^2 +g\,\hbar\dfrac{\vec{\ell}\cdot\vec{\sigma}}{r^2}
                  +\dfrac{g^2\,\hbar^2}{2} \dfrac{1}{r^2}\right) \nonumber\\
            &=& \dfrac{1}{2M} \left(-\hbar^2 \dfrac{\partial^2}{\partial\,r^2}
                  -2\hbar^2 \dfrac{1}{r} \dfrac{\partial}{\partial\,r} +\dfrac{\vec{\ell}^{\;2}}{r^2}
                  +g\,\hbar\dfrac{\vec{\ell}\cdot\vec{\sigma}}{r^2}
                  +\dfrac{g^2\,\hbar^2}{2} \dfrac{1}{r^2}\right).
      \end{eqnarray}
      One may notice that in the right-hand side of Eq. (\ref{eq:E-19}), there are two operators:
      $\vec{\ell}^{\;2}$ and $\vec{\ell}\cdot\vec{S}$. As we have pointed out above, three operators in the
      set $\{H_{\rm S}, \vec{\ell}^{\;2}, \vec{\ell}\cdot\vec{S}\}$ are mutually commutative, which hints
      us to solve the eigen-problem in this common set.
\end{remark}

      In the following, for convenience let us list some useful results that have been known in quantum mechanics textbook, and then we shall use them to solve the eigen-problem mentioned above.

      \section{The Eigen-Problem of $\{\vec{\ell}^{\; 2},\ell_z\}$}
            In the spherical coordinate $(r,\theta,\phi)$, the orbit angular momentum operator
            $\vec{\ell}=(\ell_x ,\ell_y ,\ell_z)$ is given by
            \begin{eqnarray}\label{eq:E-21}
                  \ell_x &=& {\rm i}\hbar\left(\sin\phi \dfrac{\partial}{\partial\,\theta}
                        +\cos\phi \cot\theta \dfrac{\partial}{\partial\,\phi}\right), \nonumber \\
                  \ell_y &=& {\rm i}\hbar\left(-\cos\phi \dfrac{\partial}{\partial\,\theta}
                        +\sin\phi \cot\theta \dfrac{\partial}{\partial\,\phi}\right), \nonumber \\
                  \ell_z &=& -{\rm i}\hbar\dfrac{\partial}{\partial\,\phi},
            \end{eqnarray}
            and its squared operator is given by
            \begin{eqnarray}\label{eq:E-22}
                  \vec{\ell}^{\;2} &=& -\hbar^2 \Biggl[\dfrac{1}{\sin\theta} \dfrac{\partial}{\partial\,\theta}
                              \left(\sin\theta \dfrac{\partial}{\partial\,\theta}\right)
                        +\dfrac{1}{{\sin^2}\theta} \dfrac{\partial^2}{\partial\,\phi^2}\Biggr],
            \end{eqnarray}
            or
            \begin{eqnarray}\label{eq:E-23}
                  \vec{\ell}^{\;2} &=& -\hbar^2 \left[ \dfrac{\partial^2}{\partial \theta^2}
                  +\cot\theta  \dfrac{\partial}{\partial \theta}
                  + (1+\cot^2\theta) \dfrac{\partial^2 }{\partial \phi^2} \right].
            \end{eqnarray}
            Since $[\vec{\ell}^{\;2}, \ell_z]=0$, one usually solves the eigen-problem of
            $\vec{\ell}^{\; 2}$ in the common set $\{\vec{\ell}^{\; 2}, \ell_z\}$, viz.
            \begin{eqnarray}\label{eq:E-24}
                  \vec{\ell}^{\;2} \; Y_{lm}(\theta, \phi)&=& l(l+1) \hbar^2 \;Y_{lm}(\theta, \phi), \nonumber\\
                  \ell_z\; Y_{lm}(\theta, \phi)&=& m \hbar \;Y_{lm}(\theta, \phi),
            \end{eqnarray}
            with quantum numbers $l=0, 1, 2, \cdots$, and $m=0, \pm 1, \pm 2, \cdots, \pm l$. Here
            \begin{eqnarray}\label{eq:E-25}
            Y_{lm}(\theta, \phi)&=& \sqrt{\frac{(2l+1)}{4\pi}\frac{(l-m)!}{(l+m)!}}\; P_{l}^m(\cos\theta) {\rm e}^{\mathrm{i} m \phi}
                  \end{eqnarray}
            is called the \emph{spherical harmonics function}, and the function $P_{l}^m(\cos\theta)$
            satisfies
            \begin{eqnarray}\label{eq:E-26}
                  \dfrac{1}{\sin\theta} \dfrac{{\rm d}}{{\rm d}\,\theta}\left[
                        \sin\theta \dfrac{{\rm d}\,P_{l}^m(\cos\theta)}{{\rm d}\,\theta}\right]
                  +\left[l(l+1)-\dfrac{m^2}{{\sin^2}\theta}\right]P_{l}^m(\cos\theta)=0,
            \end{eqnarray}
and
\begin{eqnarray}\label{eq:H-24-d}
P_l^m(z)=(-1)^m (1-z^2)^{m/2} \frac{{\rm d}^m P_l(z)}{{\rm d} z^m}.
		\end{eqnarray}

      \section{The Eigen-Problem of $\vec{J}^{\; 2}$}

            The situation has been changed a little bit, for the Hilbert space is enlarged by
            introducing the spin operator $\vec{S}$. The eigenstates of
            $\{\vec{J}^{\; 2},\vec{\ell}^{\;2},J_z\}$ are given as
            \begin{eqnarray}\label{eq:E-28}
                  \vec{J}^{\; 2}\; \Phi_{ljm_j}(\theta, \phi)&=& j(j+1) \hbar^2 \; \Phi_{ljm_j}(\theta, \phi), \nonumber\\
                  \vec{\ell}^{\;2}\; \Phi_{ljm_j}(\theta, \phi)&=& l(l+1) \hbar^2 \; \Phi_{ljm_j}(\theta, \phi), \nonumber\\
                  J_z\; \Phi_{ljm_j}(\theta, \phi)&=& m_j \hbar \; \Phi_{ljm_j}(\theta, \phi).
            \end{eqnarray}
            More precisely,

            \emph{Case A.---}For the quantum number $j=l+1/2$, $(m_j=m+1/2)$, the eigenstate is given by
                  \begin{eqnarray}\label{eq:E-29}
                        \Phi^A_{ljm_j}(\theta, \phi) &=& \dfrac{1}{\sqrt{2l+1}} \begin{bmatrix}
                              \sqrt{l+m+1}\:Y_{lm}(\theta, \phi) \\
                              \sqrt{l-m}\:Y_{l,m+1}(\theta, \phi) \\
                        \end{bmatrix}.
                  \end{eqnarray}

            \emph{Case B.---}For the quantum number $j=l-1/2$, $l\neq 0$, $(m_j=m+1/2)$, the eigenstate
                  is given by
                  \begin{eqnarray}\label{eq:E-30}
                        \Phi^B_{ljm_j}(\theta, \phi)&=& \dfrac{1}{\sqrt{2l+1}} \begin{bmatrix}
                              -\sqrt{l-m}\:Y_{lm}(\theta, \phi) \\
                              \sqrt{l+m+1}\:Y_{l,m+1}(\theta, \phi) \\
                        \end{bmatrix}.
                  \end{eqnarray}
Since the right-hand side of Eq. (\ref{eq:E-29}) and Eq. (\ref{eq:E-30}) appear only quantum numbers $l$ and $m$, for simplicity, without any confusion we may denote $\Phi^A_{ljm_j}(\theta, \phi)\equiv\Phi^A_{lm}(\theta, \phi)$ and $\Phi^B_{ljm_j}(\theta, \phi)\equiv\Phi^B_{lm}(\theta, \phi)$.

      \section{The Eigen-Problem of $\vec{\ell}\cdot\vec{S}$}

            As mentioned above, $\left[\vec{J}^2,\ \vec{\ell}\cdot\vec{S}\right]=0$, thus two operators
                  $\vec{\ell}\cdot \vec{S}$ and $\vec{J}^2$ share the same eigenstates.
            From \Eq{eq:SquredJ}, one obtains
            \begin{eqnarray}\label{eq:E-32}
                  \vec{\ell}\cdot \vec{S}  &=& \dfrac{1}{2}\left(\vec{J}^2 -\vec{\ell}^{\;2}-\dfrac{3}{4}\hbar^2\right).
            \end{eqnarray}
            We then have the eigenvalues and eigenstates for the operator $\vec{\ell}\cdot \vec{S}$ as follows:
            \begin{eqnarray}\label{eq:E-33}
                  \vec{\ell}\cdot \vec{S}\; \Phi_{ljm_j}(\theta, \phi) &&= \dfrac{\hbar^2}{2} \left[
                              j(j+1)-l(l+1)-\dfrac{3}{4}\right]\,\Phi_{ljm_j}(\theta, \phi) \nonumber\\
                    &&= \begin{cases}			
                        &\dfrac{l}{2} \hbar^2 \; \Phi^A_{ljm_j}(\theta, \phi), \;\;\;\;\;\;\;\;\; {\rm for}\;\; j=l+\dfrac{1}{2},\\
                        & -\dfrac{l+1}{2} \hbar^2 \; \Phi^B_{ljm_j}(\theta, \phi), \;\;\; {\rm for}\;\; j=l-\dfrac{1}{2},\; (l\neq 0),
                  \end{cases}
            \end{eqnarray}
            with $m_j=m+1/2$.

      \section{The Eigen-Problem of $H_{\rm S}$}

            Based on the aforementioned analysis, we may let the wavefunction as
            \begin{eqnarray}\label{eq:F-2}
                  \Psi(\vec{r})=\Psi(r, \theta, \phi)=R(r)\;\Phi_{ljm_j}(\theta, \phi),
            \end{eqnarray}
            where the two-component quantum state $\Phi_{ljm_j}(\theta, \phi)=\Phi^A_{ljm_j}(\theta, \phi)$,
            or $\Phi^B_{ljm_j}(\theta, \phi)$, as we have shown in Eqs. (\ref{eq:E-29}) and (\ref{eq:E-30}).
            We then have
            \begin{eqnarray}\label{eq:F-3}
                  H_{\rm S}\;R(r) \Phi_{ljm_j}(\theta, \phi) &=&E  \;R(r) \Phi_{ljm_j}(\theta, \phi),\nonumber\\
                  \vec{\ell}^{\; 2} \;R(r) \Phi_{ljm_j}(\theta, \phi) &=& l(l+1) \hbar^2  \;R(r) \Phi_{ljm_j}(\theta, \phi),\nonumber\\
                  \vec{\ell}\cdot\vec{S} \;R(r) \Phi_{ljm_j}(\theta, \phi) &=& K \hbar^2  \;R(r) \Phi_{ljm_j}(\theta, \phi),
            \end{eqnarray}
            where $K=l/2$ or $-(l+1)/2$ ($l\neq 0$), depending on $\Phi_{ljm_j}(\theta, \phi)$ takes
            $\Phi^A_{ljm_j}(\theta, \phi)$ or $\Phi^B_{ljm_j}(\theta, \phi)$. By substituting the last two
            equations of (\ref{eq:F-3}) into the first equation of (\ref{eq:F-3}), we then have the following
            (radial) eigen-equation:
            \begin{eqnarray}\label{eq:F-4}
                  \dfrac{1}{2M} \left[-\hbar^2 \dfrac{\partial^2}{\partial\,r^2}
                        -2\hbar^2 \dfrac{1}{r} \dfrac{\partial}{\partial\,r} +\dfrac{l(l+1)\hbar^2}{r^2}
                        +2\,g\,\dfrac{K\hbar^2}{r^2} +\dfrac{g^2\,\hbar^2}{2} \dfrac{1}{r^2}\right]R(r)
                  =E\,R(r),
            \end{eqnarray}
            i.e.
            \begin{eqnarray}\label{eq:F-5}
                  &&-\dfrac{\hbar^2}{2M} \Biggl\{\dfrac{\partial^2}{\partial\,r^2}
                        +\dfrac{2}{r} \dfrac{\partial}{\partial\,r}
                        -\dfrac{1}{r^2} \left[l(l+1)+2gK+\dfrac{g^2}{2}\right]\Biggr\}R(r)=E\,R(r).
            \end{eqnarray}
            Denote
            \begin{eqnarray}\label{eq:DefEpsilonKappa}
                  \epsilon &=& \dfrac{2M}{\hbar^2} E,\;\;\;\;\;\;                  \kappa = l(l+1)+2gK+\dfrac{g^2}{2},
            \end{eqnarray}
            and thereafter we do not distinguish $\dfrac{\partial}{\partial\,r}$ from
            $\dfrac{{\rm d}}{{\rm d}\,r}$, then Eq. (\ref{eq:F-5}) becomes
            \begin{eqnarray}\label{eq:ReduceRadialEq}
                  \dfrac{{\rm d}^2\,R(r)}{{\rm d}\,r^2}
                        +\dfrac{2}{r} \dfrac{{\rm d}\,R(r)}{{\rm d}\,r}
                        -\dfrac{\kappa}{r^2} R(r)+\epsilon R(r)=0.
            \end{eqnarray}

\emph{Observation 1.---}The parameter $\kappa$ in \Eq{eq:DefEpsilonKappa} is non-negative.

            \begin{proof}
                  \begin{itemize}
                        \item [(i)]. For $K=\dfrac{l}{2}$,  we have
                              \begin{equation}\label{eq:KappaK1}
                                    \kappa=l(l+1)+2gK+\dfrac{g^2}{2 }=l(l+1)+gl+\dfrac{g^2}{2}
                                    =\left(\dfrac{g}{2} +l\right)^2 +\dfrac{g^2}{4} +l\geq 0,
                              \end{equation}
                              iff $g=l=0$, the equal sign holds. When $g=2$,
                              \begin{equation}\label{eq:KappaK1N}
                                    \kappa=\left(1 +l\right)^2 +1 +l=(l+1)(l+2).
                              \end{equation}

                        \item [(ii)]. For $K=-\dfrac{l+1}{2}$, ($l\neq 0$), we have
                              \begin{equation}\label{eq:KappaK2}
                                    \begin{split}
                                          \kappa =&\ l(l+1)+2gK+\dfrac{g^2}{2} \\
                                          =&\ l(l+1)-g(l+1)+\dfrac{g^2}{2} \\
                                          =&\ \dfrac{1}{2} \Bigl[g^2 -2g(l+1)+2l(l+1)\Bigr] \\
                                          =&\ \dfrac{1}{2} \Bigl\{\bigl[g-(l+1)\bigr]^2 +2l(l+1)-(l+1)^2\Bigr\} \\
                                          =&\ \dfrac{1}{2} \Bigl\{\bigl[g-(l+1)\bigr]^2 +(l+1)(l-1)\Bigr\} \\
                                          =&\ \dfrac{1}{2} \Bigl\{\bigl[g-(l+1)\bigr]^2 +l^2 -1\Bigr\}\geq 0,
                                    \end{split}
                              \end{equation}

                              when $g=2$,
                              \begin{equation}\label{eq:KappaK2N}
                                    \kappa=\dfrac{1}{2} \bigl\{[g-(l+1)]^2 +l^2 -1\bigr\}
                                    =\dfrac{1}{2} \bigl[(l-1)^2 +l^2 -1\bigr]
                                    =l(l-1),
                              \end{equation}
                              further, iff $l=1$, the equal sign holds.
                  \end{itemize}
            \end{proof}

In the following, let us solve Eq. (\ref{eq:ReduceRadialEq}) by considering three different situations: $E< 0$, $E=0$, and $E >0$.

            \subsection{The Energy $E < 0$}

            Let us analyze the asymptotic behaviors of $R(r)$.

            \emph{Case (i).---} When $r\rightarrow 0$, Eq. (\ref{eq:ReduceRadialEq})
                  becomes
                  \begin{eqnarray}\label{eq:F-8}
                        &&\dfrac{{\rm d}^2 R(r)}{{\rm d} r^2}
                              +\dfrac{2 }{r} \dfrac{{\rm d} R(r)}{{\rm d} r}
                              -\dfrac{\kappa}{r^2} R(r)=0.
                  \end{eqnarray}
                  In the region near to $r=0$, let $R(r) \propto r^s$, after substituting it into above equation, we have
                  \begin{eqnarray}\label{eq:F-9}
                        && s(s-1)r^{s-2}+2s r^{s-2} -\kappa r^{s-2}=0,
                  \end{eqnarray}
                  which leads to
                  \begin{eqnarray}\label{eq:F-10}
                        && s(s+1) -\kappa=0.
                  \end{eqnarray}
                  Based on which, one obtains
                  \begin{eqnarray}\label{eq:F-11}
                        && s_1=\dfrac{-1+\sqrt{1+4\kappa}}{2}, \;\;\; {\rm or }\;\;\; s_2= \dfrac{-1-\sqrt{1+4\kappa}}{2}.
                  \end{eqnarray}
                  To avoid when $r\rightarrow 0$, $R(r)\rightarrow \infty$, from the
                  viewpoint of physics, we choose $s=s_1 \geq 0$. When $\kappa=0$, we specially have $s=0$.

            \emph{Case (ii).---} When $r \rightarrow \infty$, Eq.
                  (\ref{eq:ReduceRadialEq}) becomes
                  \begin{eqnarray}\label{eq:F-12}
                        &&\dfrac{{\rm d}^2 R(r)}{{\rm d}\,r^2} + \epsilon\,R(r)=0,
                  \end{eqnarray}
                  then we attain
                  \begin{eqnarray}\label{eq:F-13-01}
                        R(r)\propto {\rm e}^{\sqrt{-\epsilon}\,r}, \;\;\; {\rm or }\;\;\;
                        R(r)\propto {\rm e}^{-\sqrt{-\epsilon}\,r}.
                  \end{eqnarray}
                  To avoid when $r\rightarrow \infty$, $R(r)\rightarrow \infty$, we
                  choose the solution $R(r)\propto {\rm e}^{-\sqrt{-\epsilon}\,r}$.

                  Based on the analysis above, we may set $R(r)$ as
                  \begin{eqnarray}\label{eq:F-13-t}
                        && R(r)=r^{s} {\rm e}^{-\sqrt{-\epsilon}\,r} F(r),
                  \end{eqnarray}
                  with $s=\bigl(-1+\sqrt{1+4\kappa}\bigr)/2$. After that, we can
                  calculate
                  \begin{eqnarray}
                        \dfrac{{\rm d}\,R(r)}{{\rm d}\,r} &=&
                              s\,r^{s-1} {\rm e}^{-\sqrt{-\epsilon}\,r} F(r)
                              -\sqrt{-\epsilon}\,r^{s} {\rm e}^{-\sqrt{-\epsilon}\,r}\,F(r)
                              +r^{s} {\rm e}^{-\sqrt{-\epsilon}\,r}\,\dfrac{{\rm d}\,F(r)}{{\rm d}\,r} \nonumber \\
                        &=& r^{s} {\rm e}^{-\sqrt{-\epsilon}\,r} \Biggr[
                              \biggr(\dfrac{s}{r} -\sqrt{-\epsilon}\biggr)F(r)
                              +\dfrac{{\rm d} F(r)}{{\rm d} r}\Biggr] = r^{s} {\rm e}^{-\sqrt{-\epsilon}\,r}\,C(r),
                  \end{eqnarray}
                  thus
                  \begin{eqnarray}
                         \dfrac{{\rm d}^2\,R(r)}{{\rm d}\,r^2} &=&
                              \dfrac{{\rm d}^2\,\Bigl[
                                          r^{s} {\rm e}^{-\sqrt{-\epsilon}\,r}\,C(r)\Bigr]}
                                    {{\rm d}\,r^2} \nonumber \\
                        &=& r^{s} {\rm e}^{-\sqrt{-\epsilon}\,r} \Biggr[
                              \biggr(\dfrac{s}{r} -\sqrt{-\epsilon}\biggr)C(r)
                              +\dfrac{{\rm d}\,C(r)}{{\rm d}\,r}\Biggr] \nonumber \\
                        &=& r^{s} {\rm e}^{-\sqrt{-\epsilon}\,r} \Biggl\{
                              \biggr(\dfrac{s}{r} -\sqrt{-\epsilon}\biggr)\biggr[
                                    \biggr(\dfrac{s}{r} -\sqrt{-\epsilon}\biggr)F(r)
                                    +\dfrac{{\rm d} F(r)}{{\rm d} r}\biggr]
                              +\dfrac{{\rm d}}{{\rm d}\,r}\,\biggr[
                                    \biggr(\dfrac{s}{r} -\sqrt{-\epsilon}\biggr)F(r)
                                    +\dfrac{{\rm d} F(r)}{{\rm d} r}\biggr]\Biggr\} \nonumber \\
                        &=& r^{s} {\rm e}^{-\sqrt{-\epsilon}\,r} \Biggl\{
                              \biggr(\dfrac{s}{r} -\sqrt{-\epsilon}\biggr)\biggr[
                                    \biggr(\dfrac{s}{r} -\sqrt{-\epsilon}\biggr)F(r)
                                    +\dfrac{{\rm d} F(r)}{{\rm d} r}\biggr]
                              +\biggr[-\dfrac{s}{r^2} F(r)+\biggr(\dfrac{s}{r} -\sqrt{-\epsilon}\biggr)\dfrac{{\rm d} F(r)}{{\rm d} r}
                                    +\dfrac{{\rm d}^2 F(r)}{{\rm d}\,r^2}\biggr]\Biggr\} \nonumber \\
                        &=& r^{s} {\rm e}^{-\sqrt{-\epsilon}\,r} \Biggl\{
                              \dfrac{{\rm d}^2 F(r)}{{\rm d}\,r^2}
                              +2\biggr(\dfrac{s}{r} -\sqrt{-\epsilon}\biggr)
                                    \dfrac{{\rm d} F(r)}{{\rm d} r}
                              +\biggl[\Bigl(\dfrac{s}{r} -\sqrt{-\epsilon}\Bigr)^2
                                    -\dfrac{s}{r^2}\biggr]F(r)\Biggr\}.
                  \end{eqnarray}
                  Then,
                  \begin{eqnarray}\label{eq:F-14-01}
                        && \dfrac{{\rm d}^2\,R(r)}{{\rm d}\,r^2}
                                    +\dfrac{2}{r} \dfrac{{\rm d}\,R(r)}{{\rm d}\,r}
                                    -\dfrac{\kappa}{r^2} R(r)+\epsilon R(r) \nonumber \\
                        &&= r^{s} {\rm e}^{-\sqrt{-\epsilon}\,r} \Biggl\{
                                    \dfrac{{\rm d}^2 F(r)}{{\rm d}\,r^2}
                                    +2\biggr(\dfrac{s}{r} -\sqrt{-\epsilon}\biggr)
                                          \dfrac{{\rm d} F(r)}{{\rm d} r}
                                    +\biggl[\Bigl(\dfrac{s}{r} -\sqrt{-\epsilon}\Bigr)^2
                                          -\dfrac{s}{r^2}\biggr]F(r)\Biggr\} \nonumber \\
                              && +\dfrac{2}{r} r^{s} {\rm e}^{-\sqrt{-\epsilon}\,r} \Biggr[
                                    \biggr(\dfrac{s}{r} -\sqrt{-\epsilon}\biggr)F(r)
                                    +\dfrac{{\rm d} F(r)}{{\rm d} r}\Biggr]
                              +\left(\epsilon-\dfrac{\kappa}{r^2}\right)r^{s}
                                    {\rm e}^{-\sqrt{-\epsilon}\,r} F(r), \nonumber \\
                        &&= 0,
                  \end{eqnarray}
                  i.e.
                  \begin{eqnarray}\label{eq:F-22}
                        r\dfrac{{\rm d}^2 F(r)}{{\rm d}\,r^2}
                              +\biggr[2(s+1) -2\sqrt{-\epsilon}\;r\biggr]
                                    \dfrac{{\rm d}\,F(r)}{{\rm d}\,r}
                              -\biggr[2(s+1)\sqrt{-\epsilon}\biggr]F(r)=0,
                  \end{eqnarray}
                  viz.
                  \begin{eqnarray}\label{eq:F-28-t}
                        (\sqrt{-\epsilon})^2 r\dfrac{{\rm d}^2 F(z)}{{\rm d} (\sqrt{-\epsilon} \;r)^2}
                              +\sqrt{-\epsilon} \;\biggr[2(s+1) -2\sqrt{-\epsilon}\;r\biggr]
                                    \dfrac{{\rm d}\,F(z)}{{\rm d}\,(\sqrt{-\epsilon} \;r)}
                        -\biggr[2(s+1)\sqrt{-\epsilon}\biggr]F(z)=0.
                  \end{eqnarray}
                  Let
                  \begin{eqnarray}\label{eq:F-27-01}
                        z=\sqrt{-\epsilon} \; r,
                  \end{eqnarray}\\
                  then Eq. (\ref{eq:F-28-t}) becomes
                  \begin{eqnarray}\label{eq:F-29-01}
                        \sqrt{-\epsilon}\,z\dfrac{{\rm d}^2 F(z)}{{\rm d}\,z^2}
                                    +\sqrt{-\epsilon} \;\biggr[2(s+1) -2z\biggr]
                                          \dfrac{{\rm d}\,F(z)}{{\rm d}\,z}
                              -\biggr[2(s+1)\sqrt{-\epsilon}\biggr]F(z)=0,
                  \end{eqnarray}
                  i.e.
                  \begin{eqnarray}\label{eq:F-30-d-t}
                        \tau\dfrac{{\rm d}^2 F(\tau)}{{\rm d}\,\tau^2}
                        +\biggr[2(s+1) -\tau\biggr]
                              \dfrac{{\rm d}\,F(\tau)}{{\rm d}\,\tau}
                        -(s+1)F(\tau)=0,
                  \end{eqnarray}
                  with $\tau=2\,z$. One may notice that the
                  energy parameter $\epsilon$ does not yet appear in Eq. (\ref{eq:F-30-d-t}).

                  Remarkably, let us set
                  \begin{eqnarray}\label{eq:F-31-01}
                        \alpha &=& (s+1)=\dfrac{\sqrt{1+4\kappa}\,+1}{2}, \;\;\;\;\;\;
                        \gamma = 2(s+1)=\sqrt{1+4\kappa}\,+1,
                  \end{eqnarray}
                  then Eq.({\ref{eq:F-30-d-t}}) can be rewritten as
                  \begin{eqnarray}\label{eq:F-23-01}
                        \tau\dfrac{{\rm d}^2\,F(\tau)}{{\rm d}\,\tau^2}
                        +(\gamma-\tau)\dfrac{{\rm d}\,F(\tau)}{{\rm d}\,\tau}-\alpha\,F(\tau)=0,
                  \end{eqnarray}
                  which is a confluent hypergeometric equation with general solution
                  \begin{eqnarray}\label{eq:F-24-01}
                        F(\tau)= C_1\,F(\alpha;\,\gamma;\,\tau)
                              +C_2\,\tau^{1-\gamma}\,F(\alpha+1-\gamma;\,2-\gamma;\,\tau),
                  \end{eqnarray}
                  where $C_1$, $C_2$ are two constants, and
                  \begin{eqnarray}\label{eq:F-25}
                        F(\alpha;\gamma;\tau) &=& 1+\sum_{k=1}^{\infty}
                              \dfrac{(\alpha)_k}{k!\,(\gamma)_k}\tau^k = 1+\dfrac{\alpha}{\gamma}\tau
                              +\dfrac{\alpha(\alpha+1)}{2!\,\gamma(\gamma+1)}\tau^2
                              +\dfrac{\alpha(\alpha+1)(\alpha+2)}{3!\,\gamma(\gamma+1)(\gamma+2)}\tau^3+\cdots
                  \end{eqnarray}\\
                  is the confluent hypergeometric function, including
                  \begin{eqnarray}\label{eq:F-26}
                        (\alpha)_k &=& \alpha(\alpha+1)\cdots(\alpha+k-1), \;\;\;\;\;
                        (\gamma)_k = \gamma(\gamma+1)\cdots(\gamma+k-1).
                  \end{eqnarray}\\
                  For convenience, one can denote the two hypergeometric functions
                  as
                  \begin{eqnarray}\label{eq:F-32-t}
                        w_1 &=& F(\alpha;\gamma;\tau) = F\left(\dfrac{\sqrt{1+4\kappa}\,{+1}}{2};\,
                              \sqrt{1+4\kappa}\,+1;\,2 \sqrt{-\epsilon}\,r\right),
                  \end{eqnarray}
                  \begin{eqnarray}\label{eq:F-33-y}
                        w_2 &=& \tau^{1-\gamma}F(\alpha-\gamma+1;\,2-\gamma;\,\tau) = \left(2\sqrt{-\epsilon}\,r\right)^{1-\gamma}
                              F(\alpha-\gamma+1;\,2-\gamma;\,2 \sqrt{-\epsilon}\,r),
                  \end{eqnarray}
                  namely
                  \begin{eqnarray}
                        F(r)=C_1\,w_1 +C_2\,w_2.
                  \end{eqnarray}

                  Finally, by substituting Eq. (\ref{eq:F-32-t}) and Eq. (\ref{eq:F-33-y}) into Eq. (\ref{eq:F-13-t}), one then
                  obtains the radial function as
                  \begin{eqnarray}\label{eq:F-34-01}
                        R(r) &=& r^{s} {\rm e}^{-\sqrt{-\epsilon}\,r} F(r) = r^{s} {\rm e}^{-\sqrt{-\epsilon}\,r} (C_1\,w_1 +C_2\,w_2).
                  \end{eqnarray}
Unfortunately, from Mathematica program, one can check that: (i) for $\kappa>0$,
 \begin{eqnarray}\label{eq:F-34b-01}
 \lim_{r\rightarrow\infty } r^{s} {\rm e}^{-\sqrt{-\epsilon}r} w_i \rightarrow\infty,   \;\;\;\;\;\; (i=1, 2),
                  \end{eqnarray}
and (ii) for $\kappa=0$ (thus $s=0$),
\begin{eqnarray}\label{eq:F-34c-01}
 && \lim_{r\rightarrow\infty } r^{s} {\rm e}^{-\sqrt{-\epsilon}r} w_1=\lim_{r\rightarrow\infty } {\rm e}^{-\sqrt{-\epsilon}r} F\left(1;
                              2;\,2 \sqrt{-\epsilon}\,r\right) \rightarrow\infty, \nonumber\\
 && \lim_{r\rightarrow 0  } r^{s} {\rm e}^{-\sqrt{-\epsilon}r} w_2 =\lim_{r\rightarrow 0  }  {\rm e}^{-\sqrt{-\epsilon}r} \left(2\sqrt{-\epsilon}\,r\right)^{-1}
                              F(0;0;\,2 \sqrt{-\epsilon}\,r)=\lim_{r\rightarrow 0  }  \frac{1}{2\sqrt{-\epsilon}\,r}{\rm e}^{-\sqrt{-\epsilon}r}
                              \rightarrow\infty.
                  \end{eqnarray}
Therefore, for the case of energy $E< 0$, we do not have the physically allowable radial wavefunctions $R(r)$.

            \subsection{The Energy $E =0$}

In this case, the energy $E=0$ and $\epsilon=0$. Then, Eq. (\ref{eq:ReduceRadialEq}) becomes
            \begin{eqnarray}\label{eq:ReduceRadialEq-b}
                  \dfrac{{\rm d}^2\,R(r)}{{\rm d}\,r^2}
                        +\dfrac{2}{r} \dfrac{{\rm d}\,R(r)}{{\rm d}\,r}
                        -\dfrac{\kappa}{r^2} R(r)=0.
            \end{eqnarray}
Similarly, let us analyze the asymptotic behaviors of $R(r)$.

            \emph{Case (i).---} When $r\rightarrow 0$, Eq. (\ref{eq:ReduceRadialEq-b})
                  becomes
                  \begin{eqnarray}\label{eq:F-8-b}
                        &&\dfrac{{\rm d}^2 R(r)}{{\rm d} r^2}
                              +\dfrac{2 }{r} \dfrac{{\rm d} R(r)}{{\rm d} r}
                              -\dfrac{\kappa}{r^2} R(r)=0.
                  \end{eqnarray}
                  In the region near to $r=0$, let $R(r) \propto r^s$, after substituting it into above equation, we have
                  \begin{eqnarray}\label{eq:F-9-b}
                        && s(s-1)r^{s-2}+2s r^{s-2} -\kappa r^{s-2}=0,
                  \end{eqnarray}
                  which leads to
                  \begin{eqnarray}\label{eq:F-10-b}
                        && s(s+1) -\kappa=0.
                  \end{eqnarray}
                  Based on which, one obtains
                  \begin{eqnarray}\label{eq:F-11-b}
                        && s_1=\dfrac{-1+\sqrt{1+4\kappa}}{2}, \;\;\; {\rm or }\;\;\; s_2= \dfrac{-1-\sqrt{1+4\kappa}}{2}.
                  \end{eqnarray}
                  To avoid when $r\rightarrow 0$, $R(r)\rightarrow \infty$, from the
                  viewpoint of physics, we choose $s=s_1 \geq 0$. When $\kappa=0$, we specially have $s=0$.

            \emph{Case (ii).---} When $r \rightarrow \infty$, Eq.
                  (\ref{eq:ReduceRadialEq-b}) becomes
                  \begin{eqnarray}\label{eq:F-12-b}
                        &&\dfrac{{\rm d}^2 R(r)}{{\rm d}\,r^2} =0,
                  \end{eqnarray}
                  then we attain
                  \begin{eqnarray}\label{eq:F-13-b}
                        R(r)= c_1 r +c_2,
                  \end{eqnarray}
                  where $c_1$ and $c_2$ are constant numbers.
To avoid when $r\rightarrow \infty$, $R(r)\rightarrow \infty$, one has to choose $c_1=0$. By considering cases (i) and (ii), we finally have
 \begin{eqnarray}\label{eq:F-13-c}
                        R(r)= 1,
                  \end{eqnarray}
which corresponds to $s=0$, hence $\kappa=0$. From \emph{Observation 1}, we have known that $\kappa=0$ occurs only for the following situation:
 \begin{eqnarray}\label{eq:F-13-d}
                        g=2, \;\;\; l=1, \;\;\; K=-\frac{l+1}{2}=-1.
                  \end{eqnarray}
Thus, the wavefunction has only the angular part as
            \begin{eqnarray}\label{eq:F-2-b}
                  \Psi(\vec{r})=\Psi(r, \theta, \phi)=R(r)\;\Phi_{ljm_j}(\theta, \phi)=\Phi^B_{ljm_j}(\theta, \phi)=\dfrac{1}{\sqrt{2l+1}} \begin{bmatrix}
                              -\sqrt{l-m}\:Y_{lm}(\theta, \phi) \\
                              \sqrt{l+m+1}\:Y_{l,m+1}(\theta, \phi) \\
                        \end{bmatrix}.
            \end{eqnarray}
However, in Eq. (\ref{eq:F-2-b}), the term $Y_{l,m+1}(\theta, \phi)$ has restricted $m$ to $m=0$ and $m=-1$. Accordingly, we have
         \begin{eqnarray}\label{phi-1a-1}
                  \Phi^B_{1,\frac{1}{2},m_j=\frac{1}{2}}(\theta, \phi)&=& \dfrac{1}{\sqrt{3}}
                  \begin{bmatrix}
                        -\:Y_{10} (\theta, \phi) \\
                        \sqrt{2}\:Y_{11} (\theta, \phi)
                  \end{bmatrix}=\dfrac{1}{\sqrt{3}}
                  \begin{bmatrix}
                        -\:\frac{1}{2}\sqrt{\frac{3}{\pi}}\frac{z}{r} \\
                        \sqrt{2}\:\left(-\frac{1}{2}\sqrt{\frac{3}{2\pi}}\frac{x+{\rm i}y }{r}\right)
                  \end{bmatrix}\nonumber\\
                  &\propto& \dfrac{1}{r}
                  \begin{bmatrix}
                        z \\
                        x+{\rm i}y
                  \end{bmatrix}, \quad \quad (m_j=1/2, m=0),
            \end{eqnarray}
            \begin{eqnarray}\label{phi-1b-1}
                  \Phi^B_{1,\frac{1}{2},m_j=-\frac{1}{2}}(\theta, \phi)&=& \dfrac{1}{\sqrt{3}}
                  \begin{bmatrix}
                        -\sqrt{2}\:Y_{1,-1} (\theta, \phi) \\
                        Y_{10} (\theta, \phi)
                  \end{bmatrix}=\dfrac{1}{\sqrt{3}}
                  \begin{bmatrix}
                        -\sqrt{2}\:\left(\frac{1}{2}\sqrt{\frac{3}{2\pi}}\frac{x-{\rm i}y }{r}\right) \\
                        \frac{1}{2}\sqrt{\frac{3}{\pi}}\frac{z}{r}
                  \end{bmatrix}\nonumber\\
                  &\propto& \dfrac{1}{r}
                  \begin{bmatrix}
                        x-{\rm i}y \\
                        -z
                  \end{bmatrix}, \quad \quad (m_j=1/2, m=-1).
            \end{eqnarray}
         The linear superposition of $\Phi^B_{1,\frac{1}{2},m_j=\frac{1}{2}}(\theta, \phi)$ and $\Phi^B_{1,\frac{1}{2},m_j=-\frac{1}{2}}(\theta, \phi)$ gives the most general wavefunction as
       \begin{eqnarray}\label{phi-1a-2}
           \Psi(\vec{r})   &=&    a\; \Phi^B_{1,\frac{1}{2},m_j=\frac{1}{2}}+b\;\Phi^B_{1,\frac{1}{2},m_j=-\frac{1}{2}}=
                 a\;  \dfrac{1}{r}
                  \begin{bmatrix}
                        z \\
                        x+{\rm i}y
                  \end{bmatrix}+
                  b\;  \dfrac{1}{r}
                  \begin{bmatrix}
                        x-{\rm i}y \\
                        -z
                  \end{bmatrix}=\dfrac{1}{r}
                  \begin{bmatrix}
                        a z+b (x-{\rm i}y) \\
                        a(x+{\rm i}y)-bz
                  \end{bmatrix}.
            \end{eqnarray}

            \subsection{The Energy $E>0$}

        In this case, the energy $E>0$ and $\epsilon>0$.  After multiplying $r^2$ to the the right-hand side of \Eq{eq:ReduceRadialEq}, we have
                  \begin{eqnarray}\label{eq:F-35}
                        r^2\dfrac{{\rm d}^2 R(r)}{{\rm d}\,r^2}
                        +2r\dfrac{{\rm d}\,R(r)}{{\rm d}\,r}
                        +(\epsilon\,r^2-\kappa)R(r) =0,
                  \end{eqnarray}
                  which looks like the generalized Bessel equation in the following
                  form:
                  \begin{eqnarray}\label{eq:H-1}
                        x^2\dfrac{{\rm d}^2 y}{{\rm d}\,x^2}
                        +a\,x\dfrac{{\rm d}\,y}{{\rm d}\,x} +(b+c\,x^m)y=0,\qquad
                        c>0,\;\;m\neq 0,
                  \end{eqnarray}
                  with
                  \begin{equation}
                        a=2,\ b=-\kappa,\ c=\epsilon,\ m=2.
                  \end{equation}
                  Nevertherless, we may performed the following calculation to
                  transform Eq. (\ref{eq:H-1}) to the standard Bessel equation, so
                  does \Eq{eq:F-35}.

                  Let
                  \begin{eqnarray}\label{eq:H-3}
                        w&=&x^\alpha\;y, \;\;\;\;\;  t=\gamma\; x^\beta,
                  \end{eqnarray}
                  then
                  \begin{eqnarray}\label{eq:H-4}
                        y&=&x^{-\alpha}\;w, \;\;\;\;\;
                        \dfrac{{\rm d}\,t}{{\rm d}\,x}=\beta\gamma\; x^{\beta-1},
                  \end{eqnarray}
                  which implies
                  \begin{eqnarray}\label{eq:H-5}
                        \dfrac{{\rm d}\,y}{{\rm d}\,x} &=& -\alpha\; x^{-\alpha-1}\;w
                              + x^{-\alpha} \; \dfrac{{\rm d}\,w}{{\rm d}\,t}
                                    \dfrac{{\rm d}\,t}{{\rm d}\,x} =  -\alpha\; x^{-\alpha-1}\;w+ \beta\gamma \;
                              x^{\beta-\alpha-1} \; \dfrac{{\rm d}\,w}{{\rm d}\,t},
                  \end{eqnarray}
                  further,
                  \begin{eqnarray}\label{eq:H-6}
                        \dfrac{{\rm d}^2 y}{{\rm d}\,x^2} =\beta^2\gamma^2\;
                                    x^{2\beta-\alpha -2}\; \dfrac{{\rm d}^2 w}{{\rm d}\,t^2}
                              +\bigl[\beta\gamma(\beta-\alpha-1)-\alpha\beta\gamma\bigr]
                                    x^{\beta-\alpha-2} \; \dfrac{{\rm d}\,w}{{\rm d}\,t}
                              +\alpha(\alpha+1)\; x^{-\alpha-2}\; w.
                  \end{eqnarray}
                  By substituting Eqs. (\ref{eq:H-5}) and (\ref{eq:H-6}) into Eq.
                  (\ref{eq:H-1}), we arrive at
                  \begin{eqnarray}\label{eq:H-7}
                        \beta^2 \gamma^2\;x^{2\beta-\alpha}\;
                              \dfrac{{\rm d}^2 w}{{\rm d}\,t^2}
                        +\bigl[\beta\gamma(\beta-\alpha-1)-\alpha\beta\gamma
                                    +a \beta\gamma\bigr]x^{\beta-\alpha} \;
                              \dfrac{{\rm d}\,w}{{\rm d}\,t}
                        +\bigl[\alpha(\alpha+1)-a \alpha+b+c x^m\bigr]\; x^{-\alpha}\; w=0,
                  \end{eqnarray}
                  viz.
                  \begin{eqnarray}\label{eq:H-8}
                        \gamma^2\; x^{2\beta}\; \dfrac{{\rm d}^2 w}{{\rm d}\,t^2}
                        +\dfrac{1}{\beta}\bigl[\gamma(\beta-\alpha-1)-\alpha\gamma
                              +a \gamma\bigr]x^{\beta} \; \dfrac{{\rm d}\,w}{{\rm d}\,t}
                        +\dfrac{1}{\beta^2}\bigl[\alpha(\alpha+1)-a \alpha+b+c x^m\bigr]\; w=0.
                  \end{eqnarray}
                  Because $t=\gamma\,x^\beta$, then we obtain
                  \begin{eqnarray}\label{eq:H-9}
                        t^2\; \dfrac{{\rm d}^2 w}{{\rm d}\,t^2}
                        +\dfrac{1}{\beta} (\beta-2\alpha-1+a)t \;
                              \dfrac{{\rm d}\,w}{{\rm d}\,t}
                        +\dfrac{1}{\beta^2} \bigl[\alpha(\alpha+1)-a \alpha+b+c x^m\bigr]\; w=0.
                  \end{eqnarray}
                  Let us compare Eq. (\ref{eq:H-9}) with the following standard Bessel
                  equation
                  \begin{eqnarray}\label{eq:H-10}
                        t^2\dfrac{{\rm d}^2 w}{{\rm d}\,t^2}
                        +t\dfrac{{\rm d}\,w}{{\rm d}\,t} +(t^2-\nu^2)w=0,
                  \end{eqnarray}
                  which leads to the following conditions:
                  \begin{eqnarray}\label{eq:H-11}
                        && \dfrac{1}{\beta} (\beta-2\alpha-1+a)=1, \nonumber \\
                        && \dfrac{1}{\beta^2} c= \gamma^2,\nonumber\\
                        && 2\beta=m, \nonumber\\
                        && \dfrac{1}{\beta^2} \bigl[
                              \alpha(\alpha+1)-a \alpha+b\bigr]=-\nu^2,
                  \end{eqnarray}
                  i.e.
                  \begin{eqnarray}\label{eq:H-11-a}
                        &&\alpha=\dfrac{a-1}{2} =\dfrac{1}{2},\nonumber\\
                        && \beta=\dfrac{m}{2} =1,\nonumber\\
                        && \gamma=\sqrt{\dfrac{4c}{m^2}} =\sqrt{\epsilon}, \nonumber\\
                        &&\nu^2=\dfrac{(a-1)^2-4b}{m^2} =\dfrac{1+4\kappa}{4},
                  \end{eqnarray}
                  here we select $\nu=\sqrt{1+4\kappa}/2 \geq1/2 >0$.

                  After that,
                  \begin{eqnarray}\label{eq:H-14}
                      &&  w=r^\alpha\;R(r)=\sqrt{r} R(r), \nonumber\\
                       && t=\gamma\; r^\beta=\sqrt{\epsilon}\,r.
                  \end{eqnarray}
                  The solutions of the Bessel equation (\ref{eq:H-10}) are given by the Bessel functions
                  \begin{eqnarray}\label{eq:H-15}
                        &&J_{\pm \nu}(t)=\sum_{k=0}^{\infty} \dfrac{(-1)^k}{k!}\dfrac{1}{\Gamma(\pm \nu+k+1)}\biggr(\dfrac{t}{2}\biggr)^{2k\pm \nu},
                  \end{eqnarray}

                  \subsubsection{$g\neq 2$}

                  From \Eq{eq:KappaK1} and \Eq{eq:KappaK2}, we know $\nu=\sqrt{1+4\kappa}/2$ is not
                        an integer, thus $J_\nu(t)$
                        and $J_{-\nu}(t)$ are two linear independent functions. Then the
                        general solution of $w(t)$ is
                        \begin{eqnarray}\label{eq:H-16}
                              w(t) &=& C_1\; J_{\nu}(t) +C_2 J_{-\nu}(t) \nonumber\\
                              &=& C_1\; \sum_{k=0}^{\infty} \dfrac{(-1)^k}{k!}\dfrac{1}{\Gamma(\nu+k+1)}\biggr(\dfrac{t}{2}\biggr)^{2k+ \nu}
                                    +C_2 \sum_{k=0}^{\infty} \dfrac{(-1)^k}{k!}\dfrac{1}{\Gamma(- \nu+k+1)}\biggr(\dfrac{t}{2}\biggr)^{2k- \nu},
                        \end{eqnarray}
                        with $C_1$, and $C_2$ two constants independent of $t$. Due to Eq.
                        (\ref{eq:H-14}), we finally have the solution of the radial
                        wavefunction as
                        \begin{eqnarray}\label{eq:H-17}
                              R(r) &=& \dfrac{1}{\sqrt{r}}\;\biggr[C_1\; J_{\nu}(\sqrt{\epsilon} r) +C_2 J_{-\nu}(\sqrt{\epsilon} r) \biggr]\nonumber\\
                              &=& \dfrac{1}{\sqrt{r}}\;\Biggr[C_1\; \sum_{k=0}^{\infty} \dfrac{(-1)^k}{k!}\dfrac{1}{\Gamma(\nu+k+1)}\biggr(\dfrac{\sqrt{\epsilon} r}{2}\biggr)^{2k+ \nu}
                                    +C_2 \sum_{k=0}^{\infty} \dfrac{(-1)^k}{k!}\dfrac{1}{\Gamma(- \nu+k+1)}\biggr(\dfrac{\sqrt{\epsilon} r}{2}\biggr)^{2k- \nu}\Biggr].
                        \end{eqnarray}

                        Let us analysis the asymptotic behavior of $R(r)$.

                        \emph{Case (i).---}When $r \rightarrow 0$,
                        \begin{eqnarray}\label{eq:H-18}
                              &&J_{\nu}(\sqrt{\epsilon}\,r)\sim r^\nu, \nonumber\\
                              &&J_{-\nu}(\sqrt{\epsilon}\,r)\sim r^{-\nu},
                        \end{eqnarray}
                        thus we have
                        \begin{eqnarray}\label{eq:H-19}
                              && \dfrac{1}{\sqrt{r}} J_{\nu}(\sqrt{\epsilon} r) \sim
                                    r^{\nu-\frac{1}{2}} \rightarrow 0, \nonumber\\
                              && \dfrac{1}{\sqrt{r}} J_{-\nu}(\sqrt{\epsilon} r) \sim
                                    r^{-\nu-\frac{1}{2}} \rightarrow \infty,
                        \end{eqnarray}
                        then we have to let the coefficient $C_2=0$. In this situation, the
                        radial wavefunction reads
                        \begin{eqnarray}\label{eq:H-20}
                              R(r) &=& \dfrac{C_1}{\sqrt{r}}\;
                                    J_{\nu}(\sqrt{\epsilon}\,r) =\dfrac{C_1}{\sqrt{r}}\;\sum_{k=0}^{\infty}
                                    \dfrac{(-1)^k}{k!}\dfrac{1}{\Gamma(\nu+k+1)}\biggr(
                                          \dfrac{\sqrt{\epsilon}\,r}{2}\biggr)^{2k+ \nu}.
                        \end{eqnarray}

                        \emph{Case (ii).---}When $r \rightarrow \infty$, from Mathematica
                        computation, we can find that $R(r)\rightarrow 0$. Thus the $R(r)$ in Eq. (\ref{eq:H-20}) is a
                        physically allowable radial wave function.

                  \subsubsection{$g=2$}

            \emph{ Case (i).---}If $K=l/2$, then $\kappa=(l+1)(l+2)$, we have
                              \begin{equation}
                                    \nu=\dfrac{\sqrt{1+4\kappa}}{2}
                                    =\dfrac{\sqrt{1+4(l+1)(l+2)}}{2}
                                    =\dfrac{\sqrt{1+4(l+1)+4(l+1)^2}}{2}
                                    =l+\frac{3}{2}=\frac{3}{2}, \frac{5}{2}, \cdots.
                              \end{equation}

               \emph{ Case (ii).---}If $K=-(l+1)/2$, ($l\neq 0$), then $\kappa=l(l-1)$, we have
                              \begin{equation}
                                    \nu=\dfrac{\sqrt{1+4\kappa}}{2}
                                    =\dfrac{\sqrt{1+4\,l(l-1)}}{2}
                                    =\dfrac{\sqrt{(2\,l-1)^2}}{2}
                                    =l-\dfrac{1}{2}=\frac{1}{2}, \frac{3}{2}, \cdots.
                              \end{equation}
Namely, for $g=2$, the parameter $\nu$ is equal to a semi-odd number. Thus $J_\nu(t)$ and $J_{-\nu}(t)$ are two linear independent functions. The analysis is similar to that in the previous subsection, and the form of the radial wavefunction is the same as shown in Eq. (\ref{eq:H-20}).

 In summary, for a fixed energy $E>0$, the corresponding wavefunction $\Psi(\vec{r})$ satisfies the following eigen-equation
            \begin{eqnarray}\label{eq:H-21}
                  H_{\rm S}\:\Psi_{\rm S}(\vec{r})&=& E\:\Psi_{\rm S}(\vec{r}),
            \end{eqnarray}
            and in the common set
            $\left\{H, \vec{\ell}^2, \vec{\ell}\cdot\vec{S}\right\}$, the wavefunction
            $\Psi_{\rm}(\vec{r})$  can be expressed as (for $\kappa>0$)
            \begin{eqnarray}\label{eq:H-22}
                  \Psi_{\rm S}(\vec{r}) &=& \begin{bmatrix}
                             c_1\; \chi_1(\vec{r}) \\
                             c_2\; \chi_2(\vec{r}) \\
                        \end{bmatrix} = R(r)  \; \Phi_{ljm_j}(\theta, \phi)= \mathcal{N} \dfrac{1}{\sqrt{r}}\; J_{\nu}(\sqrt{\epsilon}\,r) \;
                        \Phi_{ljm_j} (\theta, \phi),
            \end{eqnarray}
            with
            \begin{eqnarray}\label{eq:H-23}
                  \epsilon &=& \dfrac{2M}{\hbar^2} E,\nonumber\\
                  \nu&=&\dfrac{\sqrt{1+4\kappa}}{2},\nonumber\\
                  \kappa &=& l(l+1) +2gK+\dfrac{g^2}{2},\nonumber\\
                  K&=&\dfrac{l}{2}, \;\;{\rm or}\;\; -\dfrac{l+1}{2} \;\;(l\neq 0)
            \end{eqnarray}
            and $m_j =m+1/2$,
            \begin{eqnarray}
                  \Phi^A_{ljm_j}(\theta, \phi) &=& \dfrac{1}{\sqrt{2l+1}} \begin{bmatrix}
                        \sqrt{l+m+1}\:Y_{lm}(\theta, \phi) \\
                        \sqrt{l-m}\:Y_{l,m+1}(\theta, \phi) \\
                  \end{bmatrix},\quad j=l+1/2,
            \end{eqnarray}
            \begin{eqnarray}
                  \Phi^B_{ljm_j}(\theta, \phi)&=& \dfrac{1}{\sqrt{2l+1}} \begin{bmatrix}
                        -\sqrt{l-m}\:Y_{lm}(\theta, \phi) \\
                        \sqrt{l+m+1}\:Y_{l,m+1}(\theta, \phi) \\
                  \end{bmatrix},\quad j=l-1/2.
            \end{eqnarray}

It is noted that the eigenvlaues of $H_{\rm S}$, $\vec{\ell}^2$ and $\vec{\ell}\cdot\vec{S}$ do not depend on the quantum number $m$, thus the most general solution of the eigenfuction is given by
\begin{eqnarray}\label{eq:v-9}
			\Psi_{\rm S}(\vec{r}) &=&  R(r) \sum_{m=-l}^{l} \left[ c_{m} \;\Phi^A_{lm}(\theta, \phi)
                                                   +d_{m} \Phi^B_{lm}(\theta, \phi)\right],
		\end{eqnarray}
where $c_m$, $d_m$ are complex numbers, and for simplicity we have denoted $\Phi^A_{lm}(\theta, \phi)\equiv \Phi^A_{ljm_j}(\theta, \phi) $ and
$\Phi^B_{lm}(\theta, \phi)\equiv \Phi^B_{ljm_j}(\theta, \phi)$ in the main text.

\begin{remark}
          Finally, let us consider a special case: the energy $E>0$, but $\kappa=0$. In this case, $l=1$, $g=2$, and $\nu=1/2$, we then have
                  \begin{eqnarray}
                   R(r) =&& \dfrac{1}{\sqrt{r}}\;\Bigl[
                                    C_1\; J_{\nu} (\sqrt{\epsilon} r)
                                    \Bigr] =\dfrac{C_1}{\sqrt{r}}\;J_\frac{1}{2} (\sqrt{\epsilon} r)
                                    =\dfrac{C_1}{\sqrt{r}}\;\sqrt{\dfrac{2}{\pi (\sqrt{\epsilon} r)}} \sin (\sqrt{\epsilon} r)\propto\frac{1}{r}\sin (\sqrt{\epsilon} r),
                  \end{eqnarray}
and the wavefunction $\Psi(\vec{r})$ can be written as
            \begin{eqnarray}
                  \Psi(\vec{r}) &=& \begin{bmatrix}
                             c_1\; \chi_1(\vec{r}) \\
                             c_2\; \chi_2(\vec{r}) \\
                        \end{bmatrix} = R(r)  \; \Phi_{ljm_j}(\theta, \phi)=\mathcal{N} \frac{1}{r}\sin (\sqrt{\epsilon} r)
                        \Phi^B_{1,\frac{1}{2},m_j} (\theta, \phi).
            \end{eqnarray}
Similar to Eq. (\ref{phi-1a-2}), the linear superposition of $\Phi^B_{1,\frac{1}{2},m_j=\frac{1}{2}}(\theta, \phi)$ and $\Phi^B_{1,\frac{1}{2},m_j=-\frac{1}{2}}(\theta, \phi)$ gives the most general wavefunction as
       \begin{eqnarray}\label{phi-1a-3}
           \Psi(\vec{r})   &=&   \frac{1}{r}\sin (\sqrt{\epsilon} r) \left[ a\; \Phi^B_{1,\frac{1}{2},m_j=\frac{1}{2}}+b\;\Phi^B_{1,\frac{1}{2},m_j=-\frac{1}{2}}\right]=\nonumber\\
           &&
                  \frac{1}{r}\sin (\sqrt{\epsilon} r) \left\{a\;  \dfrac{1}{r}
                  \begin{bmatrix}
                        z \\
                        x+{\rm i}y
                  \end{bmatrix}+
                  b\;  \dfrac{1}{r}
                  \begin{bmatrix}
                        x-{\rm i}y \\
                        -z
                  \end{bmatrix}\right\}=\dfrac{\sin (\sqrt{\epsilon} r)}{r^2}
                  \begin{bmatrix}
                        a z+b (x-{\rm i}y) \\
                        a(x+{\rm i}y)-bz
                  \end{bmatrix}.
            \end{eqnarray}
\end{remark}

\begin{remark}
If we let $E=\hbar^2 (k_r)^2/(2M)$, where $k_r$ is the $\hat{e}_r$-component of the vector $\vec{k}=k_r \hat{e}_r+k_\theta \hat{e}_\theta+k_\phi \hat{e}_\phi$, then we have $\epsilon=(k_r)^2=|\vec{k}\cdot \hat{e}_r|^2$, and $\sqrt{\epsilon}r=\vec{k}\cdot\vec{r}$. Then $\sin (\sqrt{\epsilon} r)$ can be further written as $\sin (\vec{k}\cdot\vec{r})$, which is eigenfunction of a free particle.

From above analysis, we have known that: (i) for $\kappa>0$, the Hamiltonian (\ref{eq:HSForm1}) has a continuous energy spectrum with $E>0$; (2) for $\kappa=0$, the Hamiltonian (\ref{eq:HSForm1}) has a continuous energy spectrum with $E\geq 0$.
\end{remark}

\newpage

\part{The Eigen-Problem of the Hamiltonian $H_\text{M}$}

Let us consider the magnetic Aharonov-Bohm Hamiltonian of an electron (with mass $M$, electric charge $-e$)
\begin{equation}\label{eq:H}
  H_{\rm M}=\dfrac{1}{2M} \left(\vec{p}+\frac{e}{c}\vec{A}_{\rm M}\right)^2,
\end{equation}
where $\vec{A}_{\rm M}$ is the magnetic vector potential of the following form \cite{2005QParadox}:
\begin{equation}\label{eq:L-15}
  \vec{A}_{\rm M} =\begin{cases}
    & \dfrac{B\,\sqrt{x^2+y^2}}{2} \hat{e}_\phi, \;\;\;\;\;\; \;\;\;\;\;\;\;\;\;\;\;\;\; (\rho< r_0)\\
    & \dfrac{B r_0^2}{2\,\rho} \hat{e}_\phi=\dfrac{\Phi_{\rm M}}{2\pi\,\sqrt{x^2+y^2}} \hat{e}_\phi,\;\; (\rho> r_0)
  \end{cases}
\end{equation}
where $\vec{B}=B \hat{z}$ is the magnetic field, $\Phi_{\rm M}=B\pi r_0^2$ is the magnetic flux, and $\hat{e}_\phi = (-\sin\phi, \cos\phi, 0)$.
Here, we only consider the case of $\rho >r_0$. In this case, the magnetic vector potential reads
\begin{eqnarray}\label{eq:L-17}
  \vec{A}_{\rm M} &=& \frac{\Phi_{\rm M}}{2\pi\sqrt{x^2+y^2}} (-\hat{e}_x \sin\phi +\hat{e}_y \cos\phi)= \frac{\Phi_{\rm M}}{2\pi(x^2+y^2)} \left(-y\,\hat{e}_x  + x\,\hat{e}_y\right).
\end{eqnarray}
Note $\vec{\nabla}\cdot\vec{A}_{\rm M} =0$, and $\vec{p}\cdot \vec{A}_{\rm M}-\vec{A}_{\rm M} \cdot\vec{p}=-{\rm i}\hbar \vec{\nabla}\cdot\vec{A}_{\rm M} =0$, and
\begin{equation}
  \vec{\nabla}\times\vec{A}_{\rm M} =\begin{cases}
    & B\,\hat{z},\quad(\rho < r_0), \\
    & 0,\qquad(\rho > r_0).
  \end{cases}
\end{equation}
The eigen-problem is given by
\begin{eqnarray}\label{eq:eigen-1}
  H_{\rm M} \Psi_{\rm M}(\vec{r})&=& E \Psi_{\rm M}(\vec{r}),
\end{eqnarray}
where $E$ is the energy and $\Psi_{\rm M}(\vec{r})$ is the eigenfunction. In the following sections, we shall use two different methods to solve the eigen-problem.

\section{The First Method}

Form Eq. (\ref{eq:H}) we have
\begin{eqnarray}\label{eq:H-a}
  H_{\rm M}&=&\dfrac{1}{2M} \left(\vec{p}+\frac{e}{c}\vec{A}_{\rm M}\right)^2\nonumber\\
  &=&\dfrac{1}{2M} \left[\vec{p}^{\; 2}+\frac{e}{c} \left(\vec{p}\cdot \vec{A}_{\rm M}+\vec{A}_{\rm M} \cdot\vec{p}\right)+\frac{{\rm e}^2}{c^2}\vec{A}_{\rm M}^2\right]\nonumber\\
  &=&\dfrac{1}{2M} \left[\vec{p}^{\;2}+\frac{2 e}{c} \left(\vec{p}\cdot \vec{A}_{\rm M}\right)+\frac{{\rm e}^2}{c^2}\vec{A}_{\rm M}^2\right]\nonumber\\
   &=&\dfrac{1}{2M} \left[\vec{p}^{\; 2}+\frac{2 e}{c} \left(\frac{\Phi_{\rm M}}{2\pi} (-p_x y+ p_y x)\frac{1}{x^2+y^2}\right)
   +\frac{{\rm e}^2}{c^2}\frac{\Phi^2_{\rm M}}{4\pi^2 (x^2+y^2)}\right]\nonumber\\
    &=&\dfrac{1}{2M} \left[\vec{p}^{\; 2}+ \frac{1}{x^2+y^2}\left( \frac{2 e}{c} \frac{\Phi_{\rm M}}{2\pi} \ell_z
   +\frac{{\rm e}^2}{c^2}\frac{\Phi^2_{\rm M}}{4\pi^2}\right)\right].
\end{eqnarray}
In the cylindrical coordinate system
\begin{eqnarray}\label{eq:H-b}
  &&\rho= \sqrt{x^2+y^2}, \;\;\; \phi=\arctan\left(\frac{y}{x}\right), \;\;\; z=z,
\end{eqnarray}
we obtains
\begin{eqnarray}\label{eq:H-c}
  &&\ell_z = -{\rm i}\hbar\dfrac{\partial}{\partial\,\phi},\nonumber\\
 && \vec{\nabla} = \hat{e}_\rho \frac{\partial }{\partial \rho}+\hat{e}_\phi \frac{1}{\rho} \frac{\partial }{\partial \phi}+\hat{e}_z \frac{\partial }{\partial z},\nonumber\\
  && \vec{\nabla}^2 u = \frac{1}{\rho} \frac{\partial }{\partial \rho}\left(\rho \frac{\partial u}{\partial \rho}\right)+\frac{1}{\rho^2} \frac{\partial^2 u }{\partial \phi^2}+ \frac{\partial^2 u}{\partial z^2}
  = \left[\frac{\partial^2 u}{\partial \rho^2}+\frac{1}{\rho}\frac{\partial u}{\partial \rho}\right]+\frac{1}{\rho^2} \frac{\partial^2 u }{\partial \phi^2}+ \frac{\partial^2 u}{\partial z^2}, \nonumber\\
  &&\vec{p}^{\;2}=-\hbar^2 \vec{\nabla}^2=-\hbar^2\left[\frac{\partial^2 }{\partial \rho^2}+\frac{1}{\rho}\frac{\partial }{\partial \rho}\right]+\frac{\ell^2_z}{\rho^2} +p_z^2,\nonumber\\
  &&p_z=-{\rm i}\hbar \frac{\partial }{\partial z}.
\end{eqnarray}
By substituting Eq. (\ref{eq:H-c}) into Eq. (\ref{eq:H-a}), we have
\begin{eqnarray}\label{eq:H-d}
  H_{\rm M}&=&\dfrac{1}{2M} \left[\vec{p}^2+ \frac{1}{x^2+y^2}\left( \frac{2 e}{c} \frac{\Phi_{\rm M}}{2\pi} \ell_z
   +\frac{{\rm e}^2}{c^2}\frac{\Phi^2_{\rm M}}{4\pi^2}\right)\right]\nonumber\\
   &=&\dfrac{1}{2M} \left[-\hbar^2\left[\frac{\partial^2 }{\partial \rho^2}+\frac{1}{\rho}\frac{\partial }{\partial \rho}\right]+\frac{\ell^2_z}{\rho^2} +p_z^2+ \frac{1}{\rho^2}\left(  \frac{e\Phi_{\rm M}}{c\pi} \ell_z
   +\frac{{\rm e}^2 \Phi^2_{\rm M}}{4 c^2\pi^2}\right)\right].
\end{eqnarray}
It is easy to check that
\begin{eqnarray}\label{eq:H-e}
  &&[H_{\rm M}, \ell_z]=0, \;\;\;\;\; [H_{\rm M}, p_z^2]=0, \;\;\;\;\; [\ell_z, p_z^2]=0,
\end{eqnarray}
hence we can solve the eigen-problem in the common set $\{H_{\rm M}, \ell_z, p_z^2\}$. We have
\begin{eqnarray}\label{eq:H-f}
   H_{\rm M} \Psi_{\rm M}(\vec{r})&=& E \;\Psi_{\rm M}(\vec{r}),\nonumber\\
   \ell_z \Psi_{\rm M}(\vec{r})&=& m\hbar \;\Psi_{\rm M}(\vec{r}),\nonumber\\
   p_z^2 \Psi_{\rm M}(\vec{r})&=& k_z^2 \hbar^2\;\Psi_{\rm M}(\vec{r}),\nonumber\\
\end{eqnarray}
with
\begin{eqnarray}\label{eq:H-g}
  \Psi_{\rm M}(\vec{r})&=& \Psi_{\rm M}(\rho, \phi, z)=R(\rho) {\rm e}^{{\rm i} m\phi} \left(
      c_1 {\rm e}^{{\rm i} k_z z}+c_2 {\rm e}^{-{\rm i} k_z z}\right).
\end{eqnarray}
After substituting Eq. (\ref{eq:H-f}) and Eq. (\ref{eq:H-g}) into Eq. (\ref{eq:H-d}), we have
\begin{eqnarray}\label{eq:H-h}
  \dfrac{1}{2M} \left[-\hbar^2\left[\frac{\partial^2 }{\partial \rho^2}+\frac{1}{\rho}\frac{\partial }{\partial \rho}\right]+\frac{\hbar^2 m^2}{\rho^2} +\hbar^2 k_z^2+ \frac{1}{\rho^2}\left(  \frac{e\Phi_{\rm M}}{c\pi} \hbar m
   +\frac{{\rm e}^2 \Phi^2_{\rm M}}{4 c^2\pi^2}\right)\right] R(\rho)=E \; R(\rho),
\end{eqnarray}
i.e.,
\begin{eqnarray}\label{eq:H-i}
  -\dfrac{\hbar^2}{2M} \left\{\left[\frac{\partial^2  R(\rho)}{\partial \rho^2}+\frac{1}{\rho}\frac{\partial  R(\rho)}{\partial \rho}\right] - \frac{R(\rho)}{\rho^2}\left( m^2+ \frac{e\Phi_{\rm M}}{\hbar c\pi}  m
   +\frac{{\rm e}^2 \Phi^2_{\rm M}}{4 \hbar^2 c^2\pi^2}\right)- k_z^2 R(\rho)\right\} =E \; R(\rho),
\end{eqnarray}
i.e.,
\begin{eqnarray}\label{eq:H-j}
   \frac{\partial^2  R(\rho)}{\partial \rho^2}+\frac{1}{\rho}\frac{\partial  R(\rho)}{\partial \rho} - \frac{R(\rho)}{\rho^2}\left( m^2+ \frac{2 e\Phi_{\rm M}}{h c}  m
   +\frac{{\rm e}^2 \Phi^2_{\rm M}}{ h^2 c^2}\right)+\left(\frac{2M}{\hbar^2}E- k_z^2\right) R(\rho) =0.
\end{eqnarray}
Let
\begin{eqnarray}\label{eq:H-k}
   &&\epsilon = \frac{2M}{\hbar^2}E- k_z^2, \;\;\;\;\; \kappa=m^2+ \frac{2 e\Phi_{\rm M}}{h c}  m
   +\frac{{\rm e}^2 \Phi^2_{\rm M}}{ h^2 c^2}=\left(m+\frac{e\Phi_{\rm M}}{h c}\right)^2\geq 0,
   \end{eqnarray}
we then have
\begin{eqnarray}\label{eq:H-L}
   \frac{\partial^2  R(\rho)}{\partial \rho^2}+\frac{1}{\rho}\frac{\partial  R(\rho)}{\partial \rho} - \kappa\frac{R(\rho)}{\rho^2}+\epsilon R(\rho) =0.
\end{eqnarray}

\subsection{The Case of $\epsilon >0$}

In this case, let us multiply $\rho^2$ for both sides of Eq. (\ref{eq:H-L}), we then have
\begin{eqnarray}\label{eq:H-m}
  \rho^2 \frac{\partial^2  R(\rho)}{\partial \rho^2}+\rho \frac{\partial  R(\rho)}{\partial \rho} +(\epsilon\rho^2-\kappa )R(\rho) =0,
\end{eqnarray}
i.e.,
\begin{eqnarray}\label{eq:H-n}
  (\sqrt{\epsilon}\rho)^2 \frac{\partial^2  R((\sqrt{\epsilon}\rho))}{\partial (\sqrt{\epsilon}\rho)^2}+(\sqrt{\epsilon}\rho) \frac{\partial  R(\rho)}{\partial (\sqrt{\epsilon}\rho)} +[(\sqrt{\epsilon}\rho)^2-\kappa ]R((\sqrt{\epsilon}\rho)) =0,
\end{eqnarray}
By comparing Eq. (\ref{eq:H-n}) with the following standard Bessel equation
 \begin{eqnarray}\label{eq:H-O}
                        t^2\dfrac{{\rm d}^2 w}{{\rm d}\,t^2}
                        +t\dfrac{{\rm d}\,w}{{\rm d}\,t} +(t^2-\nu^2)w=0,
                  \end{eqnarray}
we have
\begin{eqnarray}\label{eq:H-P}
 t&=&\sqrt{\epsilon}\rho, \nonumber\\
 \nu&=&\sqrt{\kappa}=\left|m+\frac{e\Phi_{\rm M}}{h c}\right|\geq 0.
\end{eqnarray}
The solutions of the Bessel equation (\ref{eq:H-O}) are given by the Bessel functions
                  \begin{eqnarray}\label{eq:H-15-b}
                        &&J_{\pm \nu}(t)=\sum_{k=0}^{\infty} \dfrac{(-1)^k}{k!}\dfrac{1}{\Gamma(\pm \nu+k+1)}\biggr(\dfrac{t}{2}\biggr)^{2k\pm \nu}.
                  \end{eqnarray}

\subsubsection{The Case of $\nu$ is not an Integer}

If the magnetic flux
\begin{eqnarray}\label{eq:H-Q-0}
 \frac{e\Phi_{\rm M}}{h c} \neq n, \;\;\;\;\; n\in   \mathbb{Z},
\end{eqnarray}
we have
\begin{eqnarray}\label{eq:H-Qa}
\nu&=&\left|m+\frac{e\Phi_{\rm M}}{h c}\right|\neq |m+n|,
\end{eqnarray}
i.e., $\nu$ is not an integer. Thus $J_\nu(t)$ and $J_{-\nu}(r)$ are two linear independent functions, in this case the
                  general solution of $w(t)$ is
                  \begin{eqnarray}\label{eq:H-16-b}
                        w(t) &=& C_1\; J_{\nu}(t) +C_2 J_{-\nu}(t) \nonumber\\
                        &=& C_1\; \sum_{k=0}^{\infty} \dfrac{(-1)^k}{k!}\dfrac{1}{\Gamma(\nu+k+1)}\biggr(\dfrac{t}{2}\biggr)^{2k+ \nu}
                              +C_2 \sum_{k=0}^{\infty} \dfrac{(-1)^k}{k!}\dfrac{1}{\Gamma(- \nu+k+1)}\biggr(\dfrac{t}{2}\biggr)^{2k- \nu},
                  \end{eqnarray}
  with $C_1$ and $C_2$ two constant numbers. Due to Eq. (\ref{eq:H-16-b}), we have the solution of the radial wavefunction as
                  \begin{eqnarray}\label{eq:H-17-b}
                        R(\rho) &=& \biggr[C_1\; J_{\nu}(\sqrt{\epsilon} \rho) +C_2 J_{-\nu}(\sqrt{\epsilon} \rho) \biggr].
                  \end{eqnarray}

                  Let us analysis the asymptotic behavior of $R(r)$.

                  \emph{Case (i).---}When $\rho \rightarrow 0$,
                  \begin{eqnarray}\label{eq:H-18-b}
                        &&J_{\nu}(\sqrt{\epsilon}\,\rho)\sim \rho^\nu \rightarrow 0,, \nonumber\\
                        &&J_{-\nu}(\sqrt{\epsilon}\,\rho)\sim \rho^{-\nu} \rightarrow \infty,
                  \end{eqnarray}
                  then we have to let the coefficient $C_2=0$. In this situation, the
                  radial wavefunction reads
                  \begin{eqnarray}\label{eq:H-20-b}
                        R(\rho) &=& C_1\;
                              J_{\nu}(\sqrt{\epsilon}\,\rho) =\sum_{k=0}^{\infty}
                              \dfrac{(-1)^k}{k!}\dfrac{1}{\Gamma(\nu+k+1)}\biggr(
                                    \dfrac{\sqrt{\epsilon}\,\rho}{2}\biggr)^{2k+ \nu}.
                  \end{eqnarray}

                  \emph{Case (ii).---}When $r \rightarrow \infty$, from Mathematica
                  computation, we can find that $R(\rho)\rightarrow 0$. Thus the $R(\rho)$ in Eq. (\ref{eq:H-20-b}) is a
                  physically allowable radial wave function.

\subsubsection{The Case of $\nu$ is an Integer}

If the magnetic flux
\begin{eqnarray}\label{eq:H-Q}
 \frac{e\Phi_{\rm M}}{h c} = n, \;\;\;\;\; n\in   \mathbb{Z},
\end{eqnarray}
we have
\begin{eqnarray}\label{eq:H-Qaa}
\nu&=&\left|m+\frac{e\Phi_{\rm M}}{h c}\right| = |m+n|,
\end{eqnarray}
i.e., $\nu$ is an integer. In this case, $J_\nu(t)$ and $J_{-\nu}(t)$ are no longer two linear independent functions, because
\begin{eqnarray}\label{eq:H-Qa-b}
J_{-n}(t)=(-1)^n J_{n}(t), \;\;\;\; n\in \mathbb{Z}.
\end{eqnarray}
In this case the general solution of $w(t)$ is given by
                              \begin{eqnarray}\label{eq:IntW-b}
                                    w(t) &=& C_1\; J_{\nu}(t) +C_2 N_{\nu}(t)=C_1\; J_{\nu}(t) +C_2 \left[\dfrac{J_{\nu}(t) \cos(\nu\,\pi)-J_{-\nu}(t)}{\sin(\nu\,\pi)}\right],
                              \end{eqnarray}
                              with $C_1$, and $C_2$ two constants, and
                              \begin{eqnarray}\label{eq:IntW-bb}
                              N_{\nu}(t)= \dfrac{J_{\nu}(t) \cos(\nu\,\pi)-J_{-\nu}(t)}{\sin(\nu\,\pi)},
                              \end{eqnarray}
                              is the Neumann function \cite{2013Hassani}. Due to Eq. (\ref{eq:IntW-b}), we finally have the solution of the radial
                              wavefunction as
                              \begin{eqnarray}
                              R(\rho) &=& C_1\; J_{\nu}(\sqrt{\epsilon} \rho) +C_2 N_{\nu}(\sqrt{\epsilon} \rho).
                              \end{eqnarray}

                              Let us analysis the asymptotic behavior of $R(\rho)$.

                             \emph{Case (i).---}When $\rho \rightarrow 0$,
                             \begin{eqnarray}
                                   &&\lim_{\rho\rightarrow 0} J_{\nu}(\sqrt{\epsilon}\,\rho) =  \lim_{\rho\rightarrow 0} \rho^\nu\rightarrow 0, \nonumber\\
                                  &&\lim_{\rho\rightarrow 0} N_\nu (\sqrt{\epsilon}\,\rho)\rightarrow -\infty,
                                   \end{eqnarray}
                             we then have to let the coefficient $C_2=0$. In this situation, the
                             radial wavefunction reads
                             \begin{eqnarray}\label{eq:H-20-c}
                                   R(\rho) &=& C_1\;J_{\nu}(\sqrt{\epsilon}\,\rho)
                                    =C_1\;\sum_{k=0}^{\infty} \dfrac{(-1)^k}{k!}\dfrac{1}{\Gamma(\nu+k+1)}\biggr(
                                               \dfrac{\sqrt{\epsilon}\,\rho}{2}\biggr)^{2k+ \nu}.
                             \end{eqnarray}

                             \emph{Case (ii).---}When $\rho \rightarrow \infty$, from Mathematica
                             computation, we can find that $R(\rho)\rightarrow 0$. Thus the $R(\rho)$ in Eq. (\ref{eq:H-20-c}) is a
                             physically allowable radial wave function.
\begin{remark}
    Let us consider a special case
    \begin{eqnarray}
      \nu&=&\left|m+\frac{e\Phi_{\rm M}}{h c}\right|=0,
      \end{eqnarray}
    which means the magnetic flux stastifies
      \begin{eqnarray}
            m=-\frac{e\Phi_{\rm M}}{h c}.
      \end{eqnarray}
 In this case
       \begin{eqnarray}
         R(\rho) &=& C_1\;J_{\nu=0}(\sqrt{\epsilon}\,\rho),
      \end{eqnarray}
and
\begin{eqnarray}\label{eq:H-g-b}
  \Psi_{\rm M}(\vec{r})&=& \Psi_{\rm M}(\rho, \phi, z)=J_{\nu=0}(\sqrt{\epsilon}\,\rho) {\rm e}^{{\rm i} m\phi} \left(
      c_1 {\rm e}^{{\rm i} k_z z}+c_2 {\rm e}^{-{\rm i} k_z z}\right).
\end{eqnarray}
\end{remark}

\subsection{The Case of $\epsilon =0$}

In this case, from Eq. (\ref{eq:H-L})
\begin{eqnarray}\label{eq:H-L-a}
   \frac{\partial^2  R(\rho)}{\partial \rho^2}+\frac{1}{\rho}\frac{\partial  R(\rho)}{\partial \rho} - \kappa\frac{R(\rho)}{\rho^2} =0.
\end{eqnarray}
Let us analyze the asymptotic behaviors of $R(\rho)$.

            \emph{Case (i).---} When $\rho\rightarrow 0$, Eq. (\ref{eq:H-L-a})
                  becomes
                  \begin{eqnarray}\label{eq:F-8-d}
                        &&\dfrac{{\rm d}^2 R(\rho)}{{\rm d} \rho^2}
                              +\dfrac{1 }{\rho} \dfrac{{\rm d} R(\rho)}{{\rm d} \rho}
                              -\dfrac{\kappa}{\rho^2} R(\rho)=0.
                  \end{eqnarray}
                  In the region near to $\rho=0$, let $R(\rho) \propto \rho^s$, after substituting it into above equation, we have
                  \begin{eqnarray}\label{eq:F-9-d}
                        && s(s-1)\rho^{s-2}+s \rho^{s-2} -\kappa \rho^{s-2}=0,
                  \end{eqnarray}
                  which leads to
                  \begin{eqnarray}\label{eq:F-10-d}
                        && s^2 -\kappa=0,
                  \end{eqnarray}
                  Based on which, one obtains
                  \begin{eqnarray}\label{eq:F-11-d}
                        && s_1=\sqrt{\kappa}\geq 0, \;\;\; {\rm or }\;\;\; s_2= -\sqrt{\kappa}.
                  \end{eqnarray}
                  To avoid when $\rho\rightarrow 0$, $R(\rho)\rightarrow \infty$, from the
                  viewpoint of physics, we choose $s=s_1 \geq 0$.

            \emph{Case (ii).---} When $\rho \rightarrow \infty$, Eq. (\ref{eq:H-L-a}) becomes
                  \begin{eqnarray}\label{eq:F-12-d}
                        &&\dfrac{{\rm d}^2 R(\rho)}{{\rm d}\,\rho^2}=0,
                  \end{eqnarray}
                  then we attain
                  \begin{eqnarray}\label{eq:F-13-dd}
                        R(\rho)=c_1 \rho+c_2,
                  \end{eqnarray}
                  where $c_1$ and $c_2$ are constant numbers. To avoid when $\rho\rightarrow \infty$, $R(\rho)\rightarrow \infty$, we
                  have $R(\rho)=1$. Thus we have $s=0$ and $\kappa=0$.

This implies that
\begin{eqnarray}\label{eq:H-Qb}
\nu&=&=\sqrt{\kappa}=\left|m+\frac{e\Phi_{\rm M}}{h c}\right|=0,
\end{eqnarray}
which leads to
\begin{eqnarray}\label{eq:H-Qc}
\Phi_{\rm M}&=& -\frac{mhc}{e}, \;\;\;\;\;\;{\rm or} \;\;\;\;\; m=-\frac{e\Phi_{\rm M}}{h c},
\end{eqnarray}

Substituting $R(\rho)=1$ into Eq. (\ref{eq:H-g}), for $\epsilon=0$ or
\begin{eqnarray}\label{eq:H-g-a-01}
 E=\frac{\hbar^2 k_z^2}{2M},
\end{eqnarray}
we have the wavefunction as
\begin{eqnarray}\label{eq:H-g-bb}
  \Psi_{\rm M}(\vec{r})&=& \Psi_{\rm M}(\rho, \phi, z)=R(\rho) {\rm e}^{{\rm i} m\phi} \left(
      c_1 {\rm e}^{{\rm i} k_z z}+c_2 {\rm e}^{-{\rm i} k_z z}\right)={\rm e}^{{\rm i} m\phi} \left(
            c_1 {\rm e}^{{\rm i} k_z z}+c_2 {\rm e}^{-{\rm i} k_z z}\right)\nonumber\\
  &=& \left(
      c_1 {\rm e}^{{\rm i} k_z z}+c_2 {\rm e}^{-{\rm i} k_z z}\right) {\rm e}^{{\rm i} m \left[\arctan\left(\frac{y}{x}\right)\right]}\nonumber\\
  &=& \left(
      c_1 {\rm e}^{{\rm i} k_z z}+c_2 {\rm e}^{-{\rm i} k_z z}\right) {\rm e}^{-{\rm i}  \frac{e\Phi_{\rm M}}{h c} \left[\arctan\left(\frac{y}{x}\right)\right]}.
\end{eqnarray}

\subsection{The Case of $\epsilon < 0$}

Let us analyze the asymptotic behaviors of $R(\rho)$ as shown in Eq. (\ref{eq:H-L}).


            \emph{Case (i).---} When $\rho\rightarrow 0$, Eq. (\ref{eq:H-L})
                  becomes
                  \begin{eqnarray}\label{eq:F-8-d-a}
                        &&\dfrac{{\rm d}^2 R(\rho)}{{\rm d} \rho^2}
                              +\dfrac{1 }{\rho} \dfrac{{\rm d} R(\rho)}{{\rm d} \rho}
                              -\dfrac{\kappa}{\rho^2} R(\rho)=0.
                  \end{eqnarray}
                  In the region near to $\rho=0$, let $R(\rho) \propto \rho^s$, after substituting it into above equation, we have
                  \begin{eqnarray}\label{eq:F-9-d-a}
                        && s(s-1)\rho^{s-2}+s \rho^{s-2} -\kappa \rho^{s-2}=0,
                  \end{eqnarray}
                  which leads to
                  \begin{eqnarray}\label{eq:F-10-d-a}
                        && s^2 -\kappa=0,
                  \end{eqnarray}
                  Based on which, one obtains
                  \begin{eqnarray}\label{eq:F-11-d-a}
                        && s_1=\sqrt{\kappa}\geq 0, \;\;\; {\rm or }\;\;\; s_2= -\sqrt{\kappa}.
                  \end{eqnarray}
                  To avoid when $\rho\rightarrow 0$, $R(\rho)\rightarrow \infty$, from the viewpoint of physics, we choose $s=s_1 \geq 0$.

            \emph{Case (ii).---} When $\rho \rightarrow \infty$, Eq. (\ref{eq:H-L}) becomes
                  \begin{eqnarray}\label{eq:F-12-t}
                        &&\dfrac{{\rm d}^2 R(\rho)}{{\rm d}\,\rho^2} + \epsilon\,R(\rho)=0,
                  \end{eqnarray}
                  then we attain
                  \begin{eqnarray}\label{eq:F-13-02}
                        R(\rho)\propto {\rm e}^{\sqrt{-\epsilon}\,\rho}, \;\;\; {\rm or }\;\;\;
                        R(\rho)\propto {\rm e}^{-\sqrt{-\epsilon}\,\rho}.
                  \end{eqnarray}
                  To avoid when $\rho\rightarrow \infty$, $R(\rho)\rightarrow \infty$, we
                  choose the solution $R(\rho)\propto {\rm e}^{-\sqrt{-\epsilon}\,\rho}$.

                  Based on the analysis above, we may set $R(\rho)$ as
                  \begin{eqnarray}\label{eq:F-13-03}
                        && R(\rho)=\rho^{s} {\rm e}^{-\sqrt{-\epsilon}\,\rho} F(\rho),
                  \end{eqnarray}
                  with $s=\sqrt{\kappa}$. After that, we can calculate
                  \begin{eqnarray}
                        \dfrac{{\rm d}\,R(\rho)}{{\rm d}\,\rho} &=&
                              s\,\rho^{s-1} {\rm e}^{-\sqrt{-\epsilon}\,\rho} F(\rho)
                              -\sqrt{-\epsilon}\,\rho^{s} {\rm e}^{-\sqrt{-\epsilon}\,\rho}\,F(\rho)
                              +\rho^{s} {\rm e}^{-\sqrt{-\epsilon}\,\rho}\,\dfrac{{\rm d}\,F(\rho)}{{\rm d}\,\rho} \nonumber \\
                        &=& \rho^{s} {\rm e}^{-\sqrt{-\epsilon}\,\rho} \Biggr[
                              \biggr(\dfrac{s}{\rho} -\sqrt{-\epsilon}\biggr)F(\rho)
                              +\dfrac{{\rm d} F(\rho)}{{\rm d} \rho}\Biggr] = \rho^{s} {\rm e}^{-\sqrt{-\epsilon}\,\rho}\,C(\rho),
                  \end{eqnarray}
                  thus
                  \begin{eqnarray}
                         \dfrac{{\rm d}^2\,R(\rho)}{{\rm d}\,\rho^2} &=&
                              \dfrac{{\rm d}^2\,\Bigl[
                                          \rho^{s} {\rm e}^{-\sqrt{-\epsilon}\,\rho}\,C(\rho)\Bigr]}
                                    {{\rm d}\,\rho^2} \nonumber \\
                        &=& \rho^{s} {\rm e}^{-\sqrt{-\epsilon}\,\rho} \Biggr[
                              \biggr(\dfrac{s}{\rho} -\sqrt{-\epsilon}\biggr)C(\rho)
                              +\dfrac{{\rm d}\,C(\rho)}{{\rm d}\,\rho}\Biggr] \nonumber \\
                        &=& \rho^{s} {\rm e}^{-\sqrt{-\epsilon}\,\rho} \Biggl\{
                              \biggr(\dfrac{s}{\rho} -\sqrt{-\epsilon}\biggr)\biggr[
                                    \biggr(\dfrac{s}{\rho} -\sqrt{-\epsilon}\biggr)F(\rho)
                                    +\dfrac{{\rm d} F(\rho)}{{\rm d} \rho}\biggr]
                              +\dfrac{{\rm d}}{{\rm d}\,\rho}\,\biggr[
                                    \biggr(\dfrac{s}{\rho} -\sqrt{-\epsilon}\biggr)F(\rho)
                                    +\dfrac{{\rm d} F(\rho)}{{\rm d} \rho}\biggr]\Biggr\} \nonumber \\
                        &=& \rho^{s} {\rm e}^{-\sqrt{-\epsilon}\,\rho} \Biggl\{
                              \biggr(\dfrac{s}{\rho} -\sqrt{-\epsilon}\biggr)\biggr[
                                    \biggr(\dfrac{s}{\rho} -\sqrt{-\epsilon}\biggr)F(\rho)
                                    +\dfrac{{\rm d} F(\rho)}{{\rm d} \rho}\biggr]
                              +\biggr[-\dfrac{s}{\rho^2} F(\rho)+\biggr(\dfrac{s}{\rho} -\sqrt{-\epsilon}\biggr)\dfrac{{\rm d} F(\rho)}{{\rm d} \rho}
                                    +\dfrac{{\rm d}^2 F(\rho)}{{\rm d}\,\rho^2}\biggr]\Biggr\} \nonumber \\
                        &=& \rho^{s} {\rm e}^{-\sqrt{-\epsilon}\,\rho} \Biggl\{
                              \dfrac{{\rm d}^2 F(\rho)}{{\rm d}\,\rho^2}
                              +2\biggr(\dfrac{s}{\rho} -\sqrt{-\epsilon}\biggr)
                                    \dfrac{{\rm d} F(\rho)}{{\rm d} \rho}
                              +\biggl[\Bigl(\dfrac{s}{\rho} -\sqrt{-\epsilon}\Bigr)^2
                                    -\dfrac{s}{\rho^2}\biggr]F(\rho)\Biggr\}.
                  \end{eqnarray}
                  Then,
                  \begin{eqnarray}\label{eq:F-14-02}
                        && \dfrac{{\rm d}^2\,R(\rho)}{{\rm d}\,\rho^2}
                                    +\dfrac{1}{\rho} \dfrac{{\rm d}\,R(\rho)}{{\rm d}\,\rho}
                                    -\dfrac{\kappa}{\rho^2} R(\rho)+\epsilon R(\rho) \nonumber \\
                        &&= \rho^{s} {\rm e}^{-\sqrt{-\epsilon}\,\rho} \Biggl\{
                                    \dfrac{{\rm d}^2 F(\rho)}{{\rm d}\,\rho^2}
                                    +2\biggr(\dfrac{s}{\rho} -\sqrt{-\epsilon}\biggr)
                                          \dfrac{{\rm d} F(\rho)}{{\rm d} \rho}
                                    +\biggl[\Bigl(\dfrac{s}{\rho} -\sqrt{-\epsilon}\Bigr)^2
                                          -\dfrac{s}{\rho^2}\biggr]F(\rho)\Biggr\} \nonumber \\
                              &&  \;\;\; +\dfrac{1}{\rho} \rho^{s} {\rm e}^{-\sqrt{-\epsilon}\,\rho} \Biggr[
                                    \biggr(\dfrac{s}{\rho} -\sqrt{-\epsilon}\biggr)F(\rho)
                                    +\dfrac{{\rm d} F(\rho)}{{\rm d} \rho}\Biggr]
                             +\left(\epsilon-\dfrac{\kappa}{\rho^2}\right)\rho^{s}
                                    {\rm e}^{-\sqrt{-\epsilon}\,\rho} F(\rho)\nonumber \\
                        &&= 0,
                  \end{eqnarray}
                  that is
                  \begin{eqnarray}
                      &&  \Biggl\{
                                    \dfrac{{\rm d}^2 F(\rho)}{{\rm d}\,\rho^2}
                                    +2\biggr(\dfrac{s}{\rho} -\sqrt{-\epsilon}\biggr)
                                          \dfrac{{\rm d} F(\rho)}{{\rm d} \rho}
                                    +\biggl[\Bigl(\dfrac{s}{\rho} -\sqrt{-\epsilon}\Bigr)^2
                                          -\dfrac{s}{\rho^2}\biggr]F(\rho)\Biggr\}\nonumber\\
                             && +\dfrac{1}{\rho} \Biggr[
                                    \biggr(\dfrac{s}{\rho} -\sqrt{-\epsilon}\biggr)F(\rho)
                                    +\dfrac{{\rm d} F(\rho)}{{\rm d} \rho}\Biggr]
                              +\left(\epsilon-\dfrac{\kappa}{\rho^2}\right)F(\rho)
                        = 0,
                  \end{eqnarray}
                  i.e.
                  \begin{eqnarray}
                        \dfrac{{\rm d}^2 F(\rho)}{{\rm d}\,\rho^2}
                              +\biggr(\dfrac{2s+1}{\rho} -2\sqrt{-\epsilon}\biggr)
                                    \dfrac{{\rm d} F(\rho)}{{\rm d} \rho}
                              +\biggl[\Bigl(\dfrac{s}{\rho} -\sqrt{-\epsilon}\Bigr)^2
                                    -\dfrac{s}{\rho^2}
                                    +\dfrac{1}{\rho} \biggr(\dfrac{s}{\rho} -\sqrt{-\epsilon}\biggr)
                                    +\left(\epsilon-\dfrac{\kappa}{\rho^2}\right)\biggr]F(\rho)
                        = 0,
                  \end{eqnarray}
                  i.e.
                  \begin{eqnarray}
                        \dfrac{{\rm d}^2 F(\rho)}{{\rm d}\,\rho^2}
                              +\biggr(\dfrac{2s+1}{\rho} -2\sqrt{-\epsilon}\biggr)
                                    \dfrac{{\rm d} F(\rho)}{{\rm d} \rho}
                              +\biggl[\Bigl(\dfrac{s}{\rho} -\sqrt{-\epsilon}\Bigr)^2
                                    +\dfrac{1}{\rho} \biggr(\dfrac{s}{\rho} -\sqrt{-\epsilon}\biggr)
                                    +\left(\epsilon-\dfrac{\kappa+s}{\rho^2}\right)\biggr]F(\rho)
                        = 0,
                  \end{eqnarray}
                  i.e.
                  \begin{eqnarray}
                        \dfrac{{\rm d}^2 F(\rho)}{{\rm d}\,\rho^2}
                              +\biggr(\dfrac{2s+1}{\rho} -2\sqrt{-\epsilon}\biggr)
                                    \dfrac{{\rm d} F(\rho)}{{\rm d} \rho}
                              +\biggl[\dfrac{s^2}{\rho} -2\dfrac{s}{\rho} \sqrt{-\epsilon} -\epsilon
                                    +\dfrac{1}{\rho} \biggr(\dfrac{s}{\rho}  -\sqrt{-\epsilon}\biggr)
                                    +\left(\epsilon-\dfrac{\kappa+s}{\rho^2}\right)\biggr]F(\rho)
                        = 0,
                  \end{eqnarray}
                  i.e.
                  \begin{eqnarray}
                        \dfrac{{\rm d}^2 F(\rho)}{{\rm d}\,\rho^2}
                              +\biggr(\dfrac{2s+1}{\rho} -2\sqrt{-\epsilon}\biggr)
                                    \dfrac{{\rm d} F(\rho)}{{\rm d} \rho}
                             -\dfrac{(2\,s+1)}{\rho} \sqrt{-\epsilon} F(\rho)
                        = 0,
                  \end{eqnarray}
                  i.e.
                  \begin{eqnarray}
                        \rho\,\dfrac{{\rm d}^2 F(\rho)}{{\rm d}\,\rho^2}
                              +\biggr(2s+1 -2\sqrt{-\epsilon}\,\rho\biggr)
                                    \dfrac{{\rm d} F(\rho)}{{\rm d} \rho}
                             -(2\,s+1) \sqrt{-\epsilon}\,F(\rho)
                        = 0,
                  \end{eqnarray}
                  viz.
                  \begin{eqnarray}\label{eq:F-28}
                        (\sqrt{-\epsilon})^2 \rho\dfrac{{\rm d}^2 F(z)}{{\rm d} (\sqrt{-\epsilon} \;\rho)^2}
                              +\sqrt{-\epsilon} \;\biggr[2(s+1) -2\sqrt{-\epsilon}\;\rho\biggr]
                                    \dfrac{{\rm d}\,F(z)}{{\rm d}\,(\sqrt{-\epsilon} \;\rho)}
                        -\biggr[2(s+1)\sqrt{-\epsilon}\biggr]F(z)=0.
                  \end{eqnarray}
                  Let
                  \begin{eqnarray}\label{eq:F-27-02}
                        z=\sqrt{-\epsilon} \; \rho,
                  \end{eqnarray}\\
                  then Eq. (\ref{eq:F-28}) becomes
                  \begin{eqnarray}\label{eq:F-29-02}
                        \sqrt{-\epsilon}\,z\dfrac{{\rm d}^2 F(z)}{{\rm d}\,z^2}
                                    +\sqrt{-\epsilon} \;\biggr[2(s+1) -2z\biggr]
                                          \dfrac{{\rm d}\,F(z)}{{\rm d}\,z}
                              -\biggr[2(s+1)\sqrt{-\epsilon}\biggr]F(z)=0,
                  \end{eqnarray}
                  i.e.
                  \begin{eqnarray}\label{eq:F-30-d}
                        \tau\dfrac{{\rm d}^2 F(\tau)}{{\rm d}\,\tau^2}
                        +\biggr[2(s+1) -\tau\biggr]
                              \dfrac{{\rm d}\,F(\tau)}{{\rm d}\,\tau}
                        -(s+1)F(\tau)=0,
                  \end{eqnarray}
                  with $\tau=2\,z$. One may notice that the energy parameter $\epsilon$ does not yet appear in Eq. (\ref{eq:F-30-d}).

                  Remarkably, let us set
                  \begin{eqnarray}\label{eq:F-31-02}
                        \alpha &=& (s+1)=\sqrt{\kappa}\,+1, \;\;\;\;\;\;
                        \gamma = 2(s+1)=2(\sqrt{\kappa}\,+1),
                  \end{eqnarray}
                  then Eq. (\ref{eq:F-30-d}) can be rewritten as
                  \begin{eqnarray}\label{eq:F-23-02}
                        \tau\dfrac{{\rm d}^2\,F(\tau)}{{\rm d}\,\tau^2}
                        +(\gamma-\tau)\dfrac{{\rm d}\,F(\tau)}{{\rm d}\,\tau}-\alpha\,F(\tau)=0,
                  \end{eqnarray}
                  which is a confluent hypergeometric equation with general solution
                  \begin{eqnarray}\label{eq:F-24-02}
                        F(\tau)= C_1\,F(\alpha;\,\gamma;\,\tau)
                              +C_2\,\tau^{1-\gamma}\,F(\alpha+1-\gamma;\,2-\gamma;\,\tau),
                  \end{eqnarray}
                  where $C_1$, $C_2$ are two constants. For convenience, one can denote the two hypergeometric functions
                  as
                  \begin{eqnarray}\label{eq:F-32}
                        w_1 &=& F(\alpha;\gamma;\tau) = F\left(\sqrt{\kappa}\,+1;\,
                              2(\sqrt{\kappa}\,+1);\,2 \sqrt{-\epsilon}\,\rho\right),
                  \end{eqnarray}
                  \begin{eqnarray}\label{eq:F-33}
                        w_2 &=& \tau^{1-\gamma}F(\alpha-\gamma+1;\,2-\gamma;\,\tau) = \left(2\sqrt{-\epsilon}\,\rho\right)^{1-\gamma}
                              F(\alpha-\gamma+1;\,2-\gamma;\,2 \sqrt{-\epsilon}\,\rho),
                  \end{eqnarray}
                  namely
                  \begin{eqnarray}\label{eq:F-33-t}
                        F(\rho)=C_1\,w_1 +C_2\,w_2.
                  \end{eqnarray}

                  Finally, insert \Eq{eq:F-33-t} into Eq. (\ref{eq:F-13-03}), then one
                  obtains the radial function as
                  \begin{eqnarray}\label{eq:F-34-02}
                        R(\rho) &=& \rho^{s} {\rm e}^{-\sqrt{-\epsilon}\,\rho} F(\rho) = \rho^{s} {\rm e}^{-\sqrt{-\epsilon}\,\rho} (C_1\,w_1 +C_2\,w_2).
                  \end{eqnarray}
Unfortunately, from Mathematica program, one can check that: (i) for $\kappa>0$,
 \begin{eqnarray}\label{eq:F-34b-02}
 \lim_{\rho\rightarrow\infty } \rho^{s} {\rm e}^{-\sqrt{-\epsilon}\rho} w_i \rightarrow\infty,   \;\;\;\;\;\; (i=1, 2),
                  \end{eqnarray}
and (ii) for $\kappa=0$ (thus $s=0$),
\begin{eqnarray}\label{eq:F-34c-02}
 && \lim_{\rho\rightarrow\infty } \rho^{s} {\rm e}^{-\sqrt{-\epsilon}\rho} w_1=\lim_{\rho\rightarrow\infty } {\rm e}^{-\sqrt{-\epsilon}\rho} F\left(1;
                              2;\,2 \sqrt{-\epsilon}\,\rho\right) \rightarrow\infty, \nonumber\\
 && \lim_{\rho\rightarrow 0  } \rho^{s} {\rm e}^{-\sqrt{-\epsilon}\rho} w_2 =\lim_{\rho\rightarrow 0  }  {\rm e}^{-\sqrt{-\epsilon}\rho} \left(2\sqrt{-\epsilon}\,\rho\right)^{-1}
                              F(0;0;\,2 \sqrt{-\epsilon}\,\rho)=\lim_{\rho\rightarrow 0  }  \frac{1}{2\sqrt{-\epsilon}\,\rho}{\rm e}^{-\sqrt{-\epsilon}\rho}
                              \rightarrow\infty.
                  \end{eqnarray}
Therefore, for the case of energy $E< 0$, we do not have the physically allowable radial wavefunctions $R(\rho)$.

\begin{remark}
In summary, let us denote the energy as
\begin{eqnarray}\label{eq:En-1-01}
 && E=\frac{\hbar^2 \vec{k}^2}{2M}=\frac{1}{2M}\left(k_x^2+k_y^2+k_z^2\right),
\end{eqnarray}
and then the parameter
\begin{eqnarray}\label{eq:En-1-02}
 && \epsilon = \frac{2M}{\hbar^2}E- k_z^2= k_x^2+k_y^2 \geq 0.
\end{eqnarray}
For a fixed energy $E$, the most general wavefunction is a superposition state as
\begin{eqnarray}\label{eq:H-g-a}
  \Psi_{\rm M}(\vec{r})&=& \sum_{\nu,m}C_1^{\nu, m} \left(J_{\nu}(\sqrt{\epsilon}\,\rho)  {\rm e}^{{\rm i} m\phi} \right) {\rm e}^{{\rm i} k_z z}+\sum_{\nu,m} C_2^{\nu, m} \left(J_{\nu}(\sqrt{\epsilon}\,\rho)  {\rm e}^{{\rm i} m\phi} \right)  {\rm e}^{-{\rm i} k_z z}.
\end{eqnarray}
\end{remark}

\section{The Second Method}\label{secondmethod}

When there is no vector potential, the Hamiltonian $H_{\rm M}$ reduces to the Hamiltonian of a free electron, i.e.,
\begin{equation}\label{eq:H0}
            H_0=\dfrac{1}{2M} \vec{p}^{\; 2},
      \end{equation}
with $\vec{p}=-i\hbar\vec{\nabla}$. The corresponding eigen-equation is given by
\begin{equation}\label{eq:H0-c}
  H_0\,\xi_0 (\vec{r})=E\,\xi_0(\vec{r}),
\end{equation}
where $E$ is the energy of a free electron. When there is a magnetic vector potential, the eigen-equation reads
\begin{equation}\label{eq:H0-f}
  H_{\rm M} \bigl[\xi_0(\vec{r}) \xi(\vec{r})\bigr]
  =E\bigl[\xi_0(\vec{r})\xi(\vec{r})\bigr],
\end{equation}
where $E$ is the eigen-energy of the electron, and the wave function has been written in a form as
\begin{equation}\label{eq:H0-f-a-t}
  \Psi_{\rm M} (\vec{r}) =\xi_0(\vec{r})\xi(\vec{r}).
\end{equation}
Note that the energies $E$ in Eq. (\ref{eq:H0-c}) and Eq. (\ref{eq:H0-f}) are chosen as the same.

We now expand the spin AB Hamiltonian, which gives
\begin{eqnarray}\label{eq:L-1}
  H_{\rm M} &=&\dfrac{1}{2M} \left(\vec{p}+\frac{e}{c}\vec{A}_{\rm M}\right)^2
  =\dfrac{1}{2M} \left[\vec{p}^{\; 2}+\frac{e}{c}\left(\vec{p}\cdot \vec{A}_{\rm M}+\vec{A}_{\rm M}\cdot \vec{p}\right)+\frac{{\rm e}^2}{c^2}\vec{A}_{\rm M}^{\; 2}\right]\nonumber\\
  &=& \dfrac{1}{2M} \left[\vec{p}^{\; 2}+\frac{e}{c}\left(\vec{A}_{\rm M}\cdot \vec{p}
    -{\rm i}\hbar\,\vec{\nabla}\cdot\vec{A}_{\rm M} +\vec{A}_{\rm M}\cdot \vec{p}\right)+\frac{{\rm e}^2}{c^2}\vec{A}_{\rm M}^{\; 2}\right]\nonumber\\
  &=&\dfrac{1}{2M} \left(\vec{p}^{\; 2}+ \frac{e}{c} 2 \vec{A}_{\rm M}\cdot \vec{p}
    +\frac{{\rm e}^2}{c^2}\vec{A}_{\rm M}^{\; 2}\right)\nonumber\\
  &=& H_0+ \mathcal{T},
\end{eqnarray}
with
\begin{eqnarray}\label{eq:L-2}
\mathcal{T}&=& \dfrac{1}{2M} \left[ \frac{e}{c} 2 \vec{A}_{\rm M} \cdot \vec{p}
  +\frac{{\rm e}^2}{c^2}\vec{A}_{\rm M}^{\; 2}\right].
\end{eqnarray}
Notice that the derivation of Eq. (\ref{eq:L-1}) is valid for any vector
potential satisfying $\vec{\nabla}\cdot\vec{A}$. Next we select the wavefunction $\xi_0(\vec{r})$ as the common eigenstate of
the set $\{H_0, \vec{p}\}$, i.e.,
\begin{eqnarray}\label{eq:L-3-01}
 && \xi_0(\vec{r})= \mathcal{N} {\rm e}^{\mathrm{i} \vec{k}\cdot\vec{r}}, \nonumber\\
 && H_0  \xi_0(\vec{r})= \frac{\hbar^2 \vec{k}^2}{2M} \; \xi_0(\vec{r}), \nonumber\\
 && \vec{p}\; \xi_0(\vec{r}) = \hbar \vec{k} \;\xi_0(\vec{r}).
\end{eqnarray}
Then firstly we have
\begin{eqnarray}\label{eq:L-4}
  &&H_0 \;\bigl[ \xi_0(\vec{r})\; \xi(\vec{r})\bigr]
  = \dfrac{1}{2M} \vec{p}^2 \;\bigl[ \xi_0(\vec{r})\; \xi(\vec{r})\bigr]\nonumber\\
  &&=\dfrac{-\hbar^2}{2M} \biggr\{\left[\frac{{\rm d}^2 \xi_0(\vec{r})}{{\rm d}\,x^2}+\frac{{\rm d}^2 \xi_0(\vec{r})}{{\rm d}\,y^2}+\frac{{\rm d}^2 \xi_0(\vec{r})}{{\rm d}\,z^2}\right]\; \xi(\vec{r})\biggr\}
    +\dfrac{-\hbar^2}{2M} \;2 \left[\frac{{\rm d}\,\xi_0(\vec{r})}{{\rm d}\,x}\frac{{\rm d}\,\xi(\vec{r})}{{\rm d}\,x}+\frac{{\rm d}\,\xi_0(\vec{r})}{{\rm d}\,y}\frac{{\rm d}\,\xi(\vec{r})}{{\rm d}\,y}+\frac{{\rm d}\,\xi_0(\vec{r})}{{\rm d}\,z}\frac{{\rm d}\,\xi(\vec{r})}{{\rm d}\,z}\right]\nonumber\\
    &&\;\;\;+\dfrac{-\hbar^2}{2M} \biggr\{\xi_0(\vec{r})\;\left[\frac{{\rm d}^2 \xi(\vec{r})}{{\rm d}\,x^2}+\frac{{\rm d}^2 \xi(\vec{r})}{{\rm d}\,y^2}+\frac{{\rm d}^2 \xi(\vec{r})}{{\rm d}\,z^2}\right]\biggr\}\nonumber\\
  &&=\bigl[E\; \xi_0(\vec{r})\bigr]\; \xi(\vec{r})
    +\dfrac{-\hbar^2}{2M} \; 2\xi_0(\vec{r})\;  \left[ {\rm i} k_x  \frac{{\rm d}\,\xi(\vec{r})}{{\rm d}\,x}+{\rm i} k_y  \frac{{\rm d}\,\xi(\vec{r})}{{\rm d}\,y}
    +{\rm i} k_z  \frac{{\rm d}\,\xi(\vec{r})}{{\rm d}\,z}\right]
    +\dfrac{1}{2M} \xi_0(\vec{r})\; \bigl[\vec{p}^2 \xi(\vec{r})\bigr]\nonumber\\
  &&= E\; \bigl[\xi_0(\vec{r})\; \xi(\vec{r})\bigr]+\dfrac{\hbar}{2M} \; 2 \xi_0(\vec{r})\;  \left[ ( \vec{k}\cdot  \vec{p})\xi(\vec{r}) \right]
    +\dfrac{1}{2M} \xi_0(\vec{r})\; \left[\vec{p}^2 \xi(\vec{r})\right].
\end{eqnarray}
Secondly, we have
\begin{eqnarray}\label{eq:L-5}
  && \vec{A}_{\rm M} \cdot \vec{p} \;[ \xi_0(\vec{r})\; \xi(\vec{r})]\nonumber\\
  &&= -{\rm i}\hbar \left[ A_{{\rm M}x}  \frac{{\rm d}\, \xi_0(\vec{r})}{{\rm d}\,x}+A_{{\rm M}y}  \frac{{\rm d}\, \xi_0(\vec{r})}{{\rm d}\,y}
      +A_{{\rm M}z}  \frac{{\rm d}\, \xi_0(\vec{r})}{{\rm d}\,z}\right] \xi(\vec{r})
    -{\rm i}\hbar \;\xi_0(\vec{r})\; \left[ A_{{\rm M}x}  \frac{{\rm d}\, \xi(\vec{r})}{{\rm d}\,x}+A_{{\rm M}y}  \frac{{\rm d}\, \xi(\vec{r})}{{\rm d}\,y}
      +A_{{\rm M}z}  \frac{{\rm d}\, \xi(\vec{r})}{{\rm d}\,z}\right] \nonumber\\
  &&= -{\rm i}\hbar \Bigl\{ \bigl[A_{{\rm M}x} ({\rm i} k_x)+ A_{{\rm M}y} ({\rm i} k_y)+A_{{\rm M}z} ({\rm i} k_z)\bigr]\xi_0(\vec{r})\Bigr\} \xi(\vec{r})
    -{\rm i}\hbar \;\xi_0(\vec{r})\; \left[ A_{{\rm M}x}  \frac{{\rm d}\, \xi(\vec{r})}{{\rm d}\,x}+A_{{\rm M}y}  \frac{{\rm d}\, \xi(\vec{r})}{{\rm d}\,y}
      +A_{{\rm M}z}  \frac{{\rm d}\, \xi(\vec{r})}{{\rm d}\,z}\right] \nonumber\\
  &&=\hbar (\vec{k}\cdot \vec{A}_{\rm M})[\xi_0(\vec{r})\,\xi(\vec{r})]
    + \xi_0(\vec{r})\; \left[ (\vec{A}_{\rm M} \cdot \vec{p})\,\xi(\vec{r})\right].
\end{eqnarray}
By substituting Eq. (\ref{eq:L-4}) and Eq. (\ref{eq:L-5}) into Eq. (\ref{eq:H0-f}),
we have
\begin{eqnarray}\label{eq:L-7}
  &&E\;[ \xi_0(\vec{r})\; \xi(\vec{r})]+\dfrac{\hbar}{2M} \; 2 \xi_0(\vec{r})\;
      \left[ ( \vec{k}\cdot  \vec{p})\xi(\vec{r}) \right]
    +\dfrac{1}{2M} \xi_0(\vec{r})\; \left[\vec{p}^2 \xi(\vec{r})\right] \nonumber \\
  &&+ \dfrac{1}{2M}\frac{2e}{c} \left\{\hbar (\vec{k}\cdot \vec{A}_{\rm M}) \;[\xi_0(\vec{r}) \xi(\vec{r})]
    + \xi_0(\vec{r})\; \left[ (\vec{A}_{\rm M}\cdot \vec{p}) \xi(\vec{r})\right]\right\}
    + \dfrac{1}{2M}\frac{{\rm e}^2}{c^2} \vec{A}_{\rm M}^2  \;[ \xi_0(\vec{r})\; \xi(\vec{r})]\nonumber\\
  &&= E\;[ \xi_0(\vec{r})\; \xi(\vec{r})],
\end{eqnarray}
which can be simplified as
\begin{eqnarray}\label{eq:L-8-01}
  \frac{1}{2M}\left\{ 2\hbar\;  \left[ ( \vec{k}\cdot  \vec{p})\xi(\vec{r}) \right]+\left[\vec{p}^2 \xi(\vec{r})\right]
    + \frac{2e}{c} \left\{\hbar (\vec{k}\cdot \vec{A}_{\rm M}) \;[ \xi(\vec{r})]
      + \left[ (\vec{A}_{\rm M}\cdot \vec{p}) \xi(\vec{r})\right]\right\}
    +\frac{{\rm e}^2}{c^2} \vec{A}_{\rm M}^2 \right\} \; \xi(\vec{r})= 0,
\end{eqnarray}
i.e.,
\begin{eqnarray}\label{eq:L-8-02}
  \frac{1}{2M}\left\{ 2 \hbar\;  \vec{k}\cdot
  \left(\vec{p} +\frac{e}{c} \vec{A}_{\rm M} \right) + \left(\vec{p}+\frac{e}{c} \vec{A}_{\rm M}\right)\cdot
  \left(\vec{p} +\frac{e}{c} \vec{A}_{\rm M} \right)\right\} \; \xi(\vec{r})= 0,
\end{eqnarray}
i.e.
\begin{eqnarray}\label{eq:L-12}
 \frac{1}{2M} \left[ \left(2 \hbar\;  \vec{k}+ \vec{p}+\frac{e}{c} \vec{A}_{\rm M}\right)\cdot
  \left(\vec{p} +\frac{e}{c} \vec{A}_{\rm M} \right)\right] \; \xi(\vec{r})= 0.
\end{eqnarray}
Consequently, based on Eq. (\ref{eq:L-12}), the problem of solving \Eq{eq:H0-f} is transformed into solving the
following equations:
\begin{equation}\label{eq:RelatePAM}
  \left(\vec{p} +\frac{e}{c} \vec{A}_{\rm M} \right)\; \xi(\vec{r})=0.
\end{equation}
or
\begin{equation}\label{eq:RelatePAM-2}
  \left(2 \hbar\;  \vec{k}+ \vec{p}+\frac{e}{c} \vec{A}_{\rm M}\right)\; \xi(\vec{r})=0.
\end{equation}

\subsubsection{The Case of $\left(\vec{p} +\frac{e}{c} \vec{A}_{\rm M} \right)\; \xi(\vec{r})=0$}

We now consider the case in Eq. (\ref{eq:RelatePAM}). By the way, if it is valid, we shall also have the following relation
\begin{eqnarray}\label{eq:L-12-b}
 H_{\rm M} \; \xi(\vec{r})= \frac{1}{2M} \left[ \left(\vec{p}+\frac{e}{c} \vec{A}_{\rm M}\right)\cdot
  \left(\vec{p} +\frac{e}{c} \vec{A}_{\rm M} \right)\right] \; \xi(\vec{r})=0,
\end{eqnarray}
i.e., $ \xi(\vec{r})$ is the eigenstate of $H_{\rm M}$ with zero energy. We shall check this point after $\xi(\vec{r})$ is determined.

From Eq. (\ref{eq:RelatePAM}) we obtain the following three subequaltions:
\begin{eqnarray}\label{eq:L-22-a}
&& \left(\vec{p} +\frac{e}{c} \vec{A} \right)_x \xi(\vec{r})= 0,\nonumber\\
&& \left(\vec{p} +\frac{e}{c} \vec{A} \right)_y \xi(\vec{r})= 0,\nonumber\\
&& \left(\vec{p} +\frac{e}{c} \vec{A} \right)_z \xi(\vec{r})= 0,
\end{eqnarray}
which implies that $\xi(\vec{r})$ is the common eigenstate of three operators $\left(\vec{p} +\frac{e}{c} \vec{A} \right)_x$, $\left(\vec{p} +\frac{e}{c} \vec{A} \right)_y$, and $\left(\vec{p} +\frac{e}{c} \vec{A} \right)_z$. Fortunately, for the original magnetic AB effect, these three operators are mutually commutative, thus they can have a common eigenstate $\xi(\vec{r})$. Explicitly, from Eq. (\ref{eq:L-17}) one has
\begin{eqnarray}\label{eq:L-18}
  &&\left(\vec{p} +\frac{e}{c} \vec{A}_{\rm M} \right)_x = p_x
    - \frac{y\;\Phi}{2\pi(x^2+y^2)}, \nonumber\\
 && \left(\vec{p} +\frac{e}{c} \vec{A}_{\rm M} \right)_y = p_y+
    \frac{x\;\Phi}{2\pi(x^2+y^2)}, \nonumber\\
 && \left(\vec{p} +\frac{e}{c} \vec{A}_{\rm M} \right)_z = p_z.
\end{eqnarray}
It is easy to check the following commutation relations
\begin{eqnarray}
  &&\left[p_z, p_x- \frac{y\;\Phi}{2\pi (x^2+y^2)}\right]=0, \;\;\left[p_z, p_y+ \frac{x\;\Phi}{2\pi (x^2+y^2)}\right]=0,\nonumber\\
  &&\left[p_x- \frac{y\;\Phi}{2\pi (x^2+y^2)}, p_y+ \frac{x\;\Phi}{2\pi (x^2+y^2)}\right]=  \frac{\Phi}{2\pi}\left[p_x,\frac{x}{x^2+y^2}\right] +\frac{\Phi}{2\pi}\left[p_y,\frac{y}{x^2+y^2}\right]\nonumber\\
  &&= -{\rm i}\hbar\dfrac{\Phi}{2\pi}\Biggl[\dfrac{{\rm d}}{{\rm d}\,x} \left(\dfrac{x}{x^2 +y^2}\right)
  +\dfrac{{\rm d}}{{\rm d}\,y} \left(\dfrac{y}{x^2 +y^2}\right)\Biggr]= -{\rm i}\hbar\dfrac{\Phi}{2\pi }\left[\dfrac{y^2 -x^2 +x^2 -y^2}{(x^2 +y^2)^2}\right]=0,
\end{eqnarray}
and we have the function $\xi(\vec{r})$ as
\begin{eqnarray}\label{eq:L-23}
&&\xi(\vec{r}) = {\rm e}^{-\frac{\mathrm{i} e\Phi}{2 \pi\hbar c} \left[\arctan(\frac{y}{x})\right]} = {\rm e}^{-\frac{\mathrm{i} e \Phi}{h c} \left[\arctan(\frac{y}{x})\right]}.
\end{eqnarray}
By the way, after substituting Eq. (\ref{eq:L-23}) into Eq. (\ref{eq:L-12-b}), one can find that the latter is valid. Then from Eq. (\ref{eq:H0-f-a-t}) we have the wavefunction as
\begin{equation}\label{eq:H0-f-b-02}
  \Psi_{\rm M} (\vec{r}) =\xi_0(\vec{r})\xi(\vec{r})= \mathcal{N}\; {\rm e}^{\mathrm{i} \vec{k}\cdot\vec{r}}\; {\rm e}^{-\frac{\mathrm{i} e \Phi}{h c} \left[\arctan(\frac{y}{x})\right]}.
\end{equation}

\subsubsection{The Case of $ \left(2 \hbar\;  \vec{k}+ \vec{p}+\frac{e}{c} \vec{A}_{\rm M}\right)\; \xi(\vec{r})=0$}

We now consider the case in Eq. (\ref{eq:RelatePAM-2}). Similarly, the solution is
\begin{eqnarray}\label{eq:L-23-b}
&&\xi(\vec{r}) = {\rm e}^{-2\mathrm{i} \vec{k}\cdot\vec{r}}\; {\rm e}^{-\frac{\mathrm{i} e \Phi}{h c} \left[\arctan(\frac{y}{x})\right]},
\end{eqnarray}
 and from Eq. (\ref{eq:H0-f-a-t}) we have the wavefunction as
\begin{equation}\label{eq:H0-f-c-01}
  \Psi_{\rm M} (\vec{r}) =\xi_0(\vec{r})\xi(\vec{r})= \mathcal{N}\; {\rm e}^{-\mathrm{i} \vec{k}\cdot\vec{r}}\; {\rm e}^{-\frac{\mathrm{i} e \Phi}{h c} \left[\arctan(\frac{y}{x})\right]}.
\end{equation}

\begin{remark}
By consider the superposition, for a fixed energy $E=\frac{\hbar^2 \vec{k}^2}{2M}$, we have the general wavefunction as
\begin{equation}\label{eq:H0-f-d}
  \Psi_{\rm M} (\vec{r}) = \mathcal{N}\;\left(c_1 {\rm e}^{\mathrm{i} \vec{k}\cdot\vec{r}}+ c_2 {\rm e}^{-\mathrm{i} \vec{k}\cdot\vec{r}}\right)\; {\rm e}^{-\frac{\mathrm{i} e \Phi}{h c} \left[\arctan(\frac{y}{x})\right]},
\end{equation}
where $\mathcal{N}$ is the normalized constant, $c_1, c_2$ are some complex numbers.
\end{remark}

Note that the wavefunction $ \Psi_{\rm M} (\vec{r})$ in Eq. (\ref{eq:H-g-a}) is not equal directly to the one in Eq. (\ref{eq:H0-f-d}). From the viewpoint of physics, for a fixed energy $E$, the latter can be obtained by the superpositions of the former. In the next section, we shall use the wavefunction $\Psi_{\rm M} (\vec{r})$ in Eq. (\ref{eq:H0-f-d}) to study the magnetic AB effect.

\newpage

\part{The Magnetic AB Effect and the Spin AB Effect}

To demonstrate the AB effect, one usually adopts a gedanken double-slit experiment. In Fig. \ref{fig:SpinAB}, a solenoid (or a spin) is placed behind the double-slit plate, which contributes a magnetic (or a spin) vector potential. Electrons are emitted from the electron source $O$, they travel to the point $D$ on the screen along two different paths 1 and 2. Under the influence of the magnetic (or the spin) vector potential, one will observe the interference patterns produced on the screen. The interference patterns with vector potentials are generally different from that of without a vector potential. In such a way, one demonstrates the magnetic (or spin) AB effect.

\section{The Magnetic AB Effect}

Now we come to calculate the interference patterns. We shall make a unified treatment for both the magnetic and the spin AB effects. For the magnetic AB Hamiltonian, the eigen-equation reads $H_{\rm M} \Psi_{\rm M}(\vec{r}) = E_{\rm M}  \Psi_{\rm M}(\vec{r})$,
where $E_{\rm M}=\hbar^2 k^2/{2M}\geq 0$ is the energy, and $\Psi_{\rm M}(\vec{r})$ is the eigenfunction, which is given in Eq. (\ref{eq:H0-f-d}).

To connect the wavefunction (\ref{eq:H0-f-d}) with the observable double-slit interference experiment, we need to recast the wavefunction to the following integral form
\begin{eqnarray}\label{eq:v-14}
\Psi_{\rm M}(\vec{r}) & = & \mathcal{N}\; {\rm e}^{\int^{\mathcal{L}(\vec{r})} \vec{F}(\vec{r}')\cdot {\rm d}\vec{r}'},
\end{eqnarray}
where ${\mathcal{L}(\vec{r})}$ represents the path that the electron moves, and the vector $\vec{F}=(F_x, F_y, F_z)$ is determined by
\begin{eqnarray}\label{eq:v-15a}
 &&\vec{F}=\frac{1}{\Psi_{\rm M}(\vec{r})} \left[\vec{\nabla} \Psi_{\rm M}(\vec{r}) \right],
\end{eqnarray}
or explicitly
\begin{eqnarray}\label{eq:v-15}
 &&F_x = \dfrac{ \dfrac{{\rm d} \Psi_{\rm M}(\vec{r})}{{\rm d} x}}{\Psi_{\rm M}(\vec{r})},\;\; F_y = \dfrac{ \dfrac{{\rm d} \Psi_{\rm M}(\vec{r})}{{\rm d} y}}{\Psi_{\rm M}(\vec{r})},\;\;
   F_z = \dfrac{ \dfrac{{\rm d} \Psi_{\rm M}(\vec{r})}{{\rm d} z}}{\Psi_{\rm M}(\vec{r})}.
\end{eqnarray}
By the way, in the spherical coordinate system and the cylindrical coordinate system, the vector $\vec{F}$ is respectively expressed as
\begin{eqnarray}\label{eq:v-15b}
&& \vec{F} = \hat{e}_r F_r +\hat{e}_\theta F_\theta +\hat{e}_\phi F_\phi, \nonumber\\
&& F_r = \dfrac{ \dfrac{{\rm d} \Psi_{\rm M}(\vec{r})}{{\rm d} r}}{\Psi_{\rm M}(\vec{r})},\;\; F_\theta = \dfrac{ \dfrac{1}{r}\dfrac{{\rm d} \Psi_{\rm M}(\vec{r})}{{\rm d} \theta}}{\Psi_{\rm M}(\vec{r})},\;\;
   F_\phi = \dfrac{\dfrac{1}{r\sin\theta} \dfrac{{\rm d} \Psi_{\rm M}(\vec{r})}{{\rm d} \phi}}{\Psi_{\rm M}(\vec{r})},
  \end{eqnarray}
and
  \begin{eqnarray}\label{eq:v-15c}
&& \vec{F} = \hat{e}_\rho F_r +\hat{e}_\phi F_\phi +\hat{e}_z F_z, \nonumber\\
&& F_\rho = \dfrac{ \dfrac{{\rm d} \Psi_{\rm M}(\vec{r})}{{\rm d} \rho}}{\Psi_{\rm M}(\vec{r})},\;\; F_\phi = \dfrac{ \dfrac{1}{r}\dfrac{{\rm d} \Psi_{\rm M}(\vec{r})}{{\rm d} \phi}}{\Psi_{\rm M}(\vec{r})},\;\;
   F_z = \dfrac{\dfrac{{\rm d} \Psi_{\rm M}(\vec{r})}{{\rm d} z}}{\Psi_{\rm M}(\vec{r})}.
  \end{eqnarray}

For simplicity, let $c_1=1, c_2=0$, from Eq. (\ref{eq:H0-f-d}) and Eq. (\ref{eq:v-15}) we have the vector $\vec{F}$ as
\begin{eqnarray}\label{eq:v-16}
 \vec{F} &=& {\rm i} \vec{k}- \frac{{\rm i} e}{\hbar c} \vec{A}_{\rm M},
\end{eqnarray}
i.e., the vector $(\vec{F}-{\rm i} \vec{k})$ is proportional to the magnetic vector potential $\vec{A}_{\rm M}$.

\begin{figure}[t]
\includegraphics[width=100mm]{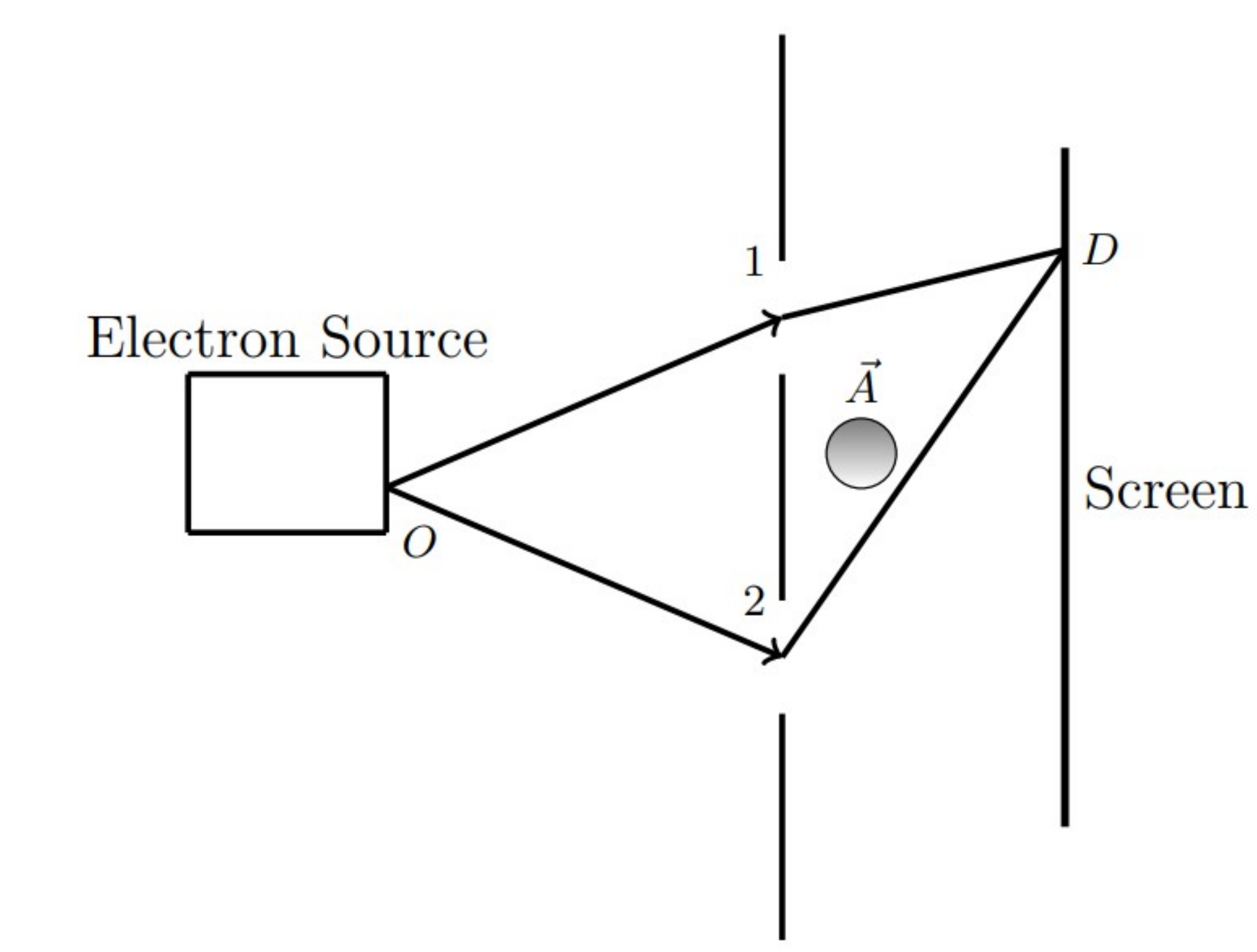}\\
	\caption{Illustration of the gedanken double-slit experiment.}\label{fig:SpinAB}
\end{figure}

Let us focus on Fig. \ref{fig:SpinAB}. The electron is initially at point $O$ and finally at point $D$, there are two different paths $\mathcal{L}_1$, $\mathcal{L}_2$ between them. Therefore, when the electron arrives at the screen, it is in the following superposition state
\begin{eqnarray}\label{eq:v-17}
\Psi(\vec{r})&=& \Psi^1_{\rm M}(\vec{r})+\Psi^2_{\rm M}(\vec{r})= \mathcal{N}\left({\rm e}^{\int^{\mathcal{L}_1(\vec{r})} \vec{F}(\vec{r}')\cdot {\rm d}\vec{r}'}+{\rm e}^{\int^{\mathcal{L}_2(\vec{r})} \vec{F}(\vec{r}')\cdot {\rm d}\vec{r}'}\right).
\end{eqnarray}
Then the probability of finding the electron at the point $D$ is given by
\begin{eqnarray}\label{eq:v-18}
			P_{\rm M}&=&|\Psi(\vec{r})|^2= \mathcal{N}^2 \left| {\rm e}^{\int^{\mathcal{L}_1(\vec{r})} \vec{F}(\vec{r}')\cdot {\rm d}\vec{r}'}+{\rm e}^{\int^{\mathcal{L}_2(\vec{r})} \vec{F}(\vec{r}')\cdot {\rm d}\vec{r}'} \right|^2\nonumber\\
&=& \mathcal{N}^2 \left| {\rm e}^{\int^{\mathcal{L}_1(\vec{r})} \vec{F}(\vec{r}')\cdot {\rm d}\vec{r}'}\left(1+{\rm e}^{\int^{\mathcal{L}_2(\vec{r})} \vec{F}(\vec{r}')\cdot {\rm d}\vec{r}'-\int^{\mathcal{L}_1(\vec{r})} \vec{F}(\vec{r}')\cdot {\rm d}\vec{r}'} \right)\right|^2 \nonumber\\
&=& \mathcal{N}^2 \left| {\rm e}^{i\int^{\mathcal{L}_1(\vec{r})} \vec{k}\cdot {\rm d}\vec{r}'}{\rm e}^{-i\int^{\mathcal{L}_1(\vec{r})} \left(\frac{e}{\hbar c} \vec{A}_{\rm M}\right)\cdot {\rm d}\vec{r}'}\left(1+{\rm e}^{\int^{\mathcal{L}_2(\vec{r})} \vec{F}(\vec{r}')\cdot {\rm d}\vec{r}'-\int^{\mathcal{L}_1(\vec{r})} \vec{F}(\vec{r}')\cdot {\rm d}\vec{r}'}\right)\right|^2 \nonumber\\
&=&\mathcal{N}^2 \left| 1+{\rm e}^{\oint \vec{F}(\vec{r}')\cdot {\rm d}\vec{r}'}\right|^2 =\mathcal{N}^2 \left| 1+{\rm e}^{\frac{\mathrm{i}}{\hbar} \oint (\hbar \vec{k}-\frac{e}{c} \vec{A}_{\rm M}(\vec{r}'))\cdot {\rm d} \vec{r}'}\right|^2\nonumber\\
&=& 2 \mathcal{N}^2 \left[1+\cos(\delta_1+\delta_2)\right],
		\end{eqnarray}
with
\begin{eqnarray}\label{eq:v-19a}
&& \delta_1=\frac{1}{\hbar} \oint (\hbar \vec{k})\cdot {\rm d} \vec{r}'=\oint \vec{k}\cdot {\rm d} \vec{l},
\end{eqnarray}
and
\begin{eqnarray}\label{eq:v-19b}
 \delta_2&=&-\frac{e}{\hbar c} \oint \vec{A}\cdot {\rm d} \vec{l}= -\frac{e}{\hbar c} {\int_S} (\vec{\nabla}\times \vec{A})\cdot {\rm d} \vec{S}\nonumber\\
  &=& {-\frac{e}{\hbar c} {\int_{S_{\text{in}}}} \left[\vec{\nabla}\times\Bigl(\dfrac{B\,\rho}{2}\Bigr)\hat{e}_\phi\right]\cdot {\rm d} \vec{S}} {-\frac{e}{\hbar c} {\int_{S_{\text{out}}}} \left[\vec{\nabla}\times\Bigl(\dfrac{\Phi_{\rm M}}{2{\pi} \rho}\Bigr)\hat{e}_\phi\right]\cdot {\rm d} \vec{S}} \nonumber \\
  &=& {-\frac{e}{\hbar c} {\int_{S_{\text{in}}}} \left[\vec{\nabla}\times\Bigl(\dfrac{B\,\rho}{2}\Bigr)\hat{e}_\phi\right]\cdot {\rm d} \vec{S}}\nonumber\\
  &=& {-\frac{e}{\hbar c} {\int_{S_{\text{in}}}}
    \Biggl\{\dfrac{1}{\rho} \dfrac{\partial}{\partial\,\rho} \left[\rho\Bigl(\dfrac{B\,\rho}{2}\Bigr)\right] \hat{z} \Biggr\}\cdot {\rm d} \vec{S}}\nonumber\\
  &=& -\frac{e}{\hbar c} {\int_{S_{\text{in}}}(B\,\hat{z})} \cdot {\rm d} \vec{S}=-\frac{e}{\hbar c} {\int_{S_{\text{in}}}} \vec{B}\cdot {\rm d} \vec{S}= -\frac{e}{\hbar c} \left(B \pi r_0^2\right)\nonumber\\
  &=& -\frac{e \Phi_{\rm M}}{\hbar c}=-2\pi \;\frac{e \Phi_{\rm M}}{h c}.
		\end{eqnarray}
Here we have used the following formula in the cylindrical coordinate system
\begin{eqnarray}\label{solution-1a}
 \vec{\nabla}\times \vec{A} &=& \left|\begin{array}{ccc}
       \dfrac{1}{\rho}\;\hat{e}_\rho & \hat{e}_\phi & \dfrac{1}{\rho}\; \hat{e}_z \\
       \dfrac{\partial}{\partial \rho} & \dfrac{\partial}{\partial \phi} & \dfrac{\partial}{\partial z} \\
       A_\rho & \rho A_\phi & A_z
     \end{array}\right|.
\end{eqnarray}

\begin{figure}[t]
      \includegraphics[width=100mm]{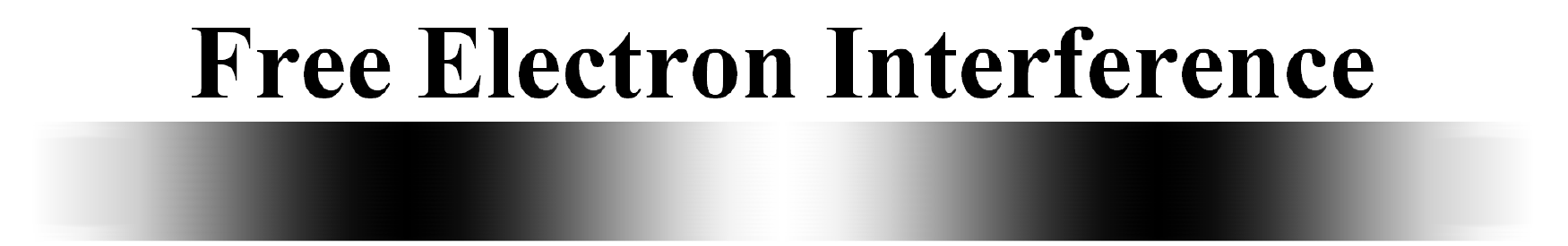}\\
      \includegraphics[width=100mm]{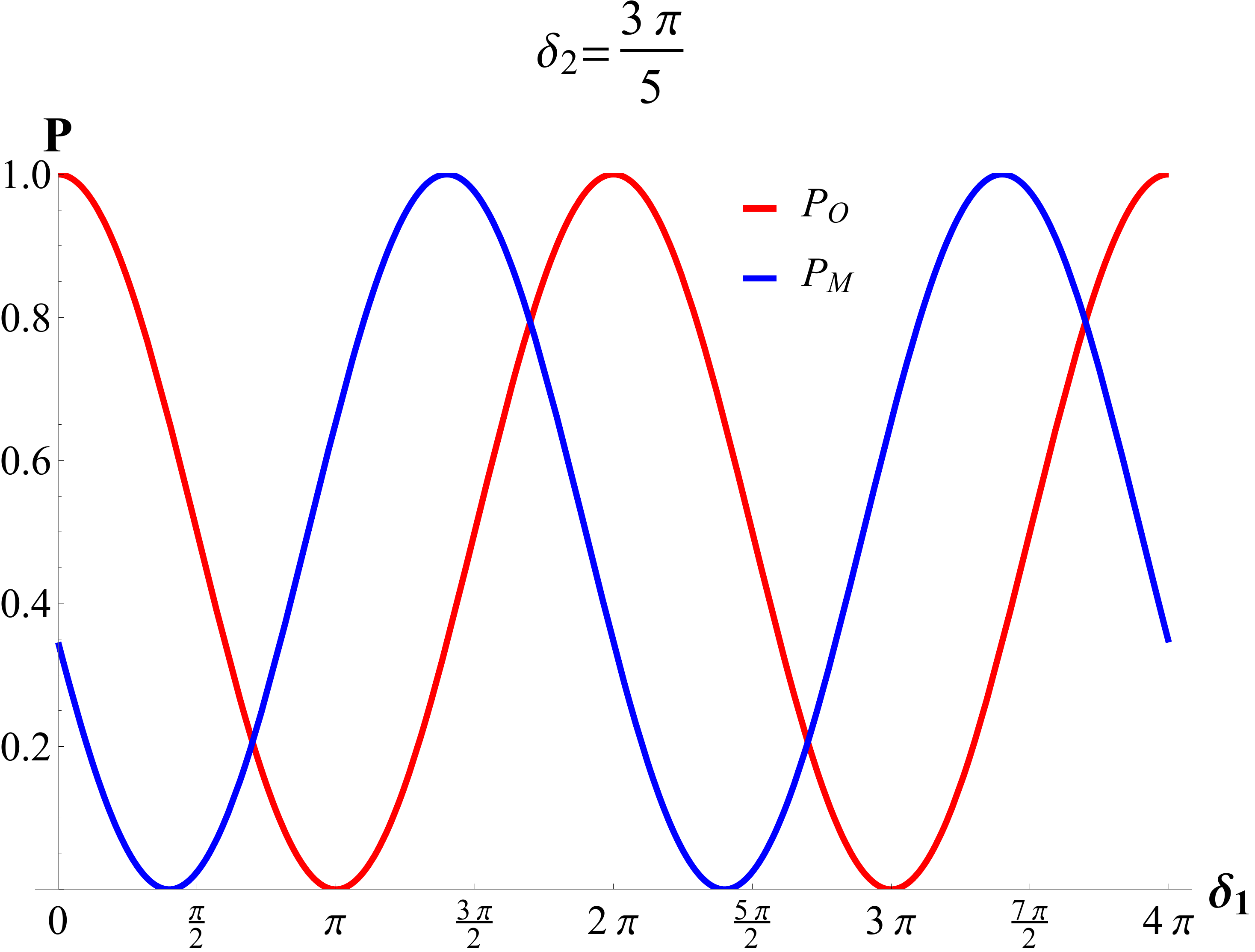}\\
      \includegraphics[width=100mm]{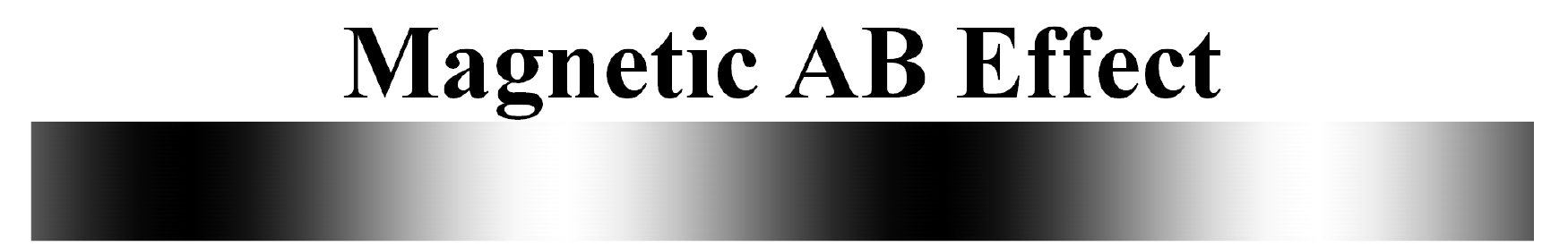}\\
            \caption{The curves for the probability $P_{\rm O}$ (red curve) and the probability $P_{\rm M}$ (blue curve). Above the curves is  the interference pattern for the free electrons, and below the curves is the interference pattern for the magnetic AB effect. For simplicity, we have set $\mathcal{N}=1/2$ and $\delta_2=3\pi/5$. One may find that, the interference fringes of the free electrons are shifted by an additional phase $\delta_2$ when one observes the magnetic AB effect.
            }\label{fig:InterDiagram}
\end{figure}

In the ``ordinary'' double-slit experiment, i.e., there is no any vector potential, it is just the phase $\delta_1$ that causes the interference fringes. In this case, in Eq. (\ref{eq:v-18}) we just let $\delta_2=0$, we then have the corresponding probability $P_{\rm O}$ for the free electron interference, namely,
\begin{eqnarray}\label{eq:v-19c}
&& P_{\rm O}=P_{\rm M}|_{\delta_2=0}= 2 \mathcal{N}^2 [1+\cos(\delta_1)].
\end{eqnarray}
In Fig. \ref{fig:InterDiagram}, we have plotted the red curve for the probability $P_{\rm O}$ as well as the corresponding interference pattern (above the curve) for the free electrons. For the interference pattern, the maximal value of $P_{\rm O}$ (i.e., $\delta_1=0$, mod $2\pi$) corresponds to the brightest region, and the minimal value of $P_{\rm O}$ (i.e., $\delta_1=\pi/2$, mod $2\pi$) corresponds to the darkest region. For the comparison,
in Fig. \ref{fig:InterDiagram}, we have also plotted the blue curve for the probability $P_{\rm M}$ as well as the corresponding interference pattern (below the curve) for the the magnetic AB effect. The interference fringes are shifted by an additional phase $\delta_2$ (in the figure we have set $\delta_2=3\pi/5$). Such an AB effect has been observed in experiments \cite{1960Chambers,1986Tonomura}.

\section{The Spin AB Effect}

Similar to the situation in Eq. (\ref{eq:v-14}), to connect the spin AB wavefunction (\ref{eq:H-22}) with the observable double-slit interference experiment, we recast it to the following integral form as
\begin{eqnarray}\label{eq:v-20}
		 \Psi_{\rm S}(\vec{r}) &=& \mathcal{N}\begin{bmatrix} c_1 \chi_1 (\vec{r}) \\ c_2 \chi_2 (\vec{r}) \end{bmatrix}
  =\mathcal{N}  \begin{bmatrix} c_1 {\rm e}^{\int^{\mathcal{L}(\vec{r})} \vec{F}_1(\vec{r}')\cdot d\vec{r}'} \\ c_2 {\rm e}^{\int^{\mathcal{L}(\vec{r})} \vec{F}_2(\vec{r}')\cdot d\vec{r}'} \end{bmatrix},
		\end{eqnarray}
where the wavefunctions $\chi_1 (\vec{r})$ and
$\chi_2 (\vec{r})$ are normalized, and $\mathcal{N}^2 (|c_1|^2 +|c_2|^2)=1$.
In the spherical coordinates, the vectors $\vec{F}_{i}$'s are expressed as
\begin{eqnarray}\label{eq:v-21}
&& \vec{F}_{i}  = \hat{e}_r F_{ir}+\hat{e}_\theta F_{i\theta}+\hat{e}_\phi F_{i\phi}, \;\;\;\;\;\;\; (i=1, 2),\nonumber\\
&& F_{ir}  = \dfrac{ \dfrac{{\rm d} \chi_i(\vec{r})}{{\rm d} r}}{\chi_i(\vec{r})},\;
  F_{i\theta} = \dfrac{ \dfrac{1}{r}\dfrac{{\rm d} \chi_i(\vec{r})}{{\rm d} \theta}}{\chi_i(\vec{r})},\;
   F_{i\phi} = \dfrac{ \dfrac{1}{r\sin\theta}\dfrac{{\rm d} \chi_i(\vec{r})}{{\rm d} \phi}}{\chi_i(\vec{r})}.
\end{eqnarray}
Similarly, by moving along two different paths $\mathcal{L}_1$ and $\mathcal{L}_2$, the electron is in a quantum superposition state
\begin{eqnarray}\label{eq:v-22}
\Psi(\vec{r})
&=& \Psi^1_{\rm S}(\vec{r})+\Psi^2_{\rm S}(\vec{r})= \mathcal{N} \; \begin{bmatrix} c_1 \Bigl({\rm e}^{\int^{\mathcal{L}_1(\vec{r})} \vec{F}_1(\vec{r}')\cdot {\rm d}\vec{r}'}+{\rm e}^{\int^{\mathcal{L}_2(\vec{r})} \vec{F}_1(\vec{r}')\cdot {\rm d}\vec{r}'}\Bigr)  \\ c_2 \Bigl({\rm e}^{\int^{\mathcal{L}_1(\vec{r})} \vec{F}_2(\vec{r}')\cdot {\rm d}\vec{r}'} +{\rm e}^{\int^{\mathcal{L}_2(\vec{r})} \vec{F}_2(\vec{r}')\cdot {\rm d}\vec{r}'}\Bigr) \end{bmatrix},
\end{eqnarray}
i.e.,
\begin{eqnarray}\label{eq:CC-1-a}
			\Psi(\vec{r})
&=& \mathcal{N}  \begin{bmatrix} c_1 \Bigl\{{\rm e}^{\int^{\mathcal{L}_1(\vec{r})} \left[{\rm Re}\vec{F_1}+{\rm i}\;{\rm Im}\vec{F}_1 \right]\cdot {\rm d}\vec{r}'} +{\rm e}^{\int^{\mathcal{L}_2(\vec{r})} \left[ {\rm Re}\vec{F}_1+{\rm i}\;{\rm Im}\vec{F}_1 \right]\cdot {\rm d}\vec{r}'}\Bigr\} \\ c_2 \Bigl\{{\rm e}^{\int^{\mathcal{L}_1(\vec{r})} \left[{\rm Re}\vec{F}_2+{\rm i}\;{\rm Im}\vec{F}_2 \right]\cdot {\rm d}\vec{r}'}+{\rm e}^{\int^{\mathcal{L}_2(\vec{r})} \left[ {\rm Re}\vec{F}_2+{\rm i}\;{\rm Im}\vec{F}_2 \right]\cdot {\rm d}\vec{r}'}\Bigr\} \end{bmatrix}\nonumber\\
&=& \mathcal{N}  \begin{bmatrix} c_1 \Bigl\{{\rm e}^{\int^{\mathcal{L}_1(\vec{r})} \left[{\rm Re}\vec{F}_1+{\rm i}\;{\rm Im}\vec{F}_1 \right]\cdot {\rm d}\vec{r}'} \times \left(1+ {\rm e}^{\oint \left[ {\rm Re}\vec{F}_1+{\rm i}\;{\rm Im}\vec{F}_1 \right]\cdot {\rm d}\vec{r}'} \right)\Bigr\}\\ c_2 \Bigl\{{\rm e}^{\int^{\mathcal{L}_1(\vec{r})} \left[{\rm Re}\vec{F}_2+{\rm i}\;{\rm Im}\vec{F}_2 \right]\cdot {\rm d}\vec{r}'} \times \left(1+ {\rm e}^{\oint \left[ {\rm Re}\vec{F}_2+{\rm i}\;{\rm Im}\vec{F}_2 \right]\cdot {\rm d}\vec{r}'} \right)\Bigr\}\end{bmatrix}\nonumber\\
&=& \mathcal{N}  \begin{bmatrix} c_1 \Bigl\{{\rm e}^{\int^{\mathcal{L}_1(\vec{r})} \left[{\rm Re}\vec{F}_1+{\rm i}\;{\rm Im}\vec{F}_1 \right]\cdot {\rm d}\vec{r}'} \times
\left(1+ {\rm e}^{\oint {\rm Re}\vec{F}_1\cdot {\rm d}\vec{r}'}\times {\rm e}^{{\rm i}\oint {\rm Im}\vec{F}_1 \cdot {\rm d}\vec{r}'} \right)\Bigr\}\\ c_2 \Bigl\{{\rm e}^{\int^{\mathcal{L}_1(\vec{r})} \left[{\rm Re}\vec{F}_2+{\rm i}\;{\rm Im}\vec{F}_2 \right]\cdot {\rm d}\vec{r}'} \times \left(1+ {\rm e}^{\oint {\rm Re}\vec{F}_2\cdot {\rm d}\vec{r}'}\times {\rm e}^{{\rm i}\oint {\rm Im}\vec{F}_2 \cdot {\rm d}\vec{r}'} \right)\Bigr\}\end{bmatrix},
		\end{eqnarray}
where ${\rm Re}\vec{F}$ and ${\rm Im}\vec{F}$ are respectively the real part and the imaginary part of the vector
\begin{eqnarray}\label{eq:v-23-01}
&&\vec{F}={\rm Re}\vec{F}+{\rm i}\; {\rm Im}\vec{F}.
\end{eqnarray}
Then we have the probability of finding the electron at the point $D$ on the screen as
\begin{eqnarray}\label{eq:v-23-02}
	P_{\rm S}&&=|\Psi(\vec{r})|^2=\left[\Psi(\vec{r})\right]^{\dagger} \Psi(\vec{r})\nonumber\\
 &&= \mathcal{N}^2\; \biggr\{|c_1|^2 {\rm e}^{2 \int^{\mathcal{L}_1(\vec{r})}\,{\rm Re}\vec{F}_1 \cdot {\rm d}\vec{r}'}
        \times \left[1+{\rm e}^{2 \oint{\rm Re}\vec{F}_1\cdot {\rm d}\vec{r}'}
      +2 \; {\rm e}^{\oint {\rm Re}\vec{F}_1\cdot {\rm d}\vec{r}'}\; \cos\biggr(\oint {\rm Im}\vec{F}_1 \cdot {\rm d}\vec{r}'\biggr) \right]\nonumber\\
      &&\;\;\;\;\;+\; |c_2|^2 {\rm e}^{2 \int^{\mathcal{L}_1(\vec{r})}\,{\rm Re}\vec{F}_2 \cdot {\rm d}\vec{r}'}
        \times \left[1+{\rm e}^{2 \oint {\rm Re}\vec{F}_2\cdot {\rm d}\vec{r}'}
      +2 \; {\rm e}^{\oint {\rm Re}\vec{F}_2\cdot {\rm d}\vec{r}'}\; \cos\biggr(\oint {\rm Im}\vec{F}_2 \cdot {\rm d}\vec{r}'\biggr) \right]\biggr\}.
\end{eqnarray}

In the following, we shall provide a concrete example on calculating the probability $P_{\rm S}$, showing that the interference pattern of the spin AB effect is different from that of the ordinary double-slit experiment. Therefore, it is very possible to observe such a spin AB effect by performing the double-slit experiment, hence confirming the existence of the spin vector potential.

Based on Eq. (\ref{eq:H-22}),  the spin AB wavefunction $\Psi_{\rm S}(\vec{r})$ is given by
\begin{eqnarray}\label{eq:MM-4}
\Psi_{\rm S}(\vec{r}) &\propto& \dfrac{J_{\nu}(\sqrt{\epsilon}\,r)}{\sqrt{r}}\;  \;
                        \Phi^A_{lm} (\theta, \phi)\propto \dfrac{J_{\nu}(\sqrt{\epsilon} r)}{\sqrt{r}} \times \dfrac{1}{\sqrt{2l+1}}\left[ \begin{array}{c}
  \sqrt{l+m+1} \;Y_{lm}(\theta, \phi) \\
 \sqrt{l-m} \;Y_{l,m+1}(\theta, \phi)\\
 \end{array}
 \right]\nonumber\\
 &\propto& \dfrac{J_{\nu}(\sqrt{\epsilon} r)}{\sqrt{r}}\;  \times{\left[\begin{array}{c}
    \tau_1\; P_l^{m}(\cos\theta){\rm e}^{{\rm i}m\phi}
       \\
    \tau_2\; P_l^{m+1}(\cos\theta){\rm e}^{{\rm i}(m+1)\phi}
    \\
    \end{array}\right]}=\mathcal{N}\begin{bmatrix}
      c_1 \chi_1(\vec{r}) \\
      c_2 \chi_2(\vec{r}) \\
\end{bmatrix},
\end{eqnarray}
with
\begin{eqnarray}\label{eq:MM-5-a}
&&   \tau_1 = \sqrt{\dfrac{(l+m +1)}{4\pi}} \sqrt{\frac{(l-m)!}{(l+m)!}}, \;\;\;\;\; \tau_2 = \sqrt{\dfrac{1}{4\pi(l+m +1)}} \sqrt{\frac{(l-m)!}{(l+m)!}},\quad
c_1 =1,\quad c_2 =\dfrac{1}{l+m +1}.
\end{eqnarray}
It is convenient to calculate the corresponding vectors $\vec{F}_1$ and $\vec{F}_2$ in the spherical coordinates.

\subsection{The Calculation of $\vec{F}_1$ and $\vec{F}_2$}

Based on Eq. (\ref{eq:v-21}), we have
\begin{eqnarray}
    F_{1r}    &=& \dfrac{ \dfrac{{\rm d} \chi_1(\vec{r})}{{\rm d} r}}{\chi_1(\vec{r})}= \dfrac{ \dfrac{{\rm d} \left[ \frac{1}{\sqrt{r}}\; J_{\nu}(\sqrt{\epsilon} r)\right]}{{\rm d} r}}{\dfrac{1}{\sqrt{r}}\; J_{\nu}(\sqrt{\epsilon} r)}= \dfrac{ \dfrac{{\rm d} \left[ \dfrac{1}{\sqrt{r}}\right]}{{\rm d} r}}{\dfrac{1}{\sqrt{r}}}+\dfrac{ \dfrac{{\rm d} \left[  J_{\nu}(\sqrt{\epsilon} r)\right]}{{\rm d} r}}{ J_{\nu}(\sqrt{\epsilon} r)}\nonumber\\
    &=&-\dfrac{1}{2 r}+\dfrac{ \sqrt{\epsilon} \;[J_{\nu-1}(\sqrt{\epsilon} r)-J_{\nu+1}(\sqrt{\epsilon} r)]}{ 2 J_{\nu}(\sqrt{\epsilon} r)},
  \end{eqnarray}
thus
\begin{eqnarray}
    {\rm Re}F_{1r}    &=& -\frac{1}{2 r}+\frac{ \sqrt{\epsilon} \;[J_{\nu-1}(\sqrt{\epsilon} r)-J_{\nu+1}(\sqrt{\epsilon} r)]}{ 2 J_{\nu}(\sqrt{\epsilon} r)},\nonumber\\
     {\rm Im}F_{1r}    &=& 0.
  \end{eqnarray}

Because
\begin{eqnarray}\label{eq:LL-10}
 \frac{{\rm d} P_l^m(\cos\theta)}{{\rm d} \theta}&=& \frac{-\sin\theta}{-\sin\theta}\frac{{\rm d} P_l^m(\cos\theta)}{d \theta}
= -\sin\theta \frac{{\rm d} P_l^m(\cos\theta)}{{\rm d} \cos\theta}=-\sin\theta \frac{{\rm d} P_l^m(z)}{{\rm d} z}\nonumber\\
&=&-\sin\theta \frac{{\rm d} }{{\rm d} z}\left[(-1)^m (1-z^2)^{m/2} \frac{{\rm d}^m P_l(z)}{{\rm d} z^m}\right]=- (-1)^m \sin\theta \frac{{\rm d} }{{\rm d} z}\left[ (1-z^2)^{m/2} \frac{{\rm d}^m P_l(z)}{{\rm d} z^m}\right]\nonumber\\
&=&- (-1)^m \sin\theta \biggr[ \frac{{\rm d} (1-z^2)^{m/2}}{{\rm d} z} \frac{{\rm d}^m P_l(z)}{{\rm d} z^m}+(1-z^2)^{m/2} \frac{{\rm d}^{m+1} P_l(z)}{{\rm d} z^{m+1}}\biggr]\nonumber\\
&=&- (-1)^m \sin\theta \biggr[(-{m}z) (1-z^2)^{\frac{m-2}{2}} \frac{{\rm d}^m P_l(z)}{{\rm d} z^m}+(1-z^2)^{m/2} \frac{{\rm d}^{m+1} P_l(z)}{{\rm d} z^{m+1}}\biggr]\nonumber\\
&=&- (-1)^m \sin\theta \biggr[ \frac{(-{m}z)}{1-z^2} (1-z^2)^{\frac{m}{2}} \; \frac{{\rm d}^m P_l(z)}{{\rm d} z^m}+\frac{1}{\sqrt{1-z^2}}(1-z^2)^{\frac{m+1}{2}}  \frac{{\rm d}^{m+1} P_l(z)}{{\rm d} z^{m+1}}\biggr]\nonumber\\
&=&- \sin\theta \biggr[ \frac{(-{m} z)}{1-z^2} \; (-1)^m\; (1-z^2)^{\frac{m}{2}} \; \frac{{\rm d}^m P_l(z)}{{\rm d} z^m}+\frac{1}{\sqrt{1-z^2}}\; (-1)\;(-1)^{m+1}\;  (1-z^2)^{\frac{m+1}{2}} \; \frac{{\rm d}^{m+1} P_l(z)}{{\rm d} z^{m+1}}\biggr]\nonumber\\
&=& -\sin\theta \biggr[ \frac{(-{m}z)}{1-z^2} \; P_l^m(\cos\theta)+\frac{-1}{\sqrt{1-z^2}}\;P_l^{m+1}(\cos\theta)\biggr]\nonumber\\
&=& \frac{{m} \sin\theta \cos\theta}{1-\cos^2\theta} \; P_l^m(\cos\theta)+\frac{\sin\theta}{\sqrt{1-\cos^2\theta}}\;P_l^{m+1}(\cos\theta)\nonumber\\
&=& \frac{{m} \cos\theta}{\sin\theta} \; P_l^m(\cos\theta)+P_l^{m+1}(\cos\theta),
\end{eqnarray}
we then have
\begin{eqnarray}\label{eq:LL-11}
F_{1\theta}    &=& {\dfrac{1}{r}}\; \dfrac{ \dfrac{{\rm d} \chi_1(\vec{r})}{{\rm d} \theta}}{\chi_1(\vec{r})}={\frac{1}{r}} \;\left[\dfrac{ \dfrac{{\rm d} P_l^m(\cos\theta)}{{\rm d} \theta}}{P_l^m(\cos\theta)}\right]={\frac{1}{r}}\;\left[\dfrac{{m} \cos\theta}{\sin\theta}+\dfrac{P_l^{m+1}(\cos\theta)}{P_l^m(\cos\theta)}\right],
\end{eqnarray}
thus
\begin{eqnarray}
    {\rm Re}F_{1\theta}    &=& {\frac{1}{r}}\;\left[\frac{{m} \cos\theta}{\sin\theta}+\frac{P_l^{m+1}(\cos\theta)}{P_l^m(\cos\theta)}\right],\nonumber\\
     {\rm Im}F_{1\theta}    &=& 0.
  \end{eqnarray}

Because
\begin{eqnarray}\label{eq:LL-12}
&& \dfrac{{\rm d} {\rm e}^{{\rm i} m \phi}}{{\rm d} \phi}= {\rm i} m {\rm e}^{{\rm i} m \phi},
\end{eqnarray}
we then have
\begin{eqnarray}\label{eq:LL-13}
&&F_{1\phi}   = { \dfrac{1}{r\sin\theta}}\;\dfrac{ \dfrac{{\rm d} \chi_1(\vec{r})}{{\rm d} \phi}}{\chi_1(\vec{r})}= { \dfrac{1}{r\sin\theta}}\; \left[\dfrac{ \dfrac{{\rm d} {\rm e}^{{\rm i} m \phi}}{{\rm d} \phi}}{{\rm e}^{{\rm i} m \phi}}\right]= \frac{{\rm i} \; m}{r\sin\theta}.
\end{eqnarray}
thus
\begin{eqnarray}
    {\rm Re}F_{1\phi}    &=& 0,\nonumber\\
     {\rm Im}F_{1\phi}    &=& \frac{m}{r\sin\theta}.
  \end{eqnarray}
Finally we obtain
\begin{eqnarray}\label{eq:LL-14}
&&\vec{F}_{1}    = \hat{e}_r F_{1r}+\hat{e}_\theta F_{1\theta}+\hat{e}_\phi F_{1\phi}={\rm Re}\vec{F}_1+{\rm i}\; {\rm Im}\vec{F}_1,\nonumber\\
&&{\rm Re}\vec{F}_1 = \hat{e}_r \left[-\frac{1}{2 r}+\frac{ \sqrt{\epsilon}\; [J_{\nu-1}(\sqrt{\epsilon} r)-J_{\nu+1}(\sqrt{\epsilon} r)]}{ 2 J_{\nu}(\sqrt{\epsilon} r)}\right]+\hat{e}_\theta { \frac{1}{r}}\;\left[\frac{{m} \cos\theta}{\sin\theta}+\frac{P_l^{m+1}(\cos\theta)}{P_l^m(\cos\theta)}\right], \nonumber\\
&&{\rm Im}\vec{F}_1 = \hat{e}_\phi { \frac{m}{r\sin\theta}}.
\end{eqnarray}
Similarly, we have
\begin{eqnarray}\label{eq:LL-14-b}
&&\vec{F}_{2}    = \hat{e}_r F_{2r}+\hat{e}_\theta F_{2\theta}+\hat{e}_\phi F_{2\phi}={\rm Re}\vec{F}_2+{\rm i}\; {\rm Im}\vec{F}_2, \nonumber\\
&&{\rm Re}\vec{F}_2 = \hat{e}_r \left[-\frac{1}{2 r}+\frac{ \sqrt{\epsilon}\; [J_{\nu-1}(\sqrt{\epsilon} r)-J_{\nu+1}(\sqrt{\epsilon} r)]}{ 2 J_{\nu}(\sqrt{\epsilon} r)}\right]+\hat{e}_\theta { \frac{1}{r}}\;\left[\frac{{(m+1)} \cos\theta}{\sin\theta}+\frac{P_l^{m+2}(\cos\theta)}{P_l^{m+1}(\cos\theta)}\right], \nonumber\\
&&{\rm Im}\vec{F}_2 = \hat{e}_\phi { \frac{m+1}{r\sin\theta}}.
\end{eqnarray}

\subsection{The Calculation of the Probability $P_{\rm S}$}

      One may notice that the up-component of the wavefunction (\ref{eq:MM-4})
      is almost the wavefunction (\ref{eq:J-21}) of an free electron with the fixed quantum numbers $l$ and $m$ (except that $\nu$ takes half-integer values for the latter), while the down-component of the wavefunction (\ref{eq:MM-4})
      is almost the wavefunction (\ref{eq:J-21}) of an free electron with the fixed quantum numbers $l$ and $m+1$ (except that $\nu$ takes half-integer values for the latter). Namely
      \begin{eqnarray}
            \Psi_{\rm S}(\vec{r}) &\propto&
            \dfrac{J_{\nu}(\sqrt{\epsilon}r)}{\sqrt{r}} \times
                  \dfrac{1}{\sqrt{2l+1}} \left[\begin{array}{c}
                              \sqrt{l+m+1} \;Y_{lm}(\theta, \phi) \\
                              \sqrt{l-m} \;Y_{l,m+1}(\theta, \phi)\\
                        \end{array}\right] \nonumber \\
            &=& \mathcal{N}\Biggl\{\sqrt{\dfrac{l+m+1}{2l+1}} R(r)
            Y_{lm}(\theta, \phi)\begin{bmatrix}
                  1 \\ 0
            \end{bmatrix} +\sqrt{\dfrac{l-m}{2l+1}} R(r) Y_{l,m+1}(\theta, \phi)\begin{bmatrix}
                  0 \\ 1
            \end{bmatrix}\Biggr\}.
            \end{eqnarray}
      This property enable us to make a comparison between the probability of the spin AB effect
      \begin{eqnarray}\label{PS-1}
            P_{\rm S}&&=|\Psi(\vec{r})|^2=\left[\Psi(\vec{r})\right]^{\dagger} \Psi(\vec{r})\nonumber\\
      &&= \mathcal{N}^2\; \biggr\{|c_1|^2 {\rm e}^{2 \int^{\mathcal{L}_1(\vec{r})}\,{\rm Re}\vec{F}_1 \cdot {\rm d}\vec{r}'}
            \times \left[1+{\rm e}^{2 \oint{\rm Re}\vec{F}_1\cdot {\rm d}\vec{r}'}
            +2 \; {\rm e}^{\oint {\rm Re}\vec{F}_1\cdot {\rm d}\vec{r}'}\; \cos\biggr(\oint {\rm Im}\vec{F}_1 \cdot {\rm d}\vec{r}'\biggr) \right]\nonumber\\
            &&\;\;\;\;\;+\; |c_2|^2 {\rm e}^{2 \int^{\mathcal{L}_1(\vec{r})}\,{\rm Re}\vec{F}_2 \cdot {\rm d}\vec{r}'}
            \times \left[1+{\rm e}^{2 \oint {\rm Re}\vec{F}_2\cdot {\rm d}\vec{r}'}
            +2 \; {\rm e}^{\oint {\rm Re}\vec{F}_2\cdot {\rm d}\vec{r}'}\; \cos\biggr(\oint {\rm Im}\vec{F}_2 \cdot {\rm d}\vec{r}'\biggr) \right]\biggr\},
      \end{eqnarray}
      and the probability of the free electron interference $P'_{\rm O}$
      \begin{eqnarray}\label{PO-1}
            P'_{\rm O}= \mathcal{(N')}^2\;{\rm e}^{2 \int^{\mathcal{L}_1(\vec{r})}\,{\rm Re}\vec{F} \cdot {\rm d}\vec{r}'}
            \times \left[1+{\rm e}^{2 \oint{\rm Re}\vec{F}\cdot {\rm d}\vec{r}'}
            +2 \; {\rm e}^{\oint {\rm Re}\vec{F}\cdot {\rm d}\vec{r}'}\; \cos\biggr(\oint {\rm Im}\vec{F} \cdot {\rm d}\vec{r}'\biggr) \right],
      \end{eqnarray}
      where $\vec{F}$ is just $\vec{F}_1$ in Eq. (\ref{eq:LL-14}) merely by taking $\nu=l+1/2$. Since the phase $\oint {\rm Im}\vec{F}_1 \cdot {\rm d}\vec{r}'$ in Eq. (\ref{PS-1}) and the phase $\oint {\rm Im}\vec{F} \cdot {\rm d}\vec{r}'$ in Eq. (\ref{PO-1}) are contributed from the same term $Y_{lm}(\theta, \phi)$ (for $m\neq 0$), and they are independent of the radial wavefunctions $R(r)$ and $R_0(r)$). Thus we may let
      \begin{equation}
            \delta\equiv\oint {\rm Im}\vec{F} \cdot {\rm d}\vec{r}'=\oint {\rm Im}\vec{F}_1 \cdot {\rm d}\vec{r}'.
      \end{equation}
      Meanwhile, by observing Eq. (\ref{eq:LL-14}) and Eq. (\ref{eq:LL-14-b}), we have
      \begin{equation}
            \oint {\rm Im}\vec{F}_2 \cdot {\rm d}\vec{r}'=\dfrac{(m+1)}{m} \delta.
      \end{equation}
      Furthermore, we denote
      \begin{eqnarray}
            &&\mu_1 =\int^{\mathcal{L}_1 (\vec{r})}\,{\rm Re}\vec{F}_1 \cdot
                  {\rm d}\vec{r}',\quad
            \mu_2 =\oint{\rm Re}\vec{F}_1\cdot {\rm d}\vec{r}',\nonumber\\
            &&\mu_3 =\int^{\mathcal{L}_2 (\vec{r})}\,{\rm Re}\vec{F}_2 \cdot
                  {\rm d}\vec{r}',\quad
            \mu_4 =\oint{\rm Re}\vec{F}_2 \cdot {\rm d}\vec{r}',\nonumber\\
            &&\mu'_1 =\int^{\mathcal{L}_1 (\vec{r})}\,{\rm Re}\vec{F} \cdot
                  {\rm d}\vec{r}',\quad
            \mu'_2 =\oint{\rm Re}\vec{F}\cdot {\rm d}\vec{r}',
      \end{eqnarray}
      which means
        \begin{eqnarray}\label{eq:PS}
             P_{\rm S} &=& \mathcal{N}^2\; \biggr\{|c_1|^2\,{\rm e}^{2\mu_1}
                              \times \left(1+{\rm e}^{2\mu_2}
                                    +2 \; {\rm e}^{\mu_2}\; \cos\delta\right)
                        +|c_2|^2\,{\rm e}^{2\mu_3}\times \left[1+{\rm e}^{2 \mu_4}
                              +2 \; {\rm e}^{\mu_4}\; \cos\delta\right]\biggr\}\nonumber \\
                       &=& \mathcal{N}^2\; \Biggr\{{\rm e}^{2\mu_1}\times \left(
                              1+{\rm e}^{2\mu_2}
                              +2 \; {\rm e}^{\mu_2}\; \cos\delta\right)
                        +\dfrac{{\rm e}^{2\mu_3}}{(l+m+1)^2}\,\times \left[
                              1+{\rm e}^{2 \mu_4}
                              +2 \; {\rm e}^{\mu_4}\; \cos\left(\dfrac{m+1}{m} \delta\right)\right]\Biggr\},
      \end{eqnarray}
       \begin{eqnarray}\label{eq:POPrime}
             P'_{\rm O} &=& \mathcal{N'}^2\;\Bigr[{\rm e}^{2\mu'_1}\times \left(
                        1+{\rm e}^{2\mu'_2}
                        +2 \; {\rm e}^{\mu'_2}\; \cos\delta\right)\Bigr]
                        \propto \left[1+\dfrac{2 {\rm e}^{\mu'_2}}{1+{\rm e}^{2\mu'_2}}\;\cos\delta \right].
      \end{eqnarray}
\begin{remark}

One may notice that the expression of $P'_{\rm O}$ in Eq. (\ref{eq:POPrime}) is the same as that of $P_{\rm O}$ in Eq. (\ref{eq:v-19c}) except a visibility factor $V=2 {\rm e}^{\mu'_2}/(1+{\rm e}^{2\mu'_2})$, thus for convenient we shall adopt the latter to do the numerical computation.
\end{remark}

\subsubsection{Numerical Simulation}

\begin{figure}[!h]
            \includegraphics[width=56mm]{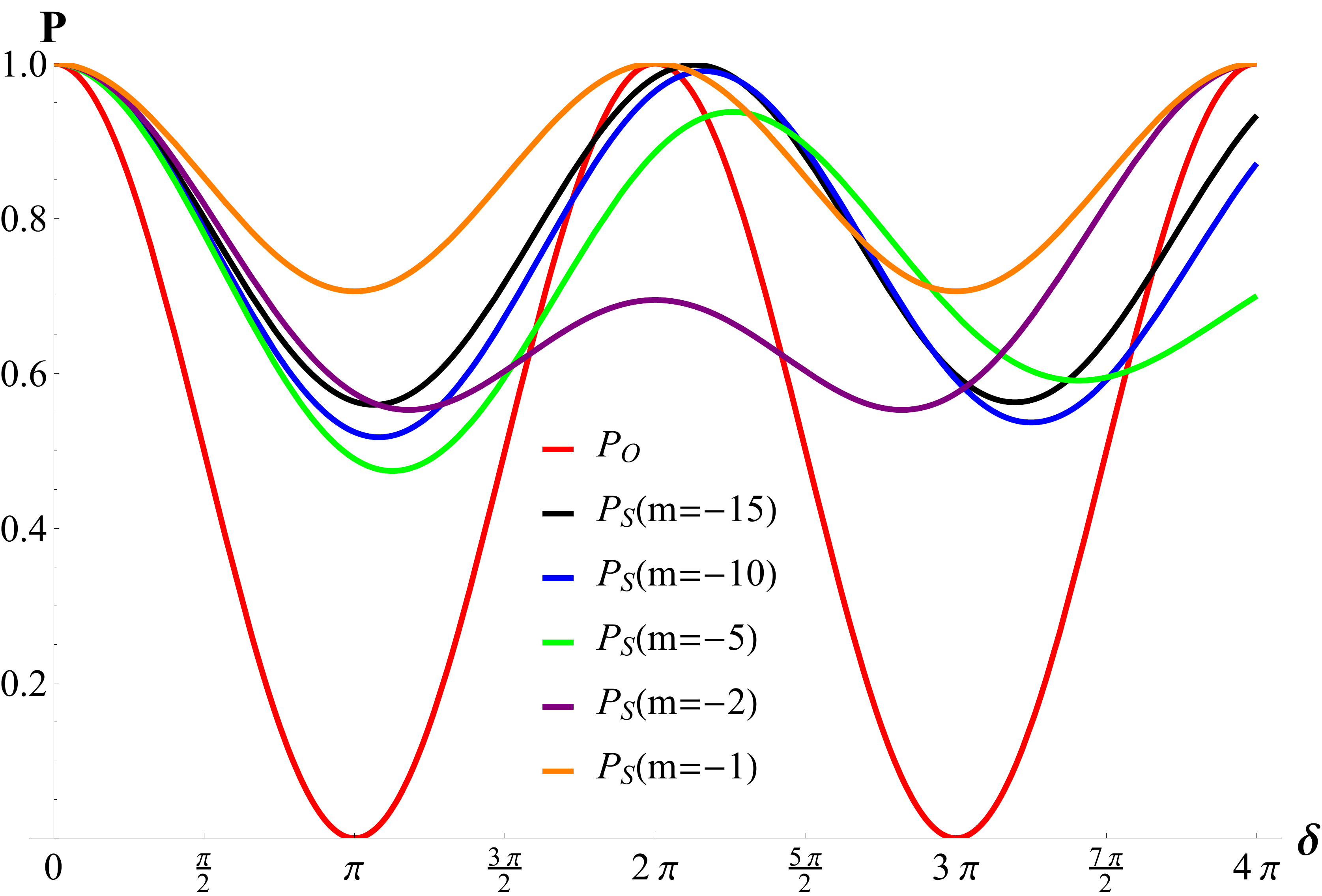}
            \includegraphics[width=56mm]{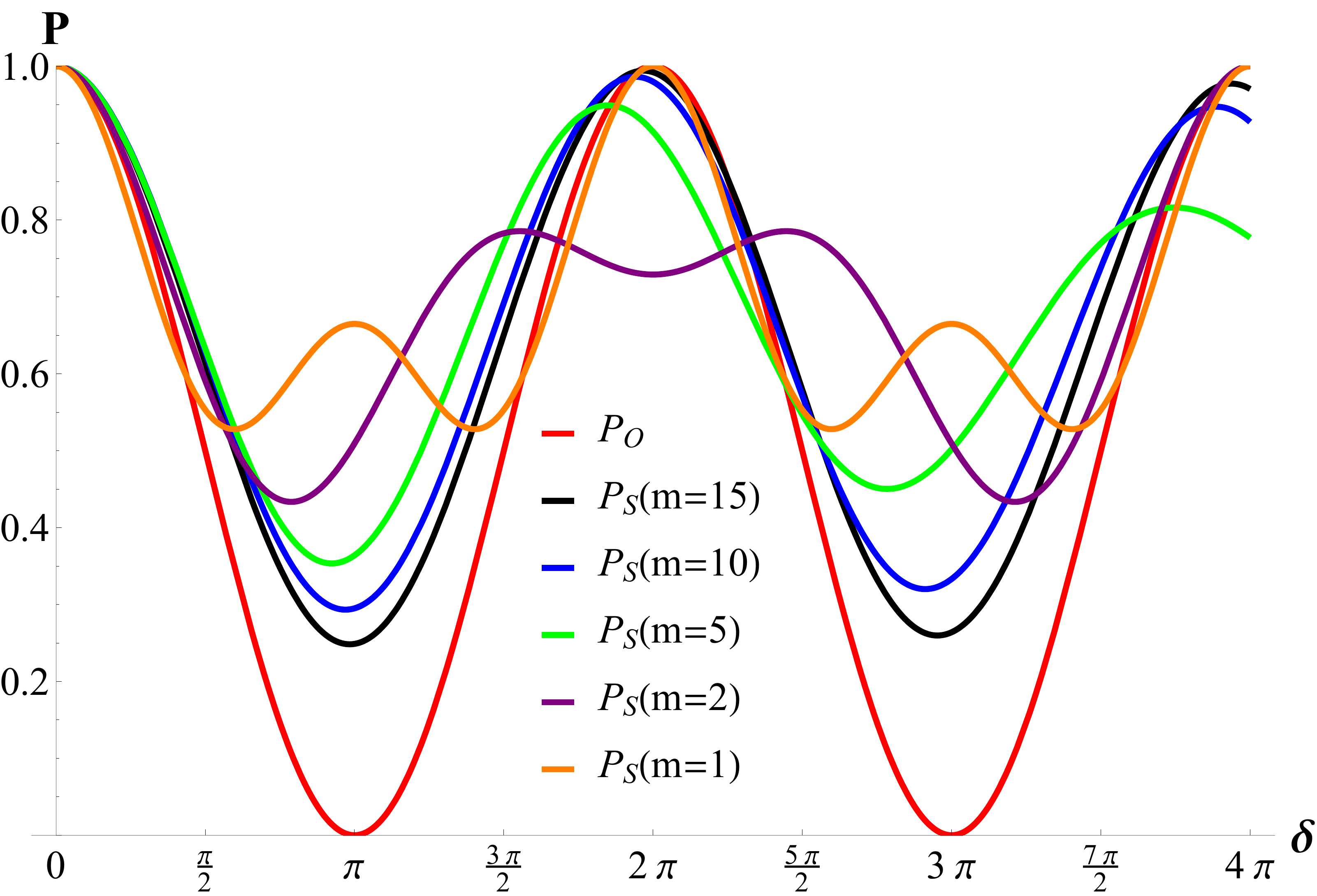}
            \includegraphics[width=56mm]{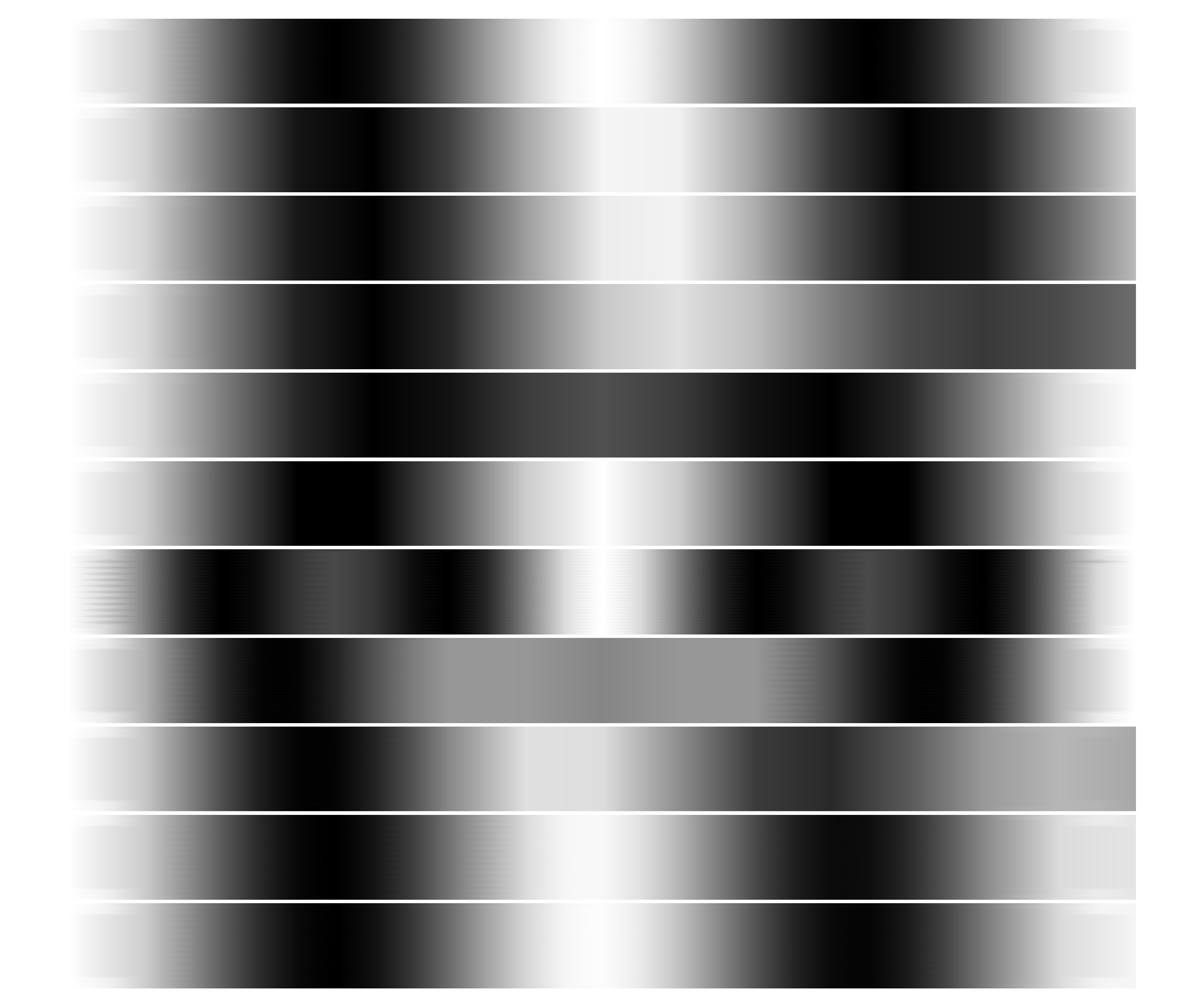}
                  \caption{Numerical simulation of the probability
                  $P_{\rm O}$ (red curve) and the probability $P_{\rm S}$, in the case of $\mu_1 =-1$, $\mu_2 =0$, $\mu_3 =1$, $\mu_4 =2$, $l=20$, and $m\in\{-15,-10,-5,-2,-1,1,2,5,10,15\}$. The interference pattern of free electrons is posed on the top of the rightmost interference figure, next to which is the situation of $m=-15$; while the other fringes are arranged from top to bottom corresponding to the same order in the set of $m$. One may find that, the amplitudes of all curves are compressed relative to the line of $P=1$. Obvious phase shift of the green lines ($m=\pm 5$) can be observed. For some values of $m$ (e.g., $m=1, 2$), their corresponding curves are evidently different from that of $P_{\rm O}$.}\label{fig:ns1m}
      \end{figure}

\begin{figure}[!h]
            \includegraphics[width=84mm]{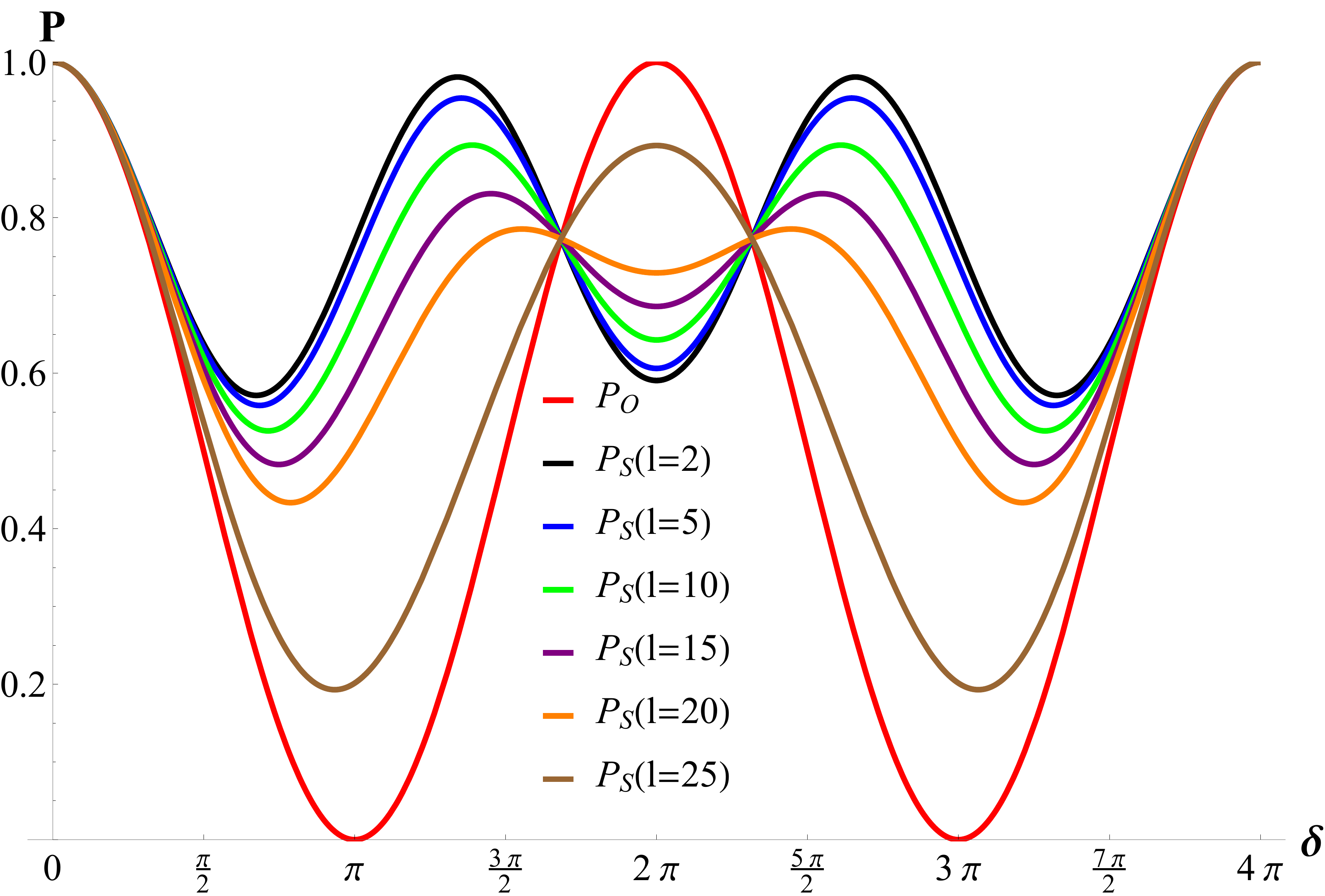}
            \includegraphics[width=84mm]{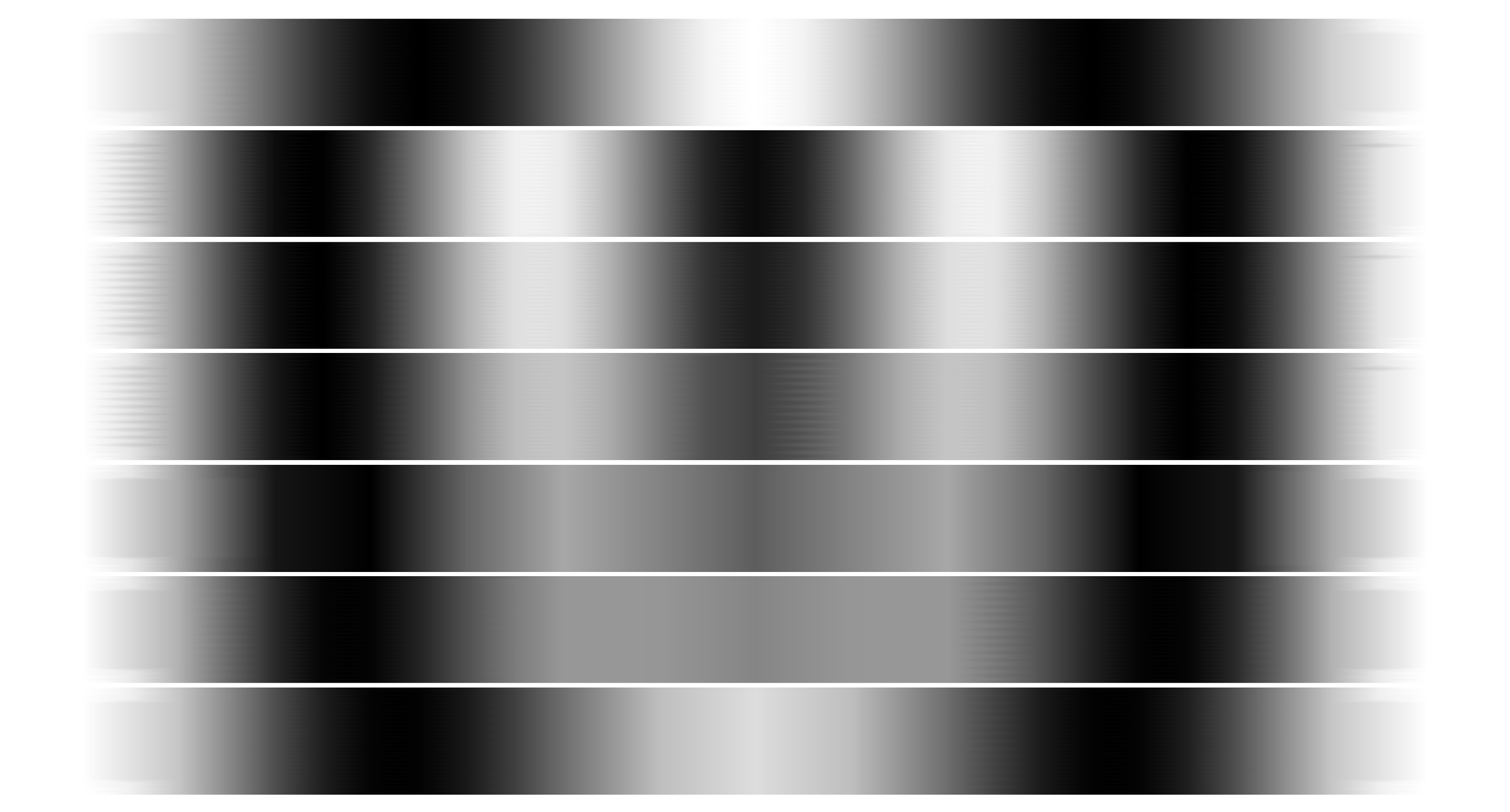}
                  \caption{Numerical simulation of the probability
                  $P_{\rm O}$ (red curve) and the probability $P_{\rm S}$, in the case of $\mu_1 =-1$, $\mu_2 =0$, $\mu_3 =1$, $\mu_4 =2$,
                  $l\in\{2,5,10,15,20,25\}$, and $m=2$. The interference pattern of free electrons is posed on the top of the rightmost interference figure, next to which is the situation of $l=2$; while the other fringes are arranged from top to bottom corresponding to the same order in the set of $l$. One may find that all interference fringes
                  are symmetric relative to the baseline of $\delta=2\pi$, and the
                  increase in $l$ shows the advolution between the patterns
                  $P_{\rm O}$ and $P_{\rm S}$.}\label{fig:ns1l}
      \end{figure}

\begin{figure}[!h]
            \includegraphics[width=56mm]{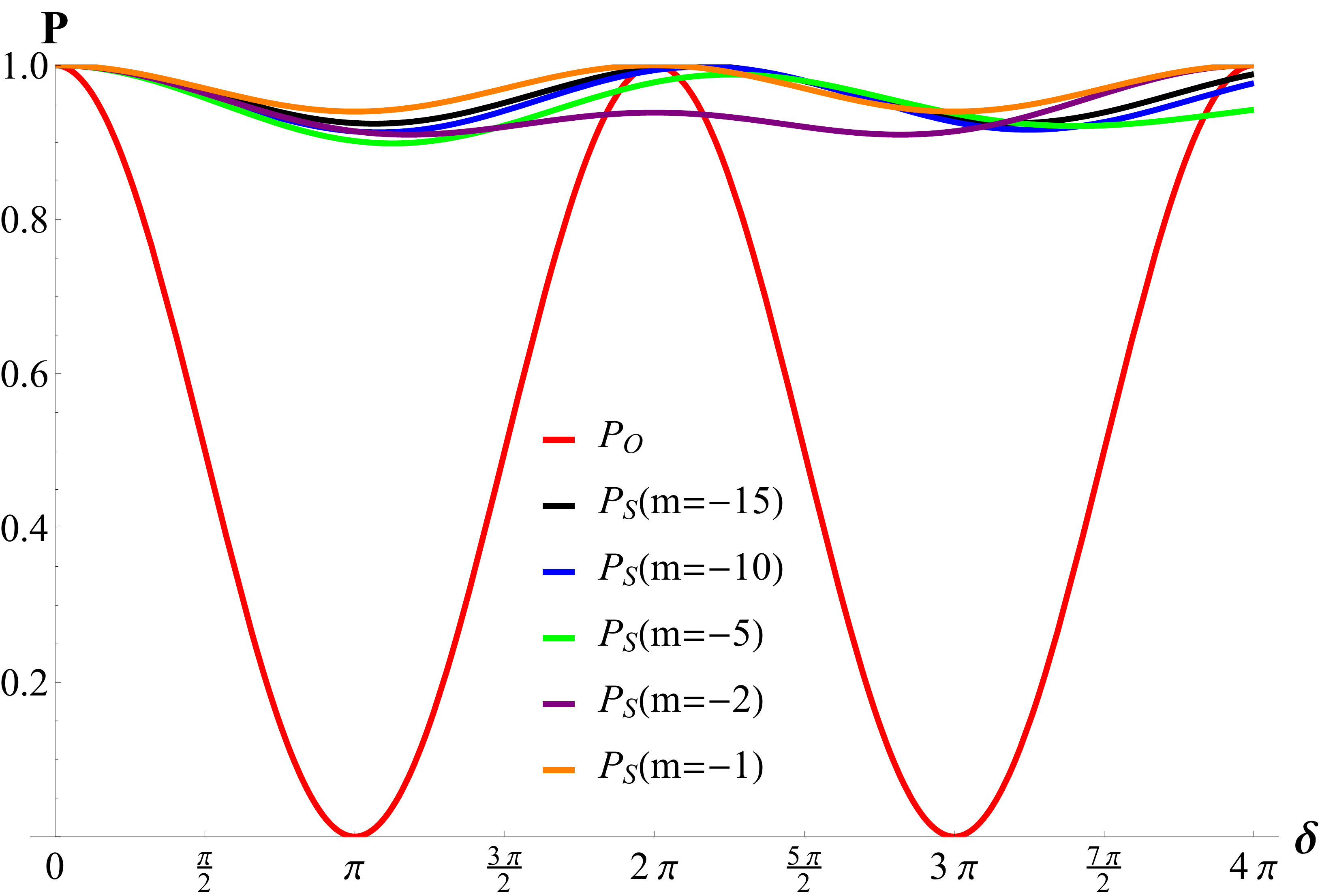}
            \includegraphics[width=56mm]{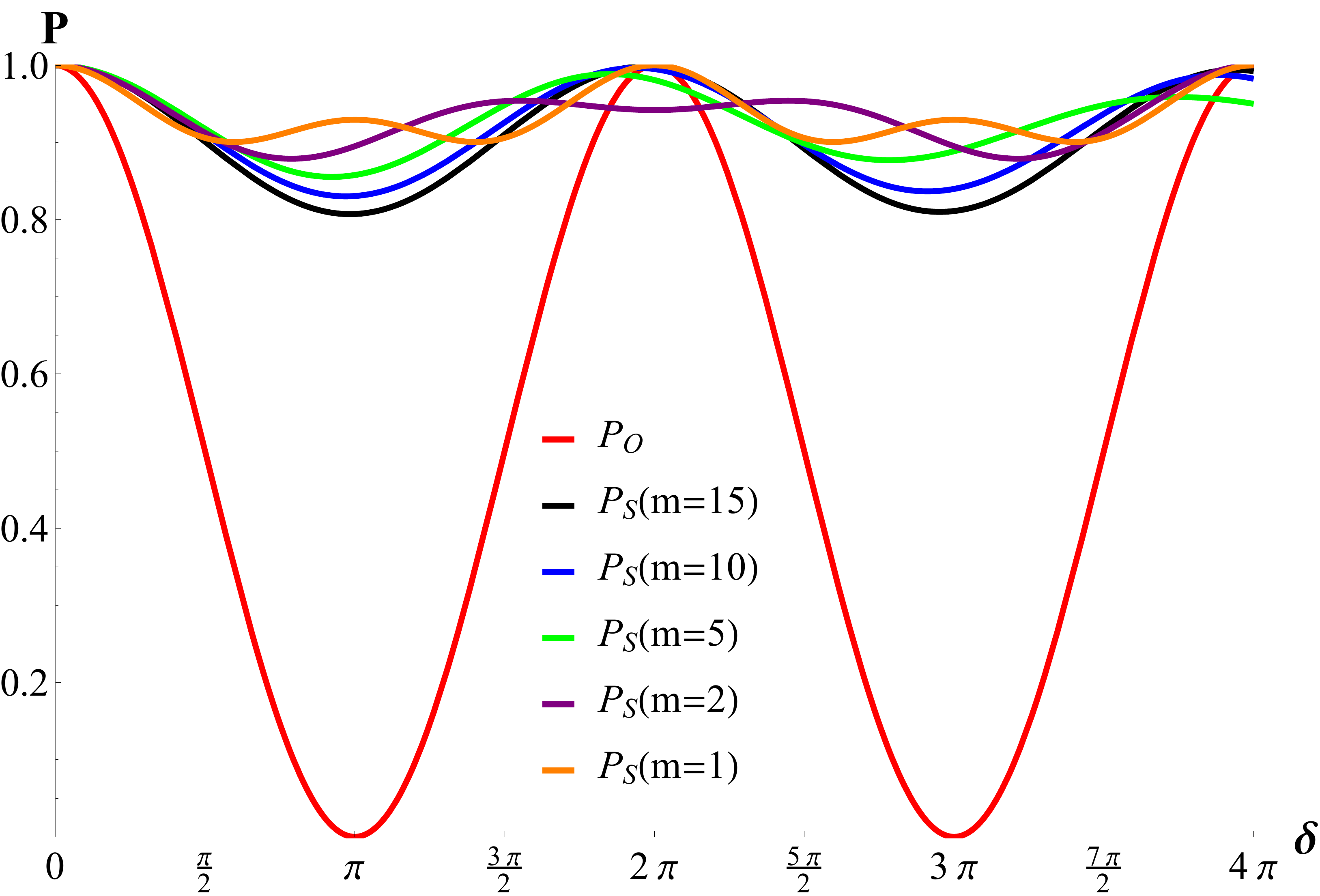}
            \includegraphics[width=56mm]{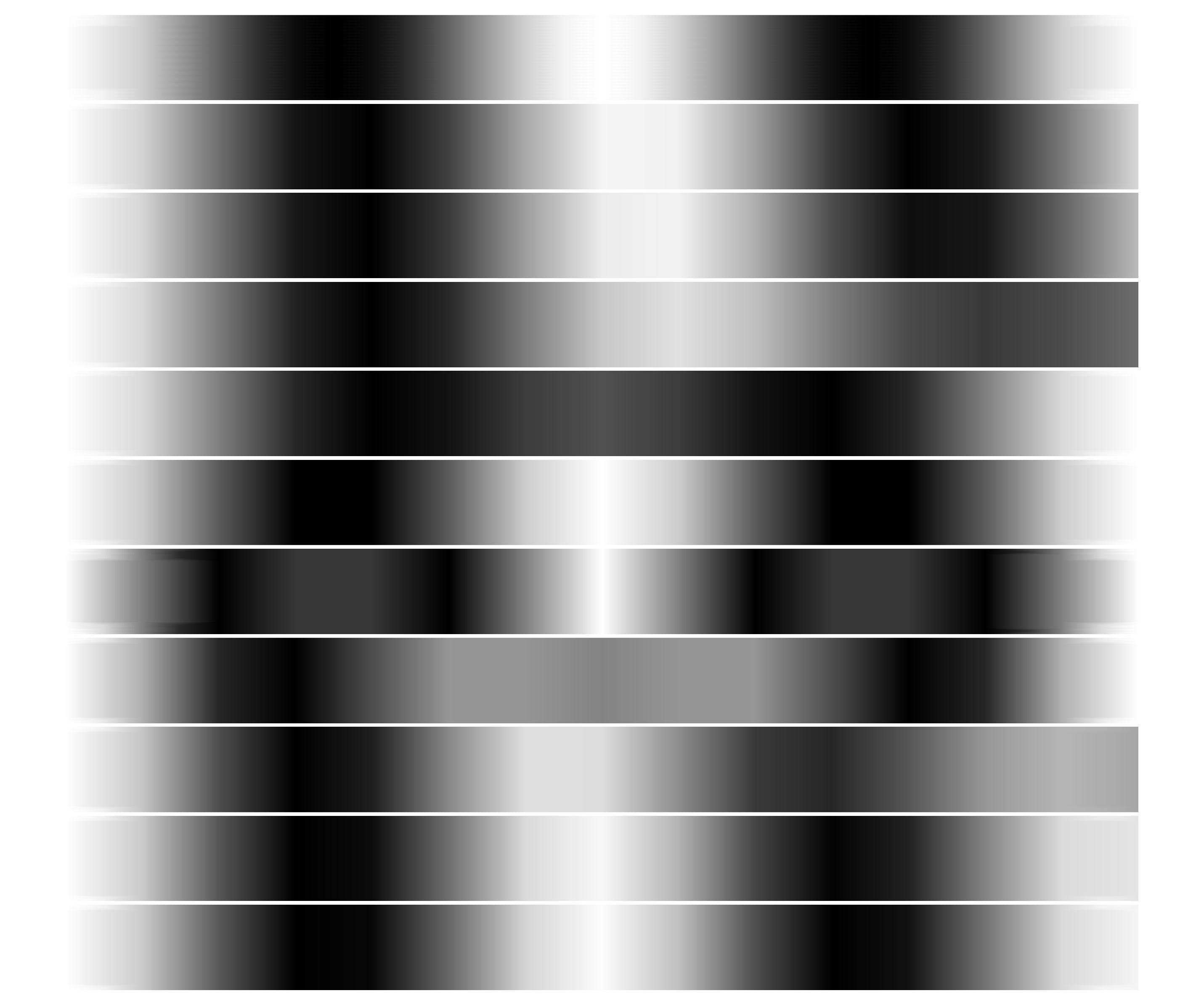}
                  \caption{Numerical simulation of the probability
                  $P_{\rm O}$ (red curve) and the probability $P_{\rm S}$, in the case of $\mu_1 =1$, $\mu_2 =2$, $\mu_3 =3$, $\mu_4 =4$, $l=20$, and
                  $m\in\{-15,-10,-5,-2,-1,1,2,5,10,15\}$. The interference pattern of free electrons is posed on the top of the rightmost interference figure, next to which is the situation of $m=-15$; while the other fringes are arranged from top to bottom corresponding to the same order in the set of $m$. One may find that,the amplitudes of all curves are compressed extremely relative to the line of $P=1$. Obvious phase shift of the green lines ($m=\pm 5$) can be observed.}\label{fig:ns2m}
      \end{figure}

\begin{figure}[!h]
            \includegraphics[width=84mm]{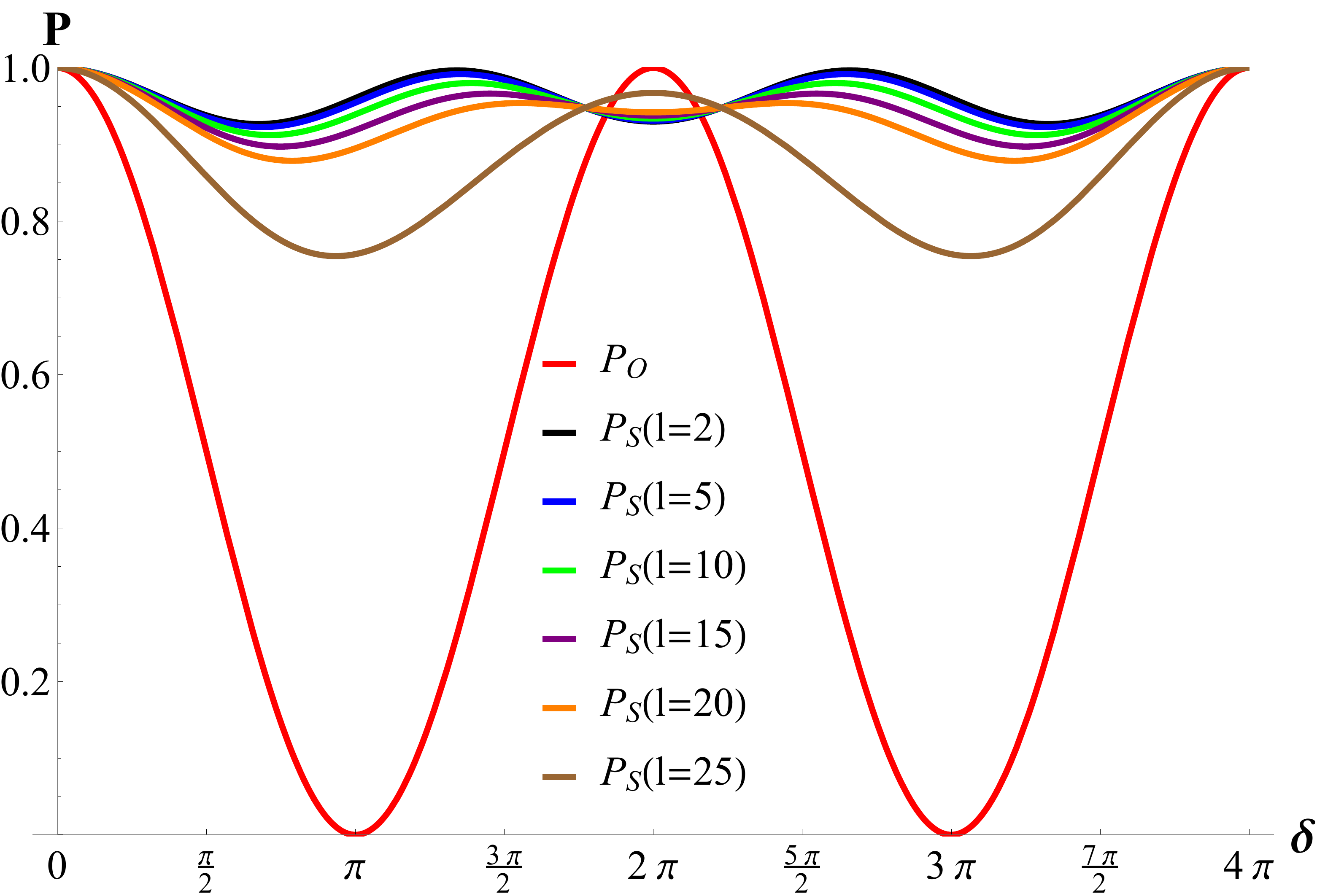}
            \includegraphics[width=84mm]{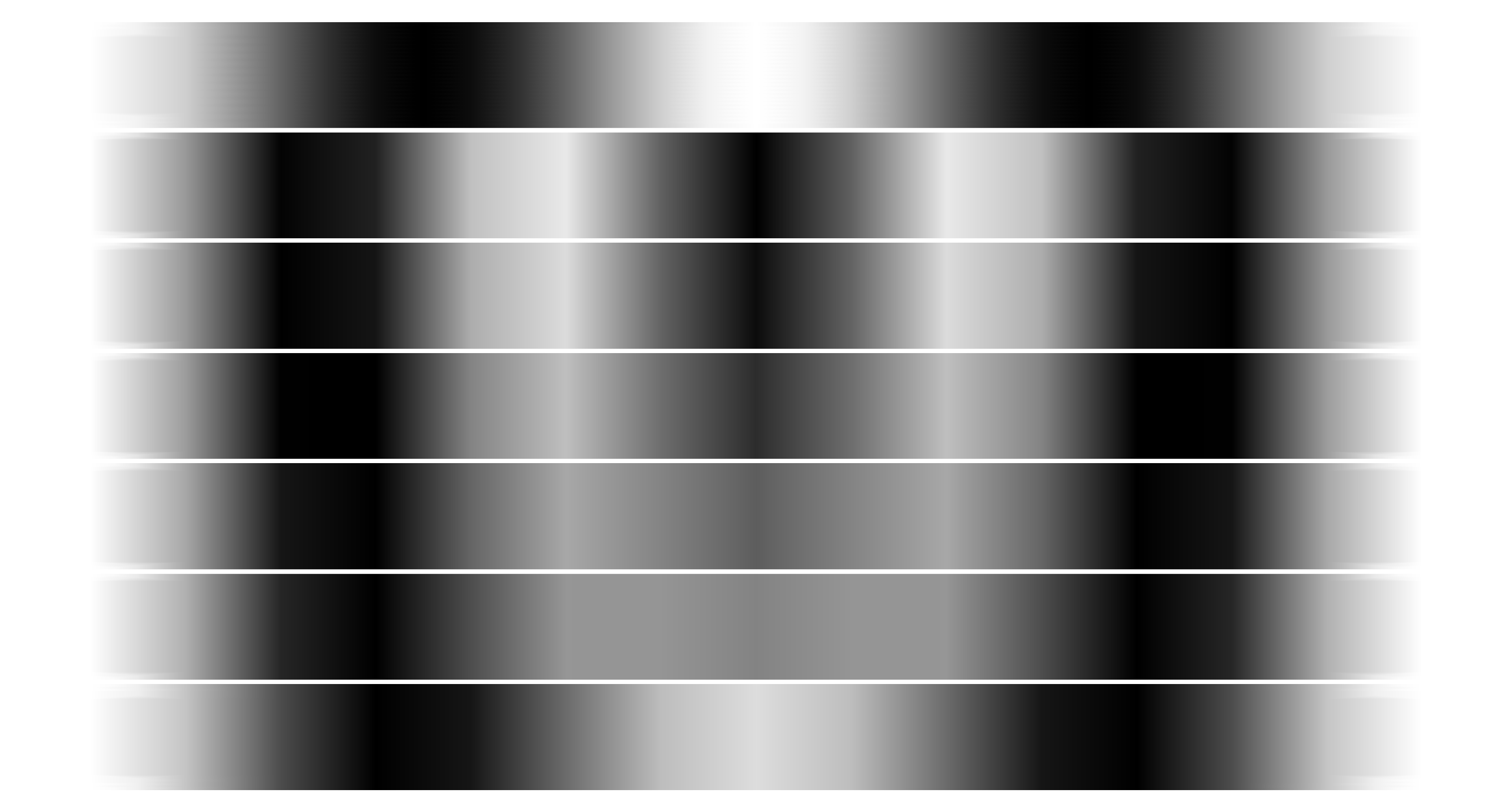}
                  \caption{Numerical simulation of the probability
                  $P_{\rm O}$ (red curve) and the probability $P_{\rm S}$, in the case of $\mu_1 =1$, $\mu_2 =2$, $\mu_3 =3$, $\mu_4 =4$,
                  $l\in\{2,5,10,15,20,25\}$, and $m=2$. The interference pattern of free electrons is posed on the top of the rightmost interference figure, next to which is the situation of $l=2$; while the other fringes are arranged from top to bottom corresponding to the same order in the set of $l$. One may find that,all interference fringes
                  are symmetric relative to the baseline of $\delta=2\pi$, and the
                  increase in $l$ shows the advolution between the patterns
                  $P_{\rm O}$ and $P_{\rm S}$.}\label{fig:ns2l}
      \end{figure}

\begin{figure}[!h]
            \includegraphics[width=56mm]{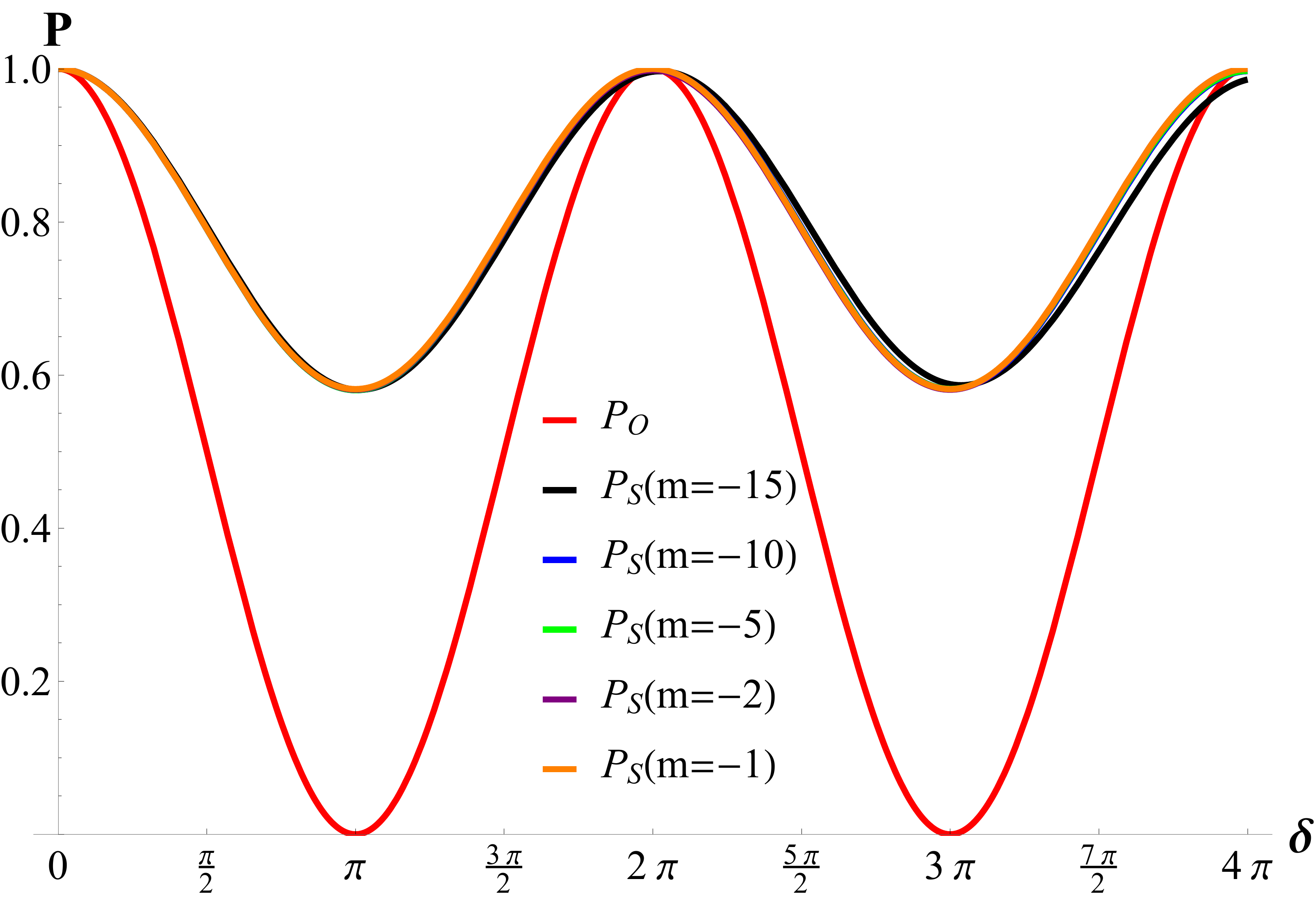}
            \includegraphics[width=56mm]{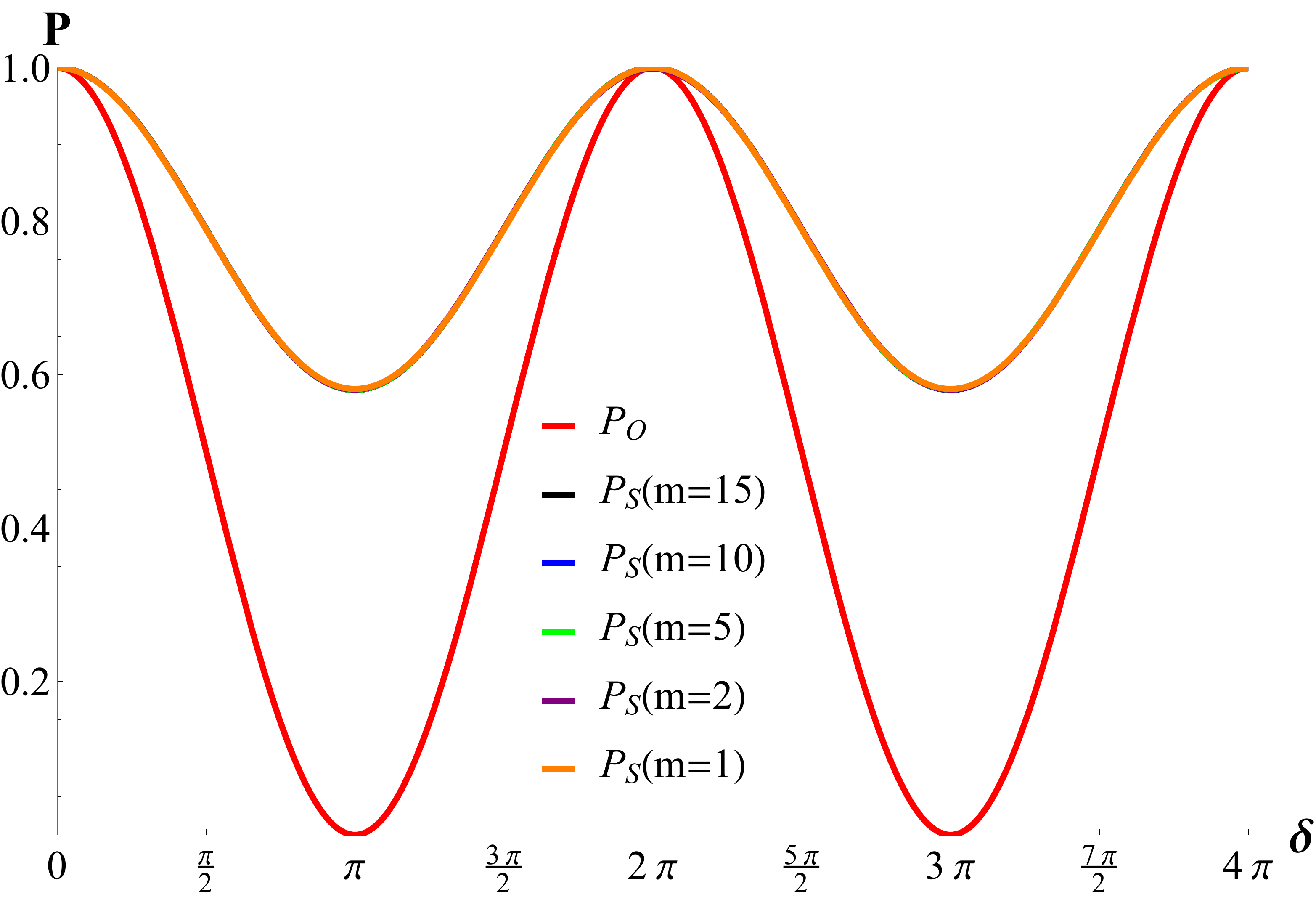}
            \includegraphics[width=56mm]{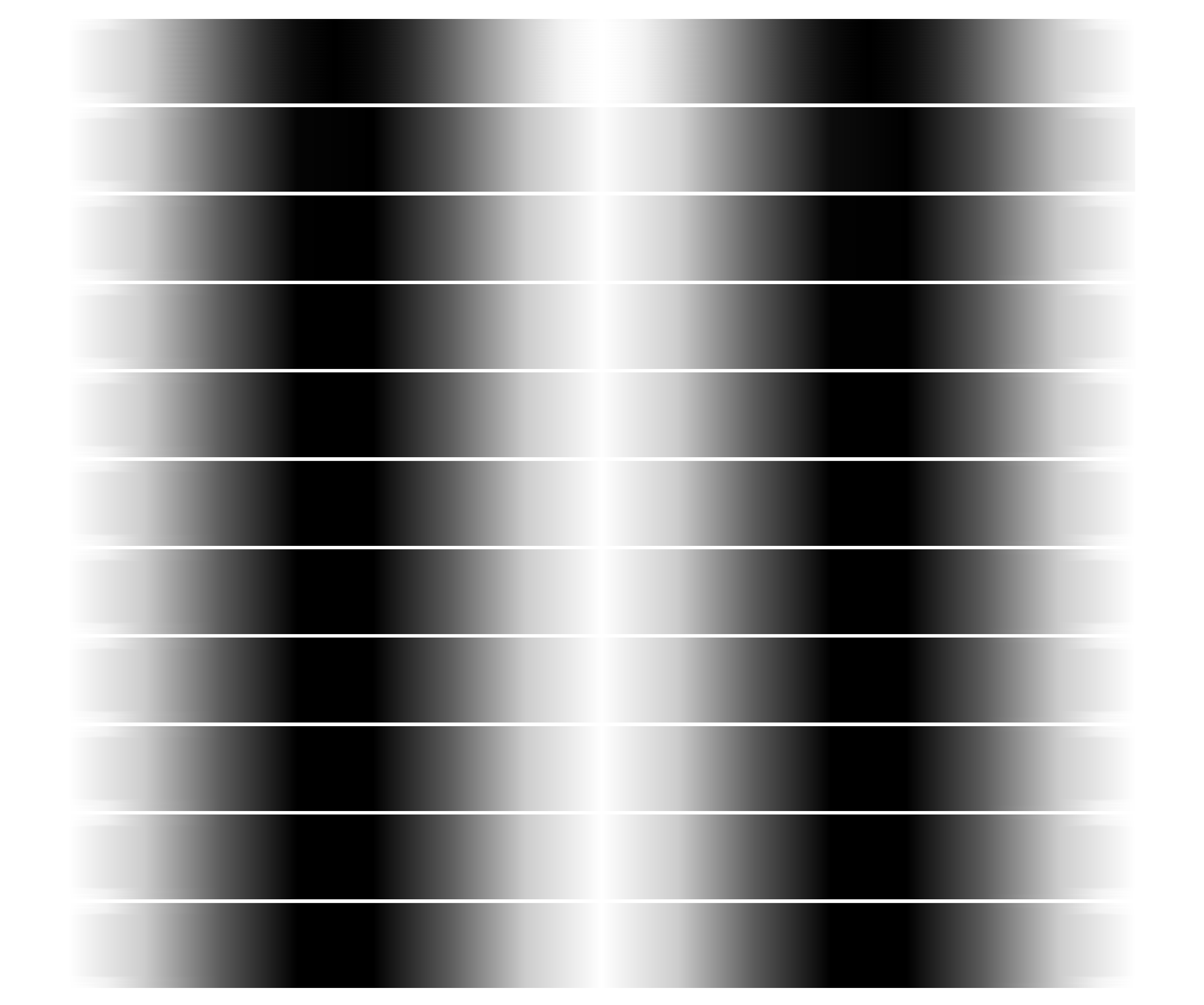} \\
            \includegraphics[width=84mm]{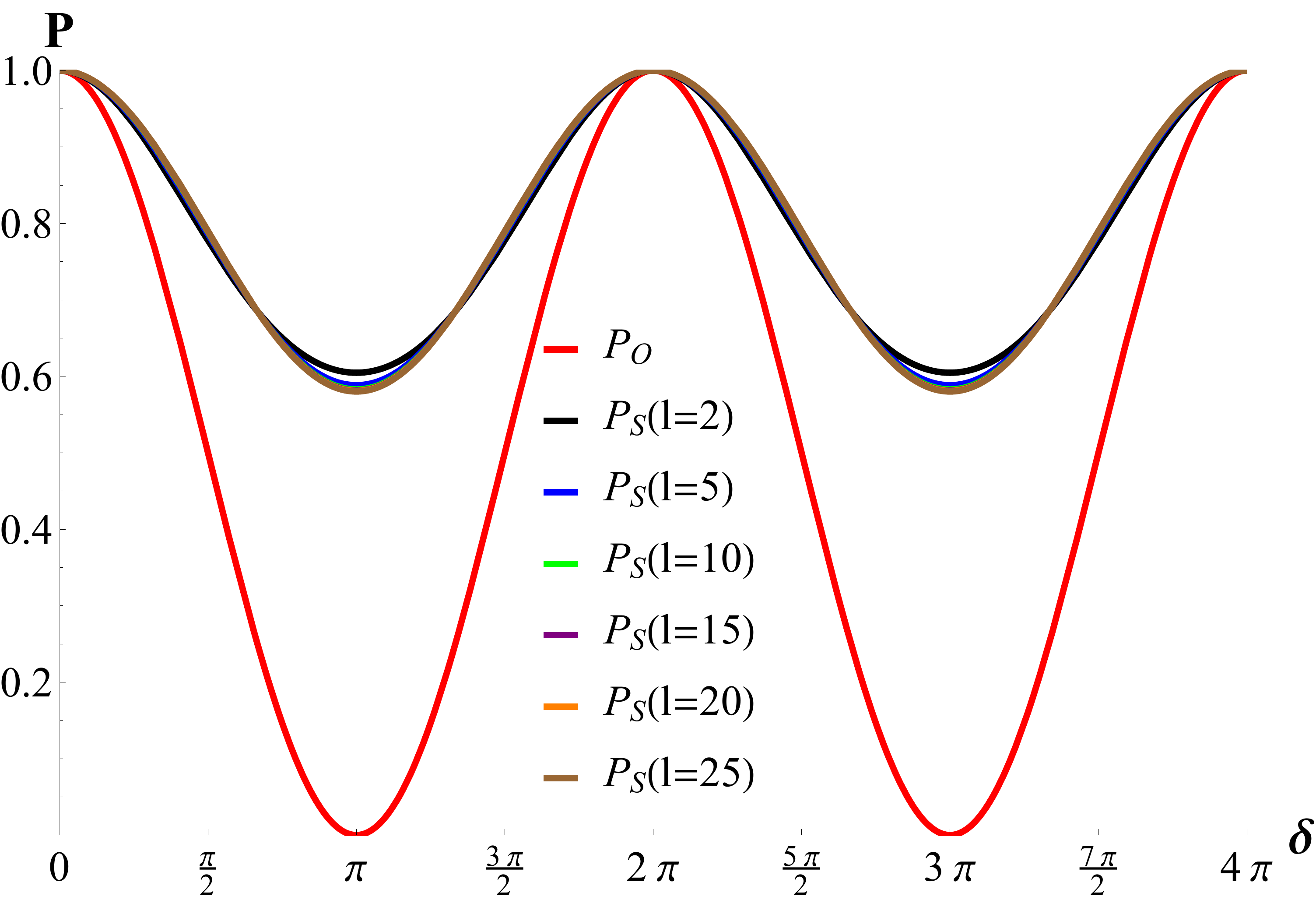}
            \includegraphics[width=84mm]{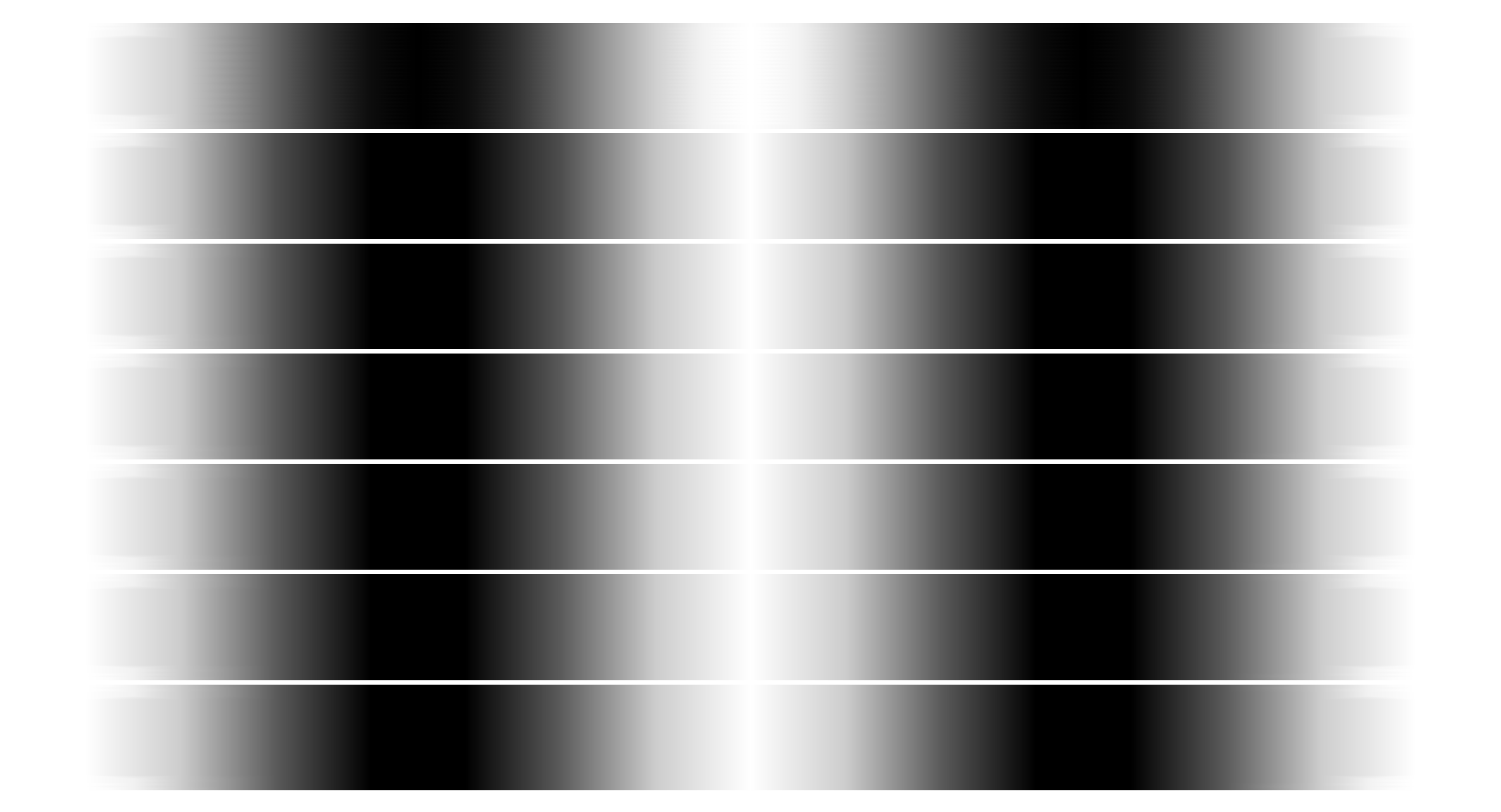}
                  \caption{Numerical simulation of the probability
                  $P_{\rm O}$ (red curve) and the probability $P_{\rm S}$. [Above] in the case of $\mu_1 =1$, $\mu_2 =2$, $\mu_3 =1$, $\mu_4 =2$,
                  (namely the weigh of $\mu_{1(2)}$ and $\mu_{3(4)}$ is equal), $l=16$, and $m\in\{-15,-10,-5,-2,-1,1,2,5,10,15\}$. The interference pattern of free electrons is posed on the top of the rightmost interference figure, next to which is the situation of $m=-15$; while the other fringes are arranged from top to bottom corresponding to the same order in the set of $m$. [Below] In the case of $\mu_1 =1$, $\mu_2 =2$, $\mu_3 =1$, $\mu_4 =2$,
                  (namely the weigh of $\mu_{1(2)}$ and $\mu_{3(4)}$ is equal),
                  $l\in\{2,5,10,15,20,25\}$, and $m=1$. The interference pattern of free electrons is posed on the top of the rightmost interference figure, next to which is the situation of $l=2$; while the other fringes are arranged from top to bottom corresponding to the same order in the set of $l$. One may find, no obvious alteration for
                  $P_{\rm S}$ will be observed relative to $P_{\rm O}$ in both cases.}\label{fig:ns3}
      \end{figure}

      In this section, we compare the curves of $P_{\rm O}$ (Eq. (\ref{eq:v-19c}))
      with $P_{\rm S}$ (Eq. (\ref{eq:PS})) under the different parameters $\mu_1$, $\mu_2$, $\mu_3$, $\mu_4$, $l$ and $m$. The curves and the corresponding interference patterns are shown from Fig. \ref{fig:ns1m} to Fig. \ref{fig:ns3}.

      \newpage


      \newpage

\newpage

\part{The Dirac Hamiltonian Involving the Spin Vector Potential}

 The spin AB Hamiltonian $H_{\rm S}$ in Eq. (\ref{eq:v-3}) is written in a non-relativistic version. In this section, let us we consider the spin AB Hamiltonian in its relativistic version. As a consequence, we find that some important types of spin interactions can be naturally emerged. Explicitly, let us consider a Dirac particle (marked `2') with charge $-ge$ and mass $M$, which moves under the spin vector potential induced by the spin $\vec{S}_1=\hbar\vec{\sigma}_1 /2$ of an electron (marked `1') with charge $-e$, namely
 \begin{eqnarray}\label{eq:v-24-a}
\vec{\mathcal{A}}=g\frac{\vec{r}\times\vec{S}_1}{r^2},
\end{eqnarray}
 where $\vec{r}$ depicts the distance vector between the Dirac particle and the electron. Then the Dirac Hamiltonian is given by $(c=1)$
\begin{eqnarray}\label{eq:v-24}
 \mathcal{H}_{\rm Dirac} &=& \vec{\alpha}_2 \cdot\left(\vec{p}-\vec{\mathcal{A}}\right)+\beta M,
\end{eqnarray}
 where
             \begin{equation}
               \vec{\alpha}_2=\begin{bmatrix}
                  0 & \vec{\sigma}_2 \\
                  \vec{\sigma}_2 & 0
               \end{bmatrix}=\sigma_x\otimes\vec{\sigma}_2,\quad  \quad \beta=\begin{bmatrix}
                  \openone & 0 \\
                  0 & -\openone
               \end{bmatrix}=\sigma_z\otimes\openone.
            \end{equation}
are the Dirac matrices, and $\openone$ is the $2\times 2$ identity matrix.

Let us expand the Dirac Hamiltonian in Eq. (\ref{eq:v-24}), i.e.,
\begin{eqnarray}\label{eq:v-25-01}
\mathcal{H}_{\rm Dirac} &=&\ \vec{\alpha}_2 \cdot\left(\vec{p}-\vec{\mathcal{A}}\right)+\beta M\nonumber\\
  &=& \sigma_x \otimes\Bigl[\vec{\sigma}_2 \cdot\left(
                           \vec{p}-\dfrac{g\hbar}{2} \dfrac{\vec{r}\times\vec{\sigma}_1}{r^2}\right)\Bigr]
                        +\beta M \nonumber\\
                     &=& \sigma_x \otimes\Bigl[\vec{\sigma}_2 \cdot\vec{p}
                           -\dfrac{g\hbar}{2} \vec{\sigma}_2 \cdot\dfrac{\vec{r}\times\vec{\sigma}_1}{r^2}\Bigr]
                        +\beta M \nonumber\\
&=&
\sigma_x \otimes\biggl[\vec{\sigma}_2 \cdot\vec{p}
                     -\dfrac{g\hbar}{2} \dfrac{\vec{r}\cdot\bigl(\vec{\sigma}_1 \times\vec{\sigma}_2\bigr)}{r^2}\biggr]
                  +\beta M,
\end{eqnarray}
thus we naturally have the Dzyaloshinsky-Moriya-like (DM) interaction \cite{1958Dzyaloshinsky,1960Moriya}
\begin{eqnarray}\label{eq:DM}
\mathcal{H}_{\rm DM} &=&\vec{r}\cdot\bigl(\vec{\sigma}_1 \times\vec{\sigma}_2\bigr),
\end{eqnarray}
which is shown in Table I.

Similarly, let us study the square operator of the Dirac Hamiltonian, after the careful calculation we have
\begin{eqnarray}\label{eq:v-26}
 \mathcal{H}^2_{\rm Dirac} &=&\openone\otimes\left[
                     \vec{p}^{\; 2} +\dfrac{g^2\,\hbar^2}{2\,r^2}+\dfrac{g\,\hbar}{r^2} (\vec{\sigma}_1 \cdot{\vec{\ell}}) - \dfrac{\hbar^2 g(g-2)}{2 r^4} \left(\vec{r}\cdot\vec{\sigma}_1\right)\left(\vec{r}\cdot\vec{\sigma}_2\right)\right]
                  +\beta^2 M^2.
\end{eqnarray}

\emph{Proof.---}From Eq. (\ref{eq:v-24}), we have
               \begin{eqnarray}\label{sub0}
                               \mathcal{H}^2_{\rm Dirac} &=& \biggl[\vec{\alpha}_2 \cdot\left(\vec{p}-\vec{\mathcal{A}}\right)+\beta M\biggr]^2 \nonumber\\
                    &=& \left[\sigma_x \otimes\Bigl[\vec{\sigma}_2 \cdot\left(
                           \vec{p}-\vec{\mathcal{A}}\right)\Bigr]\right]^2
                        +\beta^2 M^2 \nonumber\\
                     &=& \openone\otimes\biggl\{
                           \left(\vec{p}-\vec{\mathcal{A}}\right)\cdot\left(\vec{p}-\vec{\mathcal{A}}\right)
                           +{\rm i}\,\vec{\sigma}_2 \cdot\Bigl[
                              \left(\vec{p}-\vec{\mathcal{A}}\right)\times\left(\vec{p}-\vec{\mathcal{A}}\right)\Bigr]\biggr\}
                        +\beta^2 M^2 \nonumber\\
               &=& \openone\otimes\biggl\{
                           \left[\vec{p}^{\; 2}-\vec{p}\cdot\vec{\mathcal{A}}-\vec{\mathcal{A}}\cdot\vec{p}+\vec{\mathcal{A}}^2\right]
                           +{\rm i}\,\vec{\sigma}_2 \cdot\Bigl[
                              \vec{p}\times \vec{p}-\vec{p}\times\vec{\mathcal{A}}-\vec{\mathcal{A}}\times \vec{p}+\vec{\mathcal{A}}\times\vec{\mathcal{A}}\Bigr]\biggr\}
                        +\beta^2 M^2.
\end{eqnarray}
Because

\begin{eqnarray}
       \vec{p}\cdot\vec{\mathcal{A}}-\vec{\mathcal{A}}\cdot\vec{p}&=& -{\rm i}\hbar\nabla\cdot\vec{\mathcal{A}}=0, \nonumber\\
     \vec{\mathcal{A}}\cdot\vec{p} &=& \ \dfrac{g\,\hbar}{2}
                        \dfrac{(\vec{r}\times\vec{\sigma}_1)\cdot\vec{p}}{r^2} = -\dfrac{g\,\hbar}{2} \dfrac{(\vec{\sigma}_1\times\vec{r})\cdot\vec{p}}{r^2} = -\dfrac{g\,\hbar}{2} \dfrac{\vec{\sigma}_1 \cdot(\vec{r}\times\vec{p})}{r^2}= -\dfrac{g\,\hbar}{2\,r^2} \vec{\sigma}_1 \cdot\ell,\nonumber\\
    \vec{\mathcal{A}}^2 &=&\ \left(\dfrac{g\,\hbar}{2} \dfrac{\vec{r}\times\vec{\sigma}_1}{r^2}\right)\cdot
                        \left(\dfrac{g\,\hbar}{2} \dfrac{\vec{r}\times\vec{\sigma}_1}{r^2}\right) = \dfrac{g^2\,\hbar^2}{4\,r^4} (\vec{r}\times\vec{\sigma}_1)\cdot(\vec{r}\times\vec{\sigma}_1)=  \dfrac{g^2\,\hbar^2}{4\,r^4} \bigl[(\vec{r}\times\vec{\sigma}_1)\times\vec{r}\bigr]\cdot\vec{\sigma}_1 \nonumber\\
                     &=&\dfrac{g^2\,\hbar^2}{4\,r^4} \bigl[(r^2 \vec{\sigma}_1)-(\vec{\sigma}_1 \cdot\vec{r})\vec{r}\bigr]\cdot\vec{\sigma}_1 = \dfrac{g^2\,\hbar^2}{4\,r^4} \bigl[3r^2-r^2\bigr] = \dfrac{g^2\,\hbar^2}{2\,r^2},
     \end{eqnarray}

we then have
               \begin{eqnarray}\label{sub1}
               \left[\vec{p}^{\; 2} -\vec{\mathcal{A}}\cdot\vec{p}-\vec{p}\cdot\vec{\mathcal{A}}+\vec{\mathcal{A}}^2\right]
                     &=&\ \vec{p}^2 -\vec{\mathcal{A}}\cdot\vec{p}
                -\left(\vec{\mathcal{A}}\cdot\vec{p}-{\rm i}\hbar\nabla\cdot\vec{\mathcal{A}}\right)
                        +\vec{\mathcal{A}}^2 \nonumber\\
                     &=&\ \vec{p}^{\; 2} -2\vec{\mathcal{A}}\cdot\vec{p}+\vec{\mathcal{A}}^2 \nonumber\\
                     &=&\ \vec{p}^{\; 2} +\dfrac{g\,\hbar}{r^2} \vec{\sigma}_1 \cdot\vec{\ell}+\dfrac{g^2\,\hbar^2}{2\,r^2}.
               \end{eqnarray}
On the other hand, we can have
               \begin{eqnarray}
               \vec{p}\times \vec{p}&=&0,\nonumber\\
                   \vec{\mathcal{A}}\times\vec{\mathcal{A}} &=&
                        \dfrac{g^2\,\hbar^2}{4\,r^4} \left[\bigl(\vec{r}\times\vec{\sigma}_1\bigr)\times\bigl(\vec{r}\times\vec{\sigma}_1\bigr)\right] = \dfrac{g^2\,\hbar^2}{4\,r^4} \left\{\Bigl[\bigl(\vec{r}\times\vec{\sigma}_1\bigr)\cdot\vec{\sigma}_1\Bigr]\vec{r}
                        -\Bigl[\bigl(\vec{r}\times\vec{\sigma}_1\bigr)\cdot\vec{r}\Bigr]\vec{\sigma}_1\right\}\nonumber\\
                     &=& \dfrac{g^2\,\hbar^2}{4\,r^4} \Bigl[\bigl(\vec{r}\times\vec{\sigma}_1\bigr)\cdot\vec{\sigma}_1\Bigr]\vec{r} = \dfrac{g^2\,\hbar^2}{4\,r^4} \Bigl[\vec{r}\cdot\bigl(\vec{\sigma}_1 \times\vec{\sigma}_1\bigr)\Bigr]\vec{r} = \dfrac{{\rm i}\hbar^2\,g^2}{2\,r^4} \bigl(\vec{r}\cdot\vec{\sigma}_1\bigr)\vec{r}.
               \end{eqnarray}
Because
\begin{eqnarray}
\left(\vec{p}\times\vec{\mathcal{A}}\right)_z+  \left(\vec{\mathcal{A}}\times\vec{p}\right)_z &=& (p_x A_y-p_y A_x) +(A_x p_y -A_y p_x) = (p_x A_y- A_y p_x) -(p_y A_x- A_x p_y)\nonumber\\
&=& [p_x, A_y] -[p_y, A_x]= -{\rm i}\hbar \dfrac{\partial A_y}{\partial x}+ {\rm i}\hbar \dfrac{\partial A_x}{\partial y}=-{\rm i}\hbar\left(\vec{\nabla}\times \vec{\mathcal{A}}\right)_z\nonumber\\
&=& -{\rm i}\hbar g \dfrac{\partial \left( \dfrac{\vec{r}\times \vec{S}}{r^2}\right)_y}{\partial x}+ {\rm i}\hbar g \dfrac{\partial \left( \dfrac{\vec{r}\times \vec{S}}{r^2}\right)_x}{\partial y}\nonumber\\
&=& -{\rm i}\hbar g \dfrac{\partial \left( \dfrac{z S_x-x S_z}{r^2}\right)}{\partial x} +{\rm i}\hbar g \dfrac{\partial \left( \dfrac{y S_z-z S_y}{r^2}\right)}{\partial y}\nonumber\\
&=& -{\rm i}\hbar g \left( -\frac{S_z}{r^2}-\frac{2 x (-S_z x+S_x z)}{r^4}\right)+{\rm i}\hbar g \left(\frac{S_z}{r^2}-\frac{2 y (S_z y-S_y z)}{r^4}\right) \nonumber\\
&=& {\rm i}\hbar g \dfrac{1}{r^4}\left[2r^2 S_z -2 \left(x^2+y^2\right) S_z+ 2z\left(x S_x+y S_y\right)\right]\nonumber\\
&=& {\rm i}\hbar g \dfrac{2z}{r^4}\left(x S_x+y S_y+z S_z\right)={\rm i}\hbar g \dfrac{\left(\vec{r}\cdot\vec{S}\right)}{r^4} 2z\nonumber\\
&=&  {\rm i}\hbar^2 g \dfrac{\left(\vec{r}\cdot\vec{\sigma}_1\right)}{r^4} z,
\end{eqnarray}
we then have
\begin{eqnarray}
\left(\vec{p}\times\vec{\mathcal{A}}\right)+  \left(\vec{\mathcal{A}}\times\vec{p}\right)
&=& {\rm i}\hbar^2 g \dfrac{\left(\vec{r}\cdot\vec{\sigma}_1\right)}{r^4} \vec{r},
\end{eqnarray}
therefore
               \begin{eqnarray}\label{sub2}
                    {\rm i}\vec{\sigma}_2 \cdot\Bigl[\vec{p}\times\vec{p}-\vec{p}\times\vec{\mathcal{A}}
                        -\vec{\mathcal{A}}\times\vec{p}+\vec{\mathcal{A}}\times\vec{\mathcal{A}}\Bigr]
                     &=&{\rm i} \vec{\sigma}_2 \cdot\Bigl[-\left(\vec{p}\times\vec{\mathcal{A}}
                        +\vec{\mathcal{A}}\times\vec{p}\right)+\vec{\mathcal{A}}\times\vec{\mathcal{A}}\Bigr] \nonumber\\
                     &=& {\rm i} \vec{\sigma}_2 \cdot\Bigl[-\left({\rm i}\hbar^2 g \dfrac{\left(\vec{r}\cdot\vec{\sigma}_1\right)}{r^4} \vec{r}\right)+\dfrac{{\rm i}\hbar^2\,g^2}{2\,r^4} \bigl(\vec{r}\cdot\vec{\sigma}_1\bigr)\vec{r}\Bigr] \nonumber\\
                     &=&  \vec{\sigma}_2 \cdot\Bigl[\hbar^2 g \dfrac{\left(\vec{r}\cdot\vec{\sigma}_1\right)}{r^4} \vec{r}-\dfrac{\hbar^2\,g^2}{2\,r^4} \bigl(\vec{r}\cdot\vec{\sigma}_1\bigr)\vec{r}\Bigr] \nonumber\\
                     &=&  \vec{\sigma}_2 \cdot\Bigl[-\hbar^2 g(g-2) \dfrac{\left(\vec{r}\cdot\vec{\sigma}_1\right)}{2 r^4} \vec{r}\Bigr] \nonumber\\
                     &=&  -\dfrac{\hbar^2 g(g-2)}{2 r^4} \left(\vec{r}\cdot\vec{\sigma}_1\right)\left(\vec{r}\cdot\vec{\sigma}_2\right).
               \end{eqnarray}
Substituting Eq. (\ref{sub1}) and Eq. (\ref{sub2}) into Eq. (\ref{sub0}), we have
\begin{eqnarray}\label{eq:v-26b}
 \mathcal{H}^2_{\rm Dirac} &=&\openone\otimes\left[
                     \vec{p}^{\; 2} +\dfrac{g^2\,\hbar^2}{2\,r^2}+\dfrac{g\,\hbar}{r^2} (\vec{\sigma}_1 \cdot{\vec{\ell}}) - \dfrac{\hbar^2 g(g-2)}{2 r^4} \left(\vec{r}\cdot\vec{\sigma}_1\right)\left(\vec{r}\cdot\vec{\sigma}_2\right)\right]
                  +\beta^2 M^2,
\end{eqnarray}
which is just Eq. (\ref{eq:v-26}). This ends the proof.

\begin{remark}
If one observes more carefully the expansion of the operator $\mathcal{H}^2_{\rm Dirac}$ as shown in Eq. (\ref{sub0}), he may find that there exist six terms of interactions related to the spin vector potential:

\begin{eqnarray}
 && \mathbb{H}^{(1)}_{\rm Inter}=\vec{\mathcal{A}}\cdot\vec{p}= -\dfrac{g\,\hbar}{2\,r^2} \vec{\sigma}_1 \cdot\ell,
\end{eqnarray}

\begin{eqnarray}
 && \mathbb{H}^{(2)}_{\rm Inter}=\vec{p}\cdot\vec{\mathcal{A}}=\vec{\mathcal{A}}\cdot\vec{p}-{\rm i}\hbar\nabla\cdot\vec{\mathcal{A}}=
 -\dfrac{g\,\hbar}{2\,r^2} \vec{\sigma}_1 \cdot\ell=\mathbb{H}^{(1)}_{\rm Inter},
\end{eqnarray}

\begin{eqnarray}
 && \mathbb{H}^{(3)}_{\rm Inter}=\vec{\mathcal{A}}^2=\dfrac{g^2\,\hbar^2}{2\,r^2},
\end{eqnarray}

\begin{eqnarray}\label{eq:ell-1}
  \mathbb{H}^{(4)}_{\rm Inter}&=&{\rm i} \vec{\sigma}_2 \cdot \left(\vec{\mathcal{A}}\times\vec{p}\right)=
  {\rm i} \vec{\sigma}_2 \cdot \left[ \dfrac{g\,\hbar}{2} \left(
                        \dfrac{\vec{r}\times\vec{\sigma}_1}{r^2} \times\vec{p}\right) \right]
  ={\rm i} \vec{\sigma}_2 \cdot \left[ \dfrac{g\,\hbar}{2} \left( \left(\dfrac{\vec{r}}{r^2}\cdot\vec{p}\right)\vec{\sigma}_1
                        -\dfrac{\vec{r}}{r^2} (\vec{\sigma}_1 \cdot\vec{p})\right)\right]\nonumber\\
  &=& {\rm i} \vec{\sigma}_2 \cdot \left[ \dfrac{g\,\hbar}{2} \frac{1}{r^2}\left[ (\vec{r}\cdot\vec{p})\vec{\sigma}_1
                        -\vec{r} (\vec{p}\cdot \vec{\sigma}_1)\right]\right]= {\rm i} \dfrac{g\,\hbar}{2} \frac{1}{r^2}\left[ (\vec{r}\cdot\vec{p})(\vec{\sigma}_1\cdot \vec{\sigma}_2)
                        -(\vec{r}\cdot\vec{\sigma}_2)(\vec{p}\cdot \vec{\sigma}_1)\right]\nonumber\\
  &=& {\rm i} \dfrac{g\,\hbar}{2} \frac{1}{r^2}\left[ (\vec{r}\cdot\vec{p})(\vec{\sigma}_1\cdot \vec{\sigma}_2)
                        +\vec{\ell}\cdot(\vec{\sigma}_1 \times\vec{\sigma}_2) -(\vec{r}\cdot\vec{\sigma}_1)(\vec{p}\cdot\vec{\sigma}_2))\right],
\end{eqnarray}

\begin{eqnarray}
 && \mathbb{H}^{(5)}_{\rm Inter}={\rm i} \vec{\sigma}_2 \cdot \left(\vec{p}\times \vec{\mathcal{A}}\right)={\rm i} \vec{\sigma}_2 \cdot \left(-\vec{\mathcal{A}}\times \vec{p}+{\rm i}\hbar^2 g \dfrac{\left(\vec{r}\cdot\vec{\sigma}_1\right)}{r^4} \vec{r}\right)=-\mathbb{H}^{(4)}_{\rm Inter}-\hbar^2 g \dfrac{\left(\vec{r}\cdot\vec{\sigma}_1\right)(\vec{r}\cdot\vec{\sigma}_2)}{r^4},
\end{eqnarray}

\begin{eqnarray}
 && \mathbb{H}^{(6)}_{\rm Inter}={\rm i} \vec{\sigma}_2 \cdot \left(\vec{\mathcal{A}}\times \vec{\mathcal{A}}\right)={\rm i} \vec{\sigma}_2 \cdot \left( \dfrac{{\rm i}\hbar^2\,g^2}{2\,r^4} \bigl(\vec{r}\cdot\vec{\sigma}_1\bigr)\vec{r}\right)=-  \dfrac{\hbar^2\,g^2}{2\,r^4} \bigl(\vec{r}\cdot\vec{\sigma}_1\bigr)(\vec{r}\cdot \vec{\sigma}_2).
\end{eqnarray}
Here in the derivation of Eq. (\ref{eq:ell-1}) we have used the following relation
\begin{equation}
    \begin{split}
      \vec{\ell}\cdot(\vec{\sigma}_1 \times\vec{\sigma}_2) =&\
        (\vec{r}\times\vec{p})\cdot(\vec{\sigma}_1 \times\vec{\sigma}_2)= \vec{r}\cdot\bigl[\vec{p}\times(\vec{\sigma}_1 \times\vec{\sigma}_2)\bigr] = \vec{r}\cdot\bigl[(\vec{p}\cdot\vec{\sigma}_2)\vec{\sigma}_1
        -(\vec{p}\cdot\vec{\sigma}_1)\vec{\sigma}_2\bigr] \\
      =&\ (\vec{r}\cdot\vec{\sigma}_1)(\vec{p}\cdot\vec{\sigma}_2)
        -(\vec{r}\cdot\vec{\sigma}_2)(\vec{p}\cdot\vec{\sigma}_1).
    \end{split}
  \end{equation}

Based on the above calculation, we find that: (i) The operators $\mathbb{H}^{(1)}_{\rm Inter}$ and $\mathbb{H}^{(2)}_{\rm Inter}$ may contribute
the spin-orbital interaction
\begin{eqnarray}\label{eq:si-1}
  \mathcal{H}_{\rm so}&=&\vec{\sigma}_1 \cdot{\vec{\ell}}.
\end{eqnarray}
(ii) The operators $\mathbb{H}^{(3)}_{\rm Inter}$ may contribute a potential energy proportional to $1/r^2$, which does not contain spin.
(iii) The operator $\mathbb{H}^{(4)}_{\rm Inter}$ may contribute the spin-spin exchange interaction (or the well-known Heisenberg exchange interaction)
\begin{eqnarray}\label{eq:si-2}
  \mathcal{H}_{\rm sse}&=&\vec{\sigma}_1 \cdot \vec{\sigma}_2,
\end{eqnarray}
and a new-type interaction (i.e., a generalized spin-orbital interaction)
\begin{eqnarray}\label{eq:si-3}
  \mathcal{H}_{\rm gso}&=&\vec{\ell}\cdot(\vec{\sigma}_1 \times\vec{\sigma}_2).
\end{eqnarray}
(iv) The sum of $\mathbb{H}^{(4)}_{\rm Inter}$ and $\mathbb{H}^{(5)}_{\rm Inter}$, and also the operator $\mathbb{H}^{(6)}_{\rm Inter}$ may contribute the dipole-dipole interaction
\begin{eqnarray}\label{eq:si-4}
  \mathcal{H}_{\rm dd}&=& \dfrac{\left(\vec{r}\cdot\vec{\sigma}_1\right)(\vec{r}\cdot\vec{\sigma}_2)}{r^2}.
\end{eqnarray}
Therefore, based on the Dirac Hamiltonian involving the spin vector potential $\mathcal{H}_{\rm Dirac}$ and its square operator $\mathcal{H}^2_{\rm Dirac}$, we have observed that some significant interactions, such as the DM interaction, the spin-spin exchange interaction, the spin-orbital interaction, the dipole-dipole interaction can naturally appear, and we also we further predicted a new-type of spin-orbital interaction $\vec{\ell}\cdot(\vec{\sigma}_1 \times\vec{\sigma}_2)$ \cite{angular}. In Table \ref{tab:interactions}, we have listed these spin interactions.

Moreover, in the textbook of quantum mechanics \cite{1999Flugge}, the so-called tensor force between two particles 1 and 2 of spin-1/2 is defined by the interaction energy
\begin{eqnarray}\label{eq:v-25a}
 &&V=W(r)\mathcal{T}_{12}
 \end{eqnarray}
with the tensor force operator
\begin{eqnarray}\label{eq:v-25-02}
 \mathcal{T}_{12} &=& \frac{3(\vec{r}\cdot\vec{\sigma}_1)(\vec{r}\cdot\vec{\sigma}_2)}{r^2} -\vec{\sigma}_1 \cdot\vec{\sigma}_2.
\end{eqnarray}
Since the tensor operator $\mathcal{T}_{12}$ is a linear combination of the dipole-dipole interaction and the spin-spin exchange interaction, thus it is also a natural outcome of the spin vector potential. We have also listed the tensor force operator in the last line of Table \ref{tab:interactions}.
\end{remark}

\begin{table}[t]
	\centering
\caption{The types of spin interactions involving in the Dirac Hamiltonian and its square operator.}
\begin{tabular}{lr}
\hline\hline
 & {\rm The types of interactions} \\
  \hline
	$\vec{\sigma}_1 \cdot\vec{\sigma}_2$ & {\rm The spin-spin exchange interaction} \\
  \hline
	$\vec{\sigma}_1 \cdot{\vec{\ell}}$ & {\rm The spin-orbital interaction} \\
  \hline
  $\vec{r}\cdot\bigl(\vec{\sigma}_1 \times\vec{\sigma}_2\bigr)$ & \;\;\;\;\;\;\;\;\;\;\;\;{\rm The Dzyaloshinsky-Moriya-type interaction} \\
   \hline
     \vspace{0.5mm}
    $\dfrac{(\vec{r}\cdot\vec{\sigma}_1)(\vec{r}\cdot\vec{\sigma}_2)}{r^2}$ & {\rm The dipole-dipole interaction} \\
  \hline
$\vec{\ell}\cdot(\vec{\sigma}_1 \times\vec{\sigma}_2)$ & {\rm The generalized spin-orbital interaction} \\
  \hline
    \vspace{0.5mm}
 $\dfrac{3(\vec{r}\cdot\vec{\sigma}_1)(\vec{r}\cdot\vec{\sigma}_2)}{r^2} -\vec{\sigma}_1 \cdot\vec{\sigma}_2$ & {\rm The tensor force operator} \\
 \hline\hline
\end{tabular}\label{tab:interactions}
\end{table}

\newpage

\part{Some Other Calculations}

\section{The Peculiarity of $g=2$}

Let us recall the spin AB Hamiltonian as shown in Eq. (\ref{eq:v-3}), i.e.,
 \begin{eqnarray}\label{eq:v-3-b}
            H_{\rm S}=\dfrac{1}{2M} \left(\vec{p}-\vec{\mathcal{A}}\right)^2,
      \end{eqnarray}
with the vector potential
      \begin{eqnarray}\label{eq:v-4-b}
      \vec{\mathcal{A}}&=&g \dfrac{\vec{r}\times\vec{S}}{r^2}.
      \end{eqnarray}
Let us define the canonical momentum operator as
      \begin{eqnarray}\label{eq:S-1}
     \vec{\Pi}=\vec{p}-\vec{\mathcal{A}},
      \end{eqnarray}
and ask a question: When are three components $\{\Pi_x, \Pi_y, \Pi_z\}$ mutually commutative?

From Eq. (\ref{eq:S-1}) we obtain
\begin{eqnarray}\label{eq:M-2-01}
&&\Pi_x=\left(\vec{p} -\vec{\mathcal{A}}\right)_x = p_x-g \frac{(\vec{r}\times\vec{S})_x}{r^2}= p_x-g \left(\frac{y}{r^2} S_z- \frac{z}{r^2} S_y\right), \nonumber\\
&&\Pi_y=\left(\vec{p}-\vec{\mathcal{A}}\right)_y = p_y-g \frac{(\vec{r}\times\vec{S})_y}{r^2}= p_y-g \left(\frac{z}{r^2} S_x- \frac{x}{r^2} S_z\right), \nonumber\\
&&\Pi_z=\left(\vec{p} -\vec{\mathcal{A}} \right)_z = p_z-g \frac{(\vec{r}\times\vec{S})_z}{r^2}= p_z-g \left(\frac{x}{r^2} S_y- \frac{y}{r^2} S_x\right).
\end{eqnarray}
We then have
\begin{eqnarray}\label{eq:M-3-01}
&&[\mathcal{A}_x,\mathcal{A}_y] = g^2 \left[\frac{y}{r^2} S_z- \frac{z}{r^2} S_y, \frac{z}{r^2} S_x- \frac{x}{r^2} S_z \right]\nonumber\\
&&\quad \quad \quad \quad=g^2  \left\{ \frac{y}{r^2}\frac{z}{r^2}[S_z, S_x] -\frac{z}{r^2}\frac{z}{r^2} [S_y, S_x] +\frac{z}{r^2}\frac{x}{r^2}[S_y, S_z] \right\}\nonumber\\
&&\quad \quad \quad \quad =g^2 \left\{ \frac{yz}{r^4}\; {\rm i} \hbar S_y +\frac{z^2}{r^4}\; {\rm i} \hbar S_z +\frac{zx}{r^4}\; {\rm i} \hbar S_x \right\}\nonumber\\
&&\quad \quad \quad \quad =g^2  \; {\rm i} \hbar \frac{\vec{r}\cdot\vec{S}}{r^4} \; z,\nonumber\\
&&[\mathcal{A}_y,\mathcal{A}_z] = g^2  \; {\rm i} \hbar \frac{\vec{r}\cdot\vec{S}}{r^4} \; x, \nonumber\\
&&[\mathcal{A}_z,\mathcal{A}_x] = g^2  \; {\rm i} \hbar \frac{\vec{r}\cdot\vec{S}}{r^4} \; y,
\end{eqnarray}
or in a vector form as follows
\begin{eqnarray}\label{eq:M-5-01}
&& \vec{\mathcal{A}} \times \vec{\mathcal{A}}= {\rm i} \hbar \; g^2 \; \frac{\vec{r}\cdot\vec{S}}{r^4} \; \vec{r}.
\end{eqnarray}

Now, we come to calculate the following commutator
\begin{eqnarray}\label{eq:M-6}
\left[\Pi_x, \Pi_y\right]&=&\left[\left(\vec{p} -\vec{\mathcal{A}}\right)_x, \left(\vec{p} -\vec{\mathcal{A}}\right)_y\right] =\left[p_x- \mathcal{A}_x, p_y-\mathcal{A}_y\right]=[p_x, -\mathcal{A}_y]- [p_y, -\mathcal{A}_x]+  [\mathcal{A}_x, \mathcal{A}_y]\nonumber\\
&=& -  g  \left[p_x, \left(\frac{z}{r^2} S_x- \frac{x}{r^2} S_z\right)\right]+ g  \left[p_y, \left(\frac{y}{r^2} S_z- \frac{z}{r^2} S_y\right)\right]+ \left[\mathcal{A}_x, \mathcal{A}_y\right]\nonumber\\
&=& -g \;{\rm i} \hbar \left[\frac{2 z x}{r^4} S_x+\left(\frac{1}{r^2}-\frac{2x^2}{r^4}\right) S_z \right]+g (-{\rm i} \hbar)\; \left[ \left(\frac{1}{r^2}-\frac{2y^2}{r^4}\right)S_z+\frac{2zy}{r^4}S_y\right]+ g^2  \; {\rm i} \hbar \frac{\vec{r}\cdot\vec{S}}{r^4} \; z\nonumber\\
&=& - {\rm i} \hbar g  \left[\frac{2 z x}{r^4} S_x+ \frac{2zy}{r^4}S_y+ \frac{2z^2}{r^4}S_z\right]+ {\rm i} \hbar g^2  \;  \frac{\vec{r}\cdot\vec{S}}{r^4} \; z\nonumber\\
&=&- {\rm i} \hbar 2 g  \frac{\vec{r}\cdot\vec{S}}{r^4} \; z + {\rm i} \hbar g^2  \;  \frac{\vec{r}\cdot\vec{S}}{r^4} \; z\nonumber\\
&=& {\rm i} \hbar \; g(g-2)  \frac{\vec{r}\cdot\vec{S}}{r^4} \; z.
\end{eqnarray}
Similarly, we have
\begin{eqnarray}\label{eq:M-7}
\left[\Pi_y, \Pi_z\right]&=&\left[\left(\vec{p} -\vec{\mathcal{A}}\right)_y, \left(\vec{p} -\vec{\mathcal{A}}\right)_z\right] = {\rm i} \hbar \; g(g-2)  \frac{\vec{r}\cdot\vec{S}}{r^4} \; x\nonumber\\
\left[\Pi_z, \Pi_x\right]&=&\left[\left(\vec{p} -\vec{\mathcal{A}}\right)_z, \left(\vec{p} -\vec{\mathcal{A}}\right)_x\right] ={\rm i} \hbar \; g(g-2)  \frac{\vec{r}\cdot\vec{S}}{r^4} \; y,
\end{eqnarray}
or in a vector form as
\begin{eqnarray}\label{eq:M-5b}
&& \vec{\Pi} \times \vec{\Pi}= {\rm i} \hbar \; g(g-2)  \frac{\vec{r}\cdot\vec{S}}{r^4} \; \vec{r}.
\end{eqnarray}
Thus from Eq. (\ref{eq:M-6}) and Eq. (\ref{eq:M-7}), one may notice that, if $g\neq 0$, for the following special case
\begin{eqnarray}\label{eq:M-8}
&& g=2,
\end{eqnarray}
the three operators $\Pi_x$, $\Pi_y$, and $\Pi_z$ are mutually commutative.

In this special case, we can use the second method as shown in Sec. \ref{secondmethod} to solve the following eigen-problem
\begin{equation}\label{eq:H0-g}
  H_{\rm S} \bigl[\xi_0(\vec{r}) \xi(\vec{r})\bigr]
  =E\bigl[\xi_0(\vec{r})\xi(\vec{r})\bigr],
\end{equation}
where $E$ is the eigen-energy, the wave function $\Psi_{\rm S} (\vec{r})$ has been written in a form as
\begin{equation}\label{eq:H0-f-a}
  \Psi_{\rm S} (\vec{r}) =\xi_0(\vec{r})\xi(\vec{r}),
\end{equation}
and $\xi_0(\vec{r})$ is the wavefunction of a free particle, which can be chosen as the common eigenstate of
the set $\{H_0, \vec{p}\}$, i.e.
\begin{eqnarray}\label{eq:L-3}
 && \xi_0(\vec{r})= \mathcal{N} {\rm e}^{\mathrm{i} \vec{k}\cdot\vec{r}}, \;\;\;\;\; H_0  \xi_0(\vec{r})= \frac{\hbar^2 \vec{k}^2}{2M} \; \xi_0(\vec{r}), \;\;\;\;\; \vec{p}\; \xi_0(\vec{r}) = \hbar \vec{k} \;\xi_0(\vec{r}).
\end{eqnarray}

Similarly, the function $\xi(\vec{r})$ is determined by the following equation
\begin{equation}\label{eq:RelatePAM-a}
  \left(\vec{p} - \vec{\mathcal{A}} \right)\; \xi(\vec{r})=0.
\end{equation}
and also
\begin{eqnarray}\label{eq:L-12-b-a}
 H_{\rm S} \; \xi(\vec{r})= \frac{1}{2M} \left[ \left(\vec{p}- \vec{\mathcal{A}}\right)\cdot
  \left(\vec{p} - \vec{\mathcal{A}} \right)\right] \; \xi(\vec{r})=0.
\end{eqnarray}
In the next step, we shall solve Eq. (\ref{eq:RelatePAM-a}) in the spherical coordinate system. Due to
   \begin{equation}
      \vec{\nabla}=\hat{e}_r \dfrac{\partial}{\partial\,r}+\hat{e}_\theta\,\dfrac{1}{r}\dfrac{\partial}{\partial\,\theta}
         +\hat{e}_\phi\,\dfrac{1}{r\sin\theta}\dfrac{\partial}{\partial\,\phi},
   \end{equation}
We have
   \begin{eqnarray}\label{eq:WavEqXiRExpand}
        &&\left(\dfrac{\partial}{\partial\,r}-\dfrac{{\rm i}}{\hbar} \vec{\mathcal{A}}_r\right)\xi(\vec{r})=0, \nonumber\\
        &&\left(\dfrac{1}{r} \dfrac{\partial}{\partial\,\theta}-\dfrac{{\rm i}}{\hbar } \vec{\mathcal{A}}_\theta\right)\xi(\vec{r})=0, \nonumber\\
        &&\left(\dfrac{1}{r\,\sin\theta} \dfrac{\partial}{\partial\,\phi}-\dfrac{{\rm i}}{\hbar} \vec{\mathcal{A}}_\phi\right)\xi(\vec{r})=0.
    \end{eqnarray}
   By considering
   \begin{eqnarray}
   \vec{\mathcal{A}} &=& g \frac{\left(\vec{r}\times\vec{S}\right)}{r^2} = g \dfrac{\left[(r\,\hat{e}_r) \times\vec{S}\right]}{r^2}=
  \dfrac{g}{r} \left|\begin{array}{ccc}
       \hat{e}_r & \hat{e}_\theta & \hat{e}_\phi \\
       1 & 0 & 0\\
       S_r & S_\theta & S_\phi
     \end{array}\right| = \dfrac{g}{r}\left(S_\theta\,\hat{e}_\phi-S_\phi\,\hat{e}_\theta\right),
       \end{eqnarray}
   then we have
   \begin{eqnarray}
            && \mathcal{A}_r=0, \;\;\; \mathcal{A}_\theta=-g \dfrac{S_\phi}{r}, \;\;\; \mathcal{A}_\phi=g \dfrac{S_\theta}{r}.
     \end{eqnarray}
   Since $\mathcal{A}_r=0$, Eq. (\ref{eq:WavEqXiRExpand}) reduces to
   \begin{eqnarray}\label{eq:WavEqXiRExpandReduce}
         &&\dfrac{\partial\,\xi(\vec{r})}{\partial\,r}=0, \nonumber\\
        & &\left(\dfrac{1}{r} \dfrac{\partial}{\partial\,\theta}-\dfrac{{\rm i}}{\hbar} \mathcal{A}_\theta\right)\xi(\vec{r})=0, \nonumber\\
        & &\left(\dfrac{1}{r\,\sin\theta} \dfrac{\partial}{\partial\,\phi}-\dfrac{{\rm i}}{\hbar} \mathcal{A}_\phi\right)\xi(\vec{r})=0.
      \end{eqnarray}
   Besides, in the spherical coordinate system, the three components of the spin vector operator are given as
   \begin{eqnarray}\label{eq:SXYZToRThetaPhi}
             S_r&=& S_x \sin\theta\cos\phi+ S_y \sin\theta\sin\phi +S_z \cos\theta= \dfrac{\hbar}{2}\begin{bmatrix}
               \cos\theta & \sin\theta\,{\rm e}^{-{\rm i}\phi} \\
               \sin\theta\,{\rm e}^{{\rm i}\phi} & -\cos\theta
            \end{bmatrix}, \nonumber\\
              S_\theta &=& S_x \cos\theta\cos\phi+ S_y \cos\theta\sin\phi -S_z \sin\theta= \dfrac{\hbar}{2}\begin{bmatrix}
               -\sin\theta & \cos\theta\,{\rm e}^{-{\rm i}\phi} \\
               \cos\theta\,{\rm e}^{{\rm i}\phi} & \sin\theta
            \end{bmatrix}, \nonumber\\
             S_\phi &=& -S_x \sin\phi+ S_y\cos\phi= \dfrac{{\rm i}\,\hbar}{2}\begin{bmatrix}
               0 & -{\rm e}^{-{\rm i}\phi} \\
               {\rm e}^{{\rm i}\phi} & 0
            \end{bmatrix}.
   \end{eqnarray}
   From the first equation of Eq. (\ref{eq:WavEqXiRExpandReduce}), we know that $\xi(\vec{r})$ is independent of $r$, thus we have
   \begin{eqnarray}
  &&\xi(\vec{r}) \equiv \xi(\theta,\phi)
         =\begin{bmatrix} \chi_1 (\theta,\phi) \\ \chi_2 (\theta,\phi) \end{bmatrix},
   \end{eqnarray}
   and Eq. (\ref{eq:WavEqXiRExpandReduce}) can be recast as
 \begin{eqnarray}\label{eq:WavEqTheta}
&& \dfrac{1}{r} \left(\openone \dfrac{\partial}{\partial\,\theta}+\dfrac{{\rm i}\,g}{\hbar} S_\phi\right)\xi(\vec{r})=0,
   \end{eqnarray}
 \begin{eqnarray}\label{eq:WavEqphi}
&&  \dfrac{1}{r} \left(\dfrac{\openone}{\sin\theta} \dfrac{\partial}{\partial\,\phi}-\dfrac{{\rm i}\,g}{\hbar} S_\theta\right)\xi(\vec{r})=0.
   \end{eqnarray}

Let us firstly study Eq. (\ref{eq:WavEqTheta}), from which we have
   \begin{equation}
      \begin{split}
         \Biggl(\openone \dfrac{\partial}{\partial\,\theta}-\dfrac{g}{2} \begin{bmatrix}
            0 & -{\rm e}^{-{\rm i}\phi} \\
            {\rm e}^{{\rm i}\phi} & 0
         \end{bmatrix}\Biggr)\begin{bmatrix}
            \chi_1 \\ \chi_2
         \end{bmatrix} &= 0,
      \end{split}
   \end{equation}
   namely
   \begin{equation}\label{eq:WavEqThetaEnd}
      \begin{split}
         & \begin{cases}
            & \dfrac{\partial\,\chi_1}{\partial\,\theta} +\dfrac{g}{2} {\rm e}^{-{\rm i}\phi}\,\chi_2=0, \\
            &  -\dfrac{g}{2} {\rm e}^{{\rm i}\phi}\, \chi_1+\dfrac{\partial\,\chi_2}{\partial\,\theta}=0,
         \end{cases}
         \quad \quad  \stackrel{g\neq 0}{\Longrightarrow} \quad \quad
         \begin{cases}
            & {\rm e}^{{\rm i}\phi}\left(\dfrac{\partial^2\,\chi_1}{\partial\,\theta^2} +\dfrac{g^2}{4}\,\chi_1\right)=0, \\
            & {\rm e}^{-{\rm i}\phi}\left(\dfrac{\partial^2\,\chi_2}{\partial\,\theta^2} +\dfrac{g^2}{4}\,\chi_2\right)=0,
         \end{cases} \\
         & \Longrightarrow \begin{cases}
            & \chi_1=C_1 (\phi) \cos\left(\dfrac{g}{2} \theta\right)+C_2 (\phi) \sin\left(\dfrac{g}{2} \theta\right), \\
            & \\
            & \chi_2=C_3 (\phi) \cos\left(\dfrac{g}{2} \theta\right)+C_4 (\phi) \sin\left(\dfrac{g}{2} \theta\right),
         \end{cases}
      \end{split}
   \end{equation}
where $C_i (\phi)$, $(i=1,2,3,4)$, are functions depending only on $\phi$. In our case $g=2$, thus from above we have
 \begin{eqnarray}\label{eq:M-1}
 \chi_1&=& C_1 (\phi) \cos\theta +C_2 (\phi) \sin\theta,\nonumber\\
  \chi_2 &= &C_3 (\phi) \cos\theta+C_4 (\phi) \sin\theta.
   \end{eqnarray}
In the next step, we shall use Eq. (\ref{eq:WavEqphi}) to determine the four coefficients $C_i (\phi)$'s.. From Eq. (\ref{eq:WavEqphi}), we have
   \begin{eqnarray}\label{eq:M-2-02}
        &&\Biggl(\dfrac{\openone}{\sin\theta} \dfrac{\partial}{\partial\,\phi}-\dfrac{{\rm i}\,g}{2} \begin{bmatrix}
            -\sin\theta & \cos\theta\,{\rm e}^{-{\rm i}\phi} \\
            \cos\theta\,{\rm e}^{{\rm i}\phi} & \sin\theta
         \end{bmatrix}\Biggr)\begin{bmatrix}
            \chi_1 \\ \chi_2
         \end{bmatrix} = 0,\nonumber\\
         &&\Longrightarrow
         \begin{bmatrix}
            \dfrac{1}{\sin\theta} \dfrac{\partial}{\partial\,\phi}+\dfrac{{\rm i}\,g\sin\theta}{2}
            & -\dfrac{{\rm i}g}{2} \cos\theta\,{\rm e}^{-{\rm i}\phi} \\
            -\dfrac{{\rm i}g}{2} \cos\theta\,{\rm e}^{{\rm i}\phi} &
            \dfrac{1}{\sin\theta} \dfrac{\partial}{\partial\,\phi}-\dfrac{{\rm i}\,g\sin\theta}{2}
         \end{bmatrix}\begin{bmatrix}
            \chi_1 \\ \chi_2
         \end{bmatrix} = 0.
   \end{eqnarray}
   Namely
   \begin{equation}\label{eq:M-3}
      \begin{split}
         & \begin{cases}
            & \dfrac{1}{\sin\theta} \dfrac{\partial\,\chi_1}{\partial\,\phi}+\dfrac{{\rm i}\,g}{2}
               \left(\sin\theta\,\chi_1 -\cos\theta\,{\rm e}^{-{\rm i}\phi}\,\chi_2\right)=0, \\
            & \\
            & \dfrac{1}{\sin\theta} \dfrac{\partial\,\chi_2}{\partial\,\phi}-\dfrac{{\rm i}\,g}{2}
               \left(\sin\theta\,\chi_2 +\cos\theta\,{\rm e}^{{\rm i}\phi}\,\chi_1\right)=0.\\
         \end{cases}
      \end{split}
   \end{equation}

  Note that
   \begin{equation}
      \begin{cases}
         & \dfrac{\partial\,\chi_1}{\partial\,\phi}=\dfrac{{\rm e}^{-{\rm i}\phi}}{\cos\theta}
            \Biggl[\dfrac{-2\,{\rm i}}{g\sin\theta} \dfrac{\partial^2\,\chi_2}{\partial\,\phi^2}
               -\left(\dfrac{2}{g\sin\theta} +\sin\theta\right)\dfrac{\partial\,\chi_2}{\partial\,\phi}
               +{\rm i}\sin\theta\,\chi_2\Biggr] \\
         & \\
         & \dfrac{\partial\,\chi_2}{\partial\,\phi}=\dfrac{{\rm e}^{{\rm i}\phi}}{\cos\theta}
         \Biggl[\dfrac{-2\,{\rm i}}{g\sin\theta} \dfrac{\partial^2\,\chi_1}{\partial\,\phi^2}
            +\left(\dfrac{2}{g\sin\theta} +\sin\theta\right)\dfrac{\partial\,\chi_1}{\partial\,\phi}
            +{\rm i}\sin\theta\,\chi_1\Biggr],
      \end{cases}
   \end{equation}
   thus we attain two same second-order linear ordinary differential equations as follows:
    \begin{eqnarray}
      && \dfrac{1}{\cos\theta} \Biggl[\dfrac{-2\,{\rm i}}{g\sin\theta} \dfrac{\partial^2\,\chi_2}{\partial\,\phi^2}
               -\left(\dfrac{2}{g\sin\theta} +\sin\theta\right)\dfrac{\partial\,\chi_2}{\partial\,\phi}
               +{\rm i}\sin\theta\,\chi_2\Biggr]-{\rm i} \dfrac{g\,{\sin^2}\theta}{2\cos\theta}
               \left[\dfrac{2\,{\rm i}}{g\sin\theta} \dfrac{\partial\,\chi_2}{\partial\,\phi}+\sin\theta\,\chi_2\right]
            -{\rm i}\left[\dfrac{g\sin\theta\cos\theta}{2}\right]\chi_2=0, \nonumber\\
       && \dfrac{1}{\cos\theta} \Biggl[\dfrac{-2\,{\rm i}}{g\sin\theta} \dfrac{\partial^2\,\chi_1}{\partial\,\phi^2}
               +\left(\dfrac{2}{g\sin\theta} +\sin\theta\right)\dfrac{\partial\,\chi_1}{\partial\,\phi}
               +{\rm i}\sin\theta\,\chi_1\Biggr]-{\rm i}\dfrac{g\,{\sin^2}\theta}{2\cos\theta}
               \left[\dfrac{-2\,{\rm i}}{g\sin\theta} \dfrac{\partial\,\chi_1}{\partial\,\phi}+\sin\theta\,\chi_1\right]
            -{\rm i}\left[\dfrac{g\sin\theta\cos\theta}{2}\right]\chi_1=0,\nonumber\\
      \end{eqnarray}
    i.e.,
    \begin{equation}
      \begin{split}
          & \begin{cases}
            & {\rm i}\left(\dfrac{-2}{g\sin\theta}\right) \dfrac{\partial^2\,\chi_2}{\partial\,\phi^2}
               -\left(\dfrac{2}{g\sin\theta}\right)\dfrac{\partial\,\chi_2}{\partial\,\phi}
               +{\rm i}\left(1-\dfrac{g}{2}\right)\sin\theta\,\chi_2=0, \\
            & \\
            & {\rm i}\left(\dfrac{-2}{g\sin\theta}\right) \dfrac{\partial^2\,\chi_1}{\partial\,\phi^2}
               +\left(\dfrac{2}{g\sin\theta}\right)\dfrac{\partial\,\chi_1}{\partial\,\phi}
               +{\rm i}\left(1-\dfrac{g}{2}\right)\sin\theta\,\chi_1=0,
         \end{cases} \\
         & \Longrightarrow\begin{cases}
            & \dfrac{\partial^2\,\chi_2}{\partial\,\phi^2}-{\rm i}\dfrac{\partial\,\chi_2}{\partial\,\phi}
               -\dfrac{g}{2} \left(1-\dfrac{g}{2}\right){\sin^2} \theta\,\chi_2=0, \\
            & \dfrac{\partial^2\,\chi_1}{\partial\,\phi^2}+{\rm i}\dfrac{\partial\,\chi_1}{\partial\,\phi}
               -\dfrac{g}{2} \left(1-\dfrac{g}{2}\right){\sin^2} \theta\,\chi_1=0,
         \end{cases} \\
         &\Longrightarrow\begin{cases}
            & \chi_1=D_1 (\theta)\:{\rm e}^{-\frac{\phi}{2} \left[{\rm i}+\sqrt{(2-g)g\,{\sin^2} \theta-1}\right]}
               +D_2 (\theta)\:{\rm e}^{\frac{\phi}{2} \left[-{\rm i}+\sqrt{(2-g)g\,{\sin^2} \theta-1}\right]}, \\
            & \\
            & \chi_2=D_3 (\theta)\:{\rm e}^{\frac{\phi}{2} \left[{\rm i}-\sqrt{(2-g)g\,{\sin^2} \theta-1}\right]}
               +D_4 (\theta)\:{\rm e}^{\frac{\phi}{2} \left[{\rm i}+\sqrt{(2-g)g\,{\sin^2} \theta-1}\right]}.
         \end{cases}
      \end{split}
   \end{equation}
    Because $g=2$ and $\sqrt{-1}={\rm i}$, we then have
   \begin{eqnarray}\label{eq:M-4}
\chi_1&=&D_1 (\theta)\:{\rm e}^{-{\rm i}\phi}
               +D_2 (\theta), \nonumber\\
          \chi_2 &=& D_3 (\theta)
               +D_4 (\theta)\:{\rm e}^{{\rm i}\phi},
   \end{eqnarray}
  where $D_i (\theta)$, $(i=1,2,3,4)$, are functions depending only on $\theta$.

  Let us compare Eq. (\ref{eq:M-1}) and Eq. (\ref{eq:M-4}) and intend to write them into a unified form. Namely, let us observe
   \begin{equation}\label{eq:ChiTheta}
      \begin{cases}
         & \chi_1=C_1 (\phi) \cos\theta+C_2 (\phi) \sin\theta, \\
         & \\
         & \chi_2=C_3 (\phi) \cos\theta+C_4 (\phi) \sin\theta,
      \end{cases}
   \end{equation}
   and
   \begin{equation}\label{eq:ChiPhi}
      \begin{cases}
         & \chi_1=D_1 (\theta)\:{\rm e}^{-{\rm i}\phi} +D_2 (\theta), \\
         & \\
         & \chi_2=D_3 (\theta)+D_4 (\theta)\:{\rm e}^{{\rm i}\phi}.
      \end{cases}
   \end{equation}
   According to the first equation in Eq. (\ref{eq:WavEqThetaEnd}), we have
   \begin{equation}
      \begin{split}
         & \begin{cases}
            & -C_1 \sin\theta +C_2 \cos\theta +\dfrac{g}{2} {\rm e}^{-{\rm i}\phi} (C_3 \cos\theta +C_4 \sin\theta)=0, \\
            &  -\dfrac{g}{2} {\rm e}^{{\rm i}\phi} (C_1 \cos\theta +C_2 \sin\theta)-C_3 \sin\theta +C_4 \cos\theta=0,
         \end{cases} \\
         & \stackrel{g=2}{\Longrightarrow}\begin{cases}
            & C_3 \cos\theta +C_4 \sin\theta={\rm e}^{{\rm i}\phi} (C_1 \sin\theta -C_2 \cos\theta), \\
            & C_3 \sin\theta -C_4 \cos\theta=-{\rm e}^{{\rm i}\phi} (C_1 \cos\theta -C_2 \sin\theta),
         \end{cases}\Longrightarrow\begin{cases}
            & C_3=-C_2\,{\rm e}^{{\rm i}\phi}, \\
            & C_4=C_1\,{\rm e}^{{\rm i}\phi},
         \end{cases}
      \end{split}
   \end{equation}
   and based on Eq. (\ref{eq:M-3}) we have
    \begin{equation}
      \begin{split}
         & \begin{cases}
            & \dfrac{-{\rm i}}{\sin\theta} D_1\,{\rm e}^{-{\rm i}\phi} +\dfrac{{\rm i}\,g}{2} \Bigl[
               \sin\theta\left(D_1\,{\rm e}^{-{\rm i}\phi} +D_2\right)
               -\cos\theta\,{\rm e}^{-{\rm i}\phi} \left(D_3 +D_4\,{\rm e}^{{\rm i}\phi}\right)\Bigr]=0, \\
            & \\
            & \dfrac{{\rm i}}{\sin\theta} D_4\,{\rm e}^{{\rm i}\phi} -\dfrac{{\rm i}\,g}{2} \Bigl[
               \sin\theta\left(D_3 +D_4\,{\rm e}^{{\rm i}\phi}\right)
               +\cos\theta\,{\rm e}^{{\rm i}\phi} \left(D_1\,{\rm e}^{-{\rm i}\phi} +D_2\right)\Bigr]=0,
         \end{cases} \\
         & \stackrel{g=2}{\Longrightarrow}\begin{cases}
            & {\rm i}\cos\theta\,{\rm e}^{-{\rm i}\phi} \left(D_3 +D_4\,{\rm e}^{{\rm i}\phi}\right)
               ={\rm i}\left(\sin\theta-\dfrac{1}{\sin\theta}\right)\,D_1\,{\rm e}^{-{\rm i}\phi} +{\rm i}\sin\theta\,D_2, \\
            & {\rm i}\left(\dfrac{1}{\sin\theta} -\sin\theta\right)D_4\,{\rm e}^{{\rm i}\phi} -{\rm i}\sin\theta\,D_3
               ={\rm i}\cos\theta\,{\rm e}^{{\rm i}\phi} \left(D_1\,{\rm e}^{-{\rm i}\phi} +D_2\right),
         \end{cases} \\
      \end{split}
   \end{equation}
   i.e.,
   \begin{equation}
      \begin{split}
        & D_3 +D_4\,{\rm e}^{{\rm i}\phi} = \dfrac{1}{\cos\theta} \left(\sin\theta-\dfrac{1}{\sin\theta}\right)\,D_1
            +\dfrac{\sin\theta}{\cos\theta} D_2\,{\rm e}^{{\rm i}\phi} = -\dfrac{\cos\theta}{\sin\theta} D_1 +\dfrac{\sin\theta}{\cos\theta} D_2\,{\rm e}^{{\rm i}\phi}, \\
       &  -D_3 +\dfrac{{\cos^2} \theta}{{\sin^2} \theta} D_4\,{\rm e}^{{\rm i}\phi}
         = \dfrac{\cos\theta}{\sin\theta}{\rm e}^{{\rm i}\phi} \left(D_1\,{\rm e}^{-{\rm i}\phi} +D_2\right)= \dfrac{\cos\theta}{\sin\theta} \left(D_1 +D_2\,{\rm e}^{{\rm i}\phi}\right),
      \end{split}
   \end{equation}
   which leads to
   \begin{equation}
      \begin{cases}
         & D_3=-\cot\theta\,D_1, \\
         & D_4=\tan\theta\,D_2.
      \end{cases}
   \end{equation}
   Therefore, \Eq{eq:ChiTheta} and \Eq{eq:ChiPhi} can be rewritten as
   \begin{equation}
      \begin{cases}
         & \chi_1=C_1 (\phi) \cos\theta+C_2 (\phi) \sin\theta, \\
         & \\
         & \chi_2=\Bigl[C_1 (\phi) \sin\theta -C_2 (\phi) \cos\theta\Bigr]{\rm e}^{{\rm i}\phi},
      \end{cases}
   \end{equation}
   and
   \begin{equation}
      \begin{cases}
         & \chi_1=D_1 (\theta)\:{\rm e}^{-{\rm i}\phi} +D_2 (\theta), \\
         & \\
         & \chi_2=-\cot\theta\,D_1 (\theta)+\tan\theta\,D_2 (\theta)\:{\rm e}^{{\rm i}\phi}.
      \end{cases}
   \end{equation}
   Notice that
   \begin{equation}
      \begin{cases}
         & C_1 (\phi)\,\cos\theta+C_2 (\phi)\,\sin\theta=D_1 (\theta)\,{\rm e}^{-{\rm i}\phi} +D_2 (\theta), \\
         & \\
         & \Bigl[C_1 (\phi) \sin\theta -C_2  (\theta)\,\cos\theta\Bigr]{\rm e}^{{\rm i}\phi}
            =-\cot\theta\,D_1 (\theta)+\tan\theta\,D_2 (\theta)\,{\rm e}^{{\rm i}\phi},
      \end{cases}\Longrightarrow\begin{cases}
         & C_1 (\phi)=\dfrac{1}{\cos\theta} D_2 (\theta), \\
         & \\
         & C_2 (\phi)=\dfrac{{\rm e}^{-{\rm i}\,\phi}}{\sin\theta} D_1 (\theta).
      \end{cases}
   \end{equation}
   Because $C_1(\phi)$ and $C_2(\phi)$ do not depend to $\theta$, so one must have
    \begin{equation}
      \begin{cases}
         & C_1 (\phi)=a, \\
         & C_2 (\phi)=b\,{\rm e}^{-{\rm i}\phi},
      \end{cases}\qquad\begin{cases}
         & D_1 (\theta)=b\,\sin\theta, \\
         & D_2 (\theta)=a\,\cos\theta,
      \end{cases}
   \end{equation}
   where $a$, $b$ are some arbitrary complex numbers. This results in that
   \begin{eqnarray}\label{eq:M-7a}
  \xi(\vec{r}) & =&\begin{bmatrix} \chi_1 (\theta,\phi) \\ \chi_2 (\theta,\phi) \end{bmatrix}=\begin{bmatrix} a\,\cos\theta+b\,\sin\theta\:{\rm e}^{-{\rm i}\phi} \\ -b\,\cos\theta+a\,\sin\theta\:{\rm e}^{{\rm i}\phi} \end{bmatrix}.
   \end{eqnarray}
or in the rectangular coordinate system it reads
 \begin{eqnarray}\label{eq:M-8a-b}
  \xi(\vec{r}) & =&\begin{bmatrix} a\,\frac{z}{r}+b\,\frac{x-{\rm i}y}{r} \\ -b\,\frac{z}{r}+a\,\frac{x+{\rm i}y}{r} \end{bmatrix},
   \end{eqnarray}
which coincides with the wavefunction as given in Eq. (\ref{phi-1a-2}).
By substituting Eq. (\ref{eq:M-8a-b}) into Eq. (\ref{eq:L-12-b-a}), one finds that the latter is valid.
After substituting Eq. (\ref{eq:L-3}) and Eq. (\ref{eq:M-8a-b}) into Eq. (\ref{eq:H0-f-a}), we eventually have the wavefunction as
\begin{equation}\label{eq:H0-f-c-02}
  \Psi_{\rm S} (\vec{r}) =\xi_0(\vec{r})\xi(\vec{r})=\mathcal{N} {\rm e}^{\mathrm{i} \vec{k}\cdot\vec{r}} \begin{bmatrix} a\,\frac{z}{r}+b\,\frac{x-{\rm i}y}{r} \\ -b\,\frac{z}{r}+a\,\frac{x+{\rm i}y}{r} \end{bmatrix}.
\end{equation}

\begin{remark}
From Eq. (\ref{eq:v-3}) we have known that
 \begin{eqnarray}\label{g2-a}
g=\frac{Q}{q},
  \end{eqnarray}
where $q$ is the ``charge'' of the particle ``1'' and $Q$ is the ``charge'' of the particle ``2''. By taking $q=-e$ and $Q=g(-e)$, one has $g=2$.
Another alternative choice of $g=2$ is selecting $Q=-e$, but $q=-e/2$ being the half  ``charge''. Recently, the quasi-particle with half-charge has been reported in condensed matter physics \cite{2019Wang}, probably it could have an application here.
\end{remark}

\section{Generating $\vec{\mathcal{A}}$ from Some Commutators}

During the exploration of the spin vector potential, we have observed that it can be generated from some commutators. The results are listed as follows:
\begin{subequations}\label{eq:4A}
\begin{align}
&{\rm i}\,\dfrac{g}{2} \left[
                  \dfrac{\vec{\sigma} \cdot\vec{\ell}}{r^2},\ \vec{r}\right]
                  =\vec{\mathcal{A}}, \label{eq:4A-1}\\
&{\rm i}\,\dfrac{g}{2} \left[\dfrac{\vec{\ell}}{r^2},\
                  \vec{\sigma} \cdot\vec{r}\right]=\vec{\mathcal{A}},\label{eq:4A-2}\\
&-{\rm i}\,\dfrac{g\,\hbar}{4} \left[\vec{\sigma},\
                  \dfrac{\vec{\sigma} \cdot\vec{r}}{r^2}\right]
                  =\vec{\mathcal{A}},\label{eq:4A-3}\\
& g\Biggl[\dfrac{\vec{r}\cdot\vec{\sigma}}{2\,r},\
                  \left[\dfrac{\vec{r}\cdot\vec{\sigma}}{2\,r} ,\
                        \vec{p}\right]\Biggr]=\vec{\mathcal{A}},\label{eq:4A-4}
\end{align}
\end{subequations}
where $\vec{\mathcal A}$ is given by Eq. (\ref{eq:v-4-b}), namely
 \begin{equation}
            \vec{\mathcal{A}}=g\dfrac{\vec{r}\times\vec{S}}{r^2}
            =\dfrac{g\,\hbar}{2} \dfrac{\vec{r}\times\vec{\sigma}}{r^2}
            =\dfrac{g\,\hbar}{2} \dfrac{\vec{r}}{r^2} \times
                  \left(\vec{\sigma} \times\dfrac{\vec{\sigma} }{2\,{\rm i}}\right).
      \end{equation}
\emph{Proof.---}Because
      \begin{equation}
            \left[\vec{r},\ \dfrac{g\,\hbar}{r^2} \vec{\sigma} \cdot\vec{\ell}\right]
            ={\rm i}\,\hbar^2 \dfrac{g}{r^2} (\vec{\sigma}\times\vec{r})
            =-{\rm i}\,2\,\hbar\left[\dfrac{g\,\hbar}{2} \dfrac{(\vec{r}\times\vec{\sigma})}{r^2}\right]
            =-{\rm i}\,2\,\hbar\:\vec{\mathcal{A}},
      \end{equation}
      which means
      \begin{equation}
         {\rm i}\,\dfrac{g}{2} \left[\dfrac{\vec{\sigma} \cdot\vec{\ell}}{r^2},\
            \vec{r}\right]=\vec{\mathcal{A}},
      \end{equation}
thus we prove Eq. (\ref{eq:4A-1}). Similarly, by interchanging $\vec{r} \rightleftharpoons \vec{\ell}/r^2$ from Eq. (\ref{eq:4A-1}) we have Eq. (\ref{eq:4A-2}) as
      \begin{equation}
         {\rm i}\,\dfrac{g}{2} \left[\dfrac{\vec{\ell}}{r^2},\
            \vec{\sigma} \cdot\vec{r}\right]=\vec{\mathcal{A}}.
      \end{equation}

Because
\begin{equation}
         \begin{split}
            \left[\vec{\sigma},\ \dfrac{\vec{\sigma} \cdot\vec{r}}{r}\right]
            =2\,{\rm i}\,\dfrac{\vec{r}\times\vec{\sigma}}{r},
         \end{split}
\end{equation}
then we have Eq. (\ref{eq:4A-3}) as
      \begin{equation}
         -{\rm i}\,\dfrac{g\,\hbar}{4} \left[\vec{\sigma},\
            \dfrac{\vec{\sigma} \cdot\vec{r}}{r^2}\right]
         =-{\rm i}\,\dfrac{g\,\hbar}{4} 2\,{\rm i}\,
            \dfrac{\vec{r}\times\vec{\sigma}}{r^2}
         =\vec{\mathcal{A}}.
      \end{equation}

      Moreover, we have
      \begin{equation}
            \begin{split}
                  \left[\dfrac{\vec{r}\cdot\vec{\sigma}}{2\,r} ,\ \vec{p}\right]
                  =&\ -\left[\vec{p},\
                        \dfrac{\vec{r}\cdot\vec{\sigma}}{2\,r}\right]
                  =\dfrac{{\rm i}\hbar}{2} \vec{\nabla}\left(
                        \dfrac{\vec{r}\cdot\vec{\sigma}}{r}\right)
                  =\dfrac{{\rm i}\hbar}{2} \vec{\nabla}\left(
                        \dfrac{x\,\sigma_{x} +y\,\sigma_{y} +z\,\sigma_{z}}{r}\right) \\
                  =&\ \dfrac{{\rm i}\hbar}{2} \Biggl[\hat{e}_x
                        \dfrac{\partial}{\partial\,x} \left(
                              \dfrac{x\,\sigma_{x} +y\,\sigma_{1y} +z\,\sigma_{z}}{r}\right)
                        +\hat{e}_y \dfrac{\partial}{\partial\,y} \left(
                                    \dfrac{x\,\sigma_{x} +y\,\sigma_{y} +z\,\sigma_{z}}{r}\right)
                        +\hat{e}_z \dfrac{\partial}{\partial\,z} \left(
                              \dfrac{x\,\sigma_{x} +y\,\sigma_{y} +z\,\sigma_{z}}{r}\right)\Biggr] \\
                  =&\ \dfrac{{\rm i}\hbar}{2} \Biggl\{\hat{e}_x
                        \left[\dfrac{\sigma_{x} r^2
                              -(\vec{r}\cdot\vec{\sigma})x}{r^3}\right]
                        +\hat{e}_y \left[\dfrac{\sigma_{y} r^2
                              -(\vec{r}\cdot\vec{\sigma})y}{r^3}\right]
                        +\hat{e}_z \left[\dfrac{\sigma_{z} r^2
                              -(\vec{r}\cdot\vec{\sigma})z}{r^3}\right]\Biggr\} \\
                  =&\ \dfrac{{\rm i}\hbar}{2} \left[\dfrac{\vec{\sigma} r^2
                        -(\vec{r}\cdot\vec{\sigma})\vec{r}}{r^3}\right],
            \end{split}
      \end{equation}
      and note
      \begin{equation}
            \begin{split}
                  \vec{r}\times\vec{\mathcal{A}} =&\ \dfrac{g}{r^2}
                        \vec{r}\times\left(\vec{r}\times\vec{S}\right)
                  =\dfrac{g}{r^2} \biggl[
                        \left(\vec{r}\cdot\vec{S}\right)\vec{r}
                        -r^2 \vec{S}\biggr]
                  =g\Biggl[\dfrac{\bigl(\vec{r}\cdot\vec{S}\bigr)}{r^2} \vec{r}
                        -\vec{S}\Biggr]
                  =g\dfrac{\hbar}{2} \Biggl[
                        \dfrac{\bigl(\vec{r}\cdot\vec{\sigma}\bigr)}{r^2} \vec{r}
                        -\vec{\sigma}\Biggr],
            \end{split}
      \end{equation}
      which implies
      \begin{equation}
            \left[\dfrac{\vec{r}\cdot\vec{\sigma}}{2\,r} ,\ \vec{p}\right]
            =-{\rm i}\,\dfrac{1}{g} \dfrac{\vec{r}\times\vec{\mathcal{A}}}{r} .
      \end{equation}
      then
      \begin{equation}
         \begin{split}
            \Biggl[\dfrac{\vec{r}\cdot\vec{\sigma}}{2\,r},\
               \left[\dfrac{\vec{r}\cdot\vec{\sigma}}{2\,r},\ \vec{p}\right]\Biggr]
            =&\ \dfrac{{\rm i}\hbar}{2} \Biggl[
                  \dfrac{\vec{r}\cdot\vec{\sigma}}{2\,r},\
                  \dfrac{\vec{\sigma} r^2
                        -(\vec{r}\cdot\vec{\sigma})\vec{r}}{r^3}\Biggr]
            =\dfrac{{\rm i}\hbar}{4} \Biggl[
                        \dfrac{\vec{r}\cdot\vec{\sigma}}{r},\
                        \dfrac{\vec{\sigma}}{r}\Biggr]
                  -\dfrac{{\rm i}\hbar}{4} \Biggl[
                        \dfrac{\vec{r}\cdot\vec{\sigma}}{r},\
                        \dfrac{(\vec{r}\cdot\vec{\sigma})\vec{r}}{r^3}\Biggr] \\
            =&\ \dfrac{{\rm i}\hbar}{4} \Biggl[
                        \dfrac{\vec{r}\cdot\vec{\sigma}}{r},\
                        \dfrac{\vec{\sigma}}{r}\Biggr]
            =-\dfrac{{\rm i}\hbar}{4} \dfrac{1}{r}\Biggl[\vec{\sigma},\
                  \dfrac{\vec{r}\cdot\vec{\sigma}}{r}\Biggr]
            =-\dfrac{{\rm i}\hbar}{4} \dfrac{1}{r} 2\,{\rm i}\,
                  \dfrac{\vec{r}\times\vec{\sigma}}{r} \\
            =&\ \dfrac{\hbar}{2} \dfrac{\vec{r}\times\vec{\sigma}}{r^2},
         \end{split}
      \end{equation}
      then we obtain
      \begin{equation}
            g\Biggl[\dfrac{\vec{r}\cdot\vec{\sigma}}{2\,r},\
                  \left[\dfrac{\vec{r}\cdot\vec{\sigma}}{2\,r} ,\ \vec{p}\right]\Biggr]
            =g\Biggl[\dfrac{\vec{r}\cdot\vec{\sigma}}{2\,r},\
                  -{\rm i}\,\dfrac{1}{g} \dfrac{\vec{r}\times\vec{\mathcal{A}}}{r}\Biggr]
            =\Biggl[\dfrac{\vec{r}\cdot\vec{\sigma}}{2\,r},\
                  -{\rm i}\,\dfrac{\vec{r}\times\vec{\mathcal{A}}}{r}\Biggr]
            =\vec{\mathcal{A}},
      \end{equation}
thus we prove Eq. (\ref{eq:4A-4}).

\section{The ``Magnetic'' and ``Electric'' Fields and the Lorentz-Like Force}

According to the definition of field tensor \cite{1954YangMills}, we have
\begin{eqnarray}
      F_{\mu\nu}=\partial_\mu \mathbb{A}_\nu - \partial_\nu  \mathbb{A}_\mu
      {+}\dfrac{\rm i}{\hbar}[\mathbb{A}_\mu, \mathbb{A}_\nu].
\end{eqnarray}
Note that the four-vectorial notations $x=(x_0 ,x_1 ,x_2 ,x_3)$, with $x_1 \equiv x$, $x_2 \equiv y$, $x_3 \equiv z$, $x_0 \equiv c\,t$, and
   $\mathbb{A}=(\varphi,-\vec{\mathcal{A}})=(\varphi,-\mathcal{A}_x ,-\mathcal{A}_y ,-\mathcal{A}_z)$ are used (Thereafter, we set $c=1$).

The matrix form of $F_{\mu\nu}$ is given by \cite{1999Jackson}
   \begin{equation}
       F_{\mu\nu}=\begin{bmatrix}
            0 & \mathcal{E}_x & \mathcal{E}_y & \mathcal{E}_z \\
            -\mathcal{E}_x & 0 & -\mathcal{B}_z & \mathcal{B}_y \\
            -\mathcal{E}_y & \mathcal{B}_z & 0 & -\mathcal{B}_x \\
            -\mathcal{E}_z & -\mathcal{B}_y & \mathcal{B}_x & 0
      \end{bmatrix},
   \end{equation}
based on which we have the three components of ``magnetic'' field as
   \begin{equation}
      \begin{split}
            \mathcal{B}_x =&\ F_{32} =\dfrac{\partial\,\mathbb{A}_2}{\partial\,x_3}
                  -\dfrac{\partial\,\mathbb{A}_3}{\partial\,x_2}
                  +\frac{\mathrm{i}}{\hbar}[\mathbb{A}_3, \mathbb{A}_2]
            =-\dfrac{\partial\,\mathcal{A}_y}{\partial\,z}
                  +\dfrac{\partial\,\mathcal{A}_z}{\partial\,y}
                  +\frac{\mathrm{i}}{\hbar}[\mathcal{A}_z, \mathcal{A}_y], \\
            \mathcal{B}_y =&\ F_{13} =\dfrac{\partial\,\mathbb{A}_3}{\partial\,x_1}
                  -\dfrac{\partial\,\mathbb{A}_1}{\partial\,x_3}
                  +\frac{\mathrm{i}}{\hbar}[\mathbb{A}_1, \mathbb{A}_3]
            =-\dfrac{\partial\,\mathcal{A}_z}{\partial\,x}
                  +\dfrac{\partial\,\mathcal{A}_x}{\partial\,z}
                  +\frac{\mathrm{i}}{\hbar}[\mathcal{A}_x, \mathcal{A}_z], \\
            \mathcal{B}_z =&\ F_{21} =\dfrac{\partial\,\mathbb{A}_1}{\partial\,x_2}
                  -\dfrac{\partial\,\mathbb{A}_2}{\partial\,x_1}
                  +\frac{\mathrm{i}}{\hbar}[\mathbb{A}_2, \mathbb{A}_1]
            =-\dfrac{\partial\,\mathcal{A}_x}{\partial\,y}
                  +\dfrac{\partial\,\mathcal{A}_y}{\partial\,x}
                  +\frac{\mathrm{i}}{\hbar}[\mathcal{A}_y, \mathcal{A}_x].
      \end{split}
   \end{equation}
   In addition, the three components of ``electric'' field are
   \begin{equation}
      \begin{split}
            \mathcal{E}_x =&\ F_{01} =\dfrac{\partial\,\mathbb{A}_1}{\partial\,x_0}
                  -\dfrac{\partial\,\mathbb{A}_0}{\partial\,x_1}
                  +\frac{\mathrm{i}}{\hbar}[\mathbb{A}_0, \mathbb{A}_1]
            =-\dfrac{\partial\,\mathcal{A}_x}{\partial\,t}
                  -\dfrac{\partial\,\varphi}{\partial\,x}
                  -\frac{\mathrm{i}}{\hbar} [\varphi,\ \mathcal{A}_x], \\
            \mathcal{E}_y =&\ F_{02} =\dfrac{\partial\,\mathbb{A}_2}{\partial\,x_0}
                  -\dfrac{\partial\,\mathbb{A}_0}{\partial\,x_2}
                  +\frac{\mathrm{i}}{\hbar}[\mathbb{A}_0, \mathbb{A}_2]
            =-\dfrac{\partial\,\mathcal{A}_y}{\partial\,t}
                  -\dfrac{\partial\,\varphi}{\partial\,y}
                  -\frac{\mathrm{i}}{\hbar} [\varphi,\ \mathcal{A}_y], \\
            \mathcal{E}_z =&\ F_{03} =\dfrac{\partial\,\mathbb{A}_3}{\partial\,x_0}
                  -\dfrac{\partial\,\mathbb{A}_0}{\partial\,x_3}
                  +\frac{\mathrm{i}}{\hbar}[\mathbb{A}_0, \mathbb{A}_3]
            =-\dfrac{\partial\,\mathcal{A}_z}{\partial\,t}
                  -\dfrac{\partial\,\varphi}{\partial\,z}
                  -\frac{\mathrm{i}}{\hbar} [\varphi,\ \mathcal{A}_z].
      \end{split}
   \end{equation}
   Then we may define a ``magnetic'' field based on the spin vector potential as
   \begin{eqnarray}
      \vec{\mathcal{B}} &=& \vec{\nabla}\times\vec{\mathcal{A}}
            {-}\frac{\mathrm{i}}{\hbar}
                  \vec{\mathcal{A}}\times\vec{\mathcal{A}},
   \end{eqnarray}
   and a ``electric'' field as
   \begin{equation}
      \vec{\mathcal{E}} = -\dfrac{\partial\,\vec{\mathcal{A}}}{\partial\,t}
            -\vec{\nabla}\varphi
            -\frac{\mathrm{i}}{\hbar} [\varphi,\ \vec{\mathcal{A}}],
   \end{equation}
   where $\varphi=\mathbb{A}_0$ indicates the \emph{spin scalar potential}.

Due to
\begin{eqnarray}\label{eq:M-5-02}
&& \nabla\times\vec{\mathcal{A}} =  -g\,\hbar\dfrac{\left(\vec{r}\cdot\vec{\sigma} \right)}{r^4} \vec{r}, \;\;\;\;\;\;
 \vec{\mathcal{A}} \times \vec{\mathcal{A}}= {\rm i} \hbar \; g^2 \; \frac{\vec{r}\cdot\vec{S}}{r^4} \; \vec{r}
 ={\rm i} \hbar^2 \; \dfrac{g^2}{2} \; \frac{\vec{r}\cdot\vec{\sigma}}{r^4} \;
      \vec{r},
\end{eqnarray}
we have
\begin{eqnarray}
      \vec{\mathcal{B}} &=& -g\,\hbar\dfrac{\left(\vec{r}\cdot\vec{\sigma} \right)}{r^4} \vec{r}
         -\frac{\mathrm{i}}{\hbar} \dfrac{{\rm i}\hbar^2\,g^2}{2\,r^4}
            \bigl(\vec{r}\cdot\vec{\sigma}\bigr)\vec{r} = \dfrac{g(g-2)\hbar}{2} \dfrac{\left(\vec{r}\cdot\vec{\sigma} \right)}{r^4} \vec{r}= g(g-2) \dfrac{\left(\vec{r}\cdot\vec{S} \right)}{r^4} \vec{r}.
\end{eqnarray}
Interestingly, when $g=0$ and $g=2$, one has the ``magnetic'' field
$\vec{\mathcal{B}}=0$. By the direct calculation, one finds that the ``magnetic'' field satisfies the following relation
\begin{equation}
      \begin{split}
            \nabla\cdot\vec{\mathcal{B}} =&\ \dfrac{g(2-g)\hbar}{2}\,\vec{\nabla}\cdot
                  \biggl[\dfrac{\left(\vec{r}\cdot\vec{\sigma} \right)}{r^4} \vec{r}\biggr]=0.
            \end{split}
\end{equation}

We now consider the spin AB Hamiltonian
   \begin{equation}
      H=\dfrac{1}{2M} \left(\vec{p}-\vec{\mathcal{A}}\right)^2 +\varphi
            = \dfrac{1}{2M} \vec{\Pi}^2 +\varphi,
   \end{equation}
where the canonical momentum operator $\vec{\Pi}$ satisfies
\begin{eqnarray}
            \vec{\Pi} \times \vec{\Pi}= {\rm i} \hbar \; g(g-2)  \frac{\vec{r}\cdot\vec{S}}{r^4} \; \vec{r}={\rm i} \hbar \; \vec{\mathcal{B}}.
\end{eqnarray}
The Heisenberg equation of motion is given by
\begin{equation}\label{eq:HeisenbergEq}
      \begin{split}
            \dfrac{{\rm d}\,\hat{\mathcal{O}}_H (t)}{{\rm d}\,t}
            =&\ \dfrac{\partial\,\hat{U}^\dagger}{\partial\,t} \hat{\mathcal{O}} U
                  +\hat{U}^\dagger \dfrac{\partial\,\hat{\mathcal{O}}}{\partial\,t} U
                  +\hat{U}^\dagger \hat{\mathcal{O}} \dfrac{\partial\,U}{\partial\,t} \\
            =&\ -\dfrac{1}{{\rm i}\hbar} \hat{U}^\dagger HU\hat{U}^\dagger \hat     {\mathcal{O}}U
                  +\left(\dfrac{\partial\,\hat{\mathcal{O}}}{\partial\,t}\right)_H
                  +\dfrac{1}{{\rm i}\hbar} \hat{U}^\dagger \hat{\mathcal{O}}U\hat{U}^\dagger HU \\
            =&\ \dfrac{1}{{\rm i}\hbar} \left(\Bigl[\hat{\mathcal{O}},\ H\Bigr]\right)_H +\left(\dfrac{\partial\,\hat{\mathcal{O}}}{\partial\,t}\right)_H.
      \end{split}
\end{equation}
When $\hat{\mathcal{O}}$ does not apparently contain time $t$, we have
\begin{equation}
      \left(\dfrac{\partial\,\hat{\mathcal{O}}}{\partial\,t}\right)_H =0.
\end{equation}

Then based on the Heisenberg equation in \Eq{eq:HeisenbergEq}, we obtain the ``velocity'' operator as
   \begin{eqnarray}
           \vec{v}:=\dfrac{{\rm d}\,\vec{r}}{{\rm d}\,t} &=&\ \dfrac{1}{{\rm i}\,\hbar}
            \left[\vec{r},H\right] +\dfrac{\partial\,\vec{r}}{\partial\,t} \nonumber \\
            &=&  \dfrac{1}{{\rm i}\,\hbar} \left[\vec{r},H\right] \nonumber \\
            &=& \dfrac{1}{2{\rm i}\,\hbar\,M} \left[\vec{r},
                        \left(\vec{p}-\vec{\mathcal{A}}\right)^2 \right]
                  +\left[\vec{r},\ \varphi\right] \nonumber \\
            &=& \dfrac{1}{2{\rm i}\,\hbar\,M} \left[\vec{r},
                  \left(\vec{p}-\vec{\mathcal{A}}\right)^2 \right] \nonumber \\
            &=& \dfrac{1}{2{\rm i}\,\hbar\,M} \Biggl[\vec{r},\
            \left(\vec{p}^{\; 2} +\dfrac{g\,\hbar}{r^2} \vec{\sigma} \cdot\vec{\ell}
               +\dfrac{g^2\,\hbar^2} {2\,r^2}\right)\Biggr] \nonumber\\
        & =& \dfrac{1}{2{\rm i}\,\hbar\,M} \Biggl\{\left[\vec{r},\vec{p}^{\; 2}\right]
            +\left[\vec{r},\ \dfrac{g\,\hbar}{r^2} \vec{\sigma} \cdot\vec{\ell}\right]\Biggr\} = \dfrac{1}{2{\rm i}\,\hbar\,M} \Biggl\{{\rm i}\hbar\dfrac{\partial\,\vec{p}^{\; 2}}{\partial\,\vec{p}}
            +\left[\vec{r},\ \dfrac{g\,\hbar}{r^2} \vec{\sigma} \cdot\vec{\ell}\right]\Biggr\} \nonumber\\
        & =& \dfrac{1}{2{\rm i}\,\hbar\,M} \Biggl\{2\,{\rm i}\hbar\,\vec{p}
            +\left[\vec{r},\ \dfrac{g\,\hbar}{r^2} \vec{\sigma} \cdot\vec{\ell}\right]\Biggr\} = \dfrac{1}{2{\rm i}\,\hbar\,M} \Biggl\{2\,{\rm i}\hbar\,\vec{p}
            -2\,{\rm i}\hbar\,\vec{\mathcal{A}}\Biggr\} \nonumber\\
        & =& \dfrac{1}{M} \left(\vec{p}-\vec{\mathcal{A}}\right)
         \equiv \dfrac{1}{M} \;\vec{\Pi}.
   \end{eqnarray}

   Because
      \begin{eqnarray}
          \left[{\Pi}_z,\ \vec{\Pi}^2\right]
                  &=& \left[\Pi_z,\ \Pi_x^2 +\Pi_y^2 +\Pi_z^2\right]=\left[\Pi_z,\ \Pi_x^2\right] +\left[\Pi_z,\ \Pi_y^2\right] \nonumber\\
                  &=&\left[\Pi_z,\ \Pi_x\right]\Pi_x +\Pi_x \left[\Pi_z,\ \Pi_x\right]
                        +\Pi_y \left[\Pi_z,\ \Pi_y\right]+\left[\Pi_z,\ \Pi_y\right] \Pi_y \nonumber\\
                  &=& {\rm i}\hbar \mathcal{B}_y\:\Pi_x
                        + {\rm i}\hbar\, \Pi_x \mathcal{B}_y
                        -{\rm i}\hbar\,\Pi_y \mathcal{B}_x  -{\rm i}\hbar\, \mathcal{B}_x \Pi_y \nonumber\\
                  &=& {\rm i}\hbar \left[\left(\Pi_x \mathcal{B}_y-\Pi_y \mathcal{B}_x\right)-\left(\mathcal{B}_x \Pi_y-\mathcal{B}_y\:\Pi_x \right)\right] \nonumber\\
                   &=& {\rm i}\hbar \left[\left(\vec{\Pi}\times \vec{\mathcal{B}}\right)_z-\left(\vec{\mathcal{B}}\times \vec{\Pi}\right)_z\right],
      \end{eqnarray}
     which means
      \begin{eqnarray}
                  \left[\vec{\Pi},\ \vec{\Pi}^2\right] ={\rm i}\hbar \left(\vec{\Pi}\times \vec{\mathcal{B}}-\vec{\mathcal{B}}\times \vec{\Pi}\right).
      \end{eqnarray}
   After that, we can compute the ``acceleration'' operator as
      \begin{eqnarray}
                 \vec{a}:=\dfrac{{\rm d}\,\vec{v}}{{\rm d}\,t} &=&
                        \dfrac{1}{{\rm i}\,\hbar} \left[
                              \vec{v},H\right]
                        +\dfrac{\partial\,\vec{v}}{\partial\,t} \nonumber \\
                  &=& \dfrac{1}{{\rm i}\,\hbar} \left[
                              \vec{v},H\right]
                        +\dfrac{1}{M} \dfrac{\partial\,\bigl(\vec{p}-\vec{\mathcal{A}}\bigr)}{\partial\,t} \nonumber \\
                  &=& \dfrac{1}{{\rm i}\,\hbar} \left[
                              \vec{v},H\right]
                        -\dfrac{1}{M} \dfrac{\partial\,\vec{\mathcal{A}}}{\partial\,t} \nonumber \\
                  &=& \dfrac{1}{2{\rm i}\,\hbar\,M^2} \left[
                              \vec{\Pi},\vec{\Pi}^2\right]
                        +\dfrac{1}{{\rm i}\,\hbar} \left[\vec{v},\ \varphi\right]
                        -\dfrac{1}{M} \dfrac{\partial\,\vec{\mathcal{A}}}{\partial\,t}\nonumber \\
                  &=& \dfrac{1}{2{\rm i}\,\hbar\,M^2} \left[
                              \vec{\Pi},\vec{\Pi}^2\right]
                        +\dfrac{1}{M} \dfrac{1}{{\rm i}\,\hbar} \left[\vec{p},\ \varphi\right]
                        -\dfrac{1}{M} \dfrac{1}{{\rm i}\,\hbar} \left[\vec{\mathcal{A}},\ \varphi\right]
                        -\dfrac{1}{M} \dfrac{\partial\,\vec{\mathcal{A}}}{\partial\,t}\nonumber \\
                  &=& \dfrac{1}{ M} \dfrac{1}{2}\left(\vec{v}\times \vec{\mathcal{B}}-\vec{\mathcal{B}}\times \vec{v}\right)
                  -{\rm i}\hbar\dfrac{1}{M} \dfrac{1}{{\rm i}\,\hbar} \vec{\nabla}\,\varphi
                  -\dfrac{1}{M} \dfrac{1}{{\rm i}\,\hbar} \left[\vec{\mathcal{A}},\ \varphi\right]
                        -\dfrac{1}{M} \dfrac{\partial\,\vec{\mathcal{A}}}{\partial\,t}
                        \nonumber \\
                  &=& \dfrac{1}{ M} \dfrac{1}{2}\left(\vec{v}\times \vec{\mathcal{B}}-\vec{\mathcal{B}}\times \vec{v}\right)
                        -\dfrac{1}{M} \vec{\nabla}\,\varphi
                        +\dfrac{1}{M} \dfrac{{\rm i}}{\hbar} \left[\vec{\mathcal{A}},\ \varphi\right]
                        -\dfrac{1}{M} \dfrac{\partial\,\vec{\mathcal{A}}}{\partial\,t} \nonumber \\
                  &=& \dfrac{1}{M} \Biggl\{\dfrac{1}{2}\left(\vec{v}\times \vec{\mathcal{B}}-\vec{\mathcal{B}}\times \vec{v}\right)
                  +\biggl[-\dfrac{\partial\,\vec{\mathcal{A}}}{\partial\,t}
                        -\vec{\nabla}\,\varphi
                        -\dfrac{{\rm i}}{\hbar} \left[\vec{\mathcal{A}},\ \varphi\right]\biggr] \Biggr\},
      \end{eqnarray}
   therefore we have a ``force'' as
   \begin{equation}
      \mathbb{F}=M\,\vec{a}=\mathbb{F}_{\rm Lorentz} +\mathbb{F}_{\rm electric},
   \end{equation}
   with Lorentz-like force as
          \begin{eqnarray}
                 \mathbb{F}_{\rm Lorentz} =\dfrac{1}{2}\left(\vec{v}\times \vec{\mathcal{B}}-\vec{\mathcal{B}}\times \vec{v}\right),
      \end{eqnarray}
   and the electric-like force as
   \begin{equation}
      \mathbb{F}_{\rm electric} =-\dfrac{\partial\,\vec{\mathcal{A}}}{\partial\,t}
            -\vec{\nabla}\,\varphi
            -\dfrac{{\rm i}}{\hbar} \left[\varphi,\ \vec{\mathcal{A}}\right].
   \end{equation}

\begin{remark}
Let particle ``1'' has a spin operator $\vec{S}_1$, which induces a ``magnetic'' field as
   \begin{eqnarray}
      \vec{\mathcal{B}} &=&  g(g-2) \dfrac{\left(\vec{r}\cdot\vec{S}_1 \right)}{r^4} \vec{r}.
   \end{eqnarray}
Let particle ``2'' has a spin operator $\vec{S}_2$, then the interaction between the spin $\vec{S}_2$  and the ``magnetic'' field reads
   \begin{eqnarray}
      \mathcal{H}_{\rm Inter}=\vec{S}_2\cdot \vec{\mathcal{B}} &=&  g(g-2) \frac{1}{r^4}\left(\vec{r}\cdot\vec{S}_1 \right)\left(\vec{r}\cdot\vec{S}_2 \right),
   \end{eqnarray}
which connects the dipole-dipole interaction as shown in Table \ref{tab:interactions}.
\end{remark}
\begin{remark}
      From \Eq{eq:ASqure}, we have known that
      \begin{eqnarray}
            \vec{\mathcal{A}}^2 = \dfrac{g^2\,\hbar^2}{2} \dfrac{1}{r^2}.
      \end{eqnarray}
      By observing
      \begin{equation}
            \biggl(\dfrac{\vec{r}\cdot\vec{S}}{r^2}\biggr)^2
            =\dfrac{\hbar^2}{4} \dfrac{r^2}{r^4} =\dfrac{\hbar^2}{4} \dfrac{1}{r^2},
      \end{equation}
      if we define the scalar potential as
      \begin{equation}\label{eq:scalar}
            \varphi=\sqrt{2}\,g\,\dfrac{\vec{r}\cdot\vec{S}}{r^2},
      \end{equation}
      then we have
      \begin{equation}
          \mathbb{A}\cdot \mathbb{A}\equiv  \varphi^2 -\vec{\mathcal{A}}^2 =0,
      \end{equation}
which is invariant in any coordinate system $(ct, x, y, z)$.
\end{remark}

\begin{remark}
      Let the spin scalar potential take the form as shown in Eq. (\ref{eq:scalar}), then we can simplify the ``electric'' field as
      \begin{equation}
            \begin{split}
                  \mathcal{E} =&\ -\dfrac{\partial\,\vec{\mathcal{A}}}{\partial\,t}
                        -\vec{\nabla}\,\varphi
                        -\dfrac{{\rm i}}{\hbar} \left[
                              \varphi,\ \vec{\mathcal{A}}\right] \\
                  =&\ -\dfrac{\partial\,\vec{\mathcal{A}}}{\partial\,t}
                        -\sqrt{2}\,g\,\vec{\nabla}\,
                              \left(\dfrac{\vec{r}\cdot\vec{S}}{r^2}\right)
                        -\dfrac{{\rm i}}{\hbar} \left[
                              \varphi,\ \vec{\mathcal{A}}\right] \\
                  =&\ -\dfrac{\partial\,\vec{\mathcal{A}}}{\partial\,t}
                        -\sqrt{2}\,g\,\dfrac{\hbar}{2} \vec{\nabla}\,
                              \left(\dfrac{\vec{r}\cdot\vec{\sigma}}{r^2}\right)
                        -\dfrac{{\rm i}}{\hbar} \left[
                              \varphi,\ \vec{\mathcal{A}}\right] \\
                  =&\ -\dfrac{\partial\,\vec{\mathcal{A}}}{\partial\,t}
                        -\sqrt{2}\,g\,\dfrac{\hbar}{2}
                              \dfrac{\vec{\sigma} r^2
                                    -2(\vec{r}\cdot\vec{\sigma})\vec{r}}{r^4}
                        -\dfrac{{\rm i}}{\hbar} {\rm i}\dfrac{\sqrt{2}\,g^2\,\hbar^2}{2\,r^4} \biggl[
                              r^2 \vec{\sigma}-\Bigl(\vec{r}\cdot\vec{\sigma}\Bigr)\vec{r}\biggr]  \\
                  =&\ -\dfrac{\partial\,\vec{\mathcal{A}}}{\partial\,t}
                        -\sqrt{2}\,g\,\dfrac{\hbar}{2}
                              \dfrac{\vec{\sigma} r^2
                                    -2(\vec{r}\cdot\vec{\sigma})\vec{r}}{r^4}
                        +\dfrac{1}{\hbar} \dfrac{\sqrt{2}\,g^2\,\hbar^2}{2\,r^4} \biggl[
                              r^2 \vec{\sigma}-\Bigl(\vec{r}\cdot\vec{\sigma}\Bigr)\vec{r}\biggr]  \\
                  =&\ -\dfrac{\partial\,\vec{\mathcal{A}}}{\partial\,t}
                        -\sqrt{2}\,g\,\dfrac{\hbar}{2}
                              \dfrac{(\vec{r}\cdot\vec{\sigma})\vec{r}}{r^4}
                  =-\dfrac{\partial\,\vec{\mathcal{A}}}{\partial\,t}
                        -\sqrt{2}\,g\,\dfrac{(\vec{r}\cdot\vec{S})\vec{r}}{r^4}
                  =-\dfrac{\partial\,\vec{\mathcal{A}}}{\partial\,t}
                        -\varphi\dfrac{\vec{r}}{r^2}.
            \end{split}
         \end{equation}
Here we have used the following relations:
\begin{equation}
            \begin{split}
                  \vec{\nabla}\,\left(\dfrac{\vec{r}\cdot\vec{\sigma}}{r^2}\right)
                  =&\ \hat{e}_x \dfrac{\partial}{\partial\,x} \left(
                              \dfrac{x\,\sigma_{x} +y\,\sigma_{y} +z\,\sigma_{z}}{r^2}\right)
                        +\hat{e}_y \dfrac{\partial}{\partial\,y} \left(
                                    \dfrac{x\,\sigma_{x} +y\,\sigma_{y} +z\,\sigma_{z}}{r^2}\right)
                        +\hat{e}_z \dfrac{\partial}{\partial\,z} \left(
                              \dfrac{x\,\sigma_{x} +y\,\sigma_{y} +z\,\sigma_{z}}{r^2}\right) \\
                  =&\ \Biggl\{\hat{e}_x
                        \left[\dfrac{\sigma_{x} r^4
                              -2(\vec{r}\cdot\vec{\sigma})x}{r^2}\right]
                        +\hat{e}_y \left[\dfrac{\sigma_{y} r^4
                              -2(\vec{r}\cdot\vec{\sigma})y}{r^2}\right]
                        +\hat{e}_z \left[\dfrac{\sigma_{z} r^4
                              -2(\vec{r}\cdot\vec{\sigma})z}{r^2}\right]\Biggr\} \\
                  =&\ \dfrac{\vec{\sigma} r^2
                        -2(\vec{r}\cdot\vec{\sigma})\vec{r}}{r^4},
            \end{split}
\end{equation}
and
\begin{equation}
            \begin{split}
                  \left[\varphi,\ \vec{\mathcal{A}}\right]
                  =& \sqrt{2}\,g^2 \left[
                        \dfrac{\vec{r}\cdot\vec{S}}{r^2},\
                        \dfrac{\vec{r}\times\vec{S}}{r^2}\right] = \dfrac{\sqrt{2}\,g^2}{r^4} \left[\vec{r}\cdot\vec{S},\
                        \vec{r}\times\vec{S}\right] = {\rm i}\dfrac{\sqrt{2}\,g^2\,\hbar^2}{2\,r^4} \biggl[
                        r^2 \vec{\sigma}-\Bigl(\vec{r}\cdot\vec{\sigma}\Bigr)\vec{r}\biggr].
            \end{split}
         \end{equation}

\end{remark}